%% file: fcamira.tex
\documentclass[twocolumn,numberedappendix,apj]{openjournal} 

\usepackage{newtxtext,newtxmath}

\usepackage[T1]{fontenc}

\usepackage{graphicx}	
\usepackage{amsmath}     
\usepackage{amsfonts}    
\usepackage{amssymb}     
\usepackage{xcolor}      
\usepackage{hyperref}    
\hypersetup{colorlinks=true,urlcolor=[rgb]{0.5,0,0.6},filecolor=magenta,citecolor=[rgb]{0,0.2,0.7},linkcolor=magenta,}
\usepackage{orcidlink}   
\usepackage{enumitem}

\input{newcommands}

\newcommand{\percent}{\ensuremath{\%}}
\newcommand{\appropto}{\mathrel{\vcenter{
                        \offinterlineskip\halign{\hfil$##$\cr
                        \propto\cr\noalign{\kern2pt}\sim\cr\noalign{\kern-2pt}}}}}

\newcommand{\vect}[1]{\boldsymbol{\mathbf{#1}}}

\makeatletter
\def\frontmatter@title@below{%
  \vspace*{-2.63\baselineskip}%
  \vspace*{0.25in}%
}
\makeatother

\begin{document}


\title{
Weak-lensing Shear-Selected Galaxy Clusters from the Hyper Suprime-Cam Subaru Strategic Program: \\
III. A precision cosmological sample enabled by optical confirmation 
}
\shorttitle{Optical Confirmation of Weak-Lensing Shear-Selected Clusters}
\shortauthors{Chiu et al.}

\author{I-Non~Chiu$^{1}$\thanks{E-mail: inchiu@phys.ncku.edu.tw}\orcidlink{0000-0002-5819-6566}}
\author{Kai-Feng~Chen$^{2,3}$\orcidlink{0000-0002-3839-0230}}
\author{Masamune~Oguri$^{4,5}$\orcidlink{0000-0003-3484-399X}}
\author{Satoshi~Miyazaki$^{7}$\orcidlink{0000-0002-1962-904X}}
\author{Surhud~More$^{8,9}$\orcidlink{0000-0002-2986-2371}}
\author{Atsushi~J.~Nishizawa$^{10,11}$\orcidlink{0000-0002-6109-2397}}
\author{Nobuhiro~Okabe$^{12,13,14}$\orcidlink{0000-0003-2898-0728}}
\author{Ken~Osato$^{4,5,6,9}$\orcidlink{0000-0002-7934-2569}}
\author{Naomi~Ota$^{15,16}$\orcidlink{0000-0002-2784-3652}}
\author{Tomomi~Sunayama$^{17}$\orcidlink{0009-0004-6387-5784}}
\author{Sut-Ieng~Tam$^{18}$\orcidlink{0000-0002-6724-833X}}
\author{Keiichi~Umetsu$^{17}$\orcidlink{0000-0002-7196-4822}}

\affiliation{$^{1}$Department of Physics, National Cheng Kung University, No. 1, University Road, Tainan City 70101, Taiwan}
\affiliation{$^{2}$MIT Kavli Institute, Massachusetts Institute of Technology, Cambridge, MA 02139, USA}
\affiliation{$^{3}$Department of Physics, Massachusetts Institute of Technology, Cambridge, MA 02139, USA}
\affiliation{$^{4}$Center for Frontier Science, Chiba University, 1-33 Yayoi-cho, Inage-ku, Chiba 263-8522, Japan}
\affiliation{$^{5}$Department of Physics, Graduate School of Science, Chiba University, 1-33 Yayoi-cho, Inage-ku, Chiba 263-8522, Japan}
\affiliation{$^{6}$RIKEN Center for Advanced Intelligence Project, 1-4-1 Nihonbashi, Chuo, Tokyo 103-0027, Japan}
\affiliation{$^{7}$Subaru Telescope, National Astronomical Observatory of Japan, 650 North Aohoku Place, Hilo, HI 96720, USA}
\affiliation{$^{8}$The Inter-University Centre for Astronomy and Astrophysics, Post Bag 4, Ganeshkhind, Pune 411007, India}
\affiliation{$^{9}$Kavli Institute for the Physics and Mathematics of the Universe (WPI), The University of Tokyo Institutes for Advanced Study (UTIAS), The University of Tokyo, Chiba 277-8583, Japan}
\affiliation{$^{10}$Gifu Shotoku Gakuen University, 1-1 Takakuwanishi, Yanaizucho, Gifu 501-6194, Japan}
\affiliation{$^{11}$Kobayashi--Maskawa Institute, Nagoya University, Furocho, Chikusa-ku, Nagoya, Aichi 464-8602, Japan}
\affiliation{$^{12}$Physics Program, Graduate School of Advanced Science and Engineering, Hiroshima University, 1-3-1 Kagamiyama, Higashi-Hiroshima, Hiroshima 739-8526, Japan}
\affiliation{$^{13}$Hiroshima Astrophysical Science Center, Hiroshima University, 1-3-1 Kagamiyama, Higashi-Hiroshima, Hiroshima 739-8526, Japan}
\affiliation{$^{14}$Core Research for Energetic Universe, Hiroshima University, 1-3-1 Kagamiyama, Higashi-Hiroshima, Hiroshima 739-8526, Japan}
\affiliation{$^{15}$Department of Physics, Nara Women's University, Kitauoyanishimachi, Nara 630-8506, Japan}
\affiliation{$^{16}$Argelander-Institut f{\"u}r Astronomie (AIfA), Universit{\"a}t Bonn, Auf dem H{\"u}gel 71, 53121 Bonn, Germany}
\affiliation{$^{17}$Academia Sinica Institute of Astronomy and Astrophysics (ASIAA), No. 1, Section 4, Roosevelt Road, Taipei 106319, Taiwan}
\affiliation{$^{18}$Institute of Physics, National Yang Ming Chiao Tung University, No. 1001, Daxue Rd., East Dist., Hsinchu City, Taiwan}

%
%

\begin{abstract}
We develop fCAMIRA (forced-mode CAMIRA), a tool for optical cluster confirmation, and apply it to a sample of $129$ weak-lensing (WL) shear-selected galaxy clusters identified in aperture-mass maps obtained with the Hyper Suprime-Cam Three-Year (HSC-Y3) WL data sets.
fCAMIRA is built upon the CAMIRA cluster finding algorithm 
and relies on a red-sequence (RS) galaxy model that is calibrated in a data-driven way.
The RS model adopts the metallicity-luminosity relation measured in this work using X-ray-selected clusters up to redshift $\redshift\approx1.3$, followed by the calibration of color offsets using large spectroscopic samples.
With the RS model, we build two types of galaxy richness maps, one obtained with a spatial filter matched to a typical cluster size $\rdd\approx0.8\Mpch$ and the other obtained with a fixed angular-size filter identical to that used in constructing the WL aperture-mass maps.
The fCAMIRA algorithm utilizes these two richness maps, identifies all optical counterpart candidates along the line of sight of each shear-selected cluster, and measures the cluster photometric redshift \zcl\ from the highest-ranked counterpart.
The ranking is determined by the fractional lensing contribution \flens\ of each candidate.
Using available spectroscopic cluster redshifts, we quantify the mean bias and scatter in \zcl\ at levels of $\approx0.005$ and $\approx0.008$, respectively, demonstrating excellent photo-redshift performance.
We compare \zcl\ with estimates from direct positional cross-matching and find that approximately $8\percent$ of the total sample exhibits redshift discrepancies greater than $0.15$.
This outlier fraction is primarily attributed to projection effects, leading to the mis-identification of the optical counterparts.
We quantify the systematics of the resulting cluster redshifts and find that they have a negligible impact on the cosmological constraints from HSC-Y3 like samples. 
\end{abstract}


\maketitle


%
%

\section{Introduction}
\label{sec:intro}

Galaxy clusters form at peaks of the matter density field and therefore their abundance over cosmic time directly traces the growth of cosmic structures.
Measurements of the cluster halo mass function \citep{press74}, i.e., the number density of clusters as a function of halo mass, provide a powerful probe of the matter power spectrum \citep{weinberg13,huterer15}, enabling stringent constraints on cosmological parameters, such as the mean matter density \omegam\ and the amplitude of matter fluctuations \sigmaeight, and the equation of state of dark energy \w\ \citep{haiman01,allen11,miyatake25_review}. 

The advent of wide-field surveys over the last two decades, together with the progress in making accurate theoretical predictions of the halo mass function \citep{tinker08,watson13,despali16,mcclintock19,nishimichi19,bocquet20,castro23,castro25}, has enabled precision cluster cosmology using large samples constructed in X-rays \citep{vikhlinin09b,mantz15,garrel22,chiu23,ghirardini24}, at mm wavelengths \citep{PlanckCollaboration2015b,bocquet15,deHaan16,bocquet19,salvati22}, and in the optical \citep{abbott20desy1clustercosmology,costanzi21,giocoli21,sunayama24,lesci25}.
Combinations with multiple probes \citep{to21,desy3_mp_25,bocquet25}, applications of simulation-based inference \citep{tam22,tam26,salcedo26}, and extensions to cosmological models beyond \lcdm\ \citep{chiu23,artis24,vogt25,vogt26} are becoming increasingly common and powerful in cluster cosmology.

Despite the success, the selection of current cluster samples primarily relies on ``baryonic tracers'', namely X-ray emissions from intracluster medium (ICM), CMB decrements caused by the Sunyaev-Zel'dovitch effect \citep[SZE;][]{sunyaev72}, or galaxy overdensities in optical/near-infrared imaging.
Therefore, the pivotal tasks of baryon-based cluster cosmology are to establish the association between the halo mass and the baryonic observable, as the mass proxy, through an observable-mass relation, and to understand how the sample selection in the observable space affects the resulting cluster population.
Due to complex astrophysical processes, baryon-based observable-mass relations must be 
empirically calibrated using flexible functional forms of halo mass and redshift, together with the absolute calibration of halo mass from gravitational lensing \citep[e.g.,][]{umetsu20,chiu22,grandis24}.
The complexity of baryonic physics not only prevents us from assessing the ground truth of observable-mass relations but also leads to intrinsic scatter in baryonic observables at fixed mass and redshift.
To this end, common approaches are to marginalize over large numbers of nuisance parameters, thereby further weakening constraints on cosmological parameters.
Additionally, the selection function of baryonically selected cluster samples is subject to astrophysical processes.
For example, X-ray surveys preferentially select clusters with cool cores due to their enhanced central X-ray surface brightness \citep{hudson10,eckert11,lovisari15}, leading to a selection bias compared to SZE-selected samples \citep{rossetti17}.
Calibrating baryonic physics using hydrodynamical simulations is extremely challenging and leads to predictions that are systematics-dominated and can only be as accurate as the adopted sub-grid physics \citep[see discussions in][]{chiu23}.

Weak gravitational lensing (WL), on the other hand, directly probes the total matter distribution without any assumptions about baryonic physics and, therefore, provides an alternative approach to cluster cosmology through its ability to select clusters independently of baryonic tracers \citep{wittman01,white02a,hamana04,dietrich10,fan10,lin15,shirasaki15,broxterman25}.
Importantly, the ``gravity-based'' selection function can be robustly calibrated using simulations, as gravitational lensing is a geometric consequence of General Relativity and nearly insensitive to the sub-grid baryonic physics.
Moreover, the sample selection and mass calibration rely on a single WL observable, bypassing baryonic mass proxies and subsequently reducing the number of nuisance parameters in the modelling.

One of the most important factors in constructing samples of WL shear-selected clusters for cosmological analyses is the depth of the WL data, or more specifically the number density of source galaxies \citep[see reviews in][]{oguri26a}.
A high number density of WL source galaxies reduces the shape noise and thereby enables the detection of low-mass clusters at a fixed detection threshold, significantly increasing the sample size accessible for cosmological studies.
Because of the stringent requirements, it was not until the Hyper Suprime-Cam Subaru Strategic Program (HSC-SSP) that large samples containing $\gtrsim60$ WL shear-selected clusters were compiled from the HSC first-year (HSC-Y1) WL data covering an area of $\approx160~\mathrm{deg}^2$ \citep{miyazaki18b}.
Later, a larger sample containing $\approx120$ WL shear-selected clusters was built based on the same HSC-Y1 data 
by separating source galaxies into distinct tomographic bins.
Following that, \cite{oguri21} used the HSC three-year (HSC-Y3) WL data to construct even larger samples containing $\gtrsim400$ clusters over a footprint area of $\approx500~\mathrm{deg}^2$.
In particular, a highly pure sample of $129$ WL shear-selected clusters was compiled in the HSC-Y3 aperture-mass maps obtained with an optimized, core-excised spatial filter and a conservative selection of sources at $\redshift\gtrsim0.7$ with a mean redshift $\left\langle\redshift\right\rangle\approx1.3$.
With this sample, the first cosmological constraints from the abundance modelling of
WL shear-selected clusters were derived \citep{chiu24}, enabled by a novel method for determining the sample selection function \citep{chen25}.

Unlike the cluster finding in the optical, the gravity-based selection of shear-selected clusters does not estimate cluster redshifts, as WL only probes the total matter distribution integrated along the line of sight.
To obtain redshift information, it is common to cross-match WL shear-selected clusters with external cluster catalogs that provide known redshifts \citep{oguri21,chen25}; however, doing so entangles the gravity-based selection with those of the external cluster samples and further complicates subsequent cosmological modelling.
Therefore, the cosmological constraints in the previous work \cite{chiu24} were obtained by modelling only the abundance of WL shear-selected clusters above a detection threshold without the cluster redshift information, resulting in poorly constrained parameters (e.g., \omegam\ and \sigmaeight) except \seight.
For an improved cosmological analysis, it is necessary to measure the redshifts of shear-selected clusters using a consistent method \citep[cf.][]{broxterman26}.
In fact, this strategy---detecting clusters using ICM-based observables and subsequently performing optical confirmation to measure cluster redshifts---has been an essential component in cluster cosmology with clusters selected via SZE \citep{song12b,liu15,bleem15,hernandez-lang23,bleem24,klein24,bleem26} or in X-rays \citep{klein18,klein19,klein22,klein23}.

The goal of this paper is to perform the optical confirmation of the WL shear-selected clusters used in the cosmological analysis of \cite{chiu24} and measure their redshifts in a consistent manner using HSC photometric data sets.
Moreover, we compare the resulting cluster redshifts with estimates obtained from either spectroscopic data or external cross-matching, and quantify the cluster-redshift systematics and its impact on the final cosmological constraints.
We measure the redshifts of WL shear-selected clusters by identifying the ``red sequence'' (RS) population of passively evolving cluster galaxies in color-magnitude space.
The novelty of this work lies in the calibration of the RS model over a broad range of redshift using a data-driven manner.
The resulting redshift measurements in this work will extend the cosmological analysis of \cite{chiu24} 
from the abundance modelling of the WL shear-selected clusters as a function of the signal-to-noise ratio alone to that as a joint function of signal-to-noise ratio and redshift.
As demonstrated in this paper, the inclusion of the cluster redshift in the abundance modelling will significantly improve the cosmological constraints.

We organize this paper as follows.
We describe the cosmological sample of WL shear-selected clusters and the data sets used for the optical confirmation in Section~\ref{sec:data}.
The RS galaxy model used to determine the cluster redshifts is calibrated in Section~\ref{sec:rsmodel}.
The algorithm of optical confirmation for WL shear-selected clusters is fully described in Section~\ref{sec:confirmation}.
We present the results in Section~\ref{sec:results} and quantify the systematics of cluster redshift estimates on the final cosmological constraints in Section~\ref{sec:cosmology}.
Conclusions are made in Section~\ref{sec:conclusions}.
Unless stated otherwise, all uncertainties correspond to the $68\percent$ confidence levels ($1\sigma$).
Cluster masses are defined within a radius enclosing a mean density $200$ times the critical density at the cluster redshift.
We use either \mass\ or \Mtwooo\ interchangeably to denote the cluster mass.
A flat \lcdm\ model is assumed in this work with cosmological parameters of $\omegam = 0.3$, $\sigmaeight =  0.8$ and a Hubble constant $\Hnow = 70~\mathrm{km} s^{-1} {\Mpc}^{-1}$.

%
%

\section{Data and sample}
\label{sec:data}

In Section~\ref{sec:optical_data}, we introduce the data sets used in the optical confirmation, followed by the description of the cluster sample in Section~\ref{sec:cluster}.

\subsection{The HSC photometric catalog}
\label{sec:optical_data}

We use the photometric data from the Hyper Suprime-Cam Subaru Strategic Program (HSC-SSP, or the HSC survey), which is a wide-field imaging survey in five broadband filters $grizY$ carried out with the Subaru Telescope during 2014-2021 \citep{aihara18a}.
By design, the survey consists of three layers, namely Wide, Deep, and Ultra-Deep with \imag-band limiting magnitudes of $\approx26$, $\approx27$, and $\approx 27.5$, respectively.
The Wide layer covers an area of $\approx1100$~deg$^2$ with an average seeing of $\lesssim 0.6\arcsec$.
The unique combination of imaging depth, wide area, and superb imaging quality makes the HSC survey one of the premier WL data sets prior to the Stage-IV era.
To date, three Public Data Releases have been made available \citep{aihara18b,aihara19,aihara22} with the fourth and final release (PDR4) still forthcoming.

In this work, we use the s23b photometric catalog in the Wide layer from the HSC database, which will be released in PDR4 and can be accessed through the \texttt{SQL} query provided in Appendix~\ref{app:sql} \citep[see also][]{oguri26}.
We query the point-spread-function (PSF) matched aperture photometry measured within a diameter of $\approx1.5\arcsec$ on homogenized images convolved to a target seeing of $\approx1\arcsec$, thereby minimizing the bias in colors caused by deblending in crowded fields \citep{aihara18b}. 
Specifically, we use the \texttt{cmodel} magnitude in the \zmag\ band and derive the magnitudes in the other bands using the aperture photometric colors.
We remove objects flagged by \texttt{pixelflags\_edge}, \texttt{pixelflags\_interpolatedcenter} and \texttt{pixelflags\_crcenter} in each band.
Star masks are applied through the flags of \texttt{mask\_brightstar\_halo}, \texttt{mask\_brightstar\_ghost} and \texttt{mask\_brightstar\_blooming}.
The star-galaxy separation is performed by the flag \texttt{i\_extendedness\_value}.
To have a uniform imaging, we restrict the catalog to the full-depth-full-color footprint defined by 
$\mathtt{inputcount\_value} \geq 2$ in \gmag\rmag\ and 
$\geq 3$ in \imag\zmag\ymag.
We only use galaxies with the \zmag-band magnitude error below $0.1$ and magnitude errors below $1$ in the remaining bands.

\subsection{The cluster sample}
\label{sec:cluster}

We aim to optically confirm the sample of WL shear-selected clusters constructed in \cite{oguri21}.
The clusters are selected in the WL signal-to-noise-ratio maps $\snr\left(\skyloc\right)$ obtained from the HSC Three-Year (HSC-Y3) WL data sets \citep{li22}, covering six disjoint fields spanning a total area of $\approx500$~deg$^2$.
The maps $\snr\left(\skyloc\right)$ record the ratios of the aperture mass $\mkappa\left(\skyloc\right)$ and the r.m.s. uncertainty $\sigma_{\kappa}\left(\skyloc\right)$ at the sky location \skyloc. 
Below, we provide a brief summary of the cluster sample and the maps, and refer readers to \cite{oguri21} for more details.

To produce the aperture-mass maps $\mkappa\left(\skyloc\right)$, a truncated isothermal (TI) filter is used to spatially convolve the discrete observed WL shear maps.
The filter is configured by three parameters, namely $\nu_1$, $\nu_2$, and a radial threshold $\theta_{\mathrm{R}}$, where $\nu_1 < \nu_2$.
The filter is set to zero at $\theta > \theta_{\mathrm{R}}$, so that variations arising from large-scale structures in large radii are removed.
The parameter $\nu_1$ determines the effective aperture size, from which the foreground/background (hereafter background) at $\nu_2 \theta_{\mathrm{R}} \lesssim \rdd \lesssim \theta_{\mathrm{R}}$ is subtracted locally.
With the local background subtraction, the TI filter significantly suppresses local fluctuations caused by cosmic variance.
One critical feature of the TI filter is that it removes the shear signals in cluster inner regions, approximated by $\nu_1 \theta_{\mathrm{R}}$.
By doing so, it not only avoids WL modelling systematics but also mitigates contamination from cluster member galaxies, both of which are most significant near cluster cores.
We use the cluster sample constructed using the TI20 filter with $\nu_1 = 0.121$, $\nu_2 = 0.36$, and $\theta_{\mathrm{R}} = 16.6\arcmin$, which has been shown to optimize the signal-to-noise ratios for the WL peaks \citep{oguri21}.
The uncertainty maps $\sigma_{\kappa}\left(\skyloc\right)$ are derived using the same procedure on the WL catalogs of galaxies with randomized orientations.
In this work, we have $\sigma_{\kappa}\left(\skyloc\right) \approx 0.7$ with little fluctuations across the footprint.

We select the WL sources relaying on the photometric redshift estimates.
Specifically, we require that the redshift distribution $P\left(\redshift\right)$ to have a cumulative probability since $\redshift = 0.7$ greater than $95\percent$ on a per-galaxy level, resulting in an average source density of $\approx10~\mathrm{galaxies}/\mathrm{arcmin}^2$ at a mean redshift of $\approx1.3$.
This conservative source selection leads to an extremely clean aperture-mass maps for the shear-selected clusters, which primarily locate at redshift $\redshift\lesssim0.7$ with a mean value of $\redshift\approx0.3$ \citep{chen25}.
That is, the WL signals of the shear-selected clusters are free from member contaminations, as quantified in \cite{oguri21}.

The sample contains $129$ shear-selected clusters over the HSC-Y3 footprint with WL peak signal-to-noise ratios $\snr > 4.7$.
A cross-match with external cluster catalogs identified optical counterparts for the majority of the shear-selected clusters (i.e., $126$ systems), while the remaining ones can be explained by either the projection of line-of-sight structures or a large offset to low-mass halos \citep{chen25}.
This suggests that the WL shear-selected cluster sample has a purity at a level of $\gtrsim98\percent$, owing to the clean source selection and the optimized TI20 filter.
We stress that we do not use the counterparts obtained from such cross-matching, as doing so entangles both the WL and other selection functions and is difficult to model in cosmological analyses.
Instead, we directly measure the redshifts of the shear-selected clusters, as the central objective of this work.

%
%

\section{The galaxy red-sequence model}
\label{sec:rsmodel}

In this section, we describe the construction of the red-sequence (RS) model in detail.
As a unique strength of this work, the RS model is empirically constructed in a data-driven way with minimal assumptions to avoid potential bias.

The RS model is built based on the stellar population synthesis model, followed by a two-step calibration leveraging observational data sets.
By closely following the prescription in \citet[][see also \citealt{song12a} and \citealt{hennig17}]{liu15b}, we utilize a composite stellar population model formed at redshift $z_{\mathrm{f}} = 3$ with an exponentially decaying star-formation rate with an $e$-folding time scale of $\tau = 0.4~\mathrm{Gyr}$, as our baseline model before the calibration.
The baseline model has been shown to provide a good description of the passively evolving galaxy population in massive halos \citep{bleem15,bleem20}.
We use the spectral energy distribution (SED) from the \cite{bruzual03} template library.
The SEDs with six different metallicities ($Z =  0.0001, 0.0004, 0.004, 0.008, 0.02, 0.05$) are adopted to capture the ``tilt'' of RS galaxies in color-magnitude diagrams (see Section~\ref{sec:metallicity_luminosity}).
The \citet{chabrier03} initial mass function is assumed.
The configurations of the RS model are generated using the code \texttt{ezgal} \citep{mancone12b}.

The absolute calibration of the RS model is anchored to the observed luminosity functions of halos at the local Universe \citep{lan16}, which provides a precise and accurate determination of the characteristic magnitude of $M_{\ast} = -21.26\pm0.06$ in the rest-frame SDSS $r$-band at redshift $\redshift = 0.03$.

The baseline RS model is further calibrated by a two-step process, namely the calibrations of the RS tilt (Section~\ref{sec:metallicity_luminosity}) and the color offsets (Section~\ref{sec:specz_calib}).
The two-step calibration is performed in the HSC broadband $grizY$ photometric system.

\begin{figure*}
\centering
\resizebox{0.49\textwidth}{!}{
\includegraphics[scale=1]{
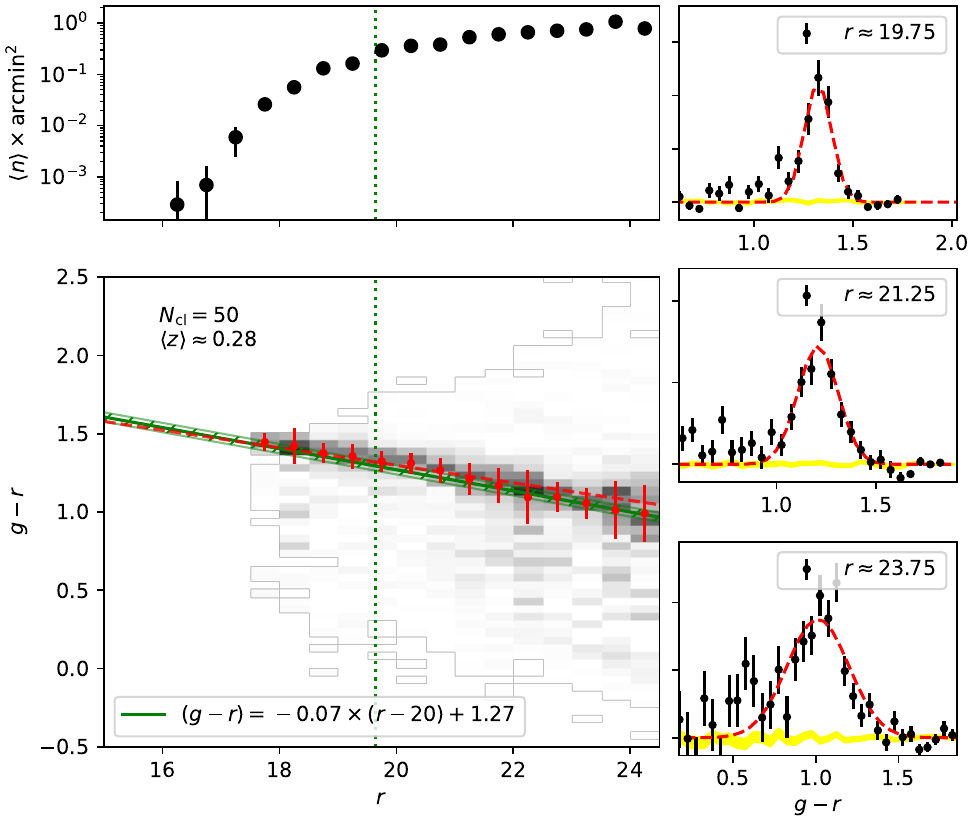
}
}
\resizebox{0.48\textwidth}{!}{
\includegraphics[scale=1]{
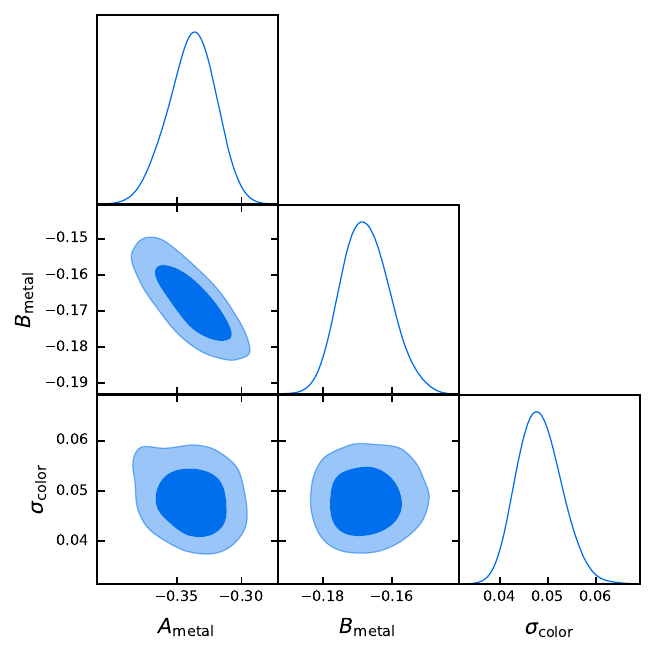
}
}
\caption{
The calibration of the metallicity-luminosity relation using the eFEDS cluster sample.
\textit{Left panel}: The background-subtracted color-magnitude diagrams of cluster galaxies for the subsample at $0.26 \lesssim z \lesssim 0.30$.
The red points and error bars show the mean colors and scatters obtained from Gaussian fits to the color distributions at fixed magnitudes, which are represented by the intensity of the underlying grey color.
The red dashed line is the observed RS relation derived from a linear fit to the data points.
The green solid line shows the prediction of the calibrated RS model at the mean cluster redshift.
The upper subplot presents the galaxy number density per square arcmin in the individual magnitude bins.
The three subplots in the right show the color distributions in the magnitude bins with the best-fit Gaussian models indicated by the red dashed lines, while the yellow regions represent the $1\sigma$ confidence levels of the null tests which are measured using random pointing.  
\textit{Right panel}: The posterior distributions of the metallicity-luminosity relation parameters $A_{\mathrm{metal}}$ and $B_{\mathrm{metal}}$ and the intrinsic RS color scatter $\sigma_{\mathrm{color}}$.
The contours indicate the $68\percent$ and $95\percent$ confidence regions.
}
\label{fig:rs_example}
\end{figure*}

\subsection{The metallicity-luminosity relation}
\label{sec:metallicity_luminosity}

The galaxy population in clusters is dominated by early-type or passively evolving galaxies, which exhibit a ``tilt'' (i.e., the RS) in the color-magnitude diagram as evidence of a luminosity-dependent color \citep{kodama97}.
The luminosity-dependent color origins from strong absorption by metallicities at around rest-frame $4000\angstrom$, indicative of a luminosity- or stellar mass-dependent metallicity.

We describe the mean luminosity-dependent metallicity $Z$ using a metallicity--luminosity ($Z-L$) relation parameterized by the observed magnitude $m$ \citep{poggianti2001},
\begin{equation}
\label{eq:metal_mag}
\left\langle \log_{10}\left(\frac{Z}{Z_{\odot}}\right) \right\rangle = A_{\mathrm{metal}} + B_{\mathrm{metal}} \times \left(\magnitude - \mstar\right) \, ,
\end{equation}
where $Z_{\odot}$ is the solar metallicity evaluated as $Z_{\odot} = 0.02$, \mstar\ is the characteristic magnitude of the cluster galaxy luminosity function in the observed frame, and $A_{\mathrm{metal}}$ and $B_{\mathrm{metal}}$ are the normalization and slope of the $Z-L$ relation, respectively.
The normalization $A_{\mathrm{metal}}$ sets the metallicity at the characteristic magnitude \mstar, while the slope $B_{\mathrm{metal}}$ determines the RS tilt.
The characteristic magnitude \mstar\ depends on redshift with its absolute calibration being anchored to that of the observed cluster luminosity functions at nearby Universe \citep[][see also Section~\ref{sec:rsmodel}]{lan16}.

We note that the luminosity-dependent metallicity in equation~(\ref{eq:metal_mag}) is expressed in terms of the magnitude offset with respect to the predicted \mstar\ of the passively evolving galaxy population at a given redshift.
By doing so, this parameterized form allows us to describe the metallicity of the passive galaxy population in clusters across a wide range of redshift using a fixed interval of the magnitude offset $\left(\magnitude - \mstar\right)$.
To describe the tilt in a color-magnitude diagram, the metallicities are evaluated at nine magnitude offsets of $\left(\magnitude - \mstar\right) = -2.5, -1.5, -1.19, -0.75, 0, 0.75, 1.19, 1.5, 2.5$, corresponding to the luminosities of $L/L_{\ast} = 10, 4, 3, 2, 1, 0.5, 0.33, 0.25, 0.1$, respectively.

In what follows, we describe the calibration of the $Z-L$ relation using the observed RS galaxies in clusters spanning a wide range of redshift.
For this purpose, we utilize the sample of X-ray-selected clusters in the \erosita\ Final Equatorial-Depth Survey \citep[eFEDS;][]{liu22}.
The eFEDS sample is ideal for studying cluster galaxy properties because 
(1) the clusters are selected in X-rays and free from any selection bias associated with galaxy properties, 
(2) each eFEDS cluster is optically confirmed with an accurate redshift estimate \citep{klein22}, 
(3) the sample spans a wide redshift range, allowing a homogeneous calibration of the RS model out to $\redshift\approx1.3$,
(4) the individual cluster mass is calibrated against weak lensing with accuracy of $\approx 6\percent$ and precision of $\approx20\percent$ \citep{chiu22}, enabling the consistent selection of cluster member galaxies within an aperture \Rfiveoo, and
(5) the eFEDS footprint is completely covered by the HSC survey, by design.
We further apply the selection of $\redshift > 0.1$ and $f_{\mathrm{cont}} < 0.3$\footnote{$f_{\mathrm{cont}}$ is the probability of observing the same richness estimate of a cluster candidate in a random line of sight.}, leading to a sample of $455$ clusters with a sample purity better than $\approx6\percent$ with halo mass \Mfiveoo\ estimated in a joint cosmological analysis of cluster abundance and WL \citep{chiu23}.

We divide the eFEDS clusters into $14$ subsamples between redshifts $0.1$ and $1.3$ using logarithmic binning in terms of $1+\redshift$. 
To straddle the $4000\angstrom$ break, we derive the galaxy color-magnitude diagram in the spaces of $\gmag-\rmag$ v.s. $\rmag$ ($\rmag-\imag$ v.s. $\imag$, $\imag-\zmag$ v.s. $\zmag$, and $\zmag-\ymag$ v.s. $\ymag$) for the clusters at $\redshift < 0.35$ ($0.35 \leq \redshift < 0.75$, $0.75 \leq \redshift < 1.12$, and $\redshift \geq 1.12$) on a per-cluster basis.
These color-magnitude diagrams are derived in the observed frame.
We use the bin widths of $0.5~\mathrm{mag}$ and $0.05~\mathrm{mag}$ for the magnitude and color binning, respectively.
For each cluster subsample, we derive the mean projected number density of galaxies in each color-magnitude cell, followed by a foreground/background (hereafter background) subtraction using a global estimate\footnote{We verify that the background removal using the local background extracted between the cluster radii of $2.5 \Mpch$ and $3.5 \Mpch$ returns statistically consistent results.} \citep{nishizawa18}.
As a result, we obtain the color-magnitude diagram of cluster member galaxies per subsample in a purely photometric way.
We note that the RS distributions are derived from the cluster galaxies within \Rfiveoo, informed by the halo mass estimates \Mfiveoo\ \citep{chiu22, chiu23}.

As an example, we show the result of the subsample at $0.26\lesssim\redshift\lesssim0.30$ in the left panel of Figure~\ref{fig:rs_example}.
We leave the results of other subsamples to Appendix~\ref{app:rs_meas}.
As seen in the figure, we detect a strong signal of the RS in the color-magnitude diagram.
We further derive the color distributions at individual magnitude bins and model each of them as a Gaussian distribution. 
The best-fit mean values $\mathfrak{c}_{\mathrm{obs}}$ and scatter $\delta\mathfrak{c}_{\mathrm{obs}}$ of the Gaussian fits at the magnitude bins are shown as the red points and errorbars in the color-magnitude diagram.
We fit a linear relation to the data points, as the parameterized RS equation shown by the red dashed line, and perform the same analysis for all subsamples.

We calibrate the $Z-L$ relation in equation~(\ref{eq:metal_mag}) with the measured RS colors at the magnitude bins over a wide redshift range. 
For each subsample at redshift \redshift, we generate a set of passively evolving galaxy SED templates at the nine magnitude offsets, $\left(\magnitude - \mstar\right) = -2.5, -1.5, -1.19, -0.75, 0, 0.75, 1.19, 1.5, 2.5$, where $\mstar\left(\redshift\right)$ is the characteristic magnitude varying with redshift with its absolute magnitude anchored to the measurements of the local clusters \citep{lan16}.
The SED templates are generated with the corresponding metallicities using equation~(\ref{eq:metal_mag}) for given parameters $\left(A_{\mathrm{metal}}, B_{\mathrm{metal}}\right)$, enabling the model prediction of the RS color $ \mathfrak{c}_{\mathrm{model}}$ with an interpolation over the magnitude range. 
Finally, we maximize the likelihood
\begin{multline}
\label{eq:rscluster_lnp}
\ln P\left(A_{\mathrm{metal}}, B_{\mathrm{metal}}, \sigma_{\mathrm{color}}\right) = \\
-
\sum\limits_{ \substack{ i\in \mathrm{subsample~bins} \\ j \in \mathrm{magnitude~bins} } }
\left[
\frac{
\left({ \mathfrak{c}_{\mathrm{obs}} }_{i,j} -  \mathfrak{c}_{\mathrm{model}, i}\left(\magnitude_{j} | A_{\mathrm{metal}}, B_{\mathrm{metal}}\right)
\right)^2
}{
2\left(
{ { \delta\mathfrak{c}_{\mathrm{obs}} }_{i, j} }^2 + \sigma_{\mathrm{color}}^2
\right)
} 
\right. \\
+
\left.
\frac{
\ln \left( { { \delta\mathfrak{c}_{\mathrm{obs}} }_{i, j} }^2 + \sigma_{\mathrm{color}}^2 \right)
}{
2
}
\right]
\, ,
\end{multline}
where ${ \mathfrak{c}_{\mathrm{obs}} }_{i,j}$ and ${ \delta\mathfrak{c}_{\mathrm{obs}} }_{i, j}$ are the mean colors and scatter of the $i$-th subsample at the $j$-th magnitude bin, respectively, and 
$\mathfrak{c}_{\mathrm{model}, i}\left(\magnitude_{j}\right)$ is the model prediction of the color for the $i$-th subsample evaluated at the magnitude $\magnitude_{j}$ using equation~(\ref{eq:metal_mag}).
In equation~(\ref{eq:rscluster_lnp}), we introduce the parameter $\sigma_{\mathrm{color}}$ to account for the intrinsic scatter in the RS color, which is assumed to be independent of redshift in this work.
The parameter space of $\left(A_{\mathrm{metal}}, B_{\mathrm{metal}}, \sigma_{\mathrm{color}}, \right)$ is explored using the code \texttt{emcee} \citep{foreman13, foreman19}, as the results are shown in the right panel of Figure~\ref{fig:rs_example}.

We constrain the parameters $\left(A_{\mathrm{metal}}, B_{\mathrm{metal}}, \sigma_{\mathrm{color}}\right)$ as $\left(-0.34 \pm 0.02, -0.17 \pm 0.01, 0.048 \pm 0.005\right)$, corresponding to metallicities $Z$ of 
$0.02$, $0.0135$, $0.0092$, $0.006$, $0.004$ at the magnitudes \magnitude\ of
$\mstar-2$, $\mstar-1$, \mstar, $\mstar + 1$, $\mstar + 2$, respectively.
We fit a linear relation to the predicted RS colors at the magnitude offset for each subsample.
The resulting RS relation of the subsample is shown as the green line in the color-magnitude diagram in the left panel of Figure~\ref{fig:rs_example}.
As seen, the best-fit RS relation (green solid line) and the observed one (red dashed line) is in good agreement, demonstrating that the resulting $Z-L$ relation well captures the observed RS tilt.

We note that the resulting metallicity-luminosity relation in equation~(\ref{eq:metal_mag}) serves an effective relation conditional on the adopted passively evolving SED template that has the formation redshift of $\redshift = 3$, the exponentially decaying star-formation rate with $\tau = 0.4~\mathrm{Gyr}$, and the \cite{chabrier03} initial mass function.
Our primary focus is to empirically model the observed RS tilts in the color-magnitude diagrams via the metallicity-luminosity relation rather than providing direct measurements of galaxy metallicities.

It is also worth noting that we do not account for blue member galaxies in the stacked color-magnitude diagrams.
However, blue galaxies are shown to be sub-dominant in the cluster galaxy population with luminosities approximately brighter than $0.15 L_{\star}$, corresponding to a magnitude range $\magnitude - \mstar \lesssim  2$ \citep{lan16,chiu16c}.
In Appendix~\ref{app:rs_meas}, we quantify the blue fraction of the eFEDS clusters in the stacked color-magnitude spaces and show that it is a sub dominant component in the cluster galaxy population.

\subsection{The color offsets}
\label{sec:specz_calib}

In the previous subsection, we calibrate the metallicity-luminosity relation using the passively evolving galaxies in eFEDS clusters, and incorporate it into the composite stellar population SED templates to reproduce the observed RS colors to first order.
However, the SED templates are imperfect and hence exhibit biases in colors with respect to those of observed galaxies.
To mitigate these biases in this subsection, we further perform an empirical calibration of the colors predicted by the RS model using spectroscopic samples of passive galaxies.
We largely follow the procedure described in \cite{oguri14}, to which we refer readers for further details.
As follows, we outline the key steps of the color calibration.

First, we collect spectroscopic samples available in the HSC database\footnote{
The spectroscopic samples compiled by the HSC survey include
zCOSMOS DR3 \citep{lilly09},
3D-HST \citep{skelton14,momcheva16},
SDSS DR15 \citep{aguado19},
GAMA DR3 \citep{baldry18},
UDSz DR1 \citep{bradshaw13,mclure13},
VANDELS DR4 \citep{garilli21},
C3R2 DR2 \citep{masters19},
VVDS Final Data Release \citep{lefevre13}
DEIMOS \citep{hasinger18},
FMOS DR2 \citep{silverman15},
LEGA-C DR2 \citep{straatman18},
PAUS+COSMOS \citep{alarcon21}
PRIMUS DR1 \citep{coil11,cool13},
VIPERS DR2 \citep{scodeggio18},
WiggleZ DR1 \citep{drinkwater10},
DEEP2 \citep{newman13}, 
and DEEP3 \citep{cooper12}
}
and from the Dark Energy Spectroscopic Instrument DR1 \citep{desi_dr1}.
We then apply a series of color selections\footnote{
The color selections are defined in Appendix 2 of \cite{oguri26}.
} to their HSC $grizY$ broadband photometry with the goal of constructing a spectroscopic RS galaxy sample with minimal contamination.
We note that the exact form of the color selections has a negligible impact on the final results, as we iteratively  remove the non-RS contaminants in a later step.

Next, we fit the RS model with its metallicity following the calibrated metallicity-luminosity relation in equation~(\ref{eq:metal_mag}) to each galaxy in the spectroscopic sample.
Specifically, we fix the redshift of the RS model to the galaxy spectroscopic redshift (hereafter spec-$z$) $\redshift_{\mathrm{spec}}$ and minimize
\begin{equation}
\label{eq:chi2_init}
\chi^2\left(\Delta\magnitude \right) =
\sum\limits_{k\in \mathrm{filters}}
\left(
\frac{ 
\magnitude_{\mathrm{obs},k} 
- \magnitude_{\mathrm{RS},k}\left( \Delta\magnitude  \right) 
}{
\sigma_{\mathrm{obs},k}
} 
\right)^2
\, ,
\end{equation}
where the index $k$ runs over the HSC filters,
the parameters $\magnitude_{\mathrm{obs},k}$ and $\sigma_{\mathrm{obs},k}$ are the observed magnitude and its uncertainty of the galaxy, respectively, and
$\magnitude_{\mathrm{RS},k}\left( \Delta\magnitude  \right)$ is the RS model magnitude prediction evaluated at the magnitude offset, $\Delta\magnitude = \magnitude - \mstar\left(\redshift_{\mathrm{spec}}\right)$.
We note that the minimization of equation~(\ref{eq:chi2_init}) only depends on one parameter $\Delta\magnitude$, as the redshift is fixed to $\redshift_{\mathrm{spec}}$.

\begin{figure}
\centering
\resizebox{0.45\textwidth}{!}{
\includegraphics[scale=1]{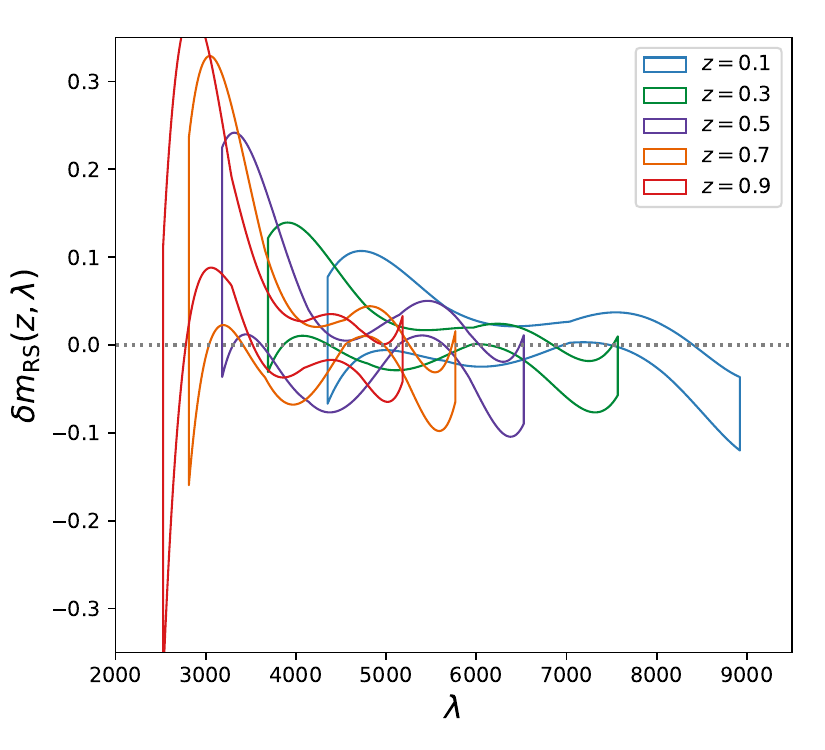}
}
\caption{
The calibration of the color offsets of the RS model.
The colored curves show the $1\sigma$ confidence levels of the magnitude biases $\delta m_{\mathrm{resid}}$ as a function of rest-frame wavelength $\lambda_{\mathrm{rf}}$ at several redshifts, whose colors are indicated in the upper-right corner.
}
\label{fig:magbias}
\end{figure}

Next, we calibrate the magnitude bias $\delta\magnitude_{\mathrm{RS}}$ and the intrinsic scatter $\sigma_{\mathrm{RS}}$ of the RS model with respect to the observed magnitude of the spectroscopic galaxies.
To do so, we divide the spectroscopic galaxy sample over the redshift  $0.01 < \redshift < 1.30$ into $100$ logarithmic bins in $1+\redshift$.
In each redshift bin with a mean redshift $\bar{ \redshift }_{\mathrm{spec}}$, we approximate the per-band magnitude distributions of RS galaxies as Gaussian and maximize the following likelihood,
\begin{multline}
\label{eq:chi2_mini}
\ln P\left( \vect{p} \right) = \\
-\frac{1}{2}
\sum\limits_{ \substack{ i\in \mathrm{galaxies} \\ k\in \mathrm{filters} } }
\left[
\left(
\frac{ 
\Delta_{i,k} - \delta\magnitude_{\mathrm{RS},k}
}{
\sigma_{\mathrm{comb},i,k}
}
\right)^2
+2
\ln\left( 
\sigma_{\mathrm{comb},i,k}
\right) 
\right]
\, ,
\end{multline}
where the term $\Delta_{i,k}$ is the residual magnitude at the $k$-th HSC broadband of the $i$-th galaxy, and 
the term $\sigma_{\mathrm{comb},i,k}$ contains both the per-galaxy measurement uncertainty and the intrinsic scatter, evaluated as 
\begin{equation}
\label{eq:rs_intrinsic_scater}
\sigma_{\mathrm{comb},i,k} = \left( {\sigma_{\mathrm{obs},i,k}}^2 + {\sigma_{\mathrm{RS},k}}^2\right)^\frac{1}{2}
\, .
\end{equation}
We model the magnitude bias of the $k$-th HSC broadband, $\delta\magnitude_{\mathrm{RS},k}$, as a polynomial function of the rest-frame wavelength, 
\begin{equation}
\label{eq:magbias_perband}
\delta\magnitude_{\mathrm{RS},k} = \sum\limits_{s = 1}^{5}
a_{s} \times \left(\lambda_{k} - \lambda_{\mathrm{piv}} \right)^s
\, ,
\end{equation}
in which $\lambda_{k}$ is the rest-frame wavelength of the $k$-th HSC broadband filter at the mean redshift $\bar{ \redshift }_{\mathrm{spec}}$, and
$\lambda_{\mathrm{piv}}$ is the pivotal wavelength of this redshift bin evaluated as 
$\lambda_{\mathrm{piv}}  = \left( 6400 - 2850 \times \bar{ \redshift }_{\mathrm{spec}} \right) \angstrom$.
Note that equation~(\ref{eq:magbias_perband}) is a fifth-degree polynomial ($\left\lbrace a_s | s = 1, 2, 3, 4, 5\right\rbrace$) because there are five filters in the HSC data set.
In addition, we set the constant term $a_0$ to be zero because we aim to calibrate the color offset while fixing the absolute luminosity calibration.
Meanwhile, we model the rest-frame RS intrinsic scatter $\left\lbrace \sigma_{\mathrm{RS},k} | k \in \mathrm{filters} \right\rbrace$ at the mean redshift $\bar{ \redshift }_{\mathrm{spec}}$.
As a result, there are ten parameters $\vect{p} = \left\lbrace a_s | s = 1, 2, 3, 4, 5\right\rbrace \cup \left\lbrace \sigma_{\mathrm{RS},k} | k \in \mathrm{filters} \right\rbrace$ in maximizing equation~(\ref{eq:chi2_mini}).

We model the magnitude bias $\delta\magnitude_{\mathrm{RS}}$ and the intrinsic scatter $\sigma_{\mathrm{RS}}$ in each redshift bin in an iterative way.
Specifically, we start with the residual magnitude using the best-fit magnitude offset in equation~(\ref{eq:chi2_init}), i.e., 
\[
\Delta_{i,k} = \magnitude_{\mathrm{obs},i,k} - \magnitude_{\mathrm{RS},i,k}\left( \Delta\magnitude  \right)
\, .
\]
Before maximizing equation~(\ref{eq:chi2_mini}), we perform a $3.5$-sigma clipping on the residual magnitude in each band.
Once the best-fit parameters $\vect{p} = \left\lbrace a_s | s = 1, 2, 3, 4, 5\right\rbrace \cup \left\lbrace \sigma_{\mathrm{RS},k} | k \in \mathrm{filters} \right\rbrace$ are found, we further remove non-RS contaminants by performing again a $3.5$-sigma clipping on the chi-square values $\left\lbrace \chi^2_i  |  i \in \mathrm{galaxies} \right\rbrace$, in which
\begin{equation}
\label{eq:chi2_updated}
\chi^2_i = \sum\limits_{ k\in \mathrm{filters} }
\left(
\frac{ 
\Delta_{i,k} - \delta\magnitude_{\mathrm{RS},k}
}{
\sigma_{\mathrm{comb},i,k}
}
\right)^2 \, .
\end{equation}
Finally, we iterate the process by updating the residual magnitude $\Delta_{i,k}$, which is achieved by fitting a new magnitude offset $\Delta\magnitude$ to minimize equation~(\ref{eq:chi2_updated}).
The process is iterated three times, after which we obtain the converged results of $ \left\lbrace a_s | s = 1, 2, 3, 4, 5\right\rbrace $ and $ \left\lbrace \sigma_{\mathrm{RS},k} | k \in \mathrm{filters} \right\rbrace$ for each redshift bin.
We present the calibration results of the magnitude bias $\left\lbrace\delta\magnitude_{\mathrm{RS},k} | k \in \mathrm{filters}\right\rbrace$ and the intrinsic scatter $\left\lbrace\sigma_{\mathrm{RS},k} | k \in \mathrm{filters}\right\rbrace$ at five redshift bins in Figure~\ref{fig:magbias}.

With the magnitude bias and the RS intrinsic scatter of the redshift bins, we are at a position to fit the RS model including the calibrated color offsets to the observed magnitudes of individual galaxies.
Moreover, the fitting is performed in terms of two parameters: the galaxy redshift \redshift, and the magnitude offset $\Delta\magnitude = \magnitude - \mstar\left(\redshift\right)$ with respect to the characteristic magnitude of the RS model at the redshift.

\begin{figure}
\centering
\resizebox{0.4\textwidth}{!}{
\includegraphics[scale=1]{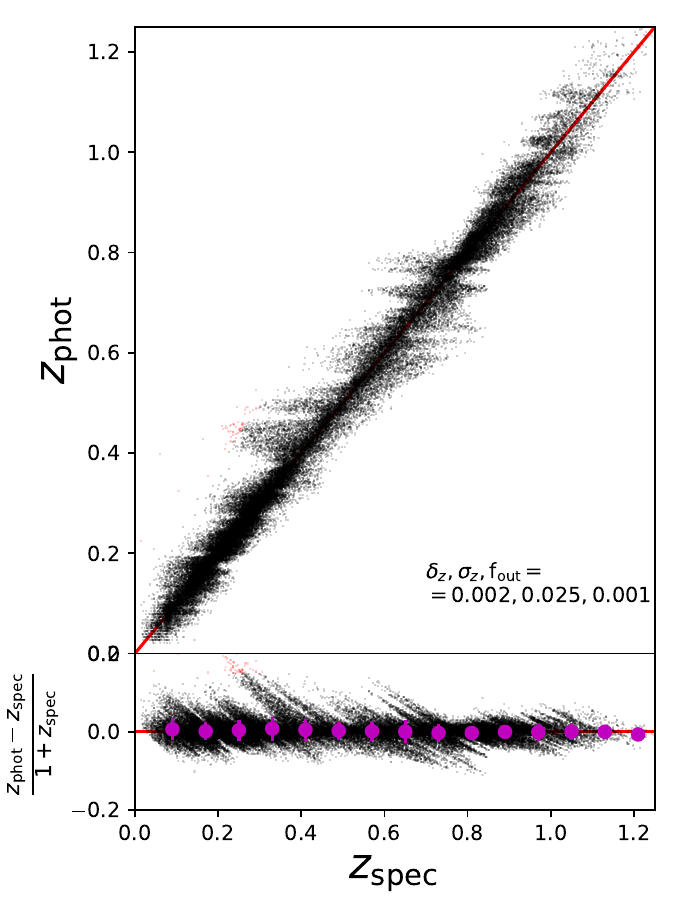}
}
\caption{
The validation of the photo-\redshift\ derived using the calibrated RS model.
\textit{Top panel}: The comparison between the RS photometric redshifts \photz\ and spectroscopic redshifts \specz.
The red dots are outliers defined by $\left(\photz - \specz\right) / \left(1 + \specz\right) > 0.15$.
The overall bias $\delta_\redshift$, scatter $\sigma_\redshift$, and the outlier fraction $f_{\mathrm{out}}$ are shown in the lower-right corner. 
\textit{Bottom panel}: The redshift residuals, $\left(\photz - \specz\right) / \left(1 + \specz\right)$, as a function of \specz.
The magenta points indicate the mean and scatter of the binned redshift intervals.
}
\label{fig:photoz_validation}
\end{figure}

\subsection{The validations}
\label{sec:validations}

In this subsection, we validate the RS model with the calibrated metallicity-luminosity relation and color offsets
using the spectroscopic galaxy sample.
As a test sample, we use $\approx67,000$ spectroscopic galaxies, which are $10\percent$ of the whole spectroscopic sample and not used in calibrating the color offsets in Section~\ref{sec:specz_calib}.
On a per-galaxy level, we fit for the photometric redshift (hereafter photo-\redshift) using the RS model by minimizing equation~(\ref{eq:chi2_updated}).

The results are presented in Figure~\ref{fig:photoz_validation}, where we show the scattering between the photo-\redshift\ and spec-\redshift\ estimates in the upper panel and the residual $\left(\redshift_{\mathrm{phot}} - \redshift_{\mathrm{spec}} \right) / \left( 1 + \redshift_{\mathrm{spec}}\right)$ in the lower panel.
As seen, the RS model achieves excellent performance (on average) with the bias of $\approx 0.2\percent$, the scatter of $\approx2.5\percent$, and an outlier rate\footnote{An outlier is identified if its $| \redshift_{\mathrm{phot}} - \redshift_{\mathrm{spec}} | / \left( 1 + \redshift_{\mathrm{spec}} \right) > 0.15$.} of $\approx0.1\percent$.
We note that only the galaxies with $\chi^2 < 20$ in equation~(\ref{eq:chi2_updated}) are used in producing Figure~\ref{fig:photoz_validation}, as those with extremely large $\chi^2$ are not considered as RS galaxies and, hence, not used in the subsequent analyses.

%
%

\section{The optical confirmation}
\label{sec:confirmation}

The goal of this section is to optically confirm weak-lensing shear-selected clusters at given peak sky locations by measuring their redshifts and RS galaxy overdensities.
This is achieved by constructing galaxy density maps at individual redshift slices, using the photo-\redshift\ estimated from the calibrated RS model (see Section~\ref{sec:rsmodel}).

The construction of the per-redshift richness maps is described in Section~\ref{sec:richness_map}.
Once the richness maps are obtained, we use the algorithm to optically confirm clusters in Section~\ref{sec:fcamira}.

\subsection{The richness maps}
\label{sec:richness_map}

We construct the richness maps in fine redshift bins by largely following the procedure described in \cite{oguri14}, to which we refer readers for further details.
Below, we summarize the key steps and provide the modifications made specifically for the purpose of the optical confirmation.

First, we compute $\chi^2\left(\redshift\right)$ for all galaxies in the HSC catalog by minimizing equation ~(\ref{eq:chi2_updated}) at each redshift in the fine binning, which has $150$ logarithmic bins in $1 + \redshift$ over $0.01 < \redshift < 1.3$.
This yields a $\chi^2\left(\redshift\right)$ array for each galaxy.
With the assumption that the RS colors are distributed as Gaussian at each band, we expect the distribution of $\chi^2\left(\redshift\right)$ at each redshift to be a chi-square distribution with $N_{\mathrm{filter}} - 1 = 4$ degrees of freedom.
In practice, we fit the degree-of-freedom parameter of a chi-square distribution to $\chi^2\left(\redshift\right)$ and find that the best-fit degree-of-freedom parameter is $\approx3.8$, which is consistent with the expectation.
With the best-fit degree-of-freedom parameter, we convert the chi-square array $\chi^2\left(\redshift\right)$ to the array of the ``number parameter'' $n\left(\redshift\right)$ as a function of redshift \redshift\ using the following equation\footnote{This is identical to equation~(5) in \cite{oguri14}, where a typo was present in the exponential term.}
\begin{equation}
\label{eq:n_chi2}
n\left(\chi^2; v \right) = \frac{ 2^{3 v / 4} }{ v^{v / 2}  U\left(v / 4, 1/2, v^2 / 8\right) }  \exp\left( -0.5 \left( \chi^2 / v \right)^2 \right) \, ,
\end{equation}
where $v$ is the degree of freedom, and $U$ is the confluent hypergeometric function of the second kind.
We note that the number parameter $n\left(\redshift\right)$ is equivalent to the effective number of the galaxy belonging to the RS population at the redshift \redshift.
With $n\left(\redshift\right)$, we are able to construct the richness maps of RS galaxies per redshift bin.

By design, non-member or non-RS galaxies have extremely large $\chi^2$ so their number parameters are down-weighted with negligible impact.
In practice, photometric uncertainties and the intrinsic scatter in galaxy colors result in foreground/background (hereafter background) contamination in the number parameter.
To address this, we adopt a spatial filter including a local background subtraction to remove the interlopers.

In this work, we use two types of spatial filters to obtain smoothed, background-subtracted richness maps, namely the CAMIRA filter $F_{\mathrm{CAMIRA}}$ and the TI20 filter $F_{\mathrm{TI20}}$.
The former is for blind cluster finding in \cite{oguri14}, which used a spatial filter matched to a typical cluster physical size of $\approx 0.8\Mpch$ while subtracting a local background at $1.2 \Mpch \lesssim R \lesssim 3\Mpch$.
On the other hand, the latter TI20 filter is used to construct the aperture-mass maps enclosed by a fixed angular scale to optimize the WL signals \citep{oguri21}.
Specifically, the TI20 filter calculates an aperture mass within $\approx5\arcmin$ subtracting background at $7\arcmin \lesssim \theta \lesssim 15\arcmin$.
We describe the applications of these richness maps obtained from the two filters in Section~\ref{sec:fcamira}.

In addition to the spatial filter, we also include magnitude weights to down-weight galaxies outside a magnitude range of interest in computing the richness maps.
Specifically, the magnitude weight $w_{\magnitude}$ is calculated in terms of the best-fit magnitude offset $\Delta\magnitude$ at each redshift bin,
\begin{equation}
\label{eq:magnitude_weight}
w_{\magnitude}\left(\Delta\magnitude\right) = \\
\exp\left(
-10^{-\frac{ 6}{2.5} \left(\Delta\magnitude + 2\right)}
-10^{ \frac{ 6}{2.5} \left(\Delta\magnitude - 2\right)}
\right) \, .
\end{equation}
With equation~(\ref{eq:magnitude_weight}), we effectively select galaxies with $-2 \lesssim \Delta\magnitude \lesssim 2$ consistently at each redshift bin in computing the richness maps. 

Finally, the richness maps at the redshift bins are given as
\begin{equation}
\label{eq:richness_map}
\rich\left(\vect{\theta} , \redshift\right) = \sum\limits_{i \in \mathrm{galaxies}} 
\frac{
n\left( \chi^2_i, \vect{\theta}_{i} | \redshift\right)
}{
f_\mathrm{mask}\left(\vect{\theta}_i, \vect{\theta} \right)  
} 
w_{\magnitude}\left(\Delta\magnitude_i\right)
F\left( | \vect{\theta}_{i} - \vect{\theta} |\right)
\, ,
\end{equation}
in which we convolve the galaxy number field by the spatial kernel $F\left(\vect{\theta}\right)$, and the factor $f_\mathrm{mask}\left(\vect{\theta}_i, \vect{\theta} \right)$ is the correction for masking, defined as
\begin{equation}
\label{eq:masking}
f_\mathrm{mask}\left(\vect{\phi}, \vect{\theta} \right)  = \left\lbrace
\begin{array}{lr}
\frac{
\int_{A_{+}} S\left(\vect{\phi}\right) F\left( | \vect{\phi} - \vect{\theta} |\right) {\dif}^2\vect{\phi} }{
\int_{A_{+}} F\left( | \vect{\phi} - \vect{\theta} |\right) {\dif}^2\vect{\phi}
} & \mathrm{if}~\vect{\phi} \in A_{+} \\
& \\
\frac{
\int_{A_{-}} S\left(\vect{\phi}\right) F\left( | \vect{\phi} - \vect{\theta} |\right) {\dif}^2\vect{\phi} }{
\int_{A_{-}} F\left( | \vect{\phi} - \vect{\theta} |\right) {\dif}^2\vect{\phi}
} & \mathrm{if}~\vect{\phi} \in A_{-} 
\end{array} 
\right.
\, ,
\end{equation}
where the sky location $A_{+}$ ($A_{-}$) is the positive (negative) region of the spatial filter defined as $F\left( | \vect{\phi} - \vect{\theta} |\right) > 0$ ($F\left( | \vect{\phi} - \vect{\theta} |\right) < 0$) at the given coordinate $\vect{\theta}$, and $S\left(\vect{\phi}\right)$ is the footprint map excluding the masks.

In practice, we use a fast Fourier transform to calculate the convolution in equation~(\ref{eq:richness_map}) in grids of $0.25\arcmin\times0.25\arcmin$. 
The footprint map $S\left(\vect{\theta}\right)$ is derived by dividing the HSC galaxy catalog into pixels on the sky using \texttt{healpix} with $\mathtt{nside} = 8192$, corresponding to a pixel resolution of $\approx0.43\arcmin$. 
We mask the pixels at $\vect{\theta}$ in the richness maps if $f_\mathrm{mask}\left(\vect{\phi}, \vect{\theta}\right) \leq 0.7$ for $\vect{\phi} \in A_{+}$ or $f_\mathrm{mask}\left(\vect{\phi}, \vect{\theta}\right) \leq 0.3$ for $\vect{\phi} \in A_{-}$.
We derive the richness-error maps $\delta \rich\left(\vect{\theta}, \redshift\right)$ by propagating the Poissonian uncertainties of all pixels in both regions, $A_{+}$ (cluster footprint) and $A_{-}$ (local background).

\begin{figure*}
\centering
\resizebox{0.9\textwidth}{!}{
\includegraphics[scale=1]{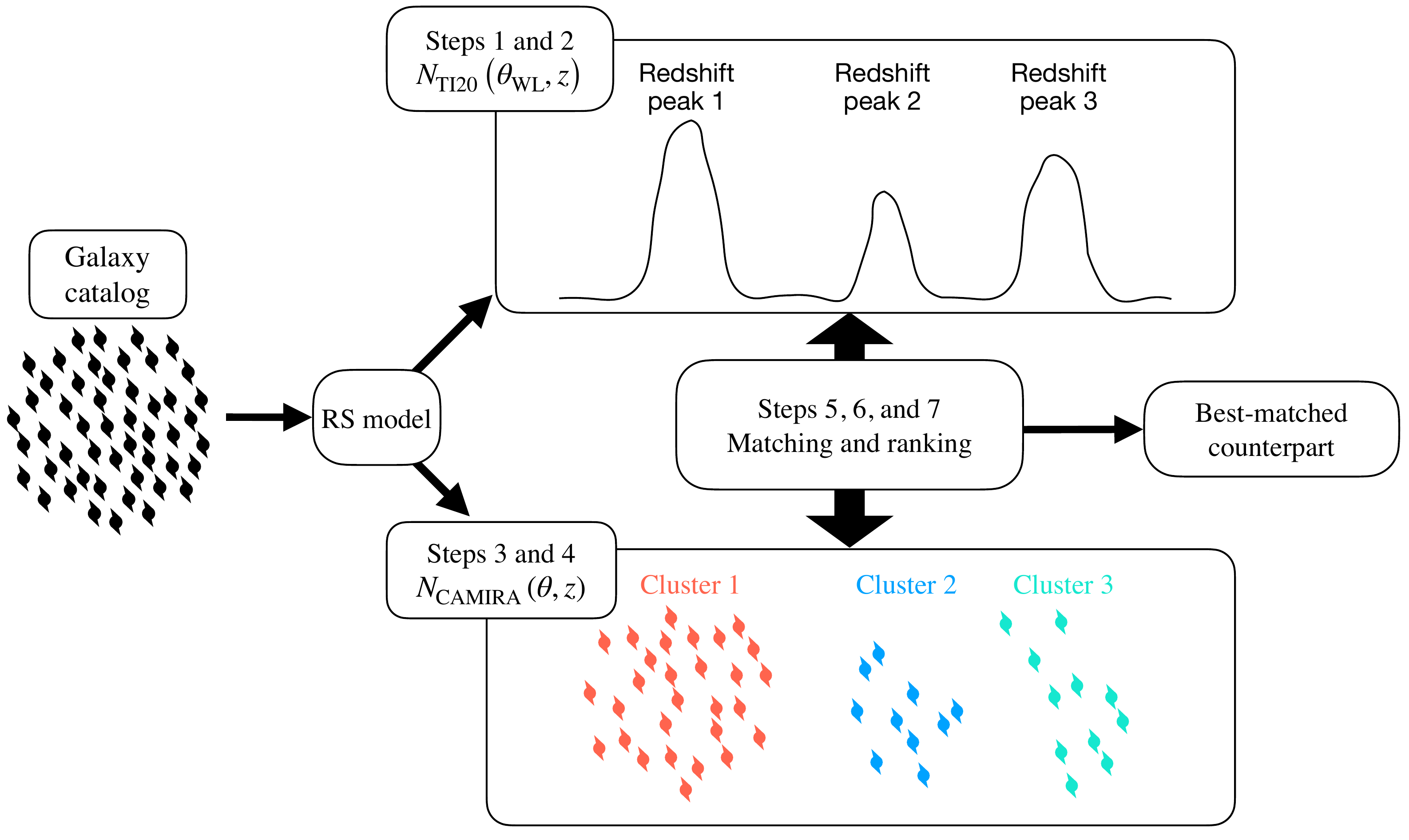}
}
\caption{The flowchart of the fCAMIRA algorithm (see details in Section~\ref{sec:fcamira}).
The number parameter $n\left(\redshift\right)$ as a function of redshift \redshift\ is inferred from the chi-square estimate $\chi^2\left(\redshift\right)$ using the RS fitting for each galaxy in the catalog, following the procedure in Section~\ref{sec:richness_map}.
The peak finding in the redshift distribution $N_{\mathrm{TI20}}\left(\mathbf{\theta}_{\mathrm{WL}}, \redshift \right)$ of the richness at the WL center $\mathbf{\theta}_{\mathrm{WL}}$ is illustrated in Steps 1 and 2.
The CAMIRA cluster finding in the richness map $N_{\mathrm{CAMIRA}}\left(\mathbf{\theta}, \redshift \right)$ and the resulting optically detected clusters are demonstrated in Steps 3 and 4.
The matching between the redshift peaks and optically detected CAMIRA clusters, the subsequent ranking based on individual lensing score $\mathcal{S}_{\mathrm{opt}}$, and the determination of the best-matched optical counterpart are carried out in Steps~5 to~7.
}
\label{fig:schematic}
\end{figure*}

\subsection{The fCAMIRA algorithm}
\label{sec:fcamira}

In this work, we aim to measure the redshifts of shear-selected clusters by locating their RS galaxy overdensities in the richness maps as optical counterparts.
There is a clear distinction between blind cluster finding and counterpart identification in richness maps.
Specifically, identifying the optical counterpart is effectively equivalent to carrying out the cluster finding algorithm in a forced mode at a given sky location, while keeping the same sample selection.
For this purpose, we develop the ``forced-mode CAMIRA'' (fCAMIRA) algorithm leveraging the existing CAMIRA cluster finder \citep{oguri14}.
In what follows, we describe the concept and the details of the fCAMIRA algorithm.

The shear-selected clusters are selected in the aperture-mass maps\footnote{Here, we assume a uniform shape noise so that the signal-to-noise-ratio maps $\snr\left(\vect{\theta}\right)$ are equivalent to the aperture-mass maps $\mkappa\left(\vect{\theta}\right)$.}, where the pixel values represent background-subtracted aperture masses enclosed within a fixed angular scale $\approx5\arcmin$.
The fixed angular scale of the TI20 filter corresponds to various physical sizes at different redshifts, e.g., 
$\approx0.4\Mpch$, $\approx0.9\Mpch$, $\approx1.4\Mpch$, and $\approx1.6\Mpch$ at $\redshift = 0.1$, $0.3$, $0.6$, and $0.9$, respectively.
Consequently, the aperture-mass peaks probe different physical scales of clusters, primarily depending on redshifts.
Moreover, the WL signals may arise from close mergers or projected pairs of physically unassociated halos along the line of sight.
As a result, direct matching between catalogs of shear-selected and optical clusters couples the two different selection functions and may lead to bias in counterpart identification.
Hence, it is of critical importance to confirm the shear-selected clusters in the richness maps obtained with the same filter.

The core concept of the fCAMIRA algorithm relies on an assumption that the galaxy richness of a halo scales linearly with the halo mass, i.e., $\rich\propto\mass$.
The linear relation between \rich\ and \mass\ can be understood from a self-similar picture of hierarchical structure formation, in which the number of member galaxies scales approximately with halo mass, neglecting satellite merging and tidal disruption.
The richness of CAMIRA clusters is evaluated within a fixed aperture size, $R\approx0.8\Mpch$, for which the observed richness-to-mass relation is well described by a power law of mass with an index broadly consistent with unity \citep{murata19,chiu20,nguyen-dang25}.
We do not expect such a linear relation over a wide range of redshift for the TI20 richness, because the corresponding physical aperture size varies with redshift.
By comparing the TI20 and CAMIRA richness estimates, we use the CAMIRA richness as a proxy for halo mass to rank the relative contributions of line-of-sight halos to the observed WL signal, and further determine the primary optical counterpart.

With the goal of a consistent assessment on both the WL signals and optical richness, we derive the richness maps using the TI20 filter, denoted as $\richti\left(\vect{\theta} , \redshift\right)$, representing the maps of enclosed galaxy numbers within an angular size $\theta \lesssim 5\arcmin$. 
Meanwhile, we also derive the richness maps $\richcamira\left(\vect{\theta} , \redshift\right)$ obtained with the CAMIRA filter, resulting in the maps of enclosed richness within a physical scale $\approx0.8\Mpch$. 

We use the \richti\ map to identify redshift peaks in the RS galaxy richness evaluated with the same WL kernel at the location of each shear-selected peak.
On the other hand, we use the \richcamira\ map to evaluate the optical properties of the halos associated with the identified redshift peaks within a consistent and physically motivated aperture.
We estimate the photometric redshifts of the shear-selected clusters by jointly utilizing both $\richti\left(\vect{\theta} , \redshift\right)$ and $\richcamira\left(\vect{\theta} , \redshift\right)$ through the following steps with a flowchart illustrated in Figure~\ref{fig:schematic}.

For each shear-selected cluster, we start from its peak location \wlcenter.
We proceed
\begin{enumerate}
\item \label{step:rich_distribution} Calculate the redshift distribution of the richness at \wlcenter\ in the map obtained with the TI20 filter, defined as ${\rich} \left(\redshift\right) \equiv \richti\left(\wlcenter, \redshift\right)$.
\item \label{step:z_wl} Find redshift peaks in $\rich\left(\redshift\right)$, resulting in a set of cluster redshift candidates $\left\lbrace {{\redshift}_{\mathrm{WL}}}_{i} | i = 1, 2, \cdots\right\rbrace$. 
\item \label{step:optical_rich} Find richness peaks in the map $\richcamira\left(\vect{\theta} , \redshift\right)$ within $7\arcmin$ of the WL peak, i.e., $|\skyloc - \wlcenter | < 7~\arcmin$. 
Label these peaks at positions
$\left\lbrace {\optcenter}_{j} | j = 1, 2, \cdots\right\rbrace$ 
with optical richness 
$\left\lbrace {\richopt}_{j} | j = 1, 2, \cdots\right\rbrace$ 
at redshifts 
$\left\lbrace {\redshift_{\mathrm{opt}}}_{j} | j = 1, 2, \cdots\right\rbrace$.
\item \label{step:lensing_score} Calculate the ``lensing scores'' $\mathcal{S}_{\mathrm{opt}}$ of the $j$-th optical-counterpart candidates identified in Step~\ref{step:optical_rich}, where
\begin{equation}
\label{eq:lensing_score}
{\mathcal{S}_{\mathrm{opt}}}_{j} = {\richopt}_{j}
\frac{
D_{\mathrm{A}} \left({\redshift_{\mathrm{opt}}}_{j}\right) D_{\mathrm{A}}\left({\redshift_{\mathrm{opt}}}_{j},\zs\right) 
}{ 
D_{\mathrm{A}} \left(\zs\right) 
}
\, ,
\end{equation}
in which $D_{\mathrm{A}}$ is the angular diameter distance, and we set $\zs = 1.3$ as the mean redshift of the WL sources.
\item \label{step:pairing} Select a single optical counterpart with the maximum lensing score $\mathcal{S}_{\mathrm{opt}}$ among candidates with $|\redshift_{\mathrm{opt}} - {{\redshift}_{\mathrm{WL}}}_{i} | < 0.1$ for each $i$-th redshift peak ${{\redshift}_{\mathrm{WL}}}_{i}$.
\item \label{step:ordering} Order the list of cluster redshift candidates $\left\lbrace {{\redshift}_{\mathrm{WL}}}_{i} \right\rbrace$ based on their lensing scores assigned in Step~\ref{step:pairing}.
\item \label{step:assign} Assign the highest-ranked candidate in the list as the most likely cluster redshift for the WL peak, corresponding to the ``best-matched'' optical counterpart at $\left(\optcenter, \redshift_{\mathrm{opt}}\right)$ with richness \richopt.
\end{enumerate}

We provide the following remarks for the above-mentioned steps.
In Step~\ref{step:z_wl} for searching for peaks in ${\rich}\left(\redshift\right)$, we require the richness of a peak to be larger than $5$ with a signal-to-noise ratio of $\rich/\delta\rich > 2$.
In addition, we restrict the peak candidates at redshifts below $1.3$, beyond which we cannot detect shear-selected clusters based on the current HSC WL data set.
The same criteria are adopted for finding peaks in the richness maps $\richcamira\left(\vect{\theta} , \redshift\right)$ in Step~\ref{step:optical_rich} in addition to the angular offset threshold of $7\arcmin$.
Beyond the scale of $7\arcmin$, the TI20 filter becomes negative so halos no longer contribute to positive WL signals, to first order.

If no redshift peak is found in ${\rich}\left(\redshift\right)$ in Step~\ref{step:z_wl}, we directly order the optical cluster candidates in Step~\ref{step:optical_rich} based on their lensing scores to assign the cluster redshift in Step~\ref{step:assign}.
If a redshift peak is found in Step~\ref{step:z_wl} but no optical counterpart candidates are obtained in Step~\ref{step:pairing}, we consider this redshift peak as a projection of line-of-sight halos that are too small to be found in Step~\ref{step:optical_rich} and do not assign a photometric redshift to it.

At this stage, a single optical counterpart is assigned to each redshift peak of the shear-selected clusters.
Based on the lensing scores, the highest-ranked redshift peaks of the individual shear-selected clusters are identified as the best-matched optical counterparts.
Following \cite{oguri14}, we further refine the redshift and richness estimates of the optical counterparts as follows.

We refine the photo-\redshift\ estimate of an optical counterpart at $\vect{\theta}_{\mathrm{opt}}$ by maximizing a weighted log-likelihood modified from equation~(\ref{eq:chi2_mini}), i.e., 
\[
-\frac{1}{2}
\sum\limits_{ \substack{ i\in \mathrm{galaxies~in}~A_{+} \\ k\in \mathrm{filters} } }
w_i
\left[
\left(
\frac{ 
\Delta_{i,k} - \delta\magnitude_{\mathrm{RS},k}
}{
\sigma_{\mathrm{comb},i,k}
}
\right)^2
+2
\ln\left( 
\sigma_{\mathrm{comb},i,k}
\right) 
\right]
\, ,
\]
where we only sum over galaxies in the region $A_{+}$ of the CAMIRA filter, and we include a per-galaxy weight $w_i$ defined in equation~(16) in \cite{oguri14} to effectively include only the top $50\percent$ galaxies contributing to the richness.
As a result, the best-fit redshift is referred to as the photometric redshift \zcl\ of the cluster optical counterpart.
The uncertainty of \zcl\ is estimated by a jackknife technique on the HSC galaxy catalog, followed by the identical redshift-refinement procedure.
In addition, the richness \rich\ of the cluster counterpart is refined by evaluating it at the photometric redshift \zcl.
As a result, we have the refined richness \rich\ at the refined redshift \zcl.

We further calculate the per-galaxy membership probability \wmem\ of a galaxy located at $\vect{\theta}$ as a member of a cluster at redshift \zcl, namely
\begin{equation}
\label{eq:wmem}
\wmem = n\left(\chi^2, \vect{\theta} \big| \zcl \right) 
w_{\magnitude}\left(\Delta\magnitude \right) F\left(|\vect{\theta} - \vect{\theta}_{\mathrm{opt}}| \right)
\, ,
\end{equation}
where $F$ is the CAMIRA filter, and $\vect{\theta}_{\mathrm{opt}}$ is the optical cluster center.
We assign \wmem\ only if $\vect{\theta} \in A_{+}$; otherwise, we set $\wmem = 0$.
With equation~(\ref{eq:wmem}), we note that a galaxy with a membership probability \wmem\ contributes an amount of $\wmem / f_\mathrm{mask}$ to the total richness, as seen in equation~(\ref{eq:richness_map}).

Below, we summarize our results.
For each shear-selected cluster, we identify the peaks in the redshift distribution of richness at the WL location \wlcenter\ in the richness map $\richti\left(\wlcenter\right)$.
Then, we pair the individual redshift peaks with the optical counterparts found in the richness map $\richcamira\left(\skyloc,\redshift\right)$ by ranking their lensing scores.
Next, we refine the redshift and richness of the optical counterparts.
Finally, the refined redshift and richness of the optical counterpart with the highest lensing score, referred to as the ``best match'', are adopted for the photo-\redshift\ and richness of the shear-selected clusters, denoted as \zcl\ and \rich, respectively.
By selecting the best-matched optical counterpart, we effectively associate each shear-selected cluster with a halo that dominates the WL signal along the line of sight, while contributions from other redshift peaks are treated as cosmic noise in the selection function modelling \citep{chen25}.

\begin{figure*}
\centering
\resizebox{0.95\textwidth}{!}{
\includegraphics[scale=1]{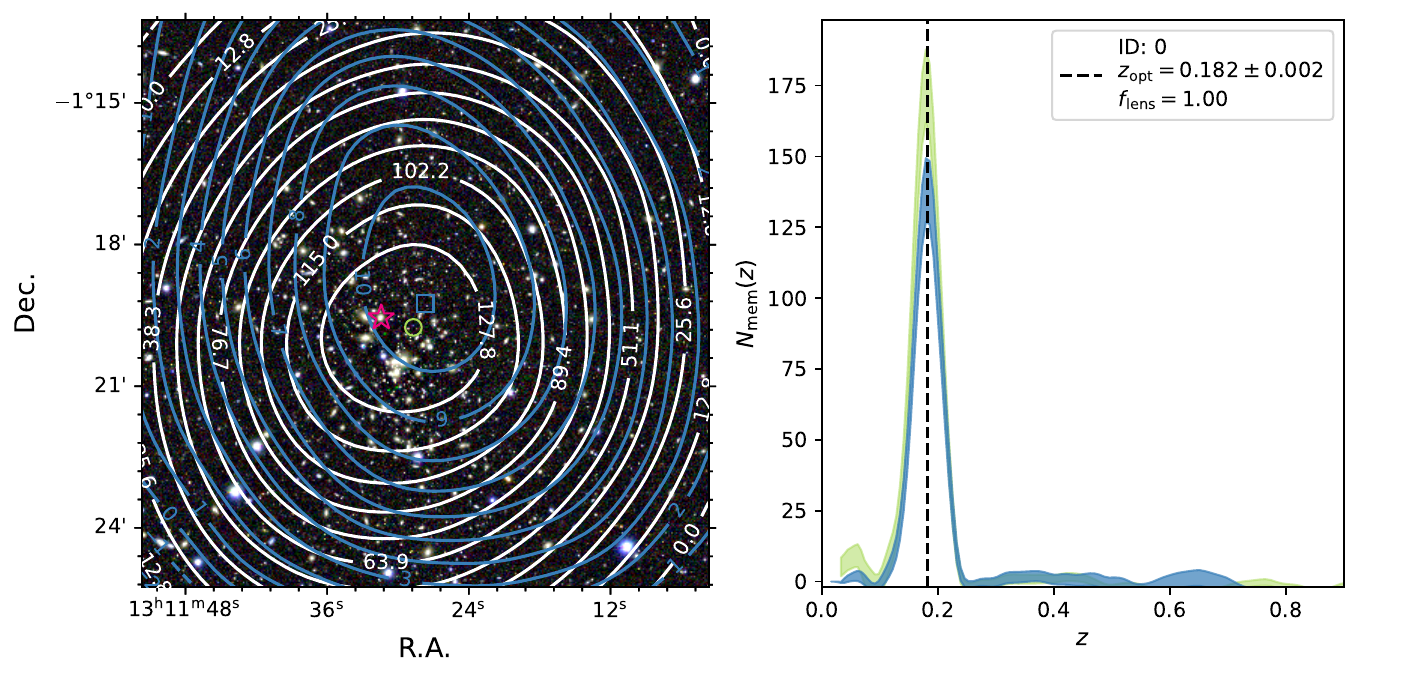}
}
\caption{
An unambiguous optical confirmation of a shear-selected cluster at $\zcl=0.182$.
\textit{Left panel}: The RGB image centered on the weak-lensing peak position \wlcenter\ (blue square), overlaid with contours of the WL signal-to-noise ratio map $\nu\left(\skyloc\right)$ in blue and the richness map $N_{\mathrm{TI20}}\left(\skyloc,\zcl\right)$ in white.
The green circle marks the optical center \optcenter, while the open star indicates the selected BCG.
\textit{Right panel}: The redshift distributions of richness measured at the WL center \wlcenter\ in the $N_{\mathrm{TI20}}$ map (blue) and at the optical center \optcenter\ in the $N_{\mathrm{CAMIRA}}$ map (green).
The single prominent peak suggests a secure optical counterpart with negligible projection effects, i.e., $\flens = 1$.
We define the quantity \flens\ in equation~(\ref{eq:flens}).
}
\label{fig:rgb_example}
\end{figure*}
%

%
%

\section{Results}
\label{sec:results}

We present the optical confirmation of the shear-selected clusters in this section, as one of the main results in this work.
By employing the fCAMIRA algorithm (see Section~\ref{sec:fcamira}) on the HSC photometric catalogs, we measure the photo-\redshift\ and richness of the optical counterparts associated with the $129$ WL shear-selected clusters.
For each WL shear-selected cluster, we record up to three most likely counterparts based on the ranking of their lensing scores.
The highest-ranked candidate is designated the ``best-matched'' optical counterpart, thereby assigning its photo-\redshift\ and richness estimates to the corresponding shear-selected cluster, denoted as \zcl\ and \rich, respectively.
For simplicity, we refer to the galaxy overdensity peak identified in the richness map $\richti\left(\vect{\theta}, \zcl\right)$ as the optical center \optcenter, while denoting the WL peak location as the WL center \wlcenter.

As an example, we present a shear-selected cluster detected with a signal-to-noise ratio
$\snr = 10.38$
at a WL center 
$\vect{\theta}_{\mathrm{WL}} = \left(13^{\mathrm{h}}11^{\mathrm{m}}27.7010^{\mathrm{s}}, -01^{\circ}19\arcmin15.1536\arcsec\right)$ 
in Figure~\ref{fig:rgb_example}.
In the left panel, we show the optical image centered at $\vect{\theta}_{\mathrm{WL}}$ (the blue square), overlaid with the contours of the WL signal-to-noise ratios \snr\ (blue contours) and the galaxy richness (white contours).
We note that the both contours are obtained with the same TI20 filter in the signal-to-noise-ratio and richness maps, $\snr\left(\skyloc\right)$ and $\richti\left(\vect{\theta}, \zcl\right)$; however, 
the former represents the overdensity of the total mass projected along the line of sight, while the latter displays a view of the RS galaxy overdensity at the cluster redshift \zcl.
As seen, the consistencies between the WL and richness contours and between \wlcenter\ (the blue square) and \optcenter\ (the green circle) suggest that the shear-selected cluster is confirmed with high confidence.
The same picture is seen in the right panel, where we show the redshift distributions of the richness 
at \wlcenter\ in the \richti\ map (blue) and at \optcenter\ in the \richcamira\ map (green), both extracted at the cluster redshift \zcl.
The shear-selected cluster shows the strong galaxy overdensity at the redshift $\zcl = 0.182 \pm 0.002$.
The single-peaked redshift distributions of both $\richti\left(\wlcenter,\redshift\right)$ and $\richcamira\left(\optcenter,\redshift\right)$ imply that the cluster does not suffer from projection effects caused by line-of-sight structures, leading to an unambiguous confirmation.
It is worth noting that the richness estimated in the \richcamira\ map (green) is higher than that in the \richti\ map (blue).
This is expected, because the matched filter $F_{\mathrm{CAMIRA}}$ with a scale of $\approx0.8\Mpch$ is larger than that $F_{\mathrm{TI}}$ at the cluster redshift $\redshift\approx0.2$.

\begin{figure*}
\centering
\resizebox{0.95\textwidth}{!}{
\includegraphics[scale=1]{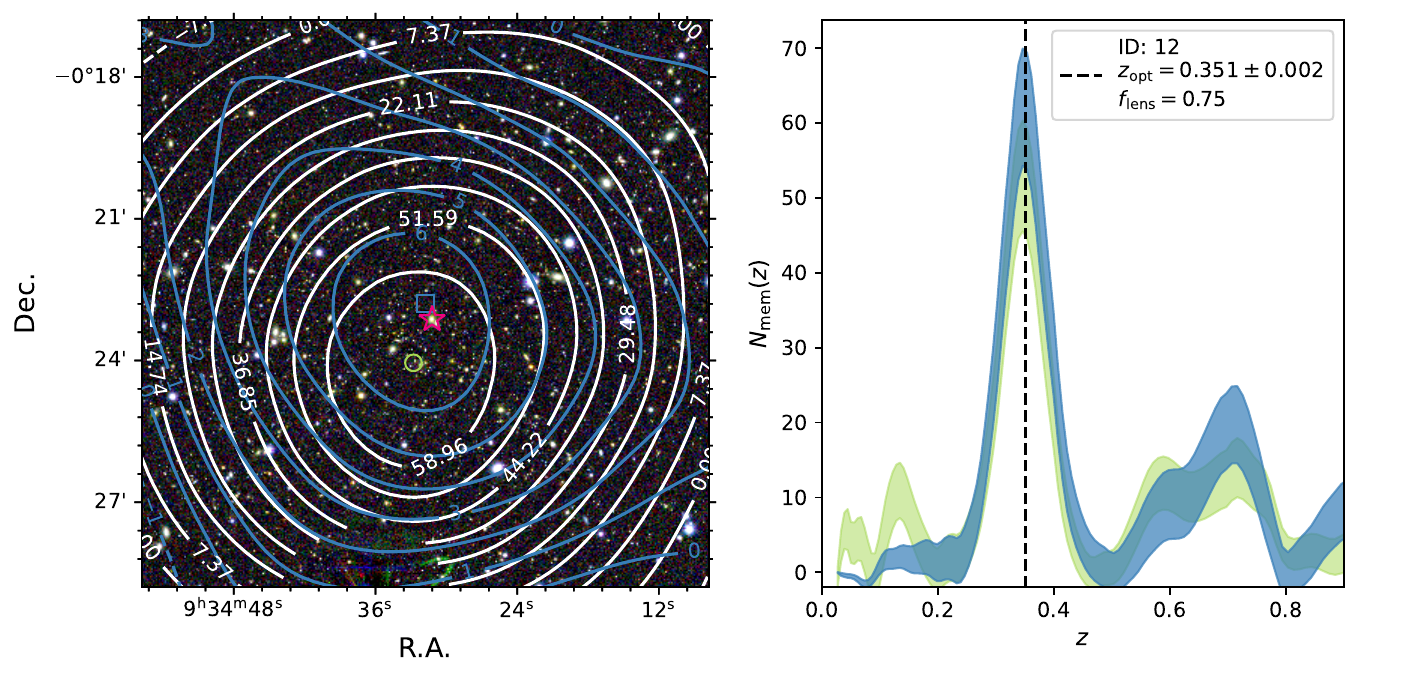}
}
\caption{
An optical confirmation of a shear-selected cluster at $\zcl\approx0.35$ with projection effects quantified as $\flens \approx 0.75$.
The same plotting style is used as in Figure~\ref{fig:rgb_example}.
}
\label{fig:rgb_example2}
\end{figure*}

In Figure~\ref{fig:rgb_example2}, we show another example of a WL shear-selected cluster with 
$\snr = 6.9$
at 
$\vect{\theta}_{\mathrm{WL}} = \left(09^{\mathrm{h}}34^{\mathrm{m}}31.7832^{\mathrm{s}},-00^{\circ}22\arcmin47.892\arcsec\right)$.
We leave the optical images of other shear-selected clusters in Appendix~\ref{app:optical_images}.
In the right panel, we find two peaks in the redshift distribution (blue shaded region) of the richness at the WL center \wlcenter\ in the \richti\ map, one at $\redshift \approx 0.35$ and the other at $\redshift\approx0.7$.
A consistent picture is indicated by the redshift distribution at the optical center \optcenter\ in the matched-filter \richcamira\ map, as shown by the green shaded curve.
We choose the halo at $\redshift\approx0.35$ as the best-matched optical counterpart of the shear-selected cluster, because its lensing score is higher than the other.
This demonstrates that (1) the WL signal of the shear-selected cluster is affected by the projection effect from a low-mass halo at $\redshift\approx0.7$, and that (2) the algorithm correctly attributes the optical counterpart to the dominant halo.
Again, we note that the difference in the richness estimates 
primarily reflects the filter choice.
Specifically, the richness in the \richti\ map is estimated with the TI20 filter with a size of $\theta \lesssim 5\arcmin$, which is larger than the size of the CAMIRA filter, $0.8\Mpch$, at redshift beyond $\approx0.25$.
This explains the richness estimated in the \richti\ map (blue) is larger than that in the \richcamira\ map (green) for the two halos at $\approx0.35$ and $\approx0.7$.

The lensing score $\mathcal{S}$ defined in equation~(\ref{eq:lensing_score}) quantifies the lensing strength of a halo at redshift \redshift\ for sources at redshift \zs, to first order.
Motivated by this, we approximate the fractional contribution \flens\ of the lensing signal from the best-matched counterpart to the observed WL \snr\ of a shear-selected cluster as its lensing score divided by the sum of the scores of the top three counterparts.
That is,
\begin{equation}
\label{eq:flens}
\flens = \frac{ \mathcal{S}_{1} }{ \mathcal{S}_1 + \mathcal{S}_2 + \mathcal{S}_3 } \, ,
\end{equation}
where the index $i$ denotes the ranking with $\mathcal{S}_1$ as the lensing score of the best-matched counterpart.
For peaks without the second and third ranked optical counterparts, we set $\mathcal{S}_2 = 0$ and $\mathcal{S}_3 = 0$, respectively.
By construction, an unambiguous optical confirmation of a shear-selected cluster leads to $\flens = 1$, as demonstrated in Figure~\ref{fig:rgb_example}.
In addition, a higher \flens\ implies smaller projection effects caused by line-of-sight halos and, hence, greater confidence in identifying the correct optical counterpart.

\begin{figure}
\centering
\resizebox{0.5\textwidth}{!}{
\includegraphics[scale=1]{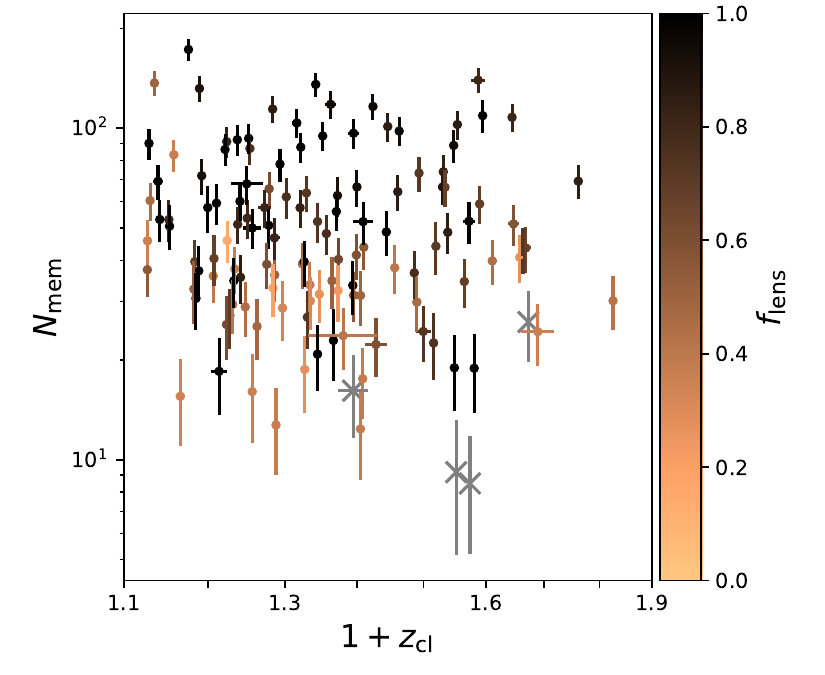}
}
\caption{
The distribution of the photometric redshifts \zcl\ and optical richnesses \rich\ of the $129$ shear-selected clusters.
The cluster redshifts \zcl\ include the correction for the bias (see Section~\ref{sec:results}).
The color scale indicates the lensing fraction \flens\ defined in equation~(\ref{eq:flens}), which approximates the fractional contribution of the best-matched optical counterpart to the observed WL signal.
Clusters with higher richness generally exhibit larger values of \flens, indicating weaker projection effects and a larger contribution from a single dominant halo (see discussion in the end of Section~\ref{sec:cosmology}).
The shear-selected clusters without redshift peaks identified at the WL center in the richness map $N\left(\skyloc, \redshift\right)$, due to heavy star masks, are indicated by the grey crosses.
}
\label{fig:rich_z}
\end{figure}

We present the distribution of the cluster photometric redshifts \zcl\ and richness \rich\ with their \flens\ indicated by colors in Figure~\ref{fig:rich_z}.
As seen, the optical richness \rich\ correlates positively with the lensing score \flens, indicating that the WL signals of the high-\snr\ shear-selected clusters are primarily produced by single massive halos.
In this work, we find that the majority ($\approx90$ systems or $\approx70\percent$) of the shear-selected clusters have $\flens \geq 0.5$, among which $\gtrsim40$ clusters are confirmed with $\flens \geq 0.9$.
Approximately $20$ clusters (corresponding to $16\percent$) have the best-matched counterparts with $\flens \leq 0.4$, with a minimum value of $\approx0.15$.
Only $4$ clusters (corresponding to $\approx3\percent$) do not find redshift peaks at the WL centers in their \richti\ maps, as their richness maps are heavily masked due to bright stars.

We stress that the quantity \flens\ approximates the expected contribution of the identified optical counterpart to the observed lensing signal; however, it does not represent a direct decomposition of that signal.

\subsection{The assessment of the cluster photometric redshifts}
\label{sec:assessment}

\begin{figure}
\centering
\resizebox{0.48\textwidth}{!}{
\includegraphics[scale=1]{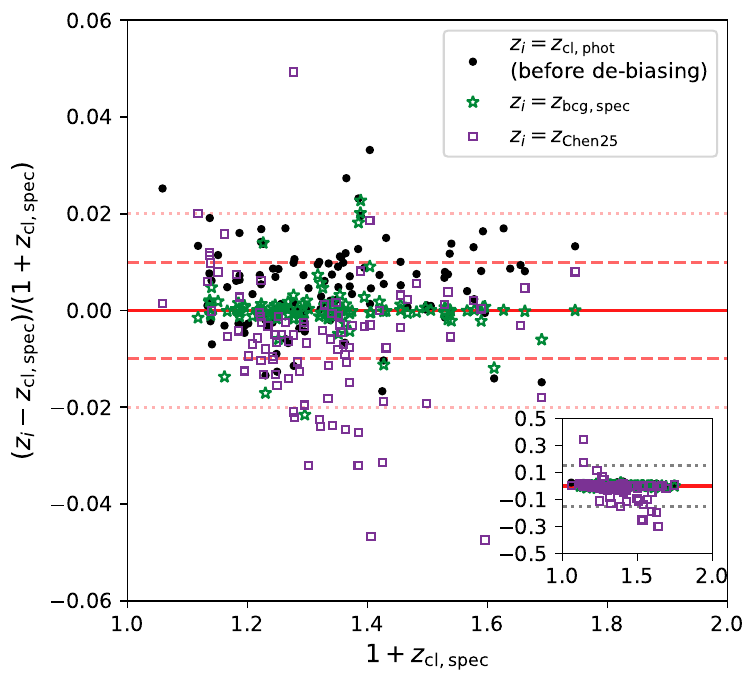}
}
\caption{
The assessment of the cluster photometric redshifts.
The black points show the comparison between the cluster photometric redshifts $\redshift_{\mathrm{cl, phot}}$ and the spectroscopic cluster redshifts $\redshift_{\mathrm{cl, spec}}$.
We find a mild bias at a level of $\left(\redshift_{\mathrm{cl, phot}} - \redshift_{\mathrm{cl, spec}} \right) / \left(1 + \redshift_{\mathrm{cl, spec}} \right) \approx 0.0047$ and correct for it in the final cluster photo-\redshift\ estimates.
The green stars compare $\redshift_{\mathrm{cl, spec}}$ and the spectroscopic redshifts $\redshift_{\mathrm{cl, bcg}}$ of the selected BCGs.
The magenta squares correspond to the comparison between $\redshift_{\mathrm{cl, spec}}$ and the redshift estimates $\redshift_{\mathrm{Chen25}}$ of the optical counterparts that obtained in \cite{chen25} by a positional cross-matching with external cluster catalogs.
The red solid, dashed, and dotted lines indicate the thresholds of $0$, $\pm0.01$ and $\pm0.02$, respectively.
The subplot embedded in the lower-right corner displays the same plot but with a wide dynamical range in the $y$-axis, suggesting an outlier fraction at a level of $\approx10\percent$ for the comparison between  $\redshift_{\mathrm{cl, spec}}$ and $\redshift_{\mathrm{Chen25}}$.
}
\label{fig:assessment_z}
\end{figure}

We assess the cluster photo-\redshift\ estimates using the spectroscopic galaxy sample, as follows.
First, we obtain the member galaxy candidates of each individual shear-selected cluster by requiring the membership probability of $\wmem > 0.1$.
Next, we match them to the spectroscopic samples (as collected in Section~\ref{sec:specz_calib}) to acquire their spectroscopic redshifts \specz, using a matching radius of $1\arcsec$.
We note that not all photometrically identified member galaxy candidates have spectroscopic counterparts, as the spectroscopic samples are not complete.
The mean (median) number of matched spectroscopic member galaxies with $\wmem > 0.1$ per shear-selected cluster is $\approx15$ ($\approx8$) with a maximum of $\approx190$.
We find $106$ out of the $129$ shear-selected clusters with the number of available spectroscopic member galaxies greater than three.
With \specz, we estimate the spectroscopic redshift $\redshift_{\mathrm{cl},\mathrm{spec}}$ of the clusters using the bi-weighted location estimator \citep[][]{beers1990}.
In practice, we use equations~(4) and (5) in \cite{kluge24} to estimate $\redshift_{\mathrm{cl},\mathrm{spec}}$ after performing a $3\sigma$ clipping on the per-cluster spectroscopic sample to remove the interlopers.
On the other hand, we obtain the redshift estimates labelled as $\redshift_{\mathrm{Chen25}}$ from \cite{chen25}, in which the redshift estimates are derived by a positional cross-matching between the shear-selected clusters and external cluster catalogs (see Section~\ref{sec:cluster}).
Finally, we compare the cluster photometric redshifts $\redshift_{\mathrm{cl},\mathrm{phot}}$  obtained using the fCAMIRA algorithm with $\redshift_{\mathrm{cl},\mathrm{spec}}$ and $\redshift_{\mathrm{Chen25}}$.
Figure~\ref{fig:assessment_z} shows the comparisons, and we discuss them below.

\begin{figure}
\centering
\resizebox{0.5\textwidth}{!}{
\includegraphics[scale=1]{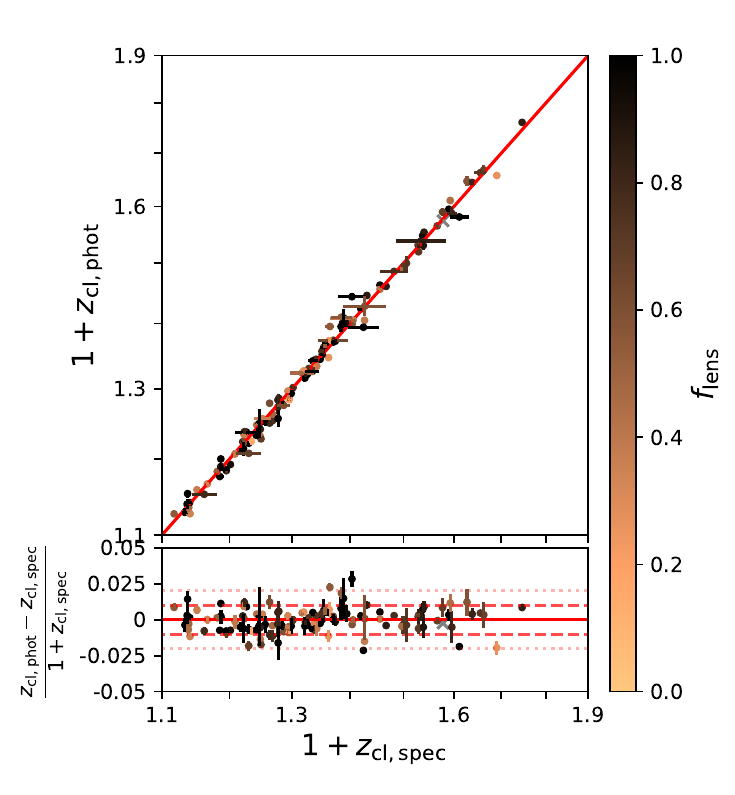}
}
\caption{
The comparison between the photometric redshifts $\redshift_{\mathrm{cl, phot}}$ obtained with the fCAMIRA algorithm and the spectroscopic cluster redshifts  $\redshift_{\mathrm{cl, spec}}$.
The cluster photometric redshifts $\redshift_{\mathrm{cl, phot}}$ include the correction for the bias (see Section~\ref{sec:results}).
The data points are colored using the same style as in Figure~\ref{fig:rich_z}.
\textit{Top panel}: 
The scatter between $1 + \redshift_{\mathrm{cl, phot}}$ and $1 + \redshift_{\mathrm{cl, spec}}$.
\textit{Bottom panel}: 
The redshift residuals $\left(\redshift_{\mathrm{cl, phot}} - \redshift_{\mathrm{cl, spec}}\right) / \left(1 + \redshift_{\mathrm{cl, spec}} \right)$ as a function of $1 + \redshift_{\mathrm{cl, spec}}$.
The red dashed and dotted lines indicate the thresholds of $\pm0.01$ and $\pm0.02$, respectively.
}
\label{fig:photo_specz}
\end{figure}

First, the black points in Figure~\ref{fig:assessment_z} present the comparisons between the cluster photometric redshifts ($\redshift_{\mathrm{cl},\mathrm{phot}}$) and spectroscopic estimates ($\redshift_{\mathrm{cl},\mathrm{spec}}$).
We find a mild mean bias at a level of $\left\langle\left(\redshift_{\mathrm{cl},\mathrm{phot}} - \redshift_{\mathrm{cl},\mathrm{spec}}\right) / \left(1 + \redshift_{\mathrm{cl},\mathrm{spec}}\right) \right\rangle\approx 0.0047\pm0.0008$ without significant dependence on the cluster redshift.
This suggests that the cluster photo-\redshift\ estimates achieve sub-percent accuracy (the red dashed line).
Nevertheless, we account for the bias in the final photo-\redshift\ estimates\footnote{The correction is applied to the cluster photometric redshift as $\photz \rightarrow \frac{\photz - \delta_\redshift}{1 + \delta_\redshift}$, where $\delta_\redshift = 0.0047$.} and present them in Figure~\ref{fig:photo_specz}.
We calculate the scatter of $\redshift_{\mathrm{cl},\mathrm{phot}}$ with respect to the spectroscopic redshifts $\redshift_{\mathrm{cl},\mathrm{spec}}$, which is found to be at a level of $\lesssim 0.008$, i.e., sub-percent precision.
We demonstrate that the precision and accuracy of the redshift estimates at such levels have negligible impact on cosmological constraints in Section~\ref{sec:cosmology}.

Second, the green stars in Figure~\ref{fig:assessment_z} compare the spectroscopic redshifts of the clusters ($\redshift_{\mathrm{cl},\mathrm{spec}}$) and the BCGs ($\redshift_{\mathrm{bcg},\mathrm{spec}}$).
We note that the BCGs are photometrically identified following the prescription in Appendix~\ref{app:bcg} and subsequently matched to the spectroscopic samples to acquire $\redshift_{\mathrm{bcg},\mathrm{spec}}$.
As seen, we find excellent agreement between $\redshift_{\mathrm{cl},\mathrm{spec}}$ and $\redshift_{\mathrm{bcg},\mathrm{spec}}$.
This indicates that the cluster member identification, which leads to the estimates of $\redshift_{\mathrm{cl},\mathrm{spec}}$, is self-consistent with the BCG identification.
It is worth mentioning that this comparison is based almost entirely on spectroscopic measurements, except for the intermediate step of identifying cluster members photometrically.
The excellent agreement is expected, because the cluster member identification relies on the RS model that is calibrated using the spectroscopic samples, which include bright galaxies as the BCGs.
Combining the comparison between $\redshift_{\mathrm{cl},\mathrm{phot}}$ and $\redshift_{\mathrm{cl},\mathrm{spec}}$, these results suggest that the mild bias at a level of $\approx0.0047$ may arise from extrapolating the RS calibration to the faint end or color spaces of the galaxy population where the spectroscopic samples are incomplete.
Future surveys including more complete spectroscopic coverage may further reduce this bias.

Finally, we compare the cluster spectroscopic redshifts $\redshift_{\mathrm{cl},\mathrm{spec}}$ and the estimates $\redshift_{\mathrm{Chen25}}$ \citep{chen25}.
We observe much larger scatter between $\redshift_{\mathrm{cl},\mathrm{spec}}$ and $\redshift_{\mathrm{Chen25}}$ broadly at a level of $\lesssim 3\percent$ with a hint of redshift-dependent bias.
The suggests that our fCAMIRA algorithm and the direct positional matching find the same optical counterparts with percent-level systematics in the redshift estimates.
The discrepancy may come from the systematics between the redshift estimate algorithms used in the external cluster catalogs, which are built using different cluster finders, or the fact that the different versions of photometric data sets and spectroscopic calibration samples are used between CAMIRA and fCAMIRA algorithms.
It is challenging to locate the cause of the redshift discrepancy, as cross-matching between cluster catalogs is an extremely non-linear process, again highlighting the importance of performing a uniform optical confirmation as in this work.
Meanwhile, as seen in the embedded plot in Figure~\ref{fig:assessment_z}, we find about $\approx 6.7 \percent$ of the shear-selected clusters have the redshift outliers, defined as $\left(\redshift_{\mathrm{cl},\mathrm{spec}} - \redshift_{\mathrm{Chen25}}\right) / \left(1 + \redshift_{\mathrm{cl},\mathrm{spec}}\right) > 0.15$.
These redshift outliers suggest that the fCAMIRA algorithm and the positional matching identify different optical counterparts for a shear-selected cluster, as evidence of projection effects along the line of sight.
The outlier fraction reaches $\approx13\percent$ and $\approx15\percent$ when defining the outlier thresholds as $0.10$ and $0.08$, respectively.

It is noteworthy that the redshift outliers may arise from physically correlated structures projected along the line of sight, in which case this effect would likely depend on the mass and/or redshift of the primary halos rather than arising from random projections.
In this work, we treat the outlier fraction as an assessment of the overall photometric redshift performance without investigating its potential dependence on other properties.
Additionally, we demonstrate in Section~\ref{sec:cosmology} that an outlier fraction at this level has a negligible impact on the final cosmological constraints. 

From the comparisons, we identify three sources of systematics in the cluster redshift estimation.
First is the bias in the cluster photometric redshift $\redshift_{\mathrm{cl},\mathrm{phot}}$, which is quantified to be at a level of $\approx0.004$ and accounted for in this work.
Second is the scatter in the estimates of $\redshift_{\mathrm{cl},\mathrm{phot}}$ with respect to the spectroscopic redshifts $\redshift_{\mathrm{cl},\mathrm{spec}}$, which is found to be at a level of $\lesssim 0.008$.
Third comes from redshift outliers, which arise from projection effects leading to the mis-identification of optical counterparts in the light-of-sight direction.
In this work, the shear-selected cluster sample suggests a redshift outlier fraction at levels of $6\percent$ to $15\percent$, depending on the outlier definition.
We assess the impacts of the cluster redshift systematics on the constraints of cosmological parameters in Section~\ref{sec:cosmology}.

%
%

\section{Cosmological impact}
\label{sec:cosmology}

In this section, we examine the impact of the shear-selected clusters' redshifts on cosmological constraints.
To this end, we generate catalogs of mock shear-selected clusters using the WL selection as the same as in observations with realistic noise, and then run the cluster-abundance analysis on them to quantify the cosmological impact.
Below, we describe the construction of the mock cluster catalogs.

We randomly sample halos with a wide mass range of $5\times10^{12} < \frac{\mass}{\Msunh} < 5\times 10^{16}$ at redshift $0.01 < \redshift < 2$, following the 
halo mass function from \cite{bocquet16}.
We include the Poisson noise when sampling the halos.
For each halo with mass \mass\ at a redshift \redshift, we sample its WL mass \mwl\ including the WL mass bias, defined as $\bwl = \mwl/\mass$, and intrinsic scatter.
This is achieved by using the functional form defined in equation~(30) in \cite{chiu24}, namely
\begin{multline}
\left\langle\ln\bwl|\mass,\redshift\right\rangle = \ln\Awl + \\
\Bwl\ln\left(\frac{\mwl}{4\times10^{14}\Msunh}\right) + 
\gammawl\ln\left(\frac{1 + \redshift}{1 + 0.45}\right)
\, , \nonumber
\end{multline}
with a constant log-normal intrinsic scatter \sigmawl.
For simplicity, we set $\left(\Awl, \Bwl, \gammawl, \sigmawl\right) = \left(1, 0.05, -0.1, 0.25\right)$, which is in broad agreement with the calibration results for X-ray-selected clusters \citep{grandis24}.
With the sampled \mwl, we calculate the projected surface mass profile using a spherical NFW model, assuming that the halo concentration \ctwooo\ follows a log-normal distribution with a mean value predicted by the \cite{diemer15} concentration-to-mass relation and a scatter of $0.3$.
We then derive the aperture mass by convolving the projected surface mass profile by the TI20 filter. 
We approximate the source redshift distribution by a skewed normal distribution $P\left(\redshift\right)$ to mimic that of the observed sources at $\left\langle\redshift\right\rangle\approx1.3$ with a high-end tail.
For each sampled halo, we compute the $P\left(\redshift\right)$-weighted critical surface mass density, which allows us to derive the dimensionless aperture aperture mass peak \mkappahat.
We do not include miscentering, i.e., the aperture mass peak \mkappahat\ is evaluated at the halo center.
To account for the WL shape noise, the observed aperture mass peak \mkappa\ is obtained by a Gaussian random sampling with the mean \mkappahat\ and the variance in the observed HSC mass maps, $\sigma_{\kappa}^2 = 0.7^2$.
The WL signal-to-noise ratios of mock halos are derived as $\snr = \mkappa / \sigma_{\kappa}$.
Finally, we adopt the selection of $\nu > \numin$ to construct the mock sample of shear-selected clusters, with $\numin = 4.7$ as in this work.

\begin{figure*}
\centering
\resizebox{0.48\textwidth}{!}{
\includegraphics[scale=1]{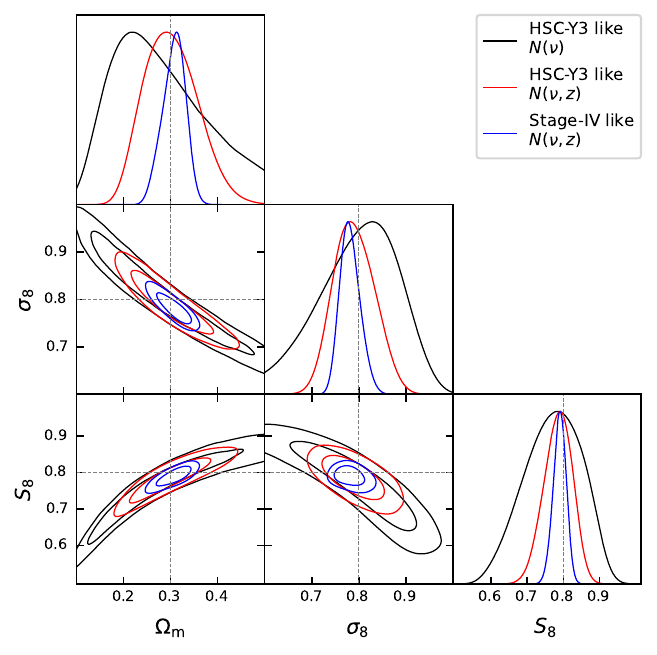}
}
\resizebox{0.48\textwidth}{!}{
\includegraphics[scale=1]{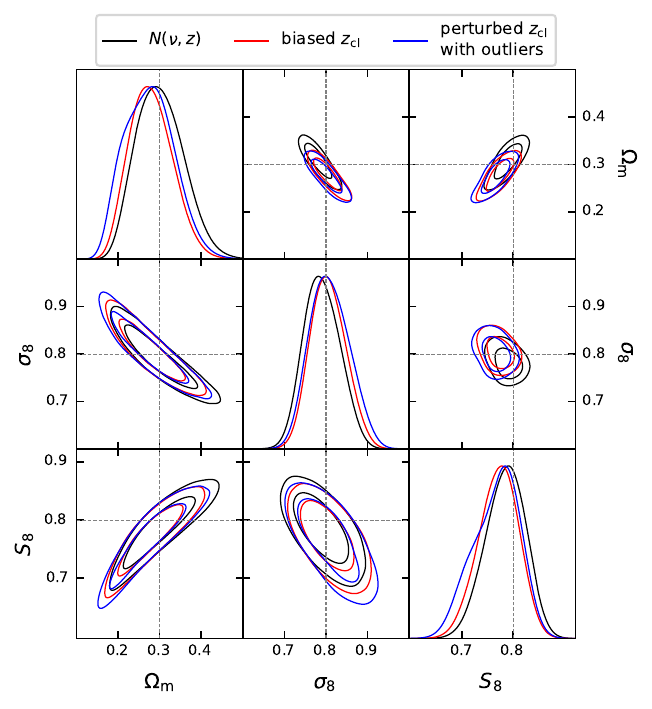}
}
\caption{
The forecast cosmological constraints from the abundance modelling of mock shear-selected clusters.
\textit{Left panel}:
The comparisons of the parameter constraints obtained with the HSC-Y3 like sample ($N_{\mathrm{cl}}\approx220$) between the modelling $N\left(\nu\right)$ without the cluster redshift in black and $N\left(\nu,\redshift\right)$ including the cluster redshift in red.
The results of the modelling $N\left(\nu,\redshift\right)$ for the Stage-IV like sample are in blue.
Including the cluster redshift modelling significantly improves the parameter constraints on \omegam, \sigmaeight, and $\seight \equiv \sigmaeight\sqrt{\omegam/0.3}$.
\textit{Right panel}:
The impact of the cluster redshift systematics on the cosmological constraints.
The black contours show the constraints obtained with the true cluster redshifts.
The red and blue colors indicate the constraints obtained in the scenarios of ``biased \zcl'' and ``perturbed \zcl\ with outliers'', respectively.
The case of  ``biased \zcl'' refers to a constant bias in the cluster redshift as $\zcl - \redshift_{\mathrm{true}} = 0.01$, while the other case ``perturbed \zcl\ with outliers'' includes a Gaussian redshift scatter and an outlier fraction of $8\percent$.
The lower-left and diagonal subplots show the parameter covariance and marginalized posteriors for the HSC-Y3 like sample ($N_{\mathrm{cl}}\approx220$), respectively.
The results obtained with the Stage-IV like sample are contained in the upper-right subplots.
Our results demonstrate that the cosmological analysis using an HSC-Y3 like sample is robust against realistic levels of cluster photo-$z$ systematics.
}
\label{fig:forecast}
\end{figure*}

The modelling of the shear-selected cluster abundance is fully described in \cite{chiu24}, to which readers are referred for further information.
In short, we model the observed cluster numbers in the parameter space of \snr\ and \redshift\ using a Poisson likelihood, in which the cluster number distribution $N\left(\snr,\redshift\right)$ is parameterized as
\begin{multline}
\label{eq:diff_n_with_selection}
\frac{ \dif N\left(\snr , \redshift | \vect{p} \right) }{ \dif\snr~\dif\redshift }  = 
 \int\dif\mass~
 \frac{ \dif N\left(\mass,\redshift | \vect{p}\right) }{ \dif\mass\dif\redshift }   \times \\
\bigg[ \int\dif\mkappahat\int\dif\thetas~ 
\comp\left(\mkappahat,\thetas\right) P\left(\snr | \snr > \numin, \mkappahat,\thetas  \right)  \\
P\left(\mkappahat,\thetas | \mass, \redshift, \vect{p} \right)
\bigg] 
\, ,
\end{multline}
where $\frac{ \dif N\left(\mass,\redshift | \vect{p}\right) }{ \dif\mass\dif\redshift }$ is the product of the halo mass function and the survey volume, the variable \thetas\ corresponds to the angular scale of the core radius and is related to the halo concentration as $\thetas = \frac{\Rtwooo}{\ctwooo D_{\mathrm{A}}\left(\redshift\right)}$, and $\vect{p}$ is the parameter vector.
In equation~(\ref{eq:diff_n_with_selection}), we account for the intrinsic scatter in the WL mass \mwl\ and the halo concentration in the probability $P\left(\mkappahat,\thetas | \mass, \redshift, \vect{p} \right)$.
Meanwhile, the probability $P\left(\snr | \snr > \numin, \mkappahat,\thetas  \right)$ takes into account the measurement uncertainty in the observed \snr\ with the presence of the selection $\nu > \numin$.
The factor $\comp\left(\mkappahat,\thetas\right)$ describes the sample completeness as a function of \mkappahat\ and \thetas.
For our mock clusters, which assumes Gaussian shape noise $\sigma_{\kappa}$, the sample completeness depends only on \mkappahat\ and can be evaluated as the cumulative probability of a unit Gaussian distribution with mean of $\mkappahat / \sigma_{\kappa}$ above the detection threshold \numin.

We generate the mock shear-selected clusters in a flat \lcdm\ cosmological model with $\left(\omegam, \omegal, \sigmaeight, h \right) = \left(0.3, 0.7, 0.8, 0.7\right)$, as the underlying true cosmological parameters.
We perform the cluster abundance modelling and examine whether the code can recover the input cosmological parameters. 
The parameter likelihood is explored using the sampler of \texttt{multinest} \citep{feroz08,feroz09,feroz19} implemented in the \texttt{CosmoSIS} framework \citep{zuntz15}.

We generate two representative mock samples of shear-selected clusters by adjusting the survey footprint.
One contains $\approx220$ clusters, comparable to the sample size obtained from the HSC-Y3 data set \citep{oguri21}, while the other consists of $\approx2400$ clusters, corresponding to a sample size expected for future Stage IV surveys \citep{oguri26}. 
We produce $20$ realizations for each mock sample.
When presenting the results, we combine the posterior chains from all $20$ realizations with equal weights to obtain the ensemble-averaged posterior distributions, marginalizing sample-to-sample fluctuations.

We further examine the impact of cluster redshift estimation on cosmological constraints under two scenarios.
In the first scenario, referred to as ``biased \zcl'', we introduce a constant bias of $0.01$ to the mock cluster redshift (i.e., $\zcl \rightarrow \zcl + 0.01$), treat the biased values as the observed cluster redshifts in the modelling, and assess its impact on the parameter constraints.
In this scenario, we examine the impact of the cluster photo-\redshift\ accuracy.
We note that the cluster photo-\redshift\ bias at the level of $0.01$ is a conservative assumption, which is generally achieved by modern RS-based algorithms \citep[e.g.,][]{klein18}, including fCAMIRA in this work.
In the second scenario, we add a Gaussian scatter\footnote{The scatter is defined in terms of $\left(\zcl - \redshift\right) / \left(1 + \redshift\right)$, therefore the scatter weakly depends on the cluster redshift, i.e., $\zcl \rightarrow \zcl + \mathcal{N} \left( 0, 0.01^2 \times \left(1 + \zcl\right)^2 \right)$.} of $0.01$ to the mock cluster redshifts and additionally introduce redshift outliers by randomly shuffling the cluster redshifts.
We set the redshift outlier fraction to be $8\percent$.
This scenario is referred to as ``perturbed \zcl\ with outliers'', in which we assess the cosmological impact of the cluster photo-\redshift\ precision and the outlier fraction.
It is worth noting that redshift outliers may arise from projection effects in shear-selected clusters, in which the optical counterparts 
may be incorrectly associated with a line-of-sight halo. 
Considering that projection effects caused by uncorrelated structures are accounted for in the selection function modelling \citep{chen25}, the mock validation presented here provides a quantitative assessment of its impact on the final cosmological constraints.

Figure~\ref{fig:forecast} shows the results of the mock validations.
In the left panel, we present the forecast of the cosmological constraints obtained from the cluster abundance modelling of the HSC-Y3 like ($N_{\mathrm{cl}}\approx220$) shear-selected samples with or without the cluster redshifts in red and black, respectively.
As seen, the constraints are improved significantly by simply including the cluster redshift modelling.
Including the redshift modelling in $N\left(\nu,\redshift\right)$, we are able to put statistically meaningful constraints on both \omegam\ and \sigmaeight\ simultaneously by using only $N_{\mathrm{cl}}\approx220$ shear-selected clusters.
This result aligns with \cite{chiu24}, where they could only constrain \seight\ by modelling $N\left(\nu\right)$.
The HSC-Y3 like sample suggests the posterior uncertainties of \omegam, \sigmaeight, and \seight\ at levels of $\approx0.06$, $\approx0.04$, and $\approx0.04$, respectively.
Increasing the sample size to $N_{\mathrm{cl}}\approx2400$ (Stage-IV like; the blue contours), we expect uncertainties at levels of $\approx0.02$, $\approx0.02$, and $\approx0.02$ for \omegam, \sigmaeight, and \seight, respectively.

In the right panel of Figure~\ref{fig:forecast}, we quantify the impacts of the cluster redshift systematics on the cosmological constraints.
The lower-left (diagonal) subplots show the parameter covariance (posteriors) when modelling $N\left(\nu,\redshift\right)$ of the HSC-Y3 like sample, where the constraints obtained with the correct cluster redshifts are in black.
Meanwhile, the results obtained with the cluster redshifts in the scenario of ``biased \zcl'' and ``perturbed \zcl\ with outliers'' are in red and blue, respectively.
As seen, the cluster redshift systematics at such levels has negligible impact on the final cosmological constraints with $N_{\mathrm{cl}}\approx220$.
On the other hand, we present the forecast results with the same colors in the upper-right subplots when increasing the mock sample size to $N_{\mathrm{cl}}\approx2400$ (Stage-IV like).
We find that both cases of ``biased \zcl'' and ``perturbed \zcl\ with outliers'' lead to shifts at $\approx1\sigma$ in the parameter contours with respect to those using the correct cluster redshifts (black). 
Our results suggest that the fCAMIRA algorithm developed in this work provides cluster redshift estimates sufficiently accurate and precise for cosmological modelling with a sample size $N_{\mathrm{cl}}\approx10^2$.
However, the cluster redshift estimates must be further improved for analyses using samples with $N_{\mathrm{cl}}\gtrsim10^3$, otherwise introducing bias comparable to the statistical uncertainty.

It is worth mentioning that these mock validations are carried out with perfect knowledge of the selection function $\comp\left(\mkappahat,\thetas\right)$ in equation~(\ref{eq:diff_n_with_selection}) and therefore represent optimistic forecasts.
In practice, the selection function modelling remains a challenging task in shear-selected cluster cosmology, due to observational systematics such as inhomogeneous survey depth, bright-star masks, and especially projection effects.
To this end, \cite{chen25} developed a data-driven and injection-based method to model the selection function, which enabled the cluster abundance modelling of $N\left(\nu\right)$ in \cite{chiu24}. 
We will present an updated modelling of $N\left(\nu, \redshift\right)$ in a forthcoming paper.

We provide a final remark on the projection effects.
For optical cluster surveys, projection effects generally refer to the preferential selection of clusters whose halo major axes are aligned close to the line of sight, as well as richness contamination from correlated structures projected along the same line of sight.
The projection effects in optically selected clusters primarily arise from the imprecise determination of distances inferred from photometric redshifts, thereby leading to biases in the sample selection, richness estimates, and weak-lensing mass calibrations \citep{costanzi19,sunayama20,wu22,zhou24,lee25,costanzi26}.
However, projection effects in shear-selected clusters are more complex.
As weak lensing probes the total matter distribution along the light path, shear-selected peaks contain the cumulative signals of all cosmic structures along the line of sight.
In addition to correlated structures \citep{marian10} and halo triaxiality \citep{hamana12}, for which the resulting bias can be calibrated through N-body simulations, the projection effects may arise when uncorrelated halos happen to lie along the line of sight to a main cluster \citep{yuan18}.
This could lead to the misidentification of optical counterparts and, hence, redshift outliers.
Moreover, the projection effects are expected to have a smaller impact for massive clusters, because it is less likely to have another massive halo with comparable mass along the line-of-sight direction.
Previous studies have demonstrated that the impact of projection effects on the identification of shear-selected clusters can be mitigated through optimized filters \citep{maturi05,marian12} and tomographic bins of WL source galaxies \citep{hennawi05,chappuis26}, although we caution that such mitigation strategies may further complicate the selection function modelling. 
We note that uncorrelated structures are not a primary source of the projection effects for optically selected clusters, as they can be identified as distinct halos through photo-\redshift\ information.

%
%

\section{Conclusions}
\label{sec:conclusions}

In this work, we develop fCAMIRA, a cluster confirmation tool built upon the cluster finder algorithm CAMIRA \citep{oguri14}, and apply it to $129$ weak-lensing shear-selected clusters identified in the HSC-Y3 aperture-mass maps \citep{oguri21}.
Using the HSC broadband photometric data, we identify the optical counterparts of the shear-selected clusters and measure their photometric redshifts \zcl\ and optical richness \rich.
The cluster redshifts estimated by fCAMIRA enable the modelling of the redshift-dependent abundance of shear-selected clusters and are expected to significantly improve the cosmological constraints compared to those obtained in \citet{chiu24}, which did not include cluster-redshift information.

The fCAMIRA algorithm relies on the RS model calibrated through a two-step, data-driven procedure.
First, we calibrate the metallicity-luminosity relation of the RS model using the observed color-magnitude diagrams at $0.1\lesssim\redshift\lesssim1.3$, which are derived from passively evolving galaxies in X-ray-selected eFEDS clusters.
The resulting metallicity-luminosity relation determines the ``tilt'' (i.e., the slope) of the RS in the color-magnitude diagrams.
Second, we further calibrate the color offsets of the RS model predictions and the intrinsic scatter of the observed RS colors using spectroscopic samples at $0.01\lesssim\redshift\lesssim1.3$.
The calibrated color offsets and intrinsic scatter account for imperfections in the theoretical SED model in an empirical way.
We validate the resulting RS model against the spectroscopic samples that are not used in the second-step calibration, and find that the resulting photo-\redshift\ estimates are sufficiently accurate and precise for the optical confirmation.

With the fCAMIRA algorithm, we derive two richness maps $N_{\mathrm{TI20}}\left(\skyloc, \redshift\right)$ and $N_{\mathrm{CAMIRA}}\left(\skyloc, \redshift\right)$ in fine redshift bins around each shear-selected cluster.
The former $N_{\mathrm{TI20}}\left(\skyloc, \redshift\right)$ is constructed using the same spatial filter as in building the WL aperture-mass maps, while the latter $N_{\mathrm{CAMIRA}}\left(\skyloc, \redshift\right)$ employs a filter matched to a typical cluster size of $R\approx0.8\Mpch$.
The fCAMIRA algorithm uses the richness map $N_{\mathrm{TI20}}\left(\skyloc, \redshift\right)$ to identify the optical counterpart candidates, followed by the redshift and richness measurements in the matched-filter richness map  $N_{\mathrm{CAMIRA}}\left(\skyloc, \redshift\right)$.
We define the lensing score of an optical counterpart candidate as the product of its richness and the WL geometric factor, which approximates the lensing strength under the assumption that the richness is linearly proportional to the halo mass (i.e., $\rich\propto\mass$).
We rank the candidates based on their lensing scores and assign the highest one as the ``best-matched'' optical counterpart of the shear-selected cluster.
Additionally, we define the lensing fraction \flens\ of the best-matched counterpart as its relative lensing score to those of the remaining candidates.
That is, the quantity \flens\ approximates the contribution of the identified optical counterpart to the total observed lensing signal.
We find that $\approx90$ out of the $129$ shear-selected clusters have $\flens\geq0.5$, indicating that the WL signals of the majority of the sample are dominated by single halos. 

We assess the systematics of the cluster photo-\redshift\ measurements by using the spectroscopic samples.
We find a mild mean bias at a level of $\approx0.005$ and correct for it in the final photo-\redshift\ estimate \zcl.
The scatter in \zcl\ with respect to the spectroscopic redshifts is quantified to be $\approx0.008$.
By comparing with the redshift estimates obtained from a positional cross-matching with external cluster catalogs, we determine the outlier fraction of the cluster photo-\redshift\ \zcl\ at a level of $\approx7\percent$ when adopting the outlier threshold as $0.15$.
The outlier fraction increases to $\approx15\percent$ if considering the redshift discrepancy $\gtrsim0.08$.
The redshift outliers indicate the presence of projection effects in the cluster sample, as the optical counterparts of shear-selected clusters may be misidentified to halos in their line-of-sight direction in a positional cross-matching.
Conversely, the fCAMIRA algorithm performs a uniform scan of all RS overdensity peaks and identifies the best-matched optical counterparts based on the relative lensing strength.

We quantify the impact of the cluster redshift measurements on cosmological constraints.
This is achieved by building mock samples of shear-selected clusters including not only the observed shape noises but also realistic cluster photo-\redshift\ systematics.
We find that the biased cluster redshifts at a level of $0.01$ result in negligible impact on the constraints on \omegam, \sigmaeight, and \seight\ in the abundance modelling of $\approx200$ shear-selected clusters.
Similarly, the cluster photo-\redshift\ scatter and outlier fraction at levels of $0.01$ and $8\percent$, respectively, do not have a significant effect on the parameter constraints.
Our results indicate that the cluster redshifts estimated by the fCAMIRA confirmation tool are sufficiently accurate for cosmological studies using an HSC-Y3 like sample.
However, we find the bias arising from the cluster photo-\redshift\ systematics at a level comparable to the statistical uncertainty of the parameter constraints when analyzing a sample size of $N_{\mathrm{cl}}\gtrsim2000$.
Future Stage-IV like samples would require further improvements in the cluster redshift measurements.

We discuss the projection effects in the context of shear-selected clusters.
Despite baryon-free and purely gravity-based selection, the projection effects in weak-lensing shear-selected clusters are more complex than those in optically selected samples.
As the weak-lensing signals of shear-selected peaks arise from the integrated matter distribution along the line of sight, the optical counterparts may be misidentified with line-of-sight halos that are uncorrelated with the main cluster, leading to redshift outliers.
In the case of two distinct halos of comparable mass projected along the same line of sight, although such cases are rare, they can be identified as separate clusters in optical observations. 
Weak lensing, however, measures the combined signal from both halos, making it difficult to disentangle their individual contributions and thereby complicating the modelling of the selection function.

This paper presents an algorithm to measure the redshifts of shear-selected clusters, as an essential ingredient towards precision cluster cosmology using WL-selected samples.
Moreover, we quantify projection effects in WL shear-selected clusters using a newly defined quantity \flens, which potentially provides a useful way to further reduce line-of-sight contaminants in a sample.
We will utilize the redshift estimates of the shear-selected clusters and update the cosmological analysis in forthcoming papers.

%
%

\section*{Acknowledgments}

We thank the anonymous referee for constructive comments that lead to improvements in this paper.
I-Non Chiu thanks Matthias Klein for useful discussions on the optical confirmation of clusters, and Yi Yang for the hospitality at Institute of Physics, Academia Sinica.
I-Non Chiu thanks the participants of the workshops ``The Most Massive Galaxies and Their Environment Across Cosmic Times'' and ``Gravitational Lensing at the turning point: Current Discoveries, New Frontiers'' held at the Sexten Center for Astrophysics Riccardo Giacconi, for insightful discussions.
This work is supported by the National Science and Technology Council in Taiwan (Grant NSTC 114-2112-M-006-017-MY3).
This work made use of the computational and storage resources in the National Center for High-Performance Computing (NCHC) in Taiwan.
This work was supported by JSPS KAKENHI Grant Numbers JP25H00662, JP25H00672, JP22K21349, JP2300108.
K.U. acknowledges support from the National Science and Technology Council of Taiwan (grants NSTC 112-2112-M-001-027-MY3 and NSTC 115-2112-M-001-027-) and the Academia Sinica Investigator Award (grant AS-IA-112-M04).

The Hyper Suprime-Cam (HSC) collaboration includes the astronomical communities of Japan and Taiwan, and Princeton University. The HSC instrumentation and software were developed by the National Astronomical Observatory of Japan (NAOJ), the Kavli Institute for the Physics and Mathematics of the Universe (Kavli IPMU), the University of Tokyo, the High Energy Accelerator Research Organization (KEK), the Academia Sinica Institute for Astronomy and Astrophysics in Taiwan (ASIAA), and Princeton University. Funding was contributed by the FIRST program from the Japanese Cabinet Office, the Ministry of Education, Culture, Sports, Science and Technology (MEXT), the Japan Society for the Promotion of Science (JSPS), Japan Science and Technology Agency (JST), the Toray Science Foundation, NAOJ, Kavli IPMU, KEK, ASIAA, and Princeton University. 

This paper makes use of software developed for Vera C. Rubin Observatory. We thank the Rubin Observatory for making their code available as free software at \url{http://pipelines.lsst.io/}.

This paper is based on data collected at the Subaru Telescope and retrieved from the HSC data archive system, which is operated by the Subaru Telescope and Astronomy Data Center (ADC) at NAOJ. Data analysis was in part carried out with the cooperation of Center for Computational Astrophysics (CfCA), NAOJ. We are honored and grateful for the opportunity of observing the Universe from Maunakea, which has the cultural, historical and natural significance in Hawaii. 

The Pan-STARRS1 Surveys (PS1) and the PS1 public science archive have been made possible through contributions by the Institute for Astronomy, the University of Hawaii, the Pan-STARRS Project Office, the Max Planck Society and its participating institutes, the Max Planck Institute for Astronomy, Heidelberg, and the Max Planck Institute for Extraterrestrial Physics, Garching, The Johns Hopkins University, Durham University, the University of Edinburgh, the Queen’s University Belfast, the Harvard-Smithsonian Center for Astrophysics, the Las Cumbres Observatory Global Telescope Network Incorporated, the National Central University of Taiwan, the Space Telescope Science Institute, the National Aeronautics and Space Administration under grant No. NNX08AR22G issued through the Planetary Science Division of the NASA Science Mission Directorate, the National Science Foundation grant No. AST-1238877, the University of Maryland, Eotvos Lorand University (ELTE), the Los Alamos National Laboratory, and the Gordon and Betty Moore Foundation.

This work is possible because of the efforts in the Vera C. Rubin Observatory \citep{juric17,ivezic19} and PS1 \citep{chambers16, schlafly12, tonry12, magnier13}, and in the HSC \citep{aihara18a} developments including the deep imaging of the COSMOS field \citep{tanaka17}, the on-site quality-assurance system \citep{furusawa18}, the Hyper Suprime-Cam \citep{miyazaki15, miyazaki18, komiyama18}, the design of the filters \citep{kawanomoto18},  the data pipeline \citep{bosch18}, the design of bright-star masks \citep{coupon18}, the characterization of the photometry by the code \texttt{Synpipe} \citep{huang18}, the photometric redshift estimation \citep{tanaka18}, the shear calibration \citep{mandelbaum18}, and the public data releases \citep{aihara18b, aihara19}.

This work made use of the IPython package \citep{ipython}, \texttt{SciPy} \citep{scipy}, \texttt{TOPCAT} \citep{topcat1,topcat2}, \texttt{matplotlib} \citep{matplotlib}, \texttt{Astropy} \citep{astropy}, \texttt{NumPy} \citep{van2011numpy}, and \texttt{Pathos} \citep{pathos}.
Cosmology-related quantities are computed using \texttt{pyccl} \citep{chisari19}.
The software management used in this work leverages the \texttt{conda-forge} project \citep{conda_forge_community_2015_4774216}.


%
%

\section*{Data Availability}

The HSC photometric catalog used in this work will be released in the Fourth Public Data Release (PDR4) of the HSC survey.
The results of the optical confirmation of the $129$ shear-selected clusters are available via \url{https://github.com/inonchiu/hsc_shear_selected_clusters}.
The other data products underlying this article will be shared upon a reasonable request to the corresponding author.

%
%

\bibliographystyle{aasjournal}
\bibliography{literature} 

%
%

\onecolumngrid

\appendix

\section{The SQL query in the HSC database}
\label{app:sql}

In what follows, we provide the SQL script to query the $grizY$ aperture photometry from the HSC PDR4 database.
In the resulting catalog, we additionally exclude $\mathtt{i\_extendedness\_value} = 0$ if $\mathtt{i\_extendedness\_value}$ is measured.

\begin{verbatim}
SELECT
photo.object_id,
photo.ra,
photo.dec,
photo.i_extendedness_value,
(afbun.[g,r,i,y]_undeblended_convolvedflux_3_15_mag -
 afbun.z_undeblended_convolvedflux_3_15_mag) + photo.z_cmodel_mag -
photo.a_[g,r,i,y] - offsets.[g,r,i,y]_mag_offset        as [g,r,i,y]mag_aper,
photo.z_cmodel_mag - photo.a_z - offsets.z_mag_offset   as zmag_aper,
photo.[g,r,i,z,y]_cmodel_magerr                         as [g,r,i,z,y]mag_err,
FROM s23b_wide.forced                           as photo
LEFT JOIN s23b_wide.forced5                     as afbun           USING (object_id)
LEFT JOIN s23b_wide.masks                       as masks           USING (object_id)
LEFT JOIN s23b_wide.stellar_sequence_offsets    as offsets         USING (skymap_id)
WHERE
photo.isprimary                                 is True  AND
photo.[g,r,i,z,y]_pixelflags_edge               is False AND
photo.[g,r,i,z,y]_pixelflags_interpolatedcenter is False AND
photo.[g,r,i,z,y]_pixelflags_crcenter           is False AND
masks.[g,r,i,z,y]_mask_brightstar_halo          is False AND
masks.[g,r,i,z,y]_mask_brightstar_ghost         is False AND
masks.[g,r,i,z,y]_mask_brightstar_blooming      is False AND
photo.[g,r]_inputcount_value                    >= 2     AND
photo.[i,z,y]_inputcount_value                  >= 3     AND
photo.[g,r,i,y]_cmodel_magerr                   < 1.0    AND
photo.z_cmodel_magerr                           < 0.1    AND
(photo.z_cmodel_mag - photo.a_z - offsets.z_mag_offset) < 25.0
ORDER BY photo.object_id
\end{verbatim}

\section{The measurements of red-sequence galaxies and the blue fraction}
\label{app:rs_meas}

\begin{figure*}
\centering
\resizebox{0.31\textwidth}{!}{\includegraphics[scale=1]{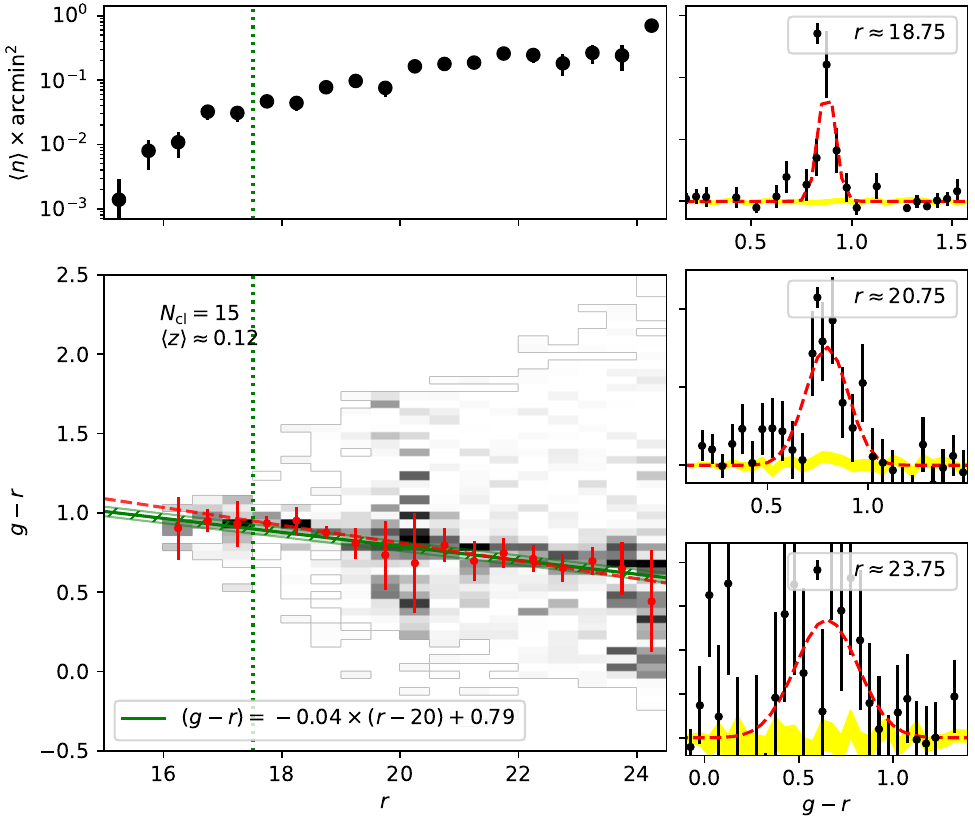}}
\resizebox{0.31\textwidth}{!}{\includegraphics[scale=1]{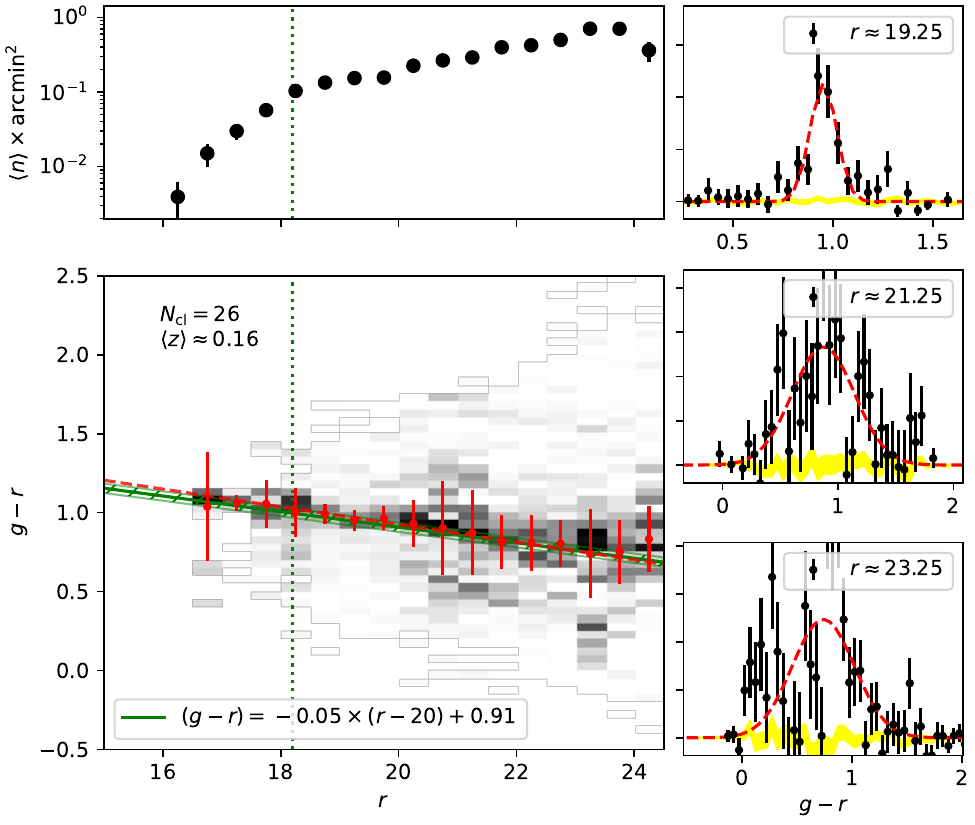}}
\resizebox{0.31\textwidth}{!}{\includegraphics[scale=1]{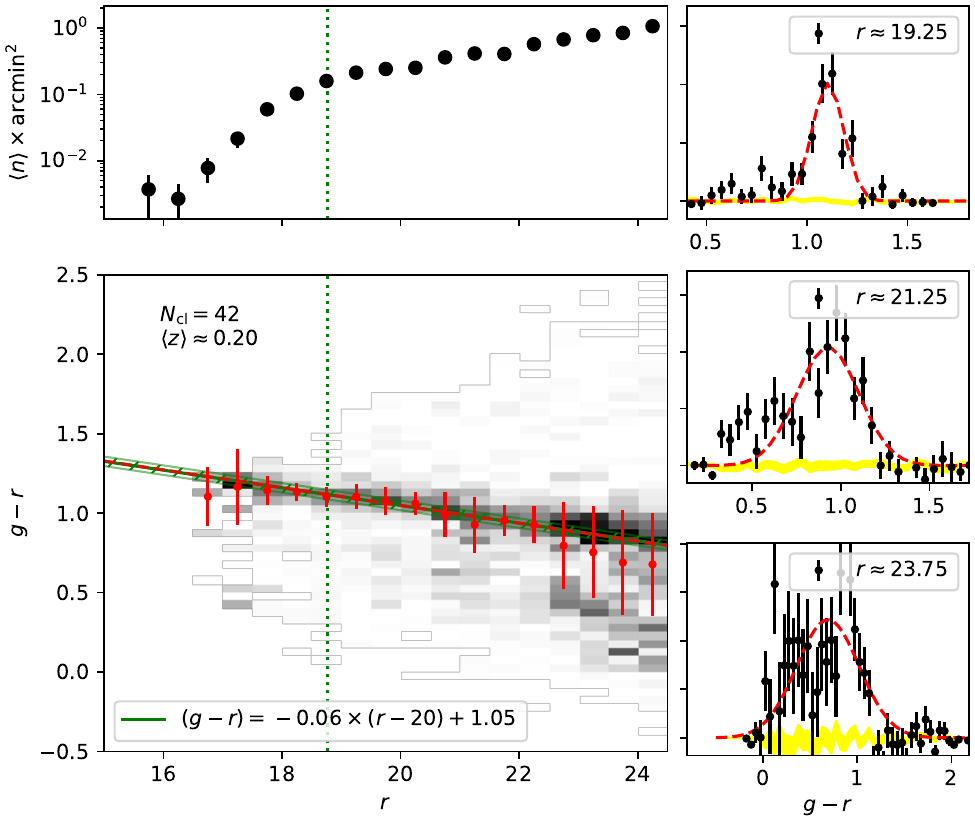}}
\resizebox{0.31\textwidth}{!}{\includegraphics[scale=1]{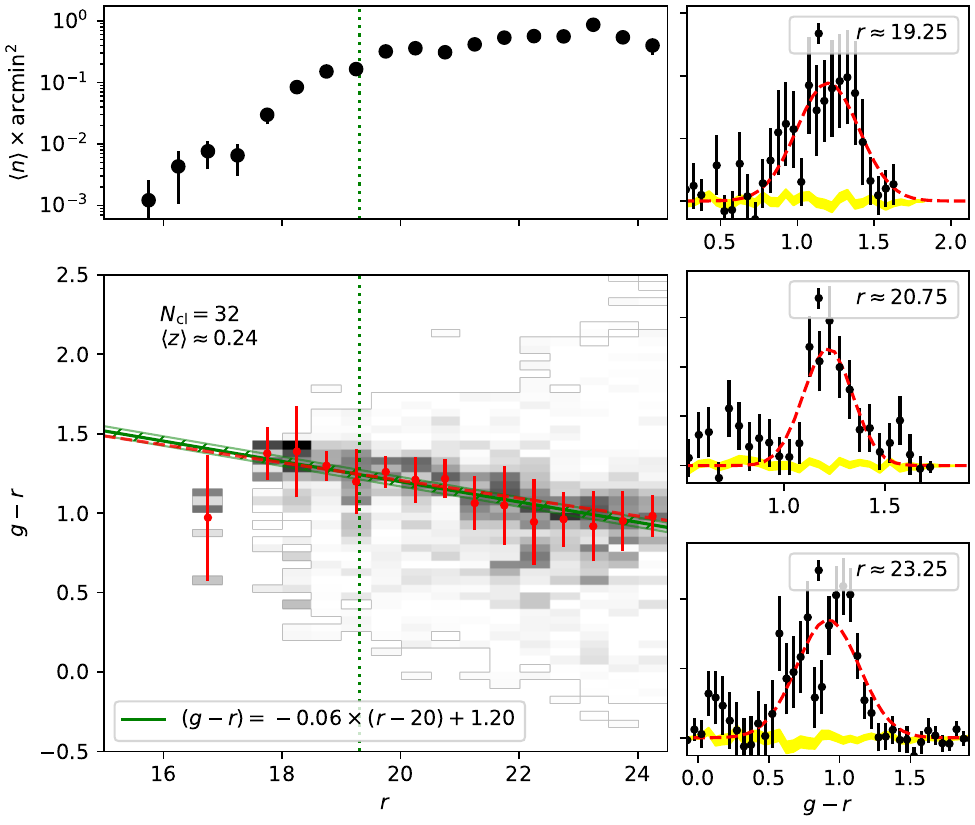}}
\resizebox{0.31\textwidth}{!}{\includegraphics[scale=1]{pplots/rs/rs_4.pdf}}
\resizebox{0.31\textwidth}{!}{\includegraphics[scale=1]{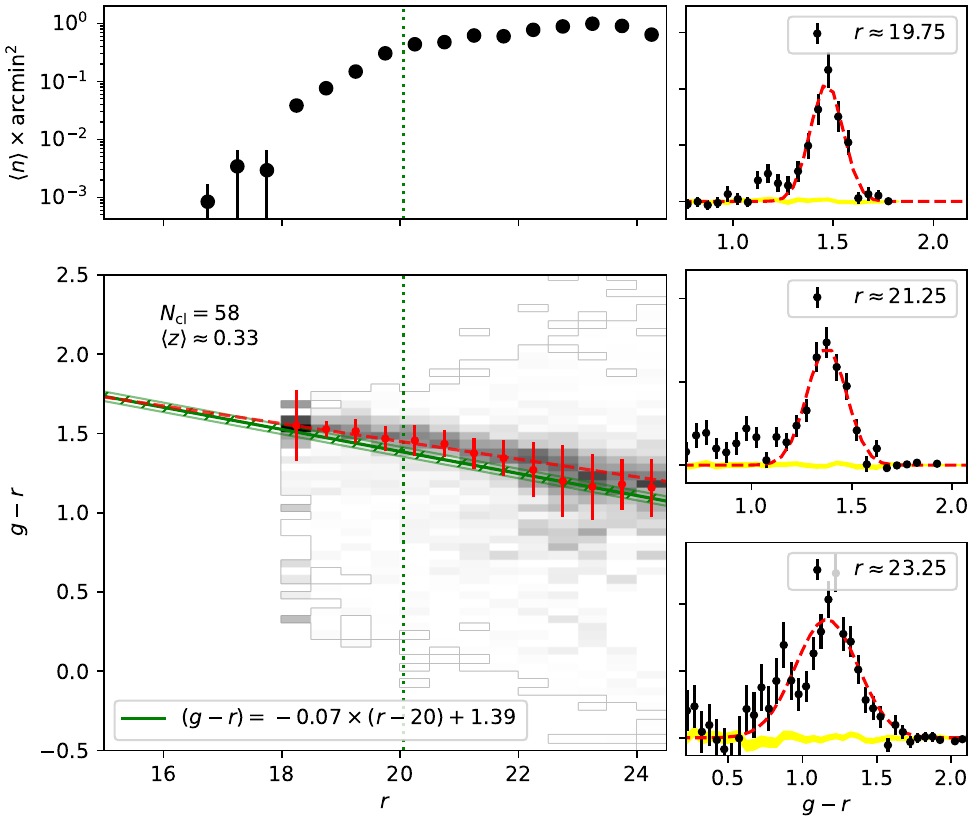}}
\resizebox{0.31\textwidth}{!}{\includegraphics[scale=1]{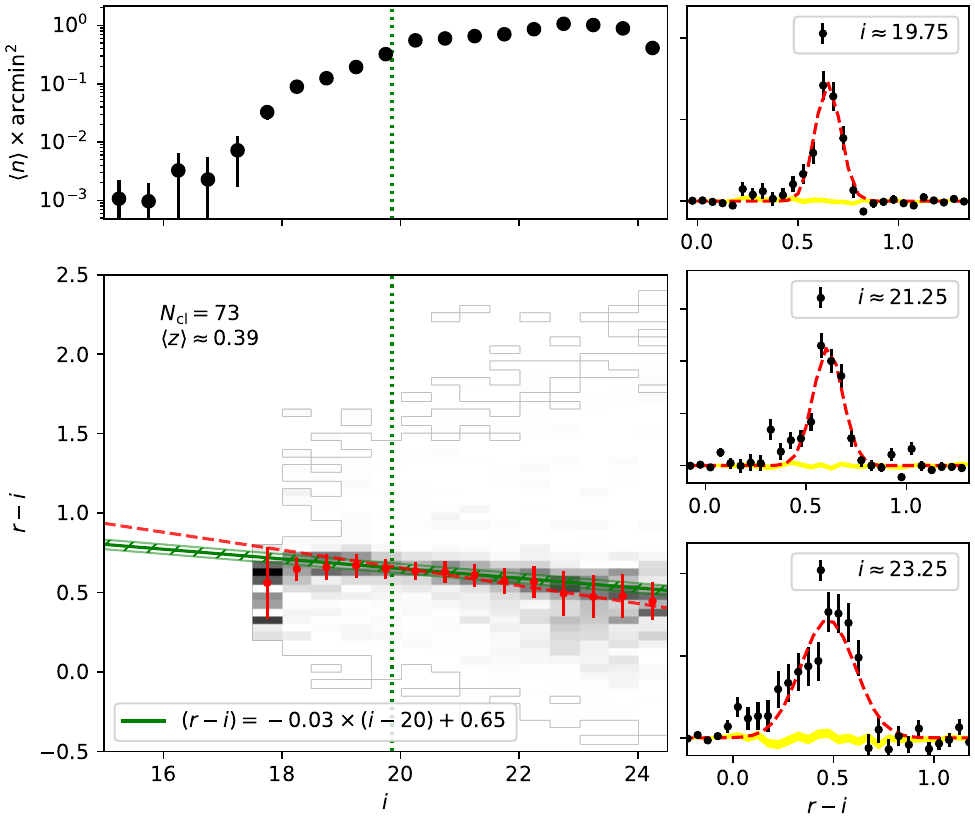}}
\resizebox{0.31\textwidth}{!}{\includegraphics[scale=1]{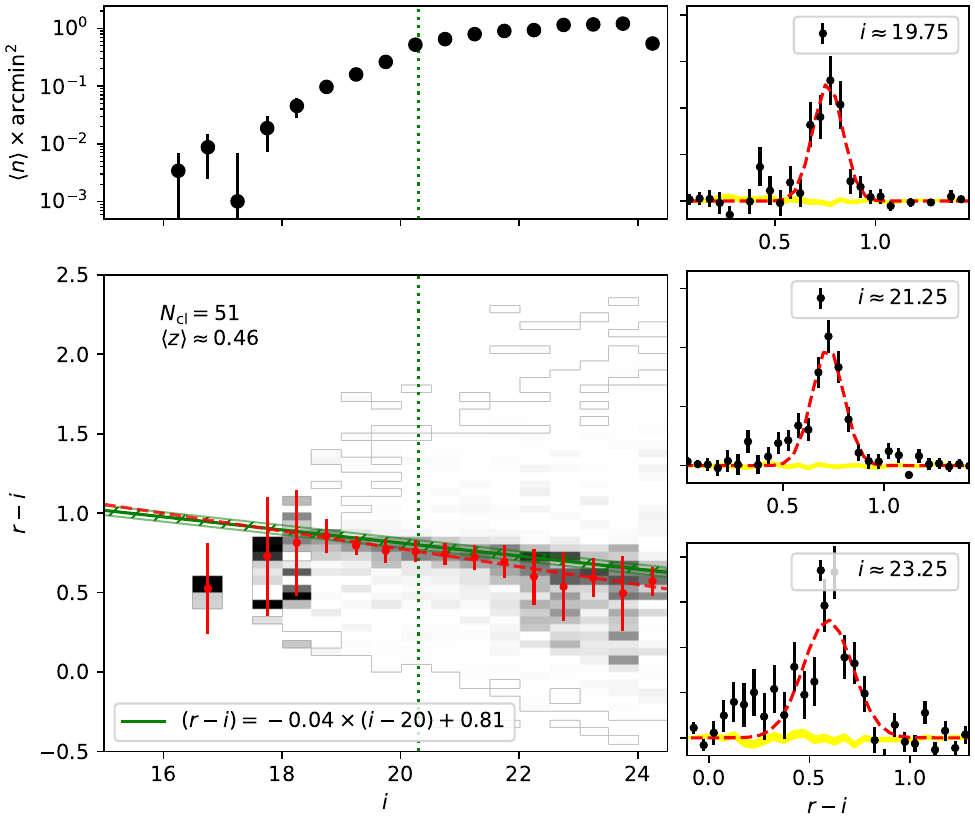}}
\resizebox{0.31\textwidth}{!}{\includegraphics[scale=1]{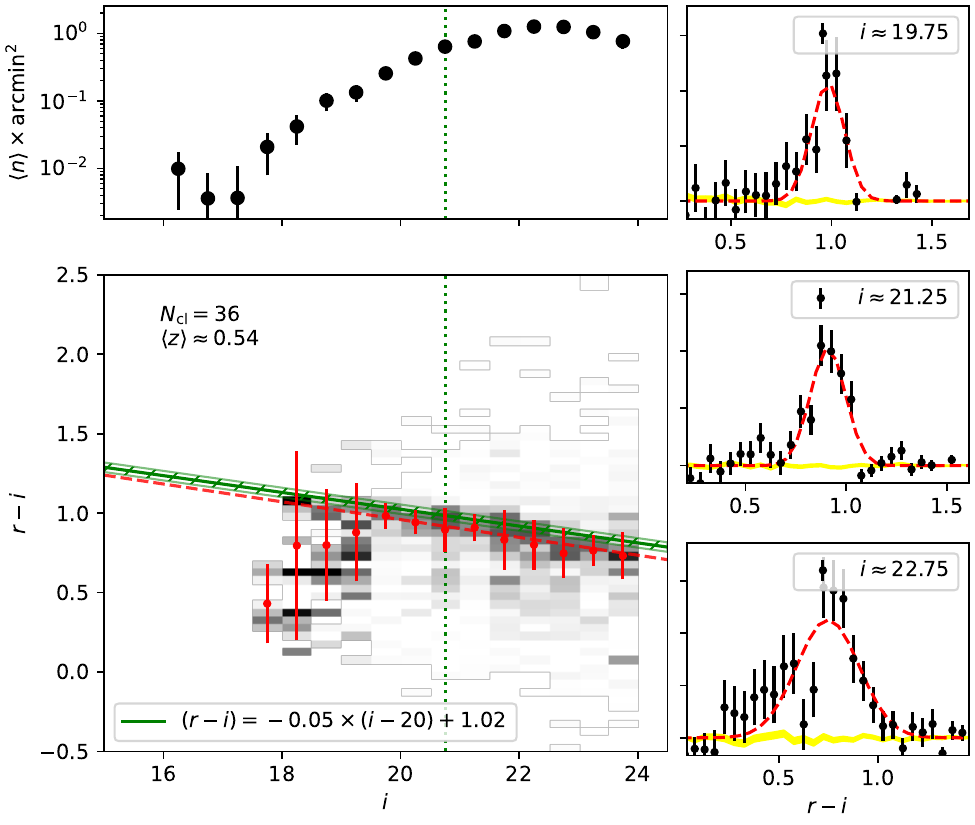}}
\resizebox{0.31\textwidth}{!}{\includegraphics[scale=1]{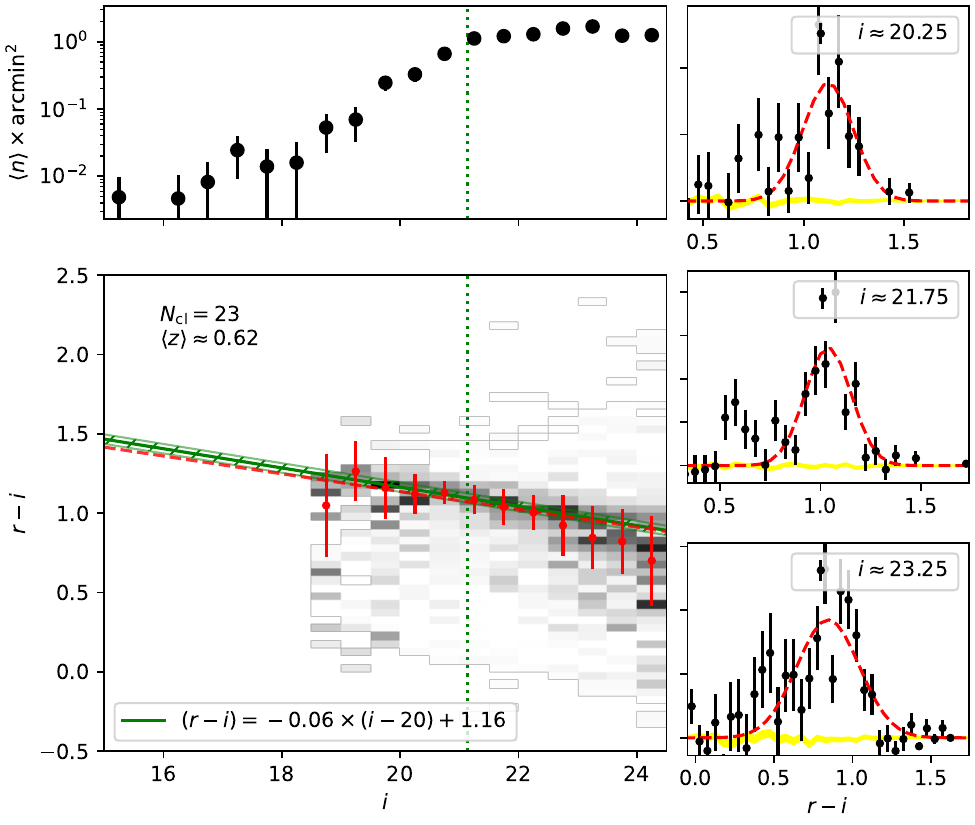}}
\resizebox{0.31\textwidth}{!}{\includegraphics[scale=1]{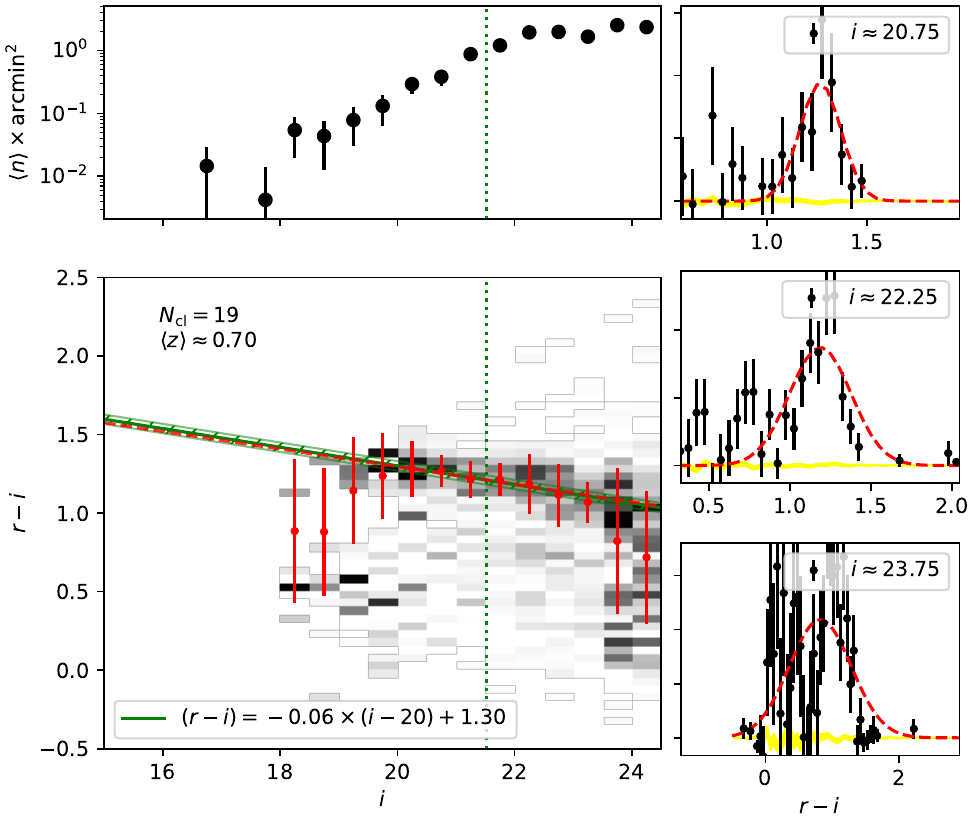}}
\resizebox{0.31\textwidth}{!}{\includegraphics[scale=1]{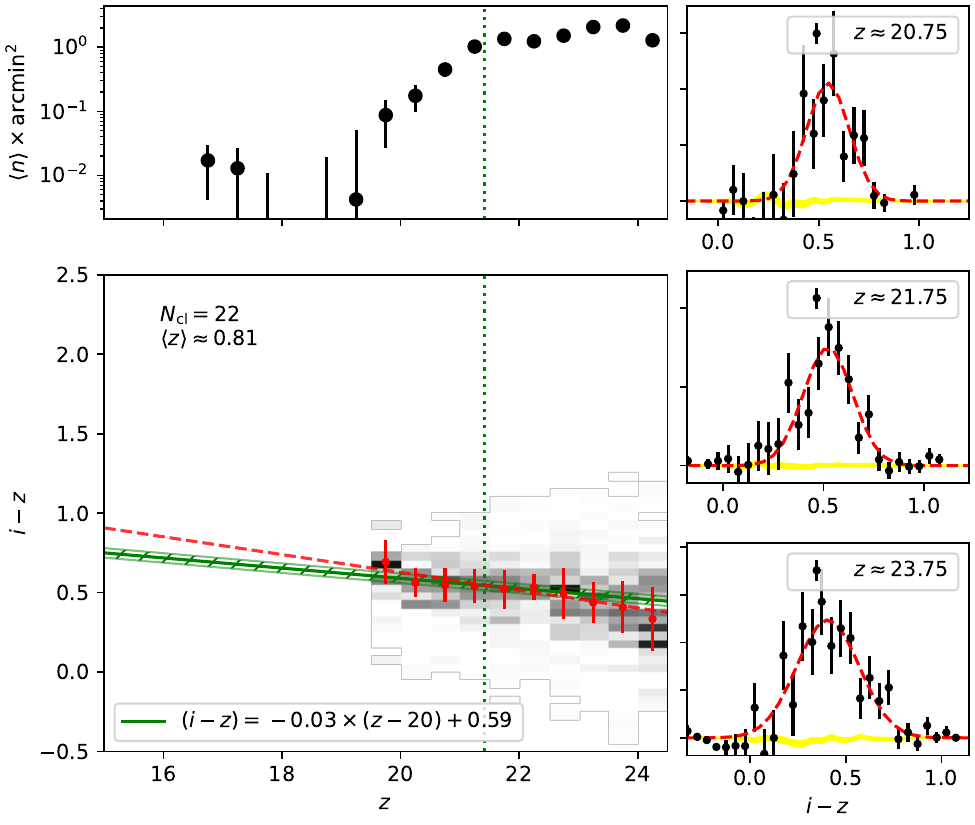}}
\resizebox{0.31\textwidth}{!}{\includegraphics[scale=1]{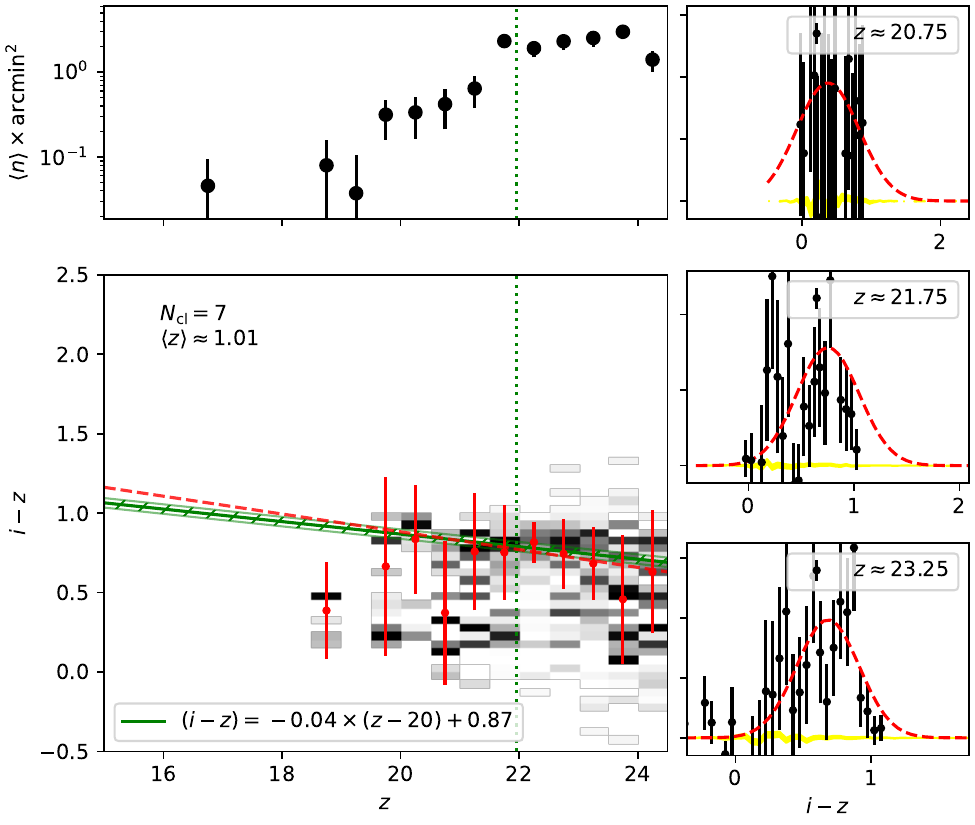}}
\resizebox{0.31\textwidth}{!}{\includegraphics[scale=1]{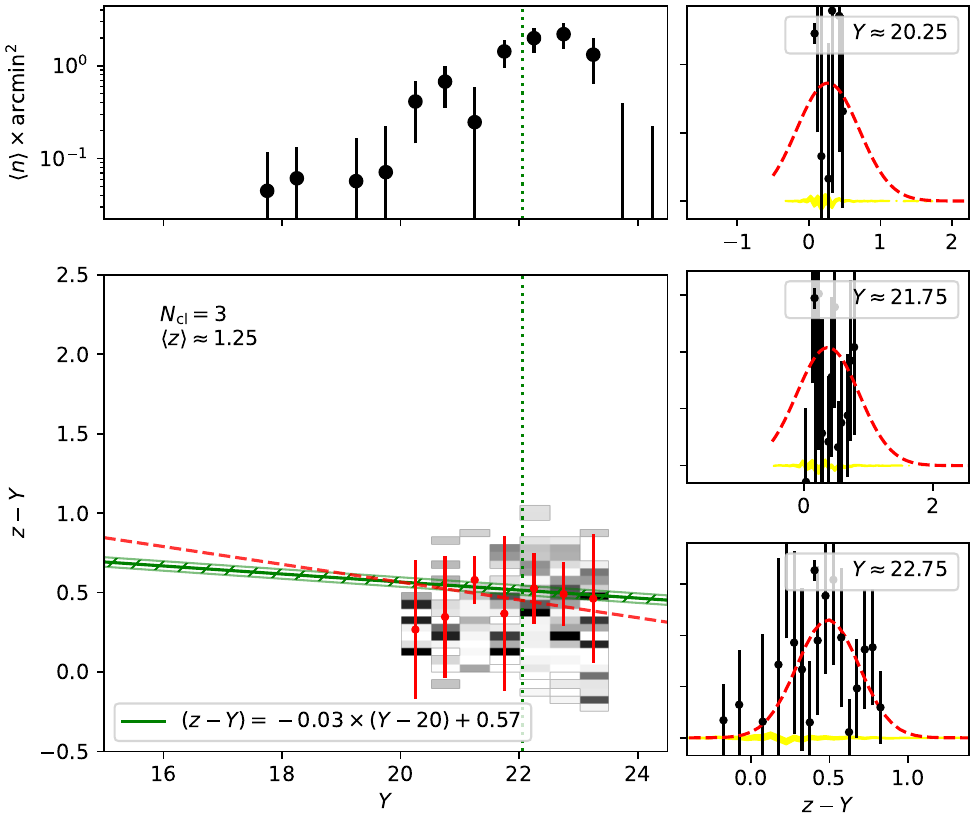}}
\caption{
The RS distribution of the passively evolving galaxies inside the eFEDS clusters in the redshift bins.
Each subplot represents a stacked result in a redshift bin and is produced in the identical way as in the left panel of Figure~\ref{fig:rs_example}.
}
\label{fig:rs_all_examples}
\end{figure*}

We present the stacked RS distribution of the eFEDS clusters measured in the color-magnitude space in Figure~\ref{fig:rs_all_examples}.

\begin{figure}
\centering
\resizebox{0.5\textwidth}{!}{
\includegraphics[scale=1]{
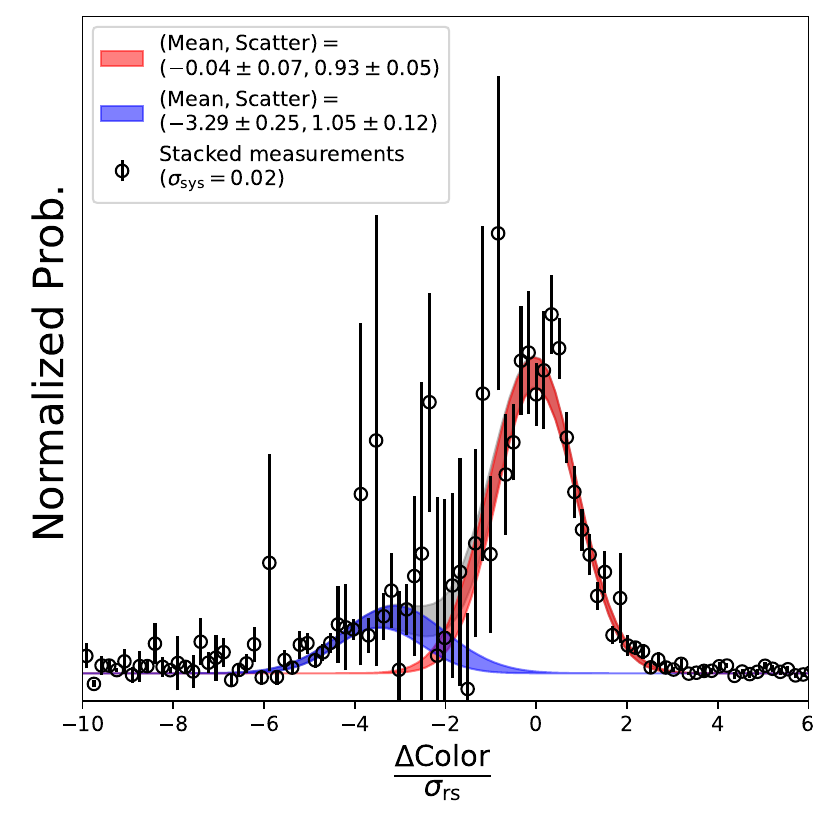
}
}
\caption{
The stacked color distribution of the member galaxies in the eFEDS clusters.
The open circles represent the distribution of the color offset $\Delta \mathrm{Color}$ with respect to the predicted RS color normalized by the RS color scatter $\sigma_{\mathrm{RS}}$.
We include a systematic uncertainty  (at a level of $0.02$) in the color offset in predicting the RS color, which arises from the linear fit to the amplitude of the Gaussian distributions; see text in Appendix~\ref{app:rs_meas} for more details. 
The red and blue regions indicate the $1\sigma$ confidence levels of the best-fit color distributions of the RS and blue populations, respectively.
Our results indicate that the blue population of the eFEDS clusters is sub-dominant compared to the RS galaxies and thereby has a negligible impact on the calibration of the RS $Z$-$L$ relation.
}
\label{fig:blue_fraction}
\end{figure}

We quantify the blue fraction as follows.
For each redshift bin, we model the color distributions at individual magnitude bins as Gaussian with two free parameters, the mean color and the color scatter.
The resulting Gaussian distributions are shown as the red data points in Figure~\ref{fig:rs_all_examples}.
Then, we fit separate linear functions of magnitude, $a(\magnitude-20)+b$, to the amplitudes and color scatters of the Gaussian distributions.
The resulting linear models enable the prediction of the amplitude and scatter as a function of magnitude at each redshift bin.
Using the predicted mean colors and color scatters, we normalize the observed color distributions in each magnitude and redshift bin to unit Gaussians, such that one unit on the horizontal axis corresponds to the predicted $1\sigma$ scatter of the RS population.
Next, we stack the normalized color distributions, each scaled to a unit Gaussian, from all magnitude and redshift bins.
We then derive the inverse-variance-weighted mean color distribution, where the weights are determined by the numbers of available clusters in the corresponding redshift bins.
The open circles in Figure~\ref{fig:blue_fraction} show the stacked normalized color distribution, where the color offset is expressed in units of the predicted RS color scatter.

As seen in Figure~\ref{fig:blue_fraction}, the stacked color distribution can be well described by a unit Gaussian distribution, suggesting that the cluster galaxy population is dominated by RS galaxies.
However, we observe a color distribution extending to the low-offset end, i.e., the blue end, indicating the presence of a sub-dominant population of blue galaxies.
Motivated by this, we fit a model containing two Gaussian distributions to the stacked color distribution.
We obtain the mean color offset and color scatter of $\left(-0.04\pm0.07, 0.93\pm0.05\right)$ for the RS population, respectively.
This indicates that the color distribution of the RS galaxies is well described by a Gaussian model with the mean color predicted by the RS model.
Meanwhile, we obtain the mean color offset and color scatter of $\left(-3.29\pm0.25, 1.05\pm0.12\right)$ for the blue population, respectively.
Our results imply that the blue galaxy population can be described by a secondary Gaussian component whose mean color is approximately $3.3\sigma$ bluer than that of the RS population.
Therefore, we conclude that contamination from blue galaxies has a negligible impact on the calibration of the RS metallicity-luminosity relation.


\section{The identification of brightest cluster galaxies (BCGs)}
\label{app:bcg}

We identify the BCGs in an iterative way, which we briefly describe below.
We start from refined photometric redshift \zcl\ of the cluster counterpart.
Then, we identify the BCG in the richness map evaluated at the refined redshift by identifying the $i$-th galaxy which maximizes the log-likelihood modified from equation~(19) in \cite{oguri14},
\[
\ln L_{\mathrm{BCG}} = -\frac{1}{2}\left( \frac{ {\Delta\magnitude}_{i} - \Delta\magnitude_{\mathrm{BCG}}}{ \sigma_{\mathrm{m},\mathrm{BCG}}} \right)^2 + \ln\left(n\left( \chi^2_i, \vect{\theta}_{i} \right)\right) - \left(\frac{R_i}{\sigma_{R}}\right)^2 + \mathrm{Const.}
\, ,
\]
where the first term includes the weight arising from the predicted BCG magnitude offset $\Delta\magnitude_{\mathrm{BCG}}$,
the second term restricts the BCG candidates to those RS galaxies, and the last term describes the BCG radial offset $R$ to the center.
We use $\Delta\magnitude_{\mathrm{BCG}} = -2$, $\sigma_{\mathrm{m},\mathrm{BCG}} = 0.3$, and $\sigma_{\mathrm{R},\mathrm{BCG}} = 0.3 ~\Mpch$ in this work.
Next, we refine the cluster redshift using the identified BCG candidate as the center, and repeat the BCG identification again at the newly refined redshift.
The process is iterated until the result converges.

\section{The optical images}
\label{app:optical_images}

We present the optical cutout images of the $129$ WL shear-selected clusters in Figures~\ref{fig:optical_images_0} to ~\ref{fig:optical_images_3}.

\begin{figure*}
\centering
\resizebox{0.245\textwidth}{!}{\includegraphics[scale=1]{pplots/rgb/0_image.pdf}}
\resizebox{0.245\textwidth}{!}{\includegraphics[scale=1]{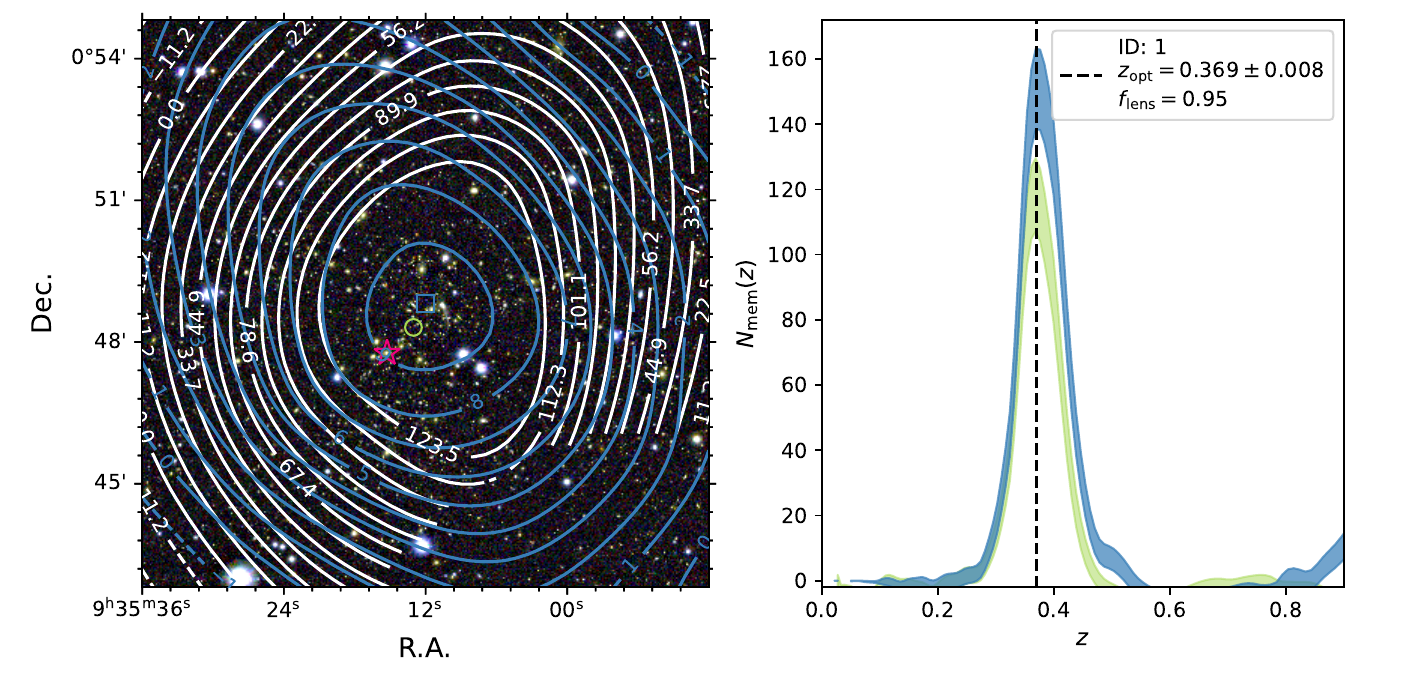}}
\resizebox{0.245\textwidth}{!}{\includegraphics[scale=1]{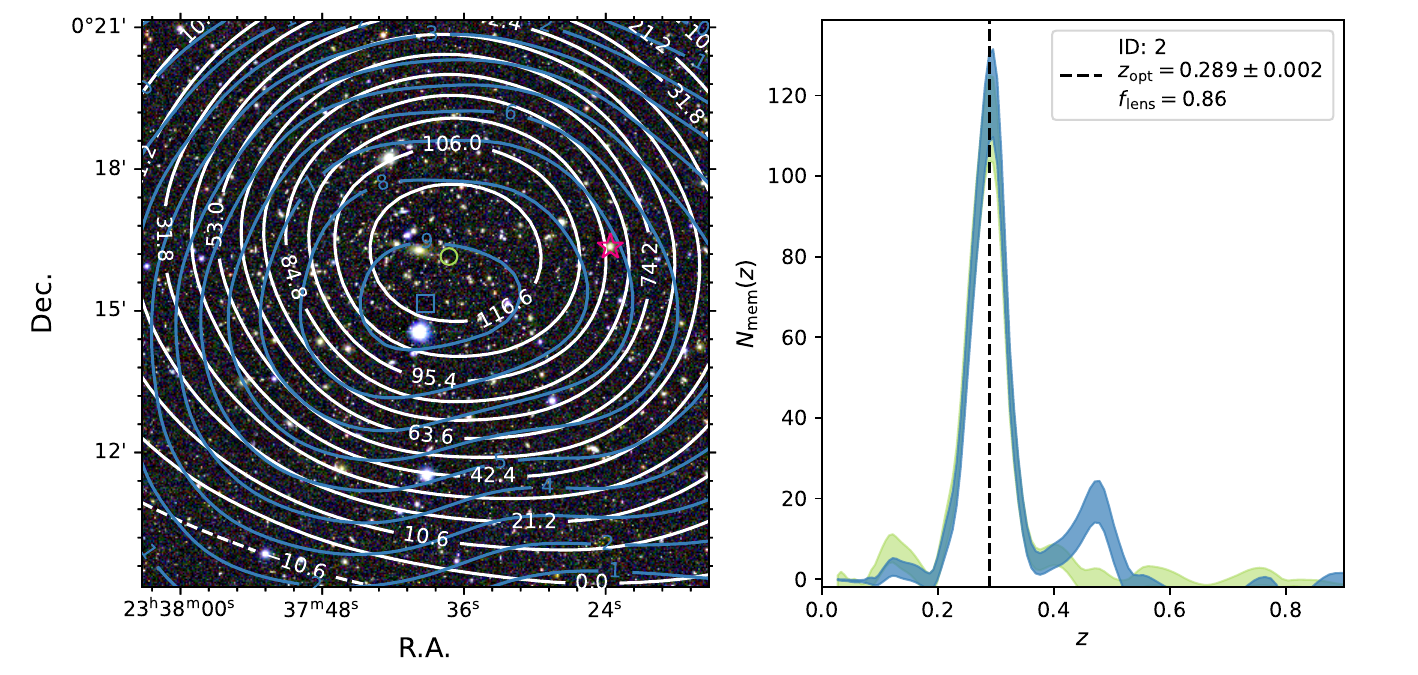}}
\resizebox{0.245\textwidth}{!}{\includegraphics[scale=1]{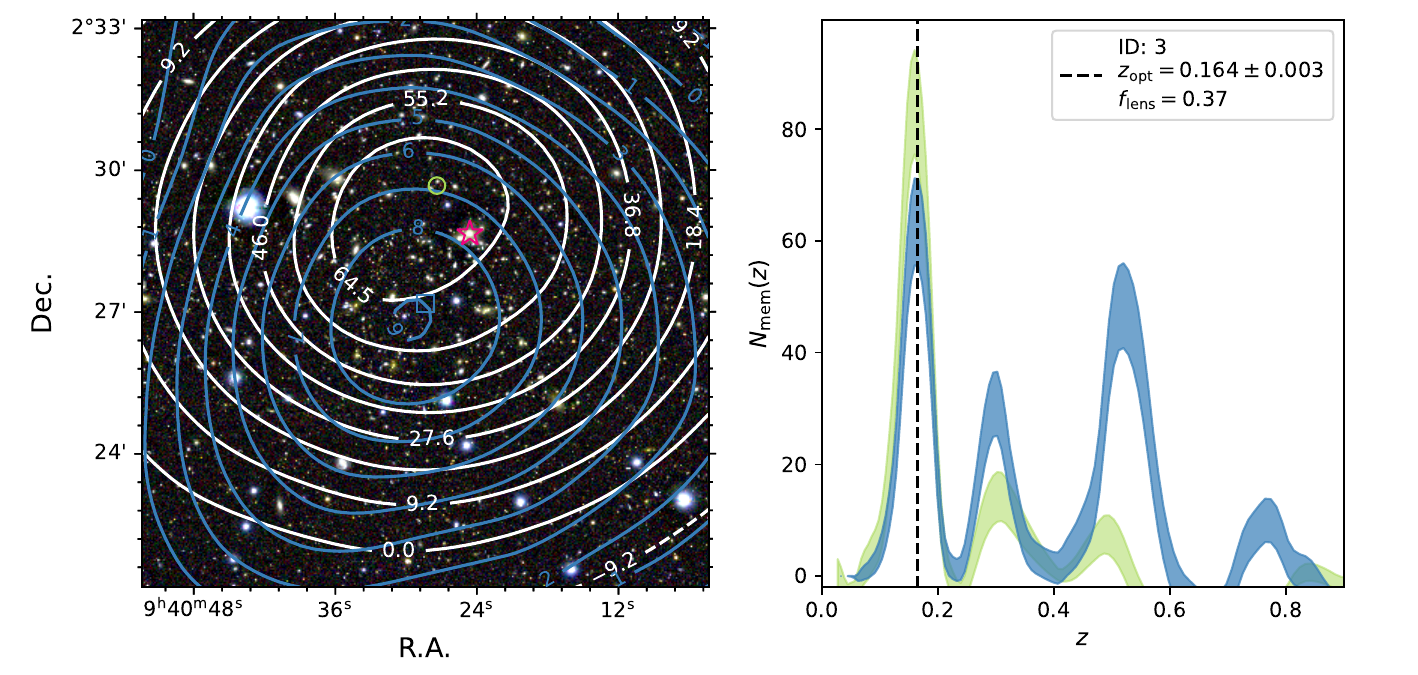}}
\resizebox{0.245\textwidth}{!}{\includegraphics[scale=1]{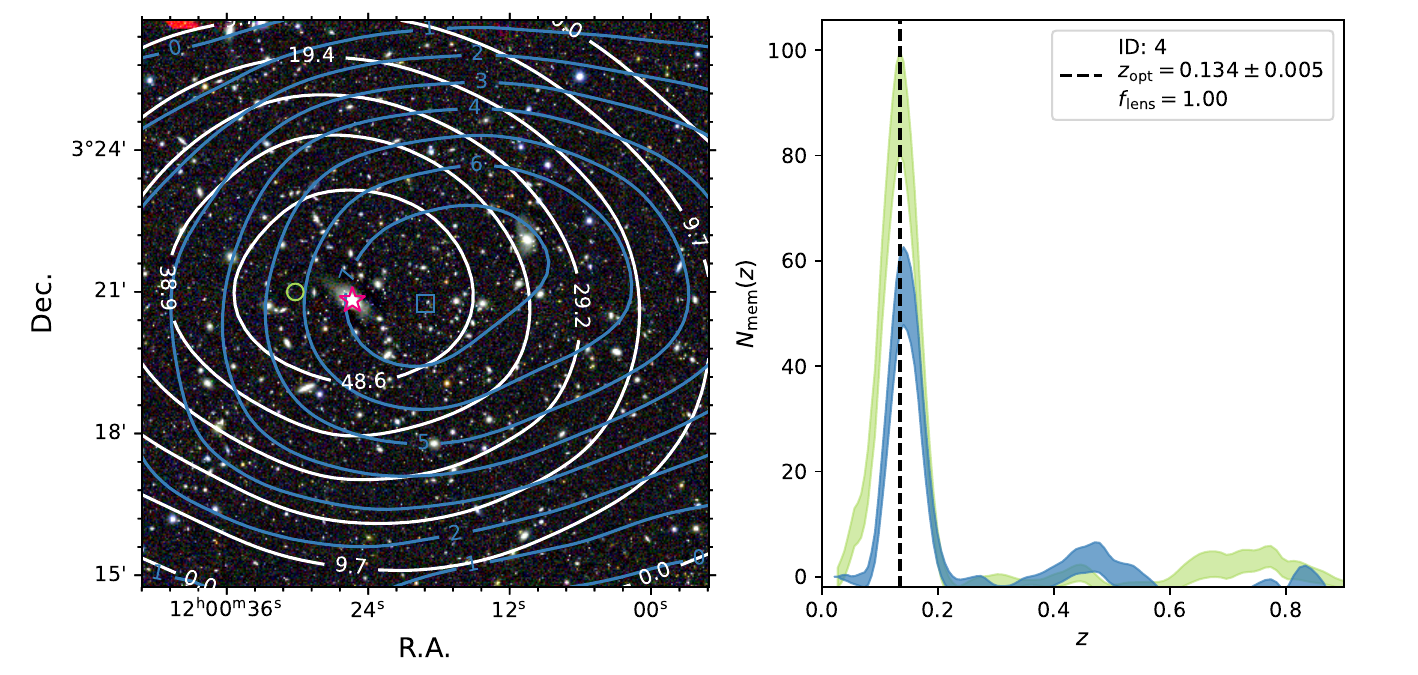}}
\resizebox{0.245\textwidth}{!}{\includegraphics[scale=1]{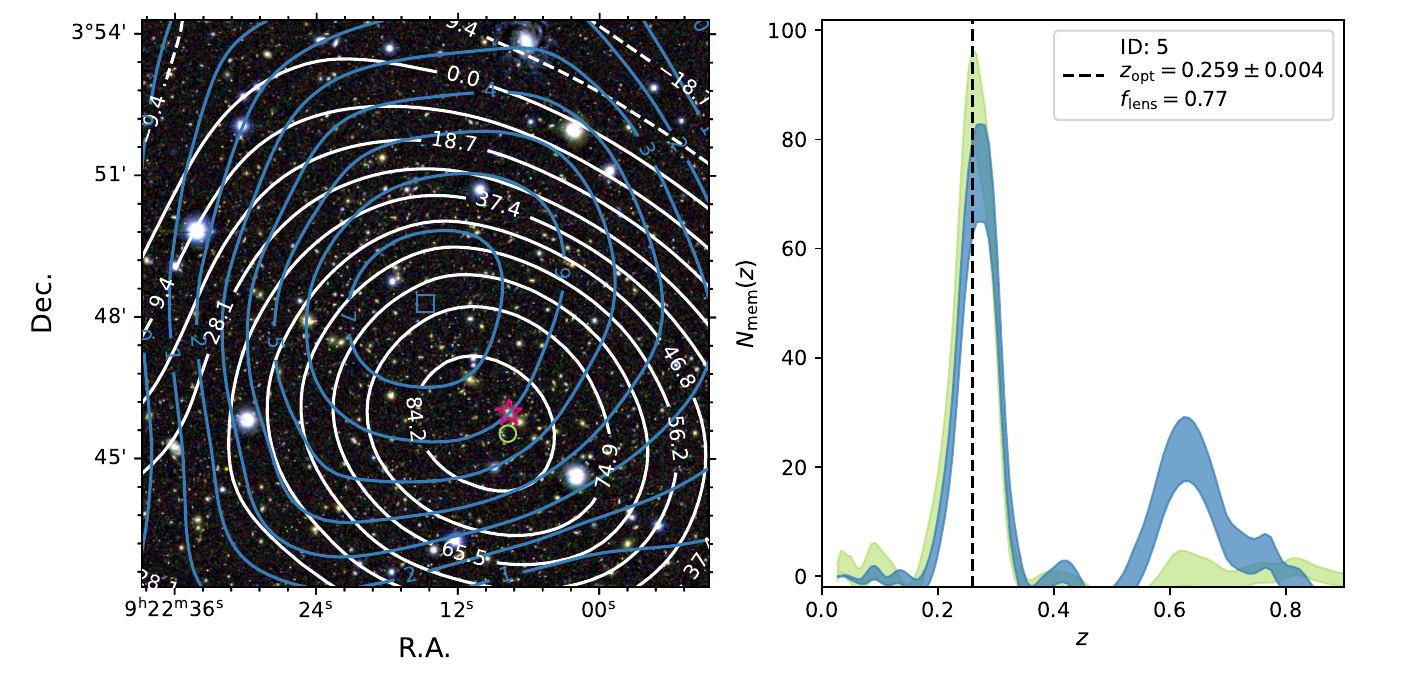}}
\resizebox{0.245\textwidth}{!}{\includegraphics[scale=1]{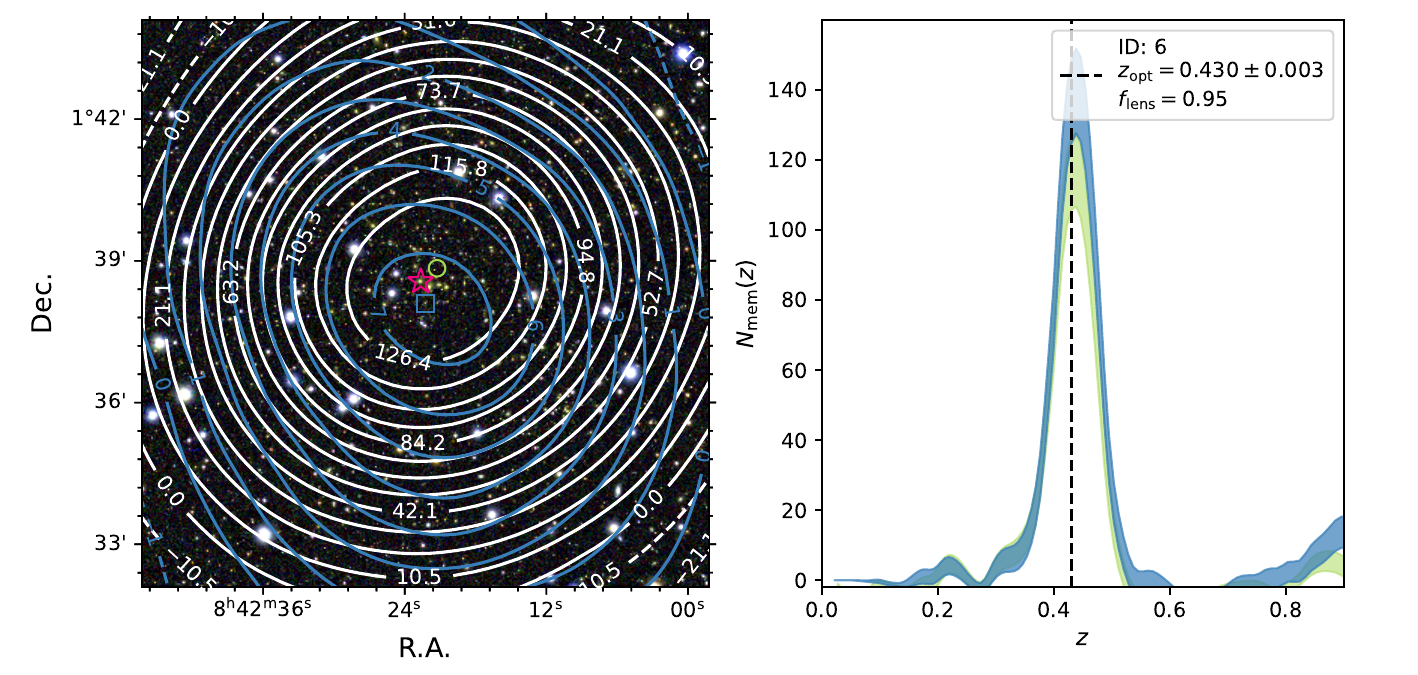}}
\resizebox{0.245\textwidth}{!}{\includegraphics[scale=1]{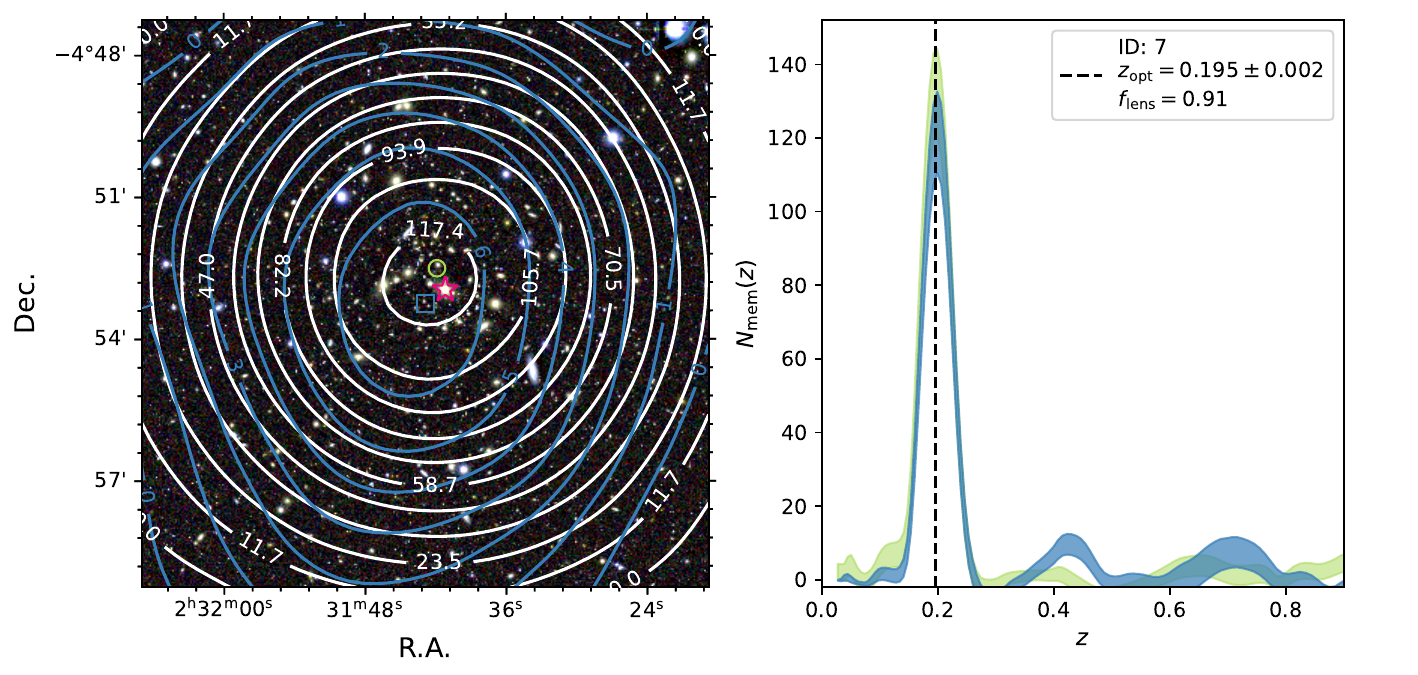}}
\resizebox{0.245\textwidth}{!}{\includegraphics[scale=1]{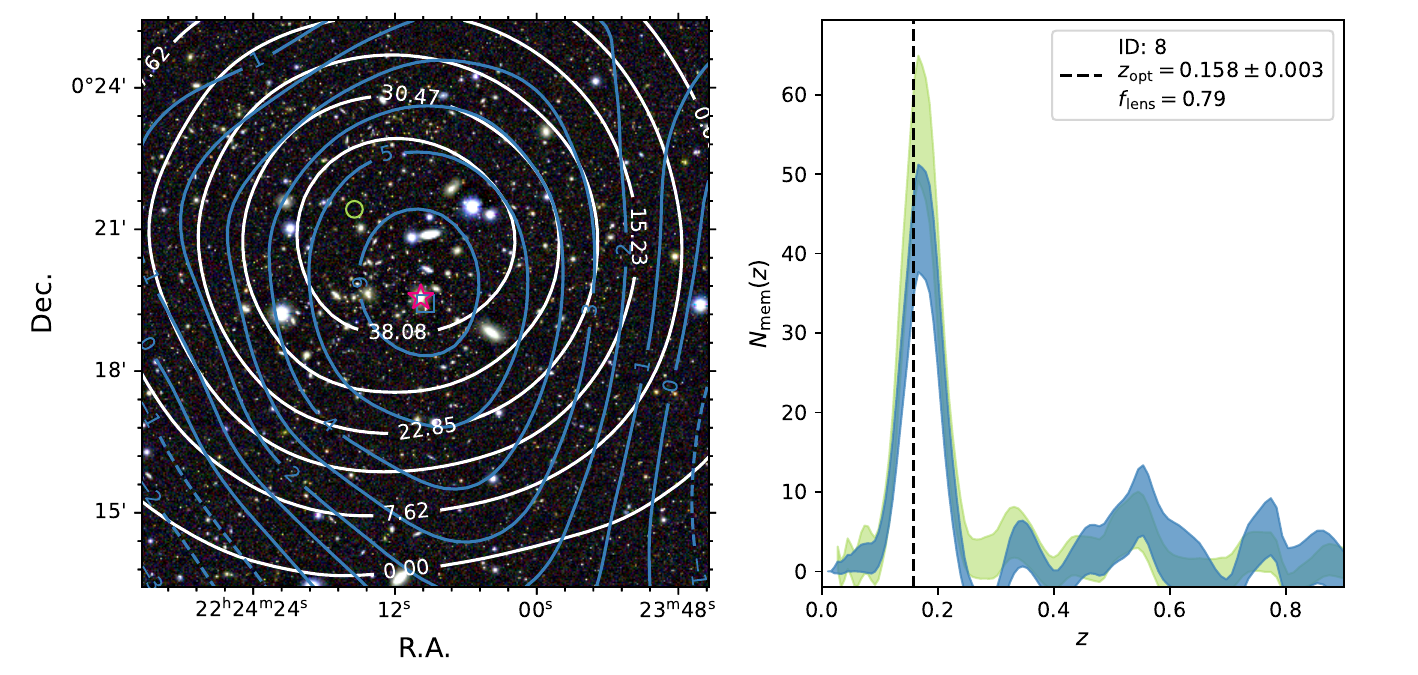}}
\resizebox{0.245\textwidth}{!}{\includegraphics[scale=1]{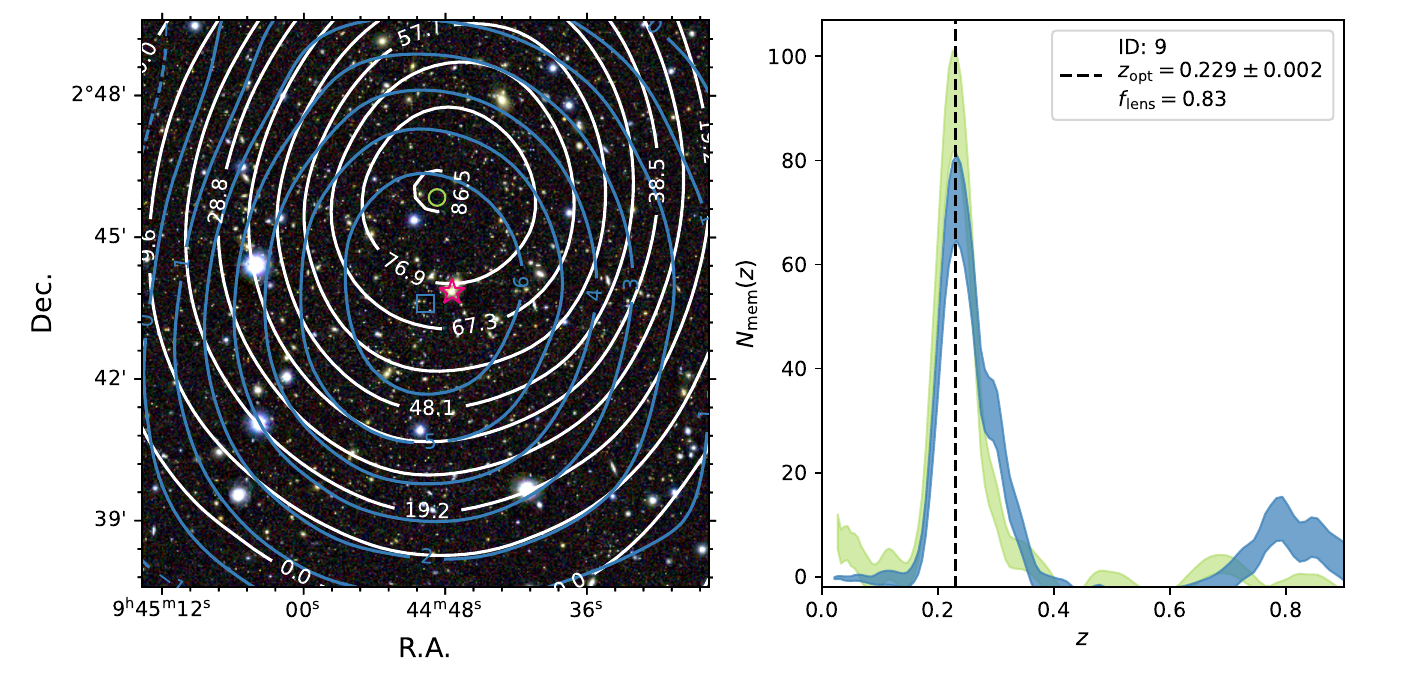}}
\resizebox{0.245\textwidth}{!}{\includegraphics[scale=1]{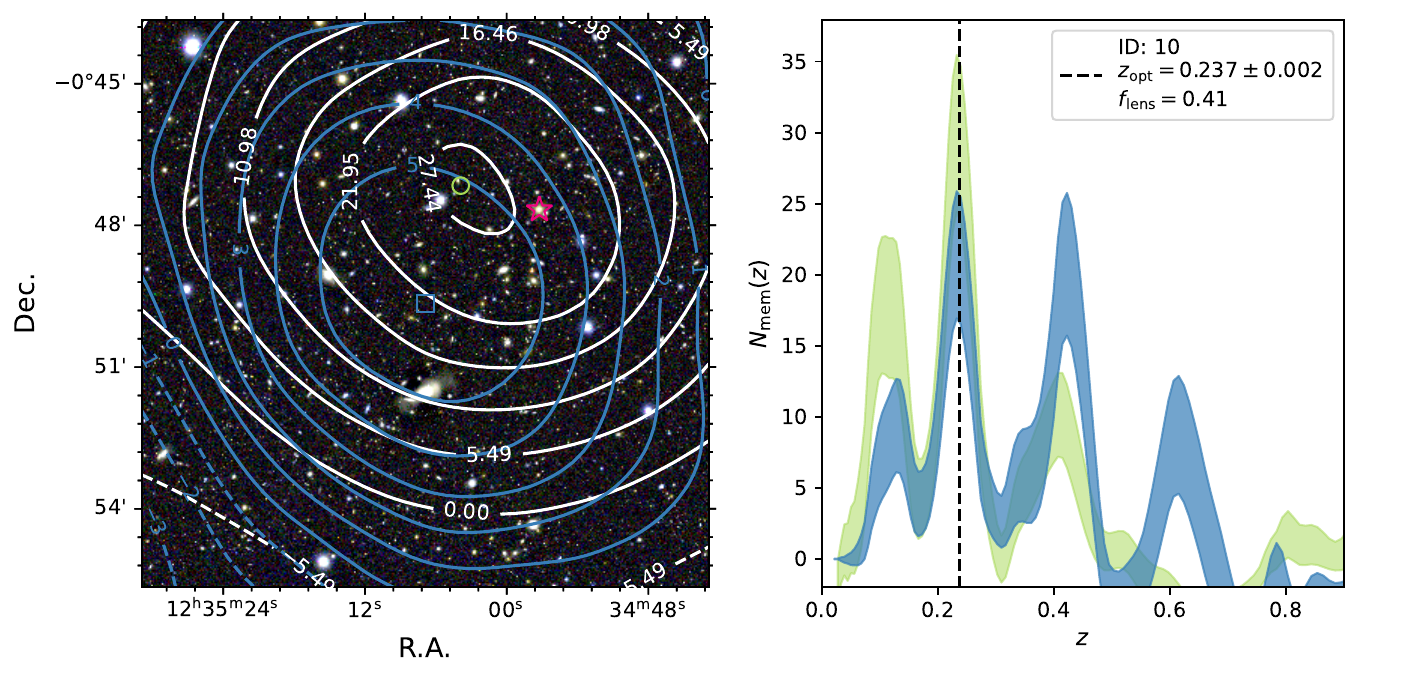}}
\resizebox{0.245\textwidth}{!}{\includegraphics[scale=1]{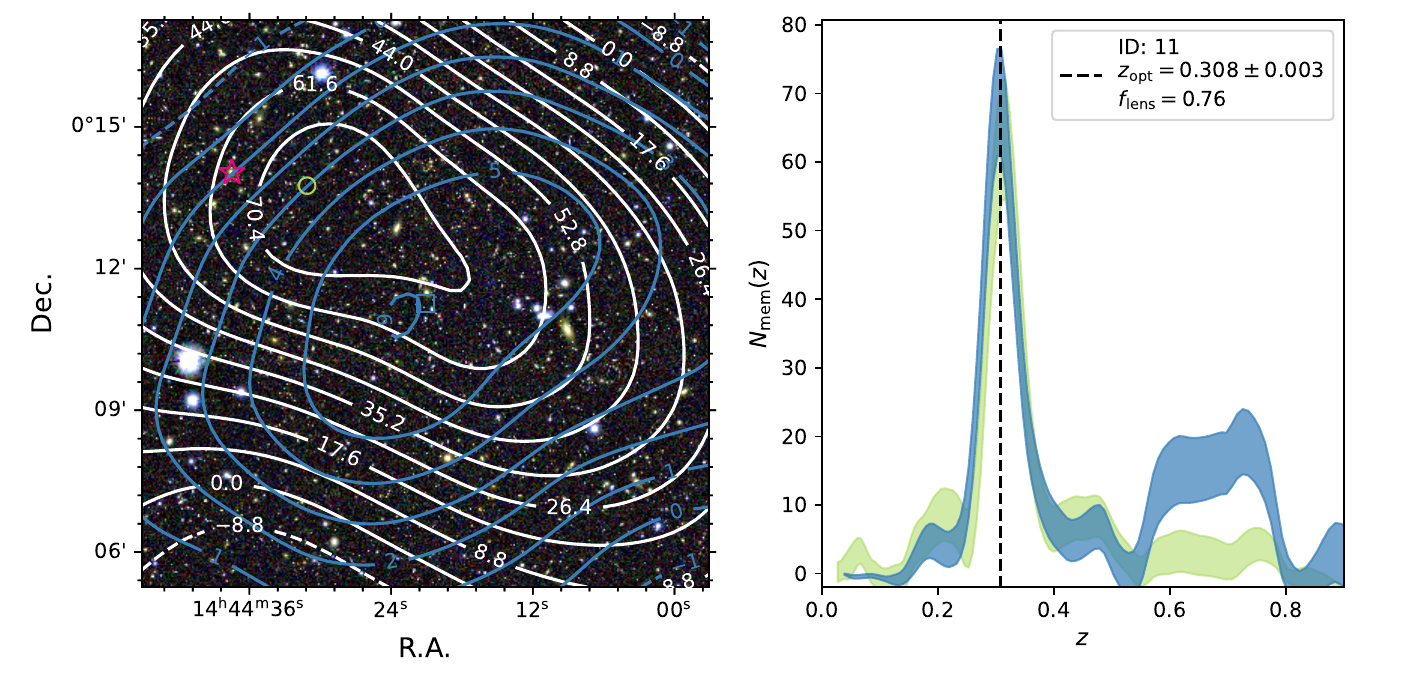}}
\resizebox{0.245\textwidth}{!}{\includegraphics[scale=1]{pplots/rgb/12_image.pdf}}
\resizebox{0.245\textwidth}{!}{\includegraphics[scale=1]{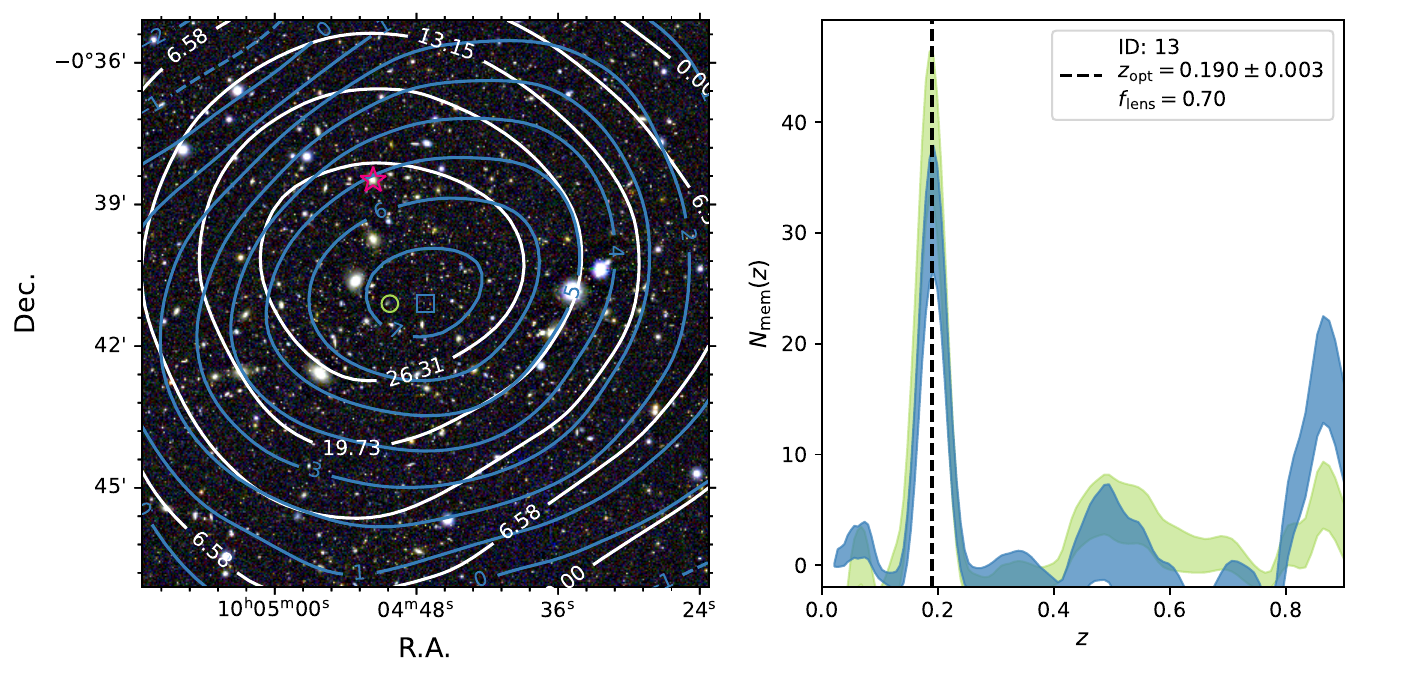}}
\resizebox{0.245\textwidth}{!}{\includegraphics[scale=1]{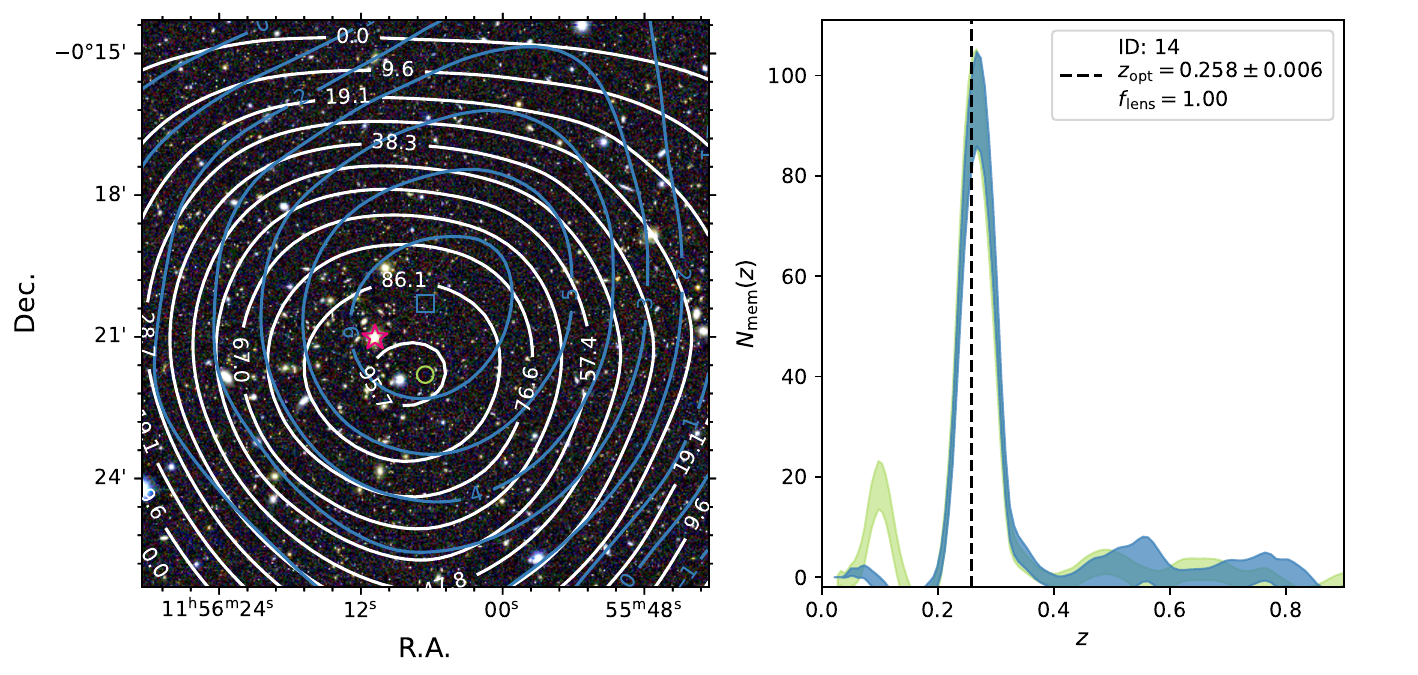}}
\resizebox{0.245\textwidth}{!}{\includegraphics[scale=1]{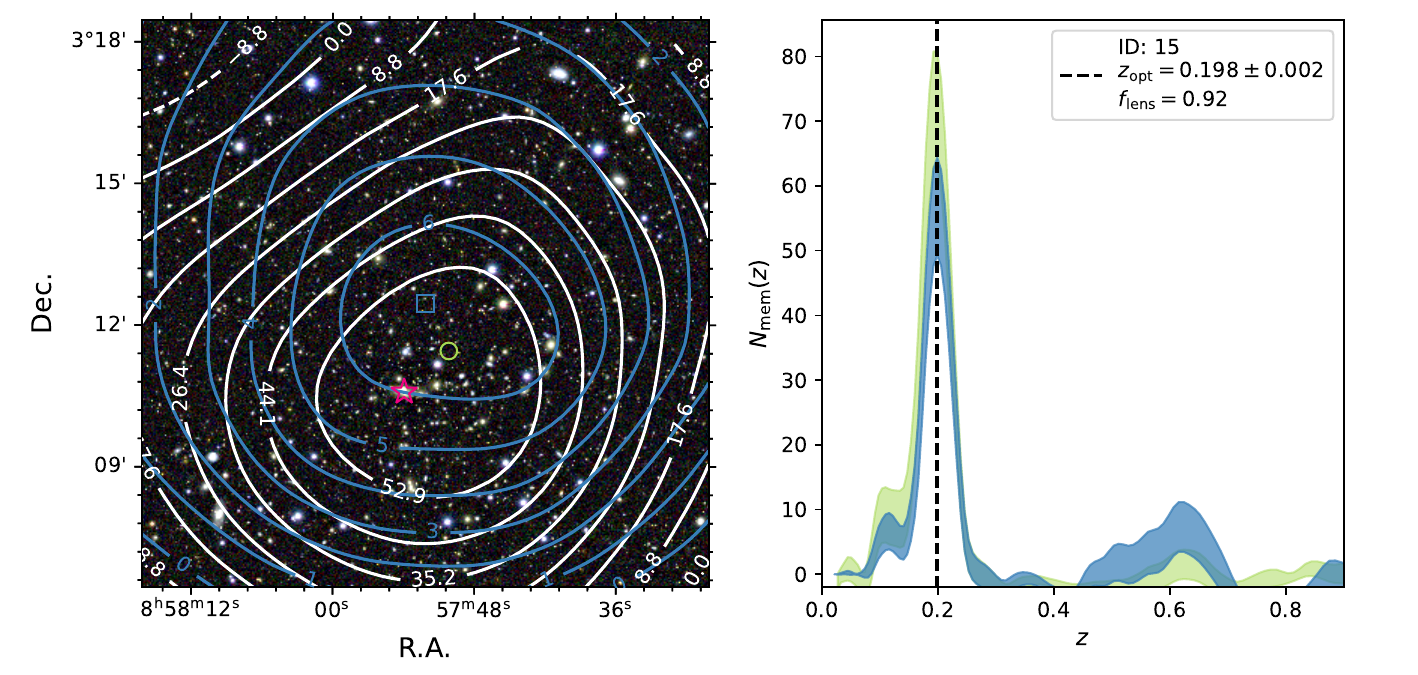}}
\resizebox{0.245\textwidth}{!}{\includegraphics[scale=1]{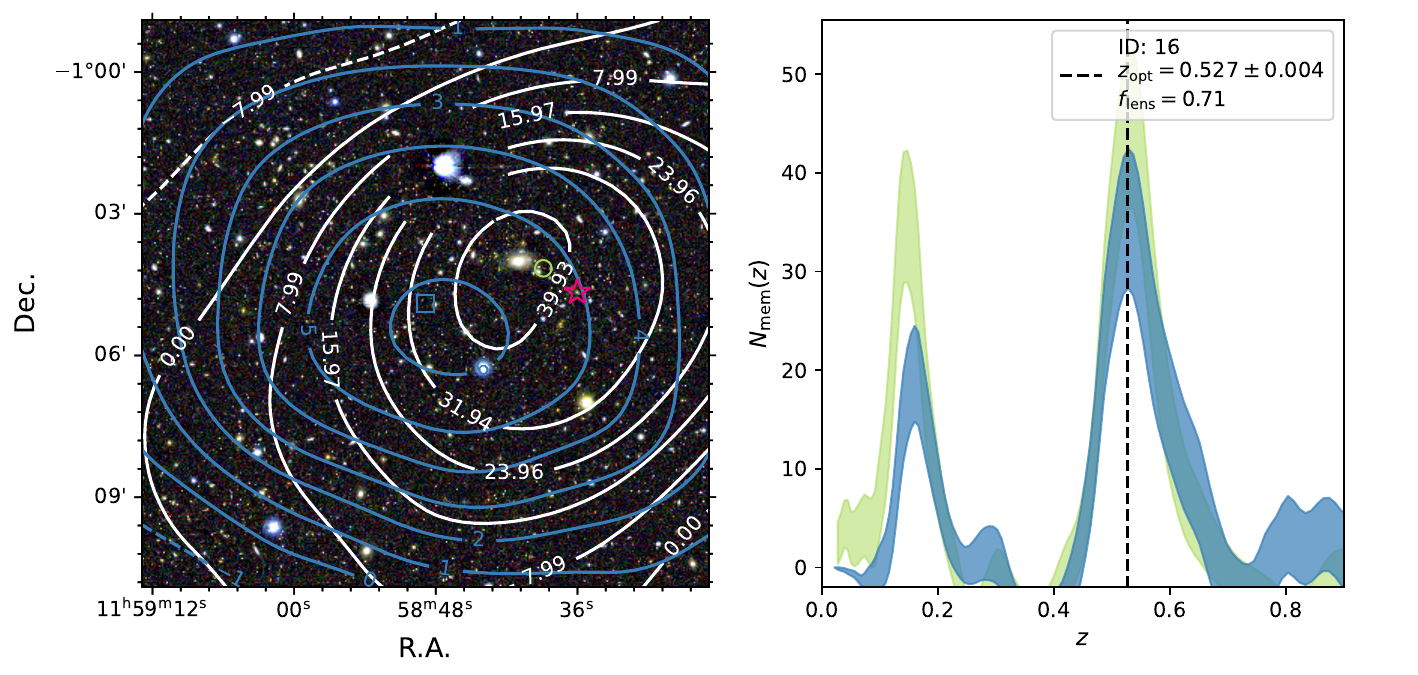}}
\resizebox{0.245\textwidth}{!}{\includegraphics[scale=1]{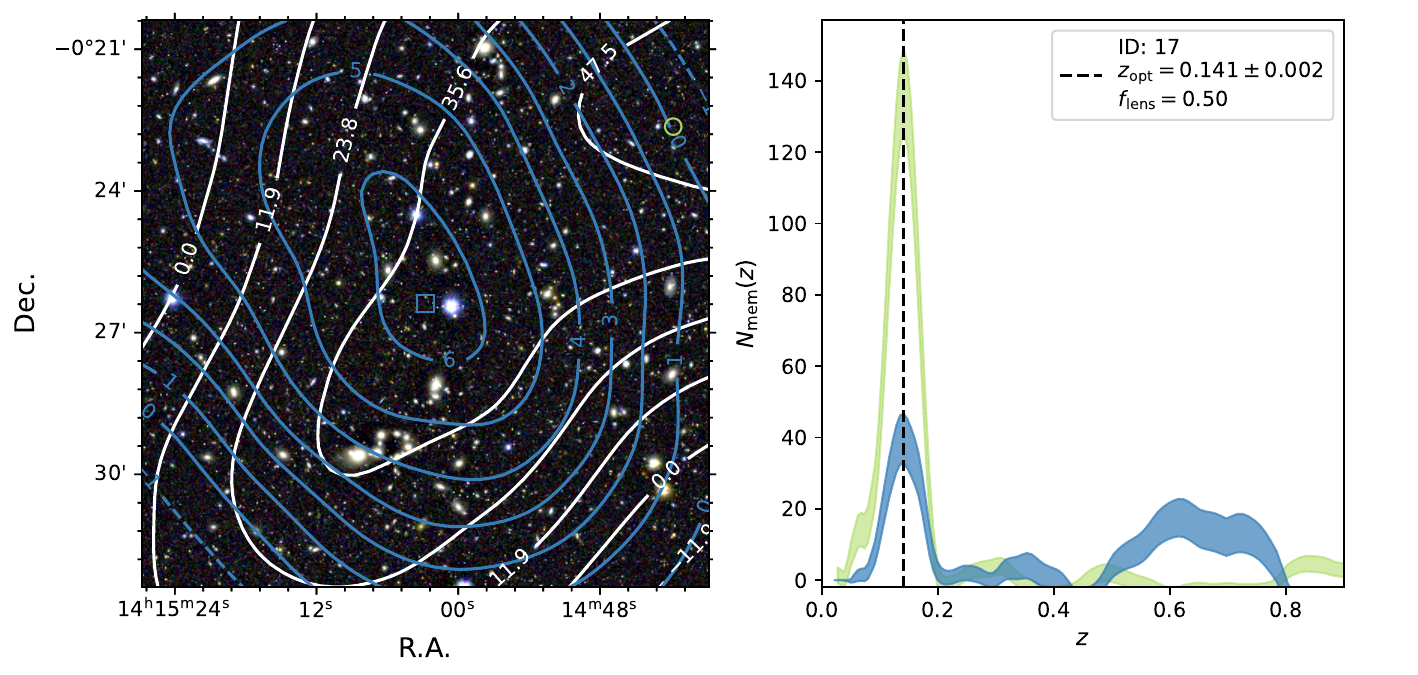}}
\resizebox{0.245\textwidth}{!}{\includegraphics[scale=1]{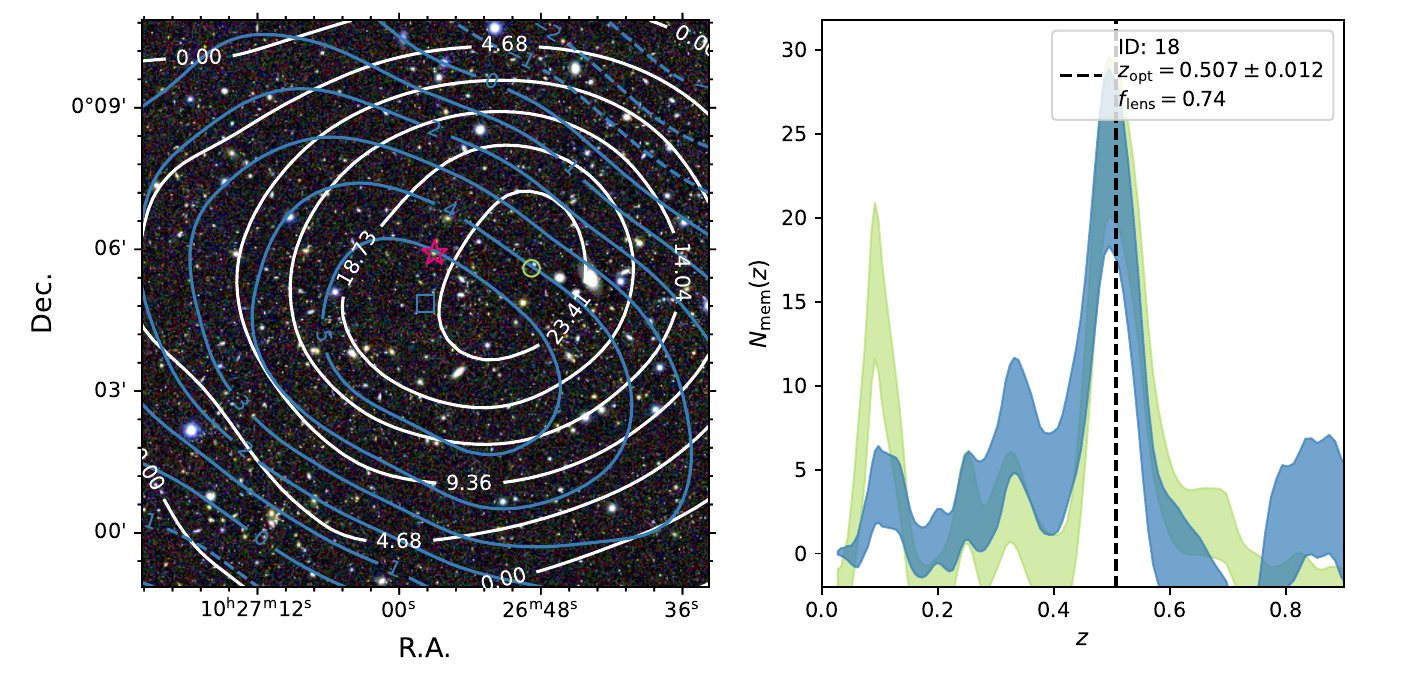}}
\resizebox{0.245\textwidth}{!}{\includegraphics[scale=1]{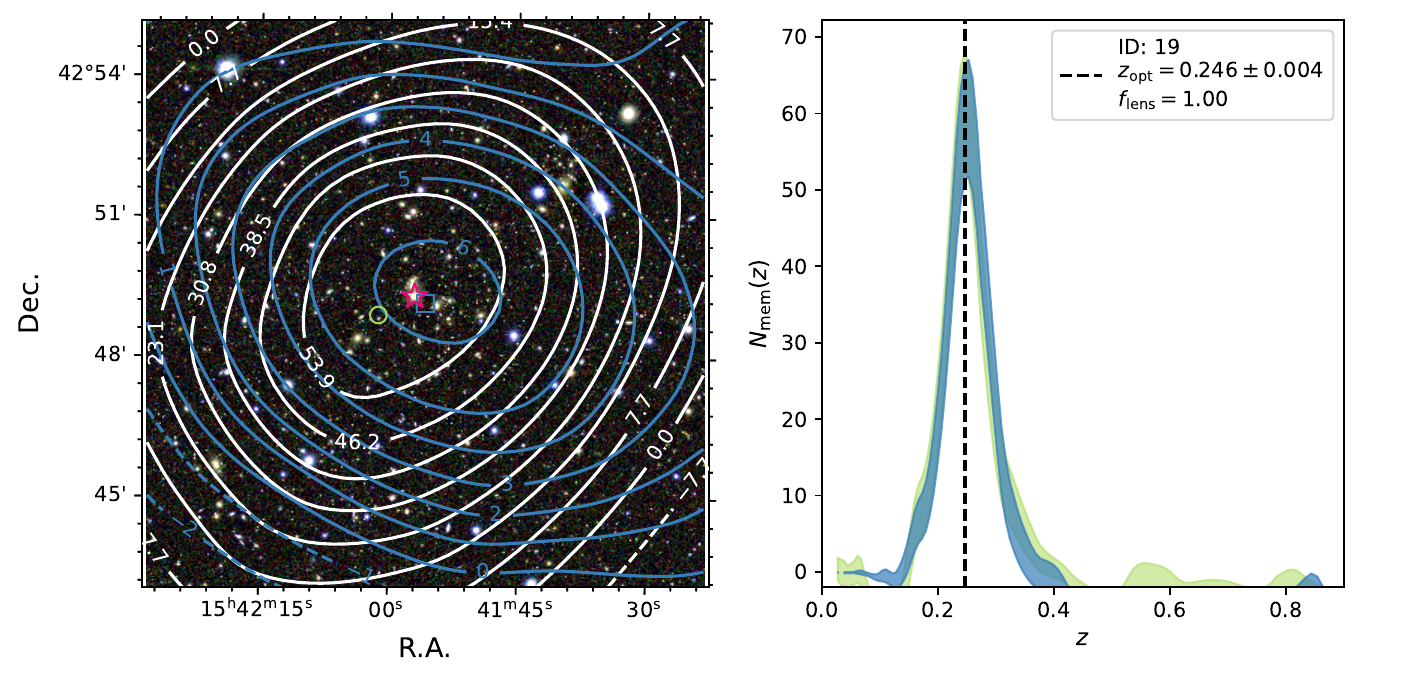}}
\resizebox{0.245\textwidth}{!}{\includegraphics[scale=1]{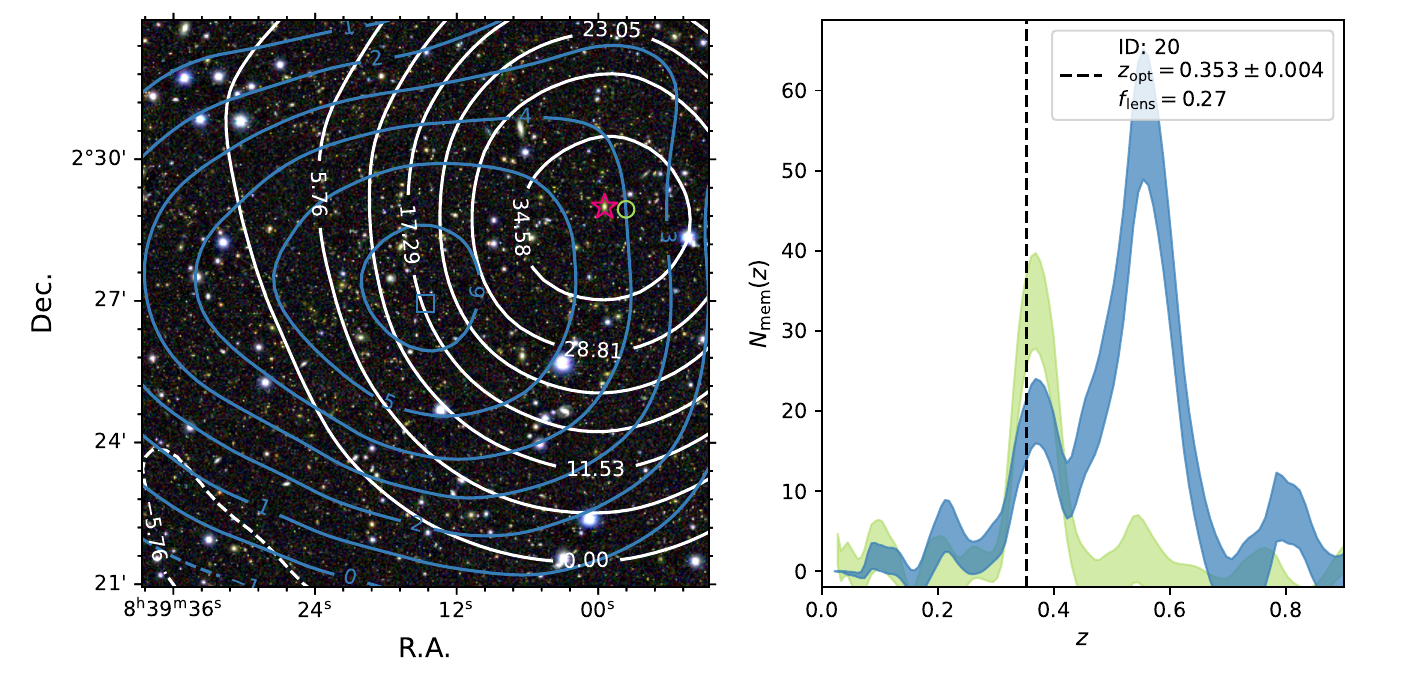}}
\resizebox{0.245\textwidth}{!}{\includegraphics[scale=1]{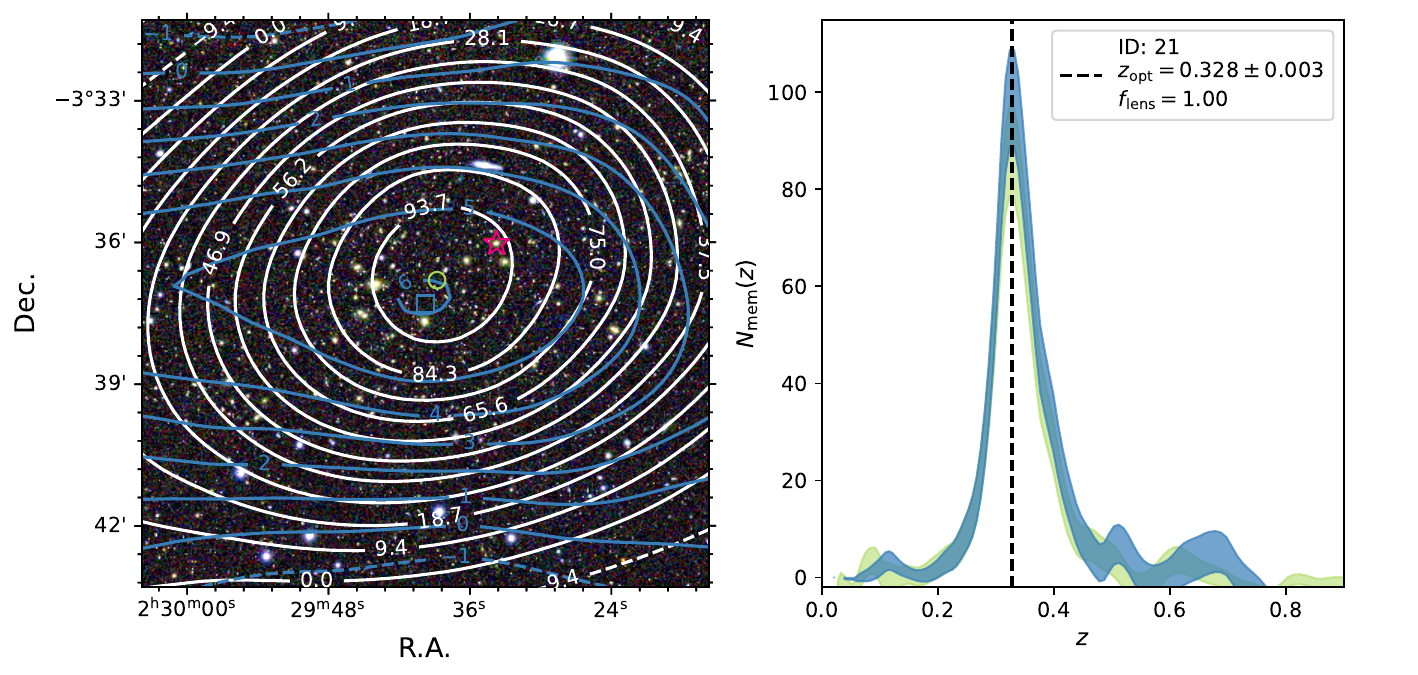}}
\resizebox{0.245\textwidth}{!}{\includegraphics[scale=1]{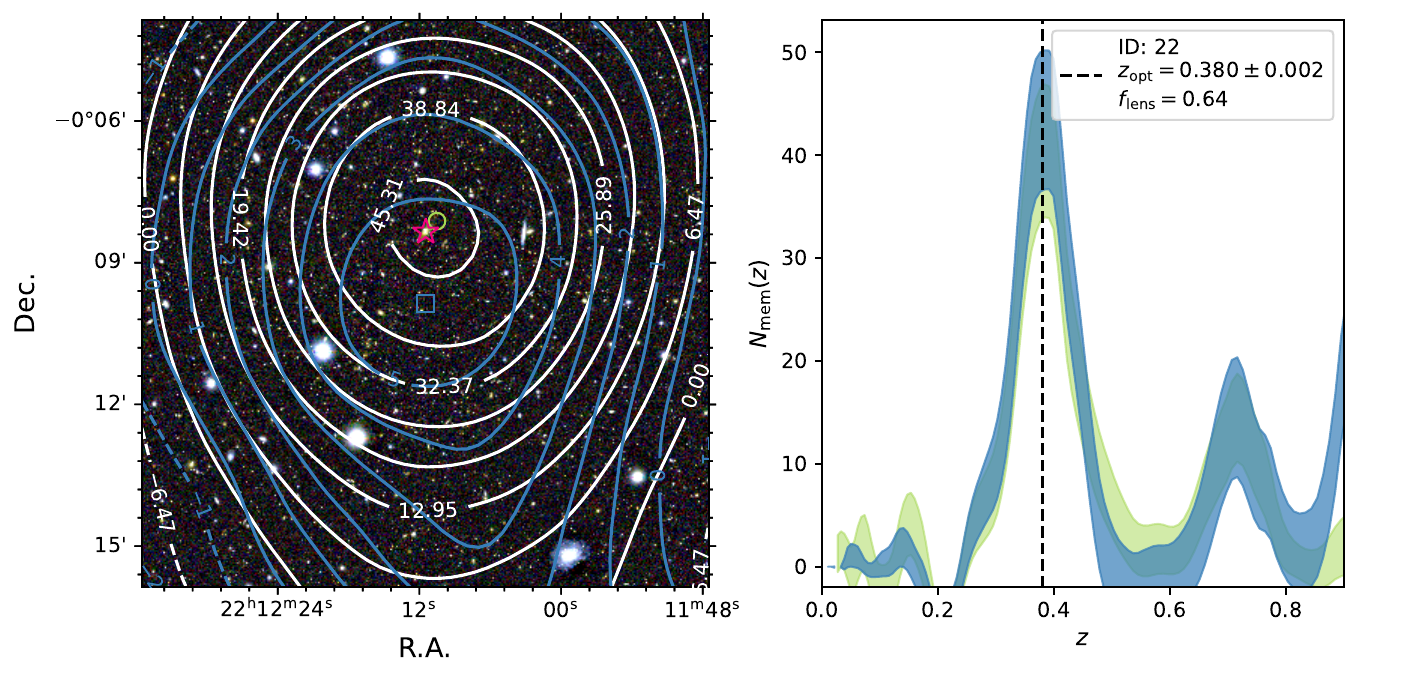}}
\resizebox{0.245\textwidth}{!}{\includegraphics[scale=1]{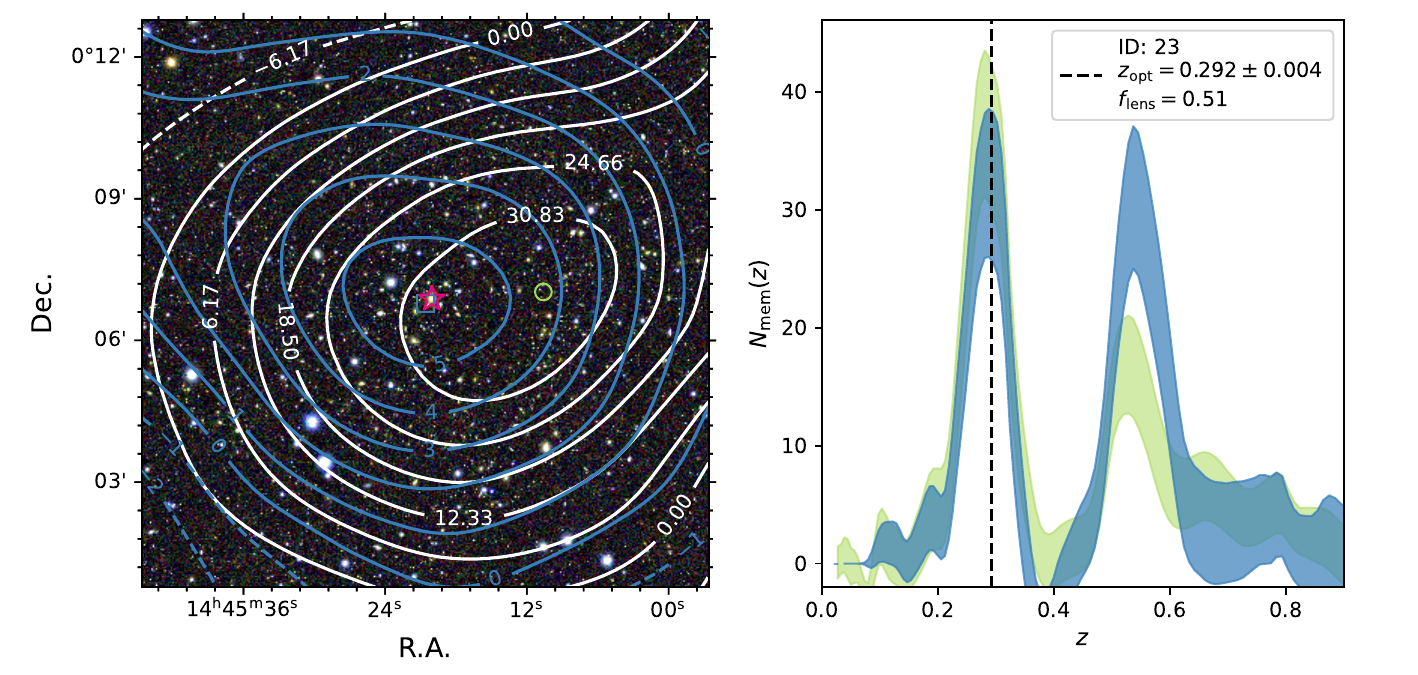}}
\resizebox{0.245\textwidth}{!}{\includegraphics[scale=1]{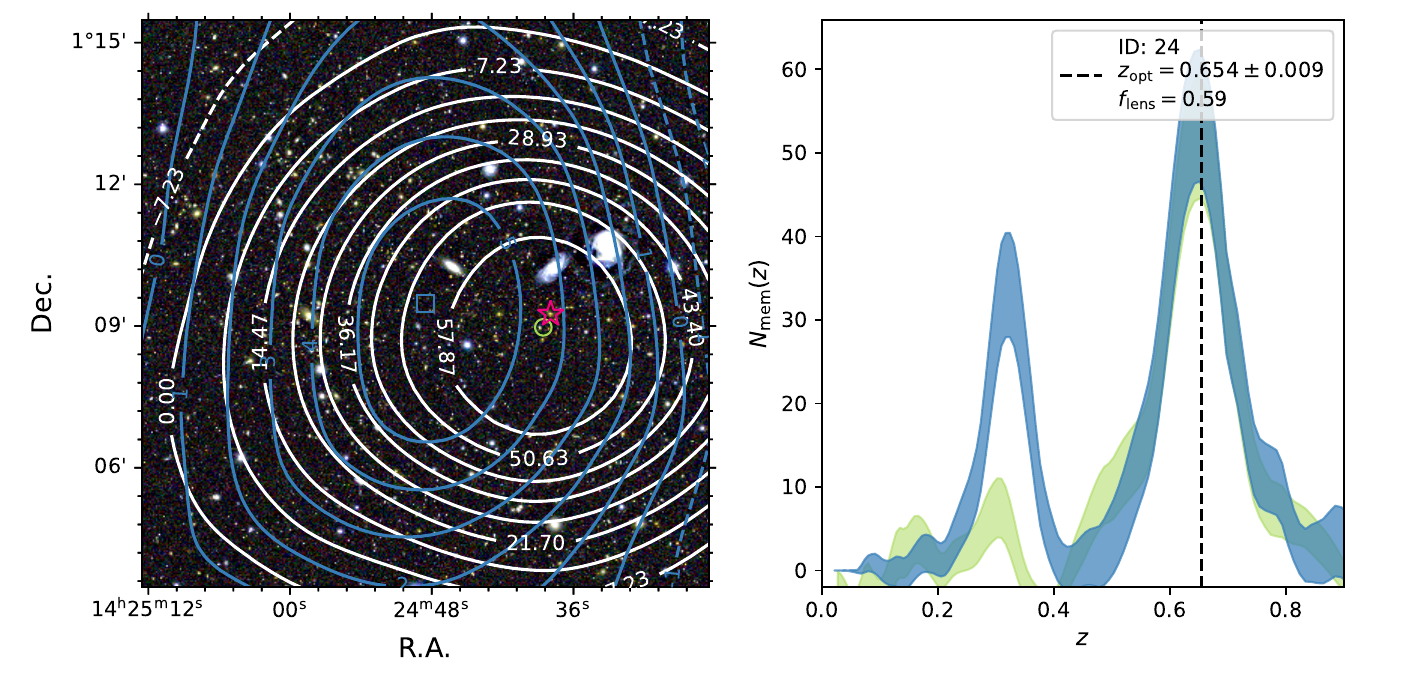}}
\resizebox{0.245\textwidth}{!}{\includegraphics[scale=1]{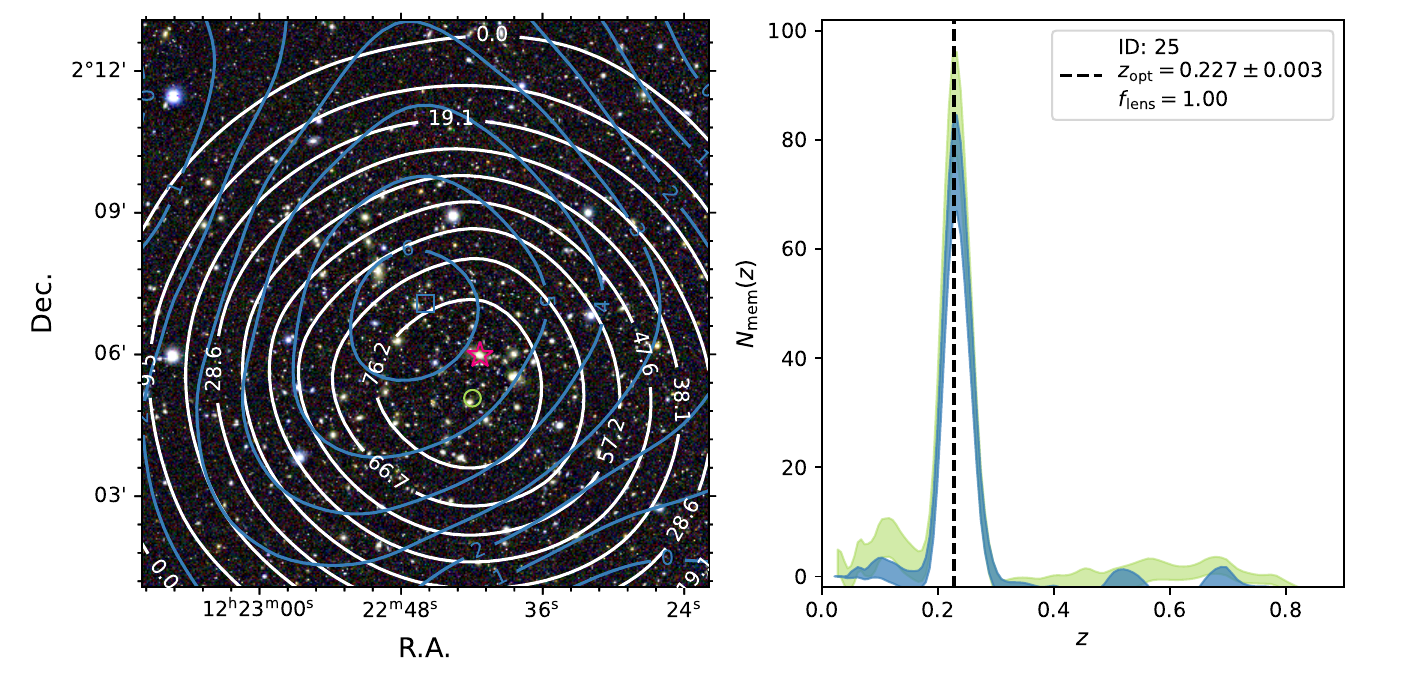}}
\resizebox{0.245\textwidth}{!}{\includegraphics[scale=1]{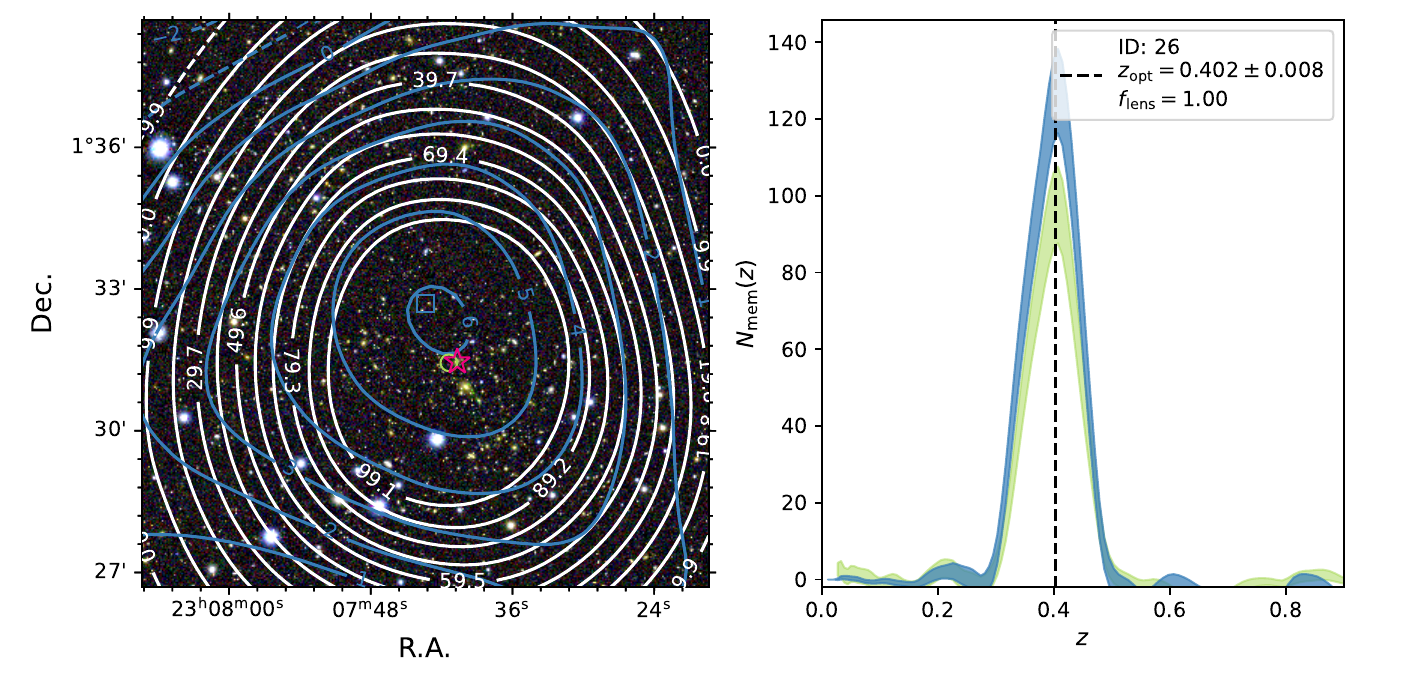}}
\resizebox{0.245\textwidth}{!}{\includegraphics[scale=1]{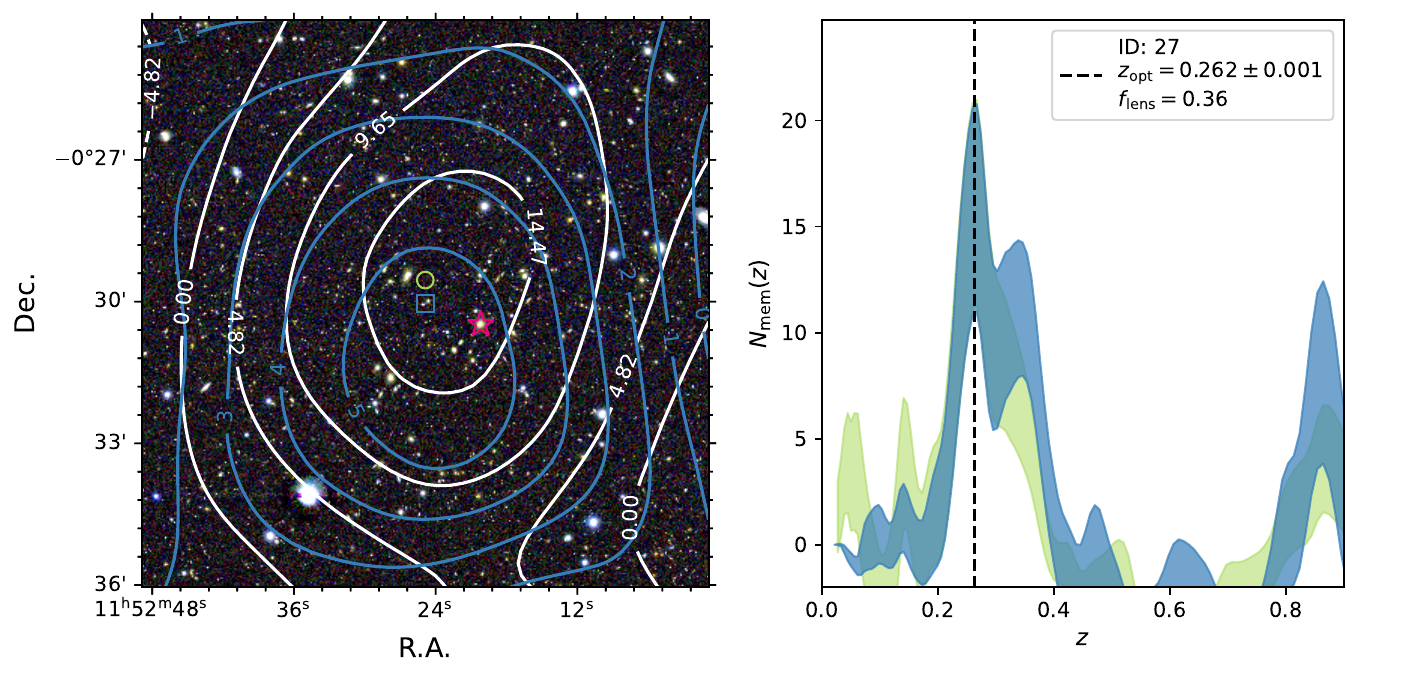}}
\resizebox{0.245\textwidth}{!}{\includegraphics[scale=1]{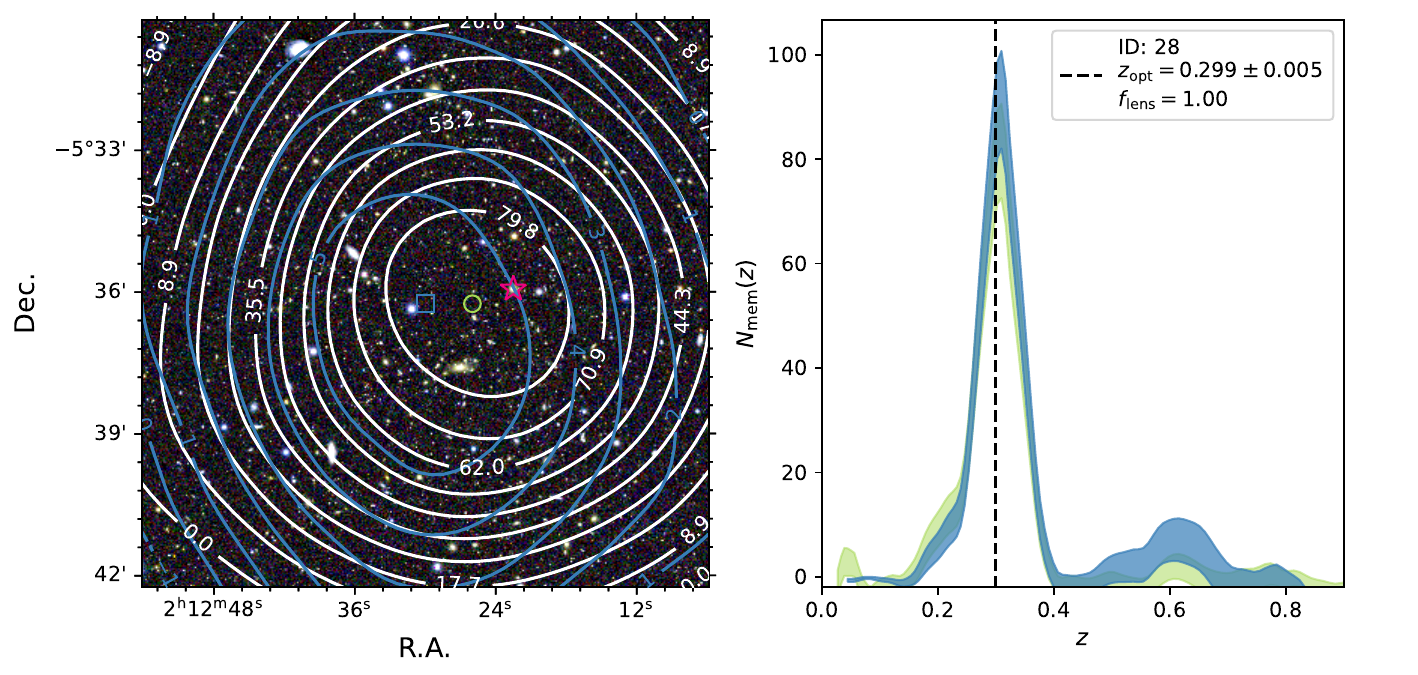}}
\resizebox{0.245\textwidth}{!}{\includegraphics[scale=1]{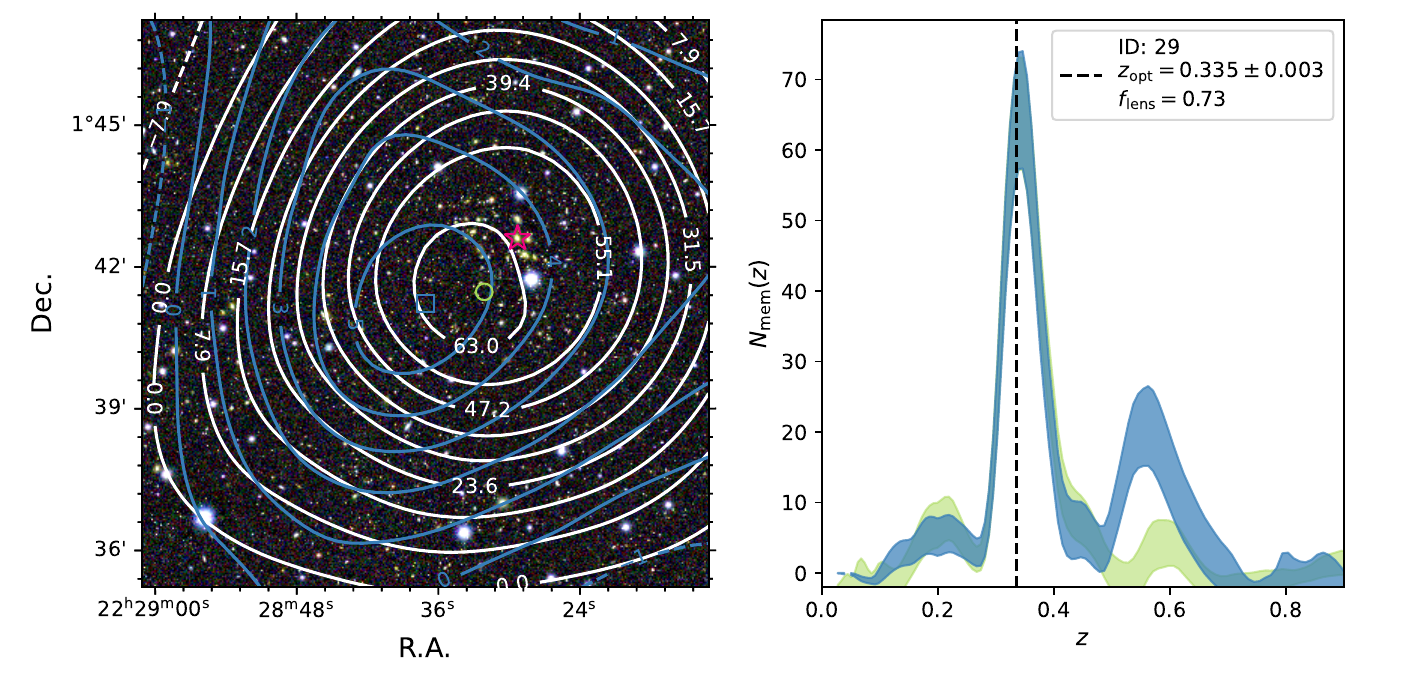}}
\resizebox{0.245\textwidth}{!}{\includegraphics[scale=1]{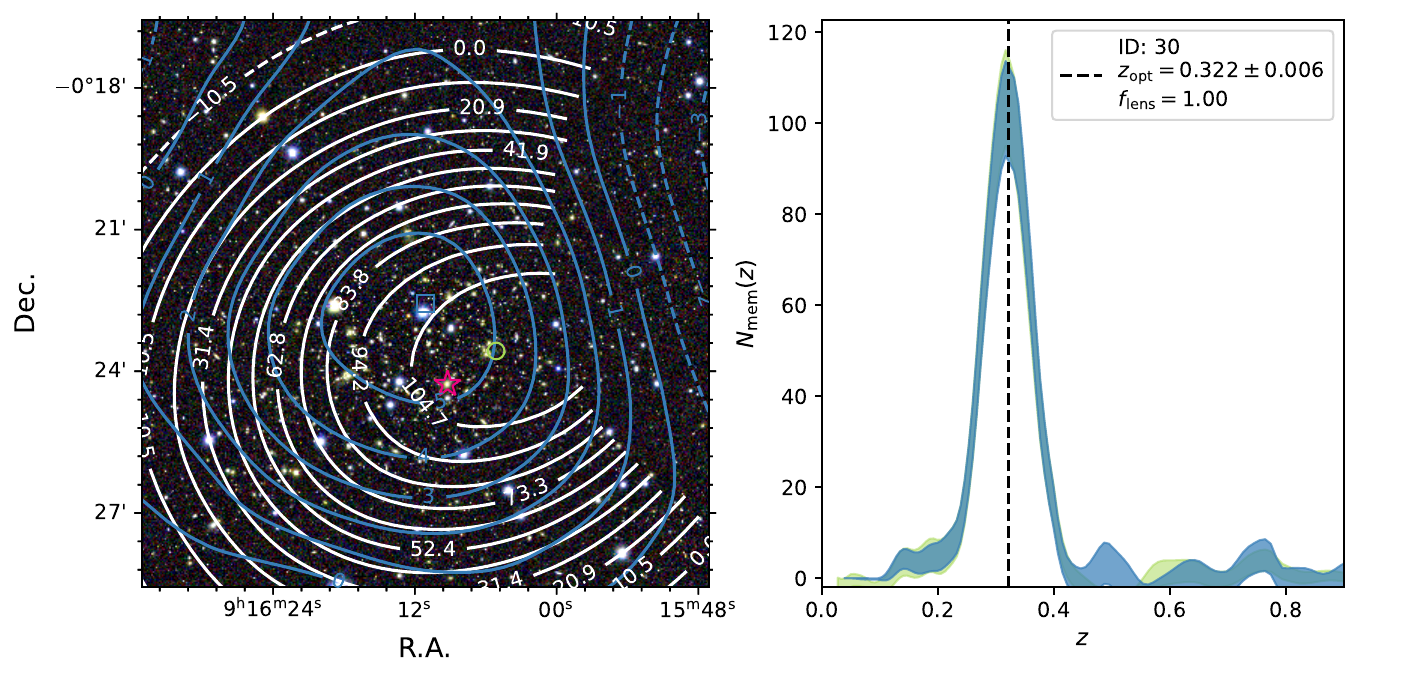}}
\resizebox{0.245\textwidth}{!}{\includegraphics[scale=1]{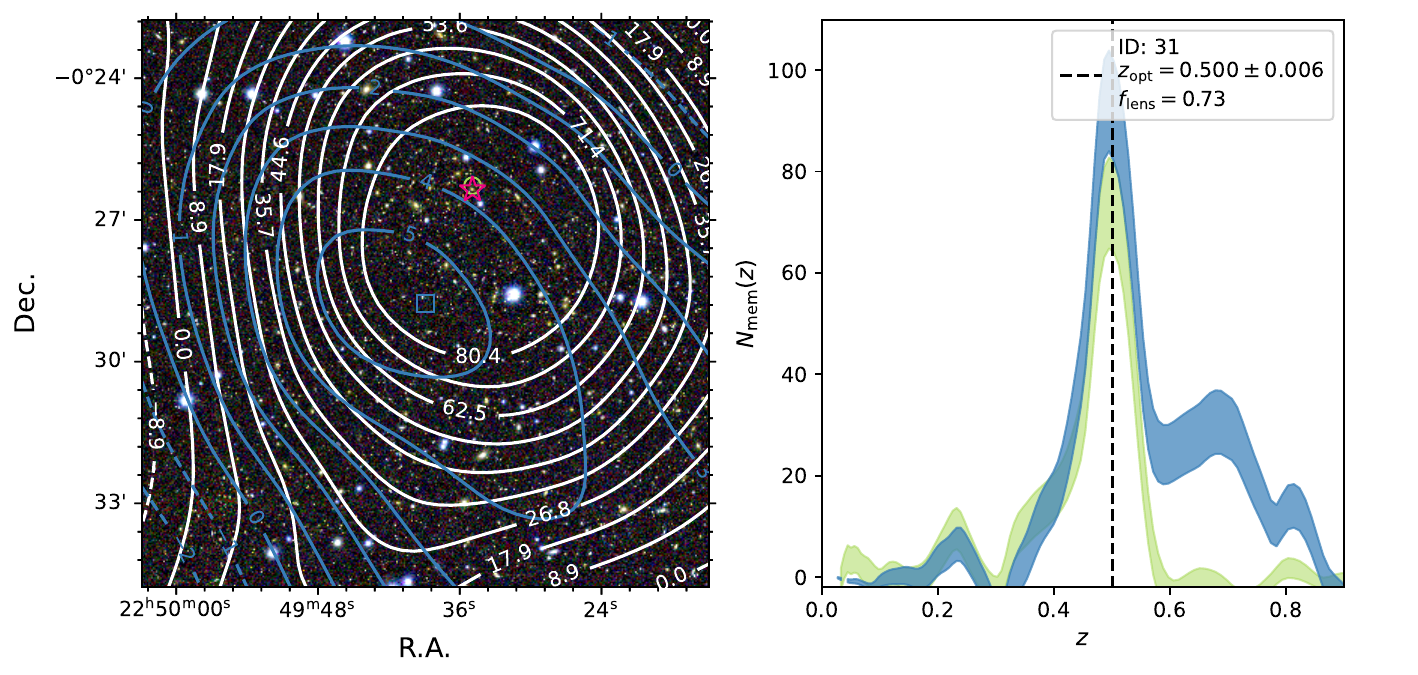}}
\resizebox{0.245\textwidth}{!}{\includegraphics[scale=1]{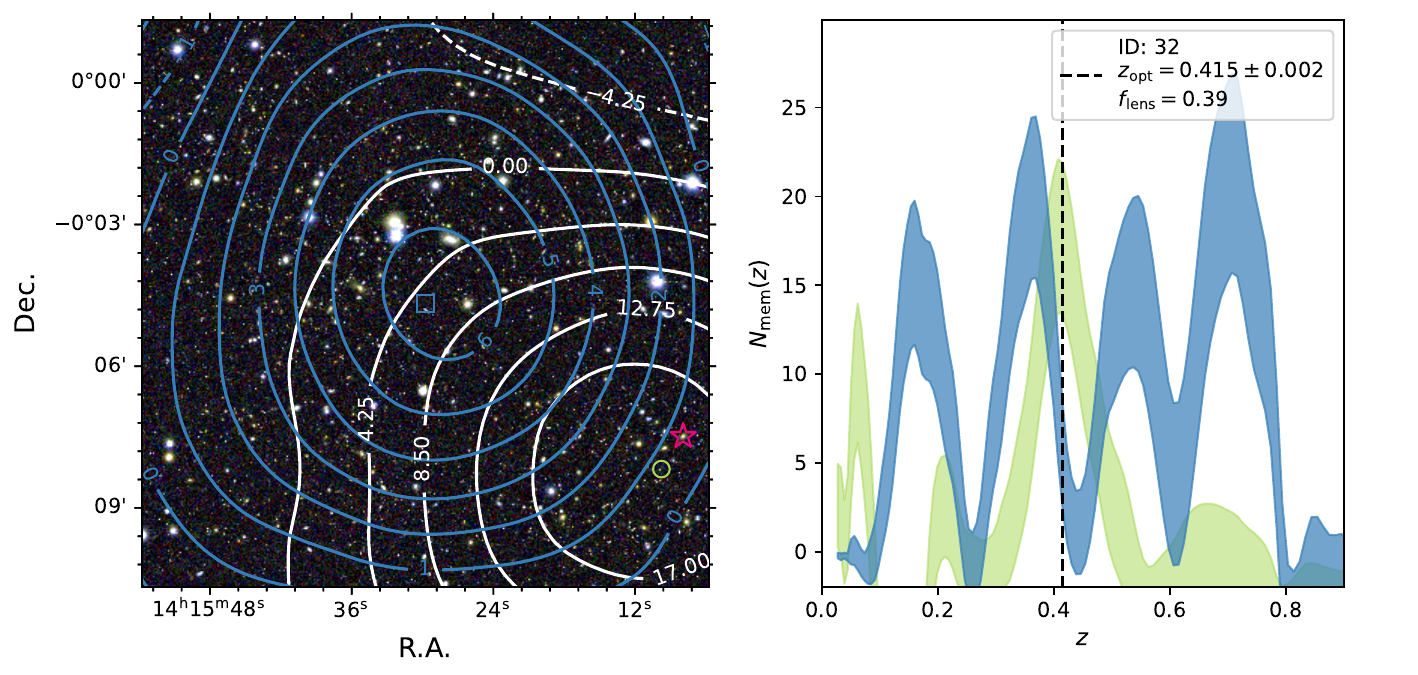}}
\resizebox{0.245\textwidth}{!}{\includegraphics[scale=1]{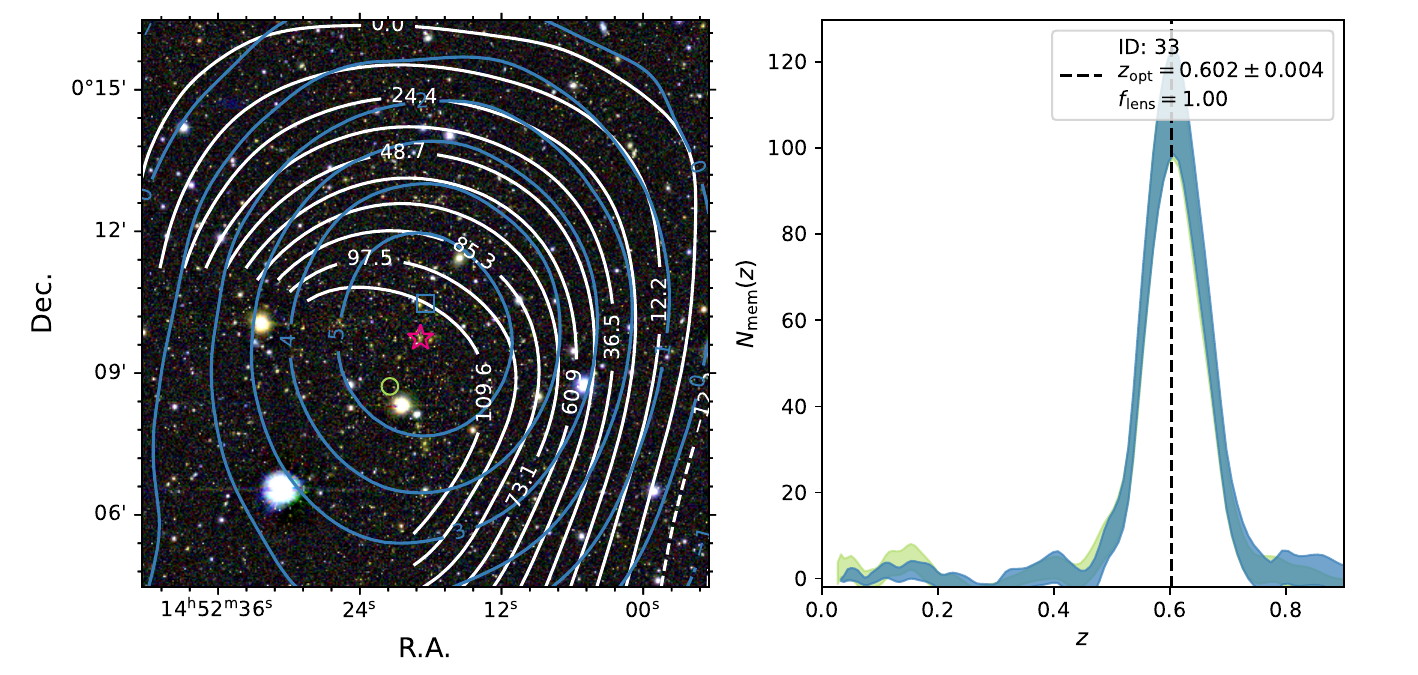}}
\resizebox{0.245\textwidth}{!}{\includegraphics[scale=1]{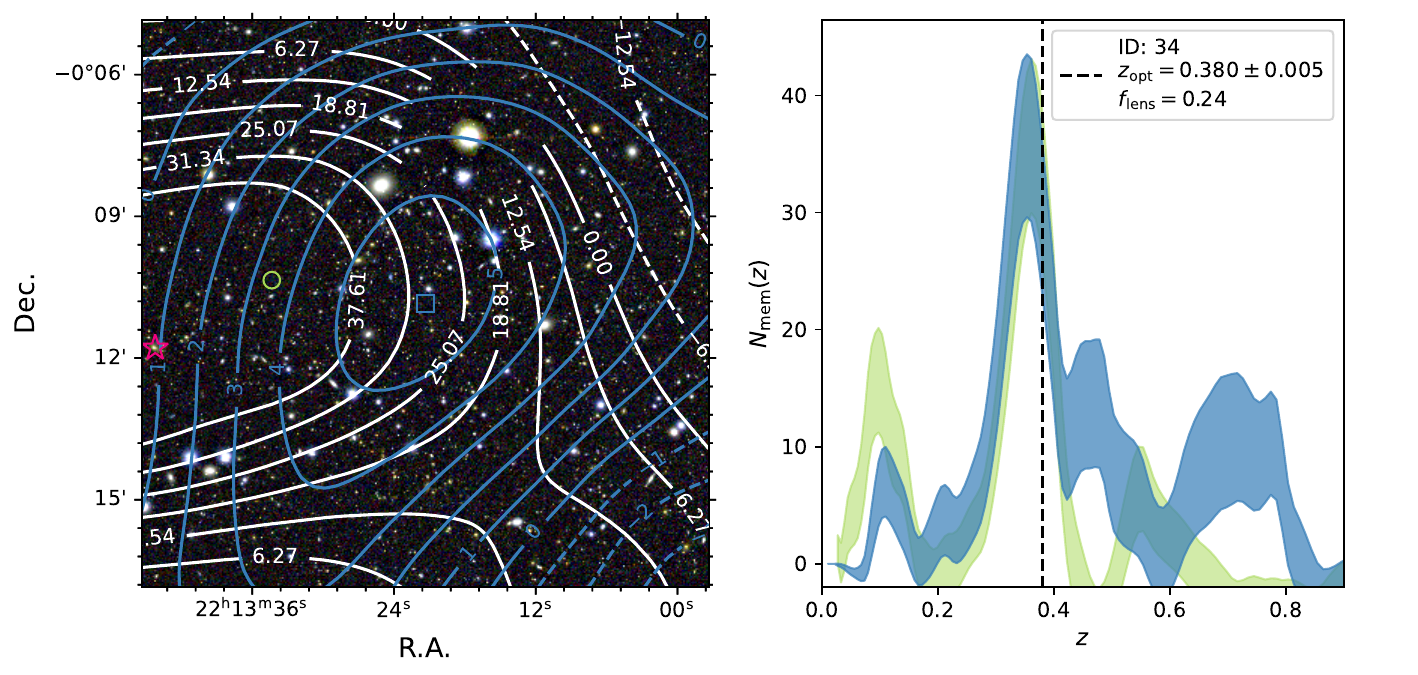}}
\resizebox{0.245\textwidth}{!}{\includegraphics[scale=1]{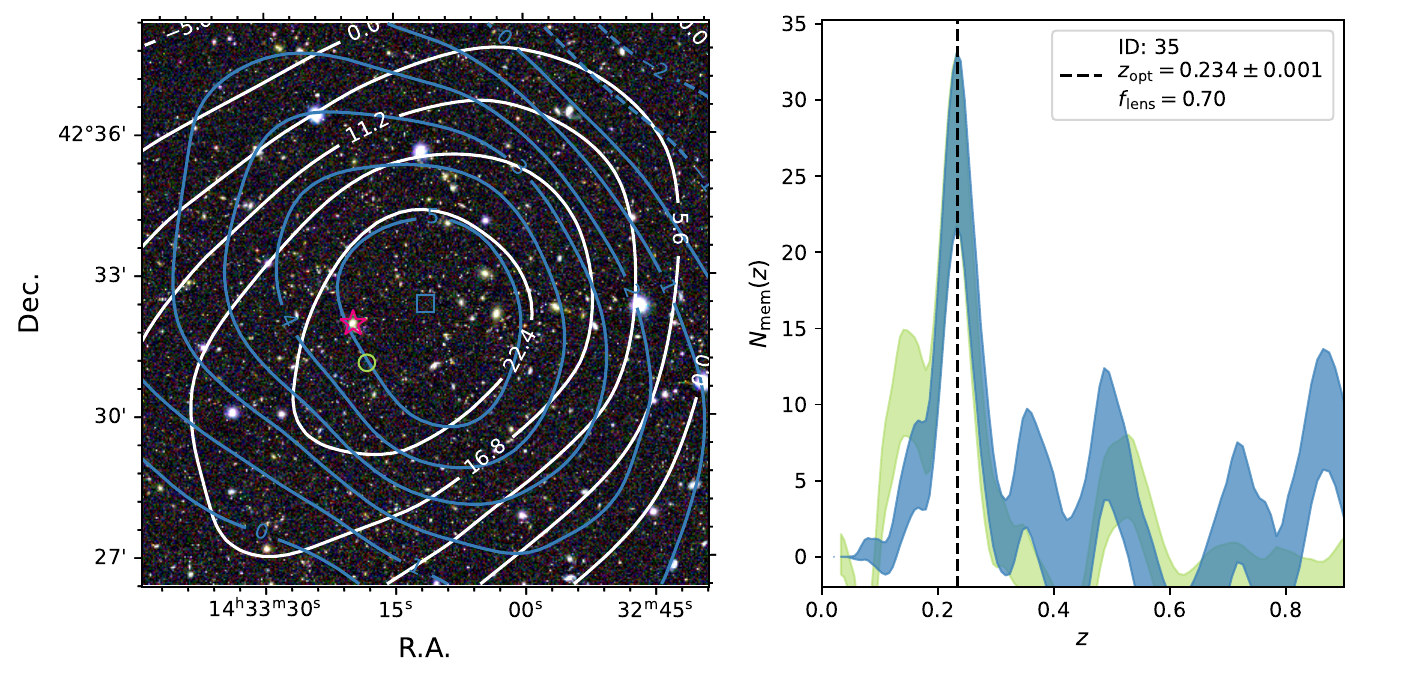}}
\resizebox{0.245\textwidth}{!}{\includegraphics[scale=1]{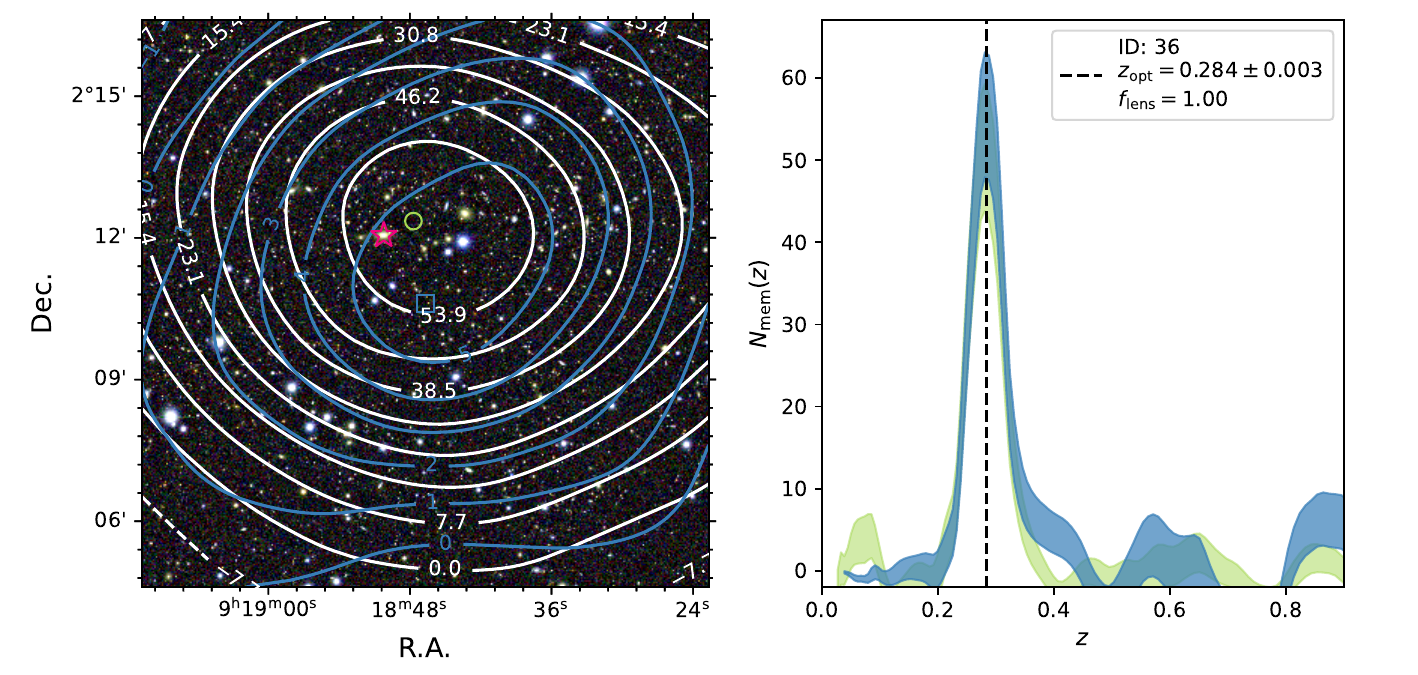}}
\resizebox{0.245\textwidth}{!}{\includegraphics[scale=1]{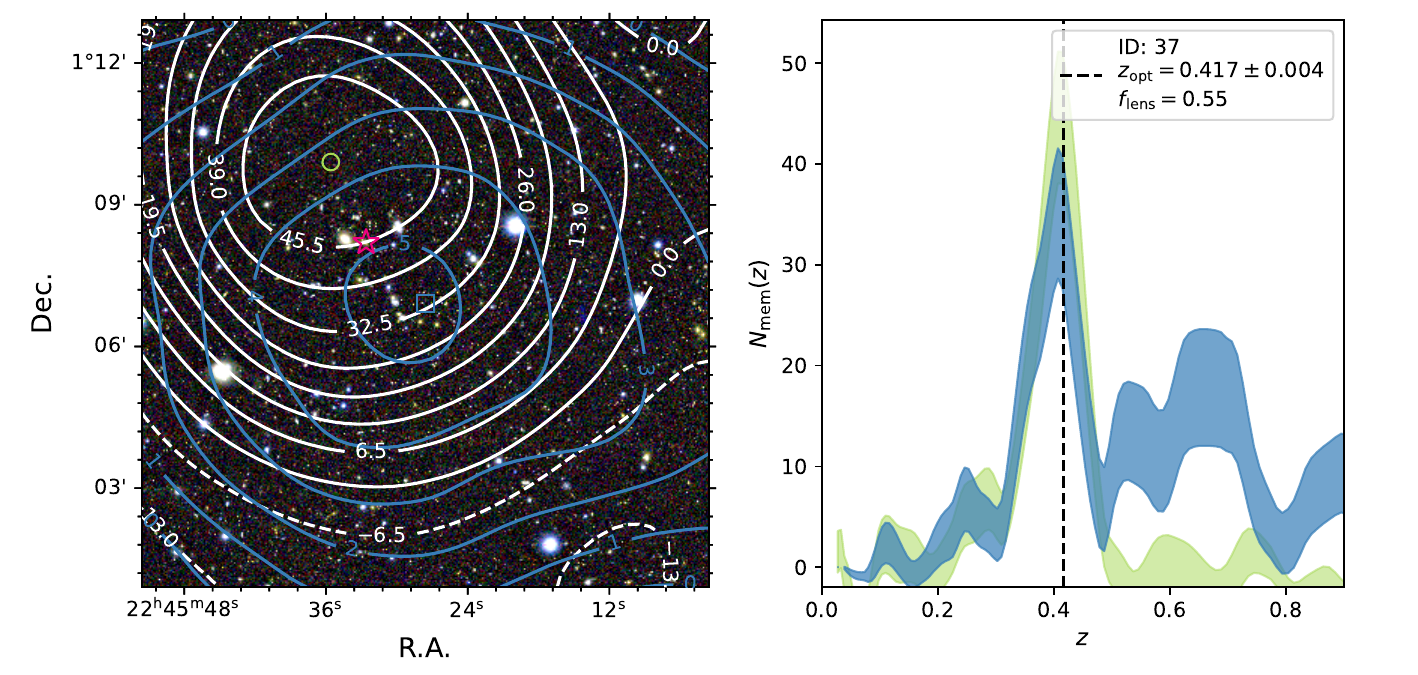}}
\resizebox{0.245\textwidth}{!}{\includegraphics[scale=1]{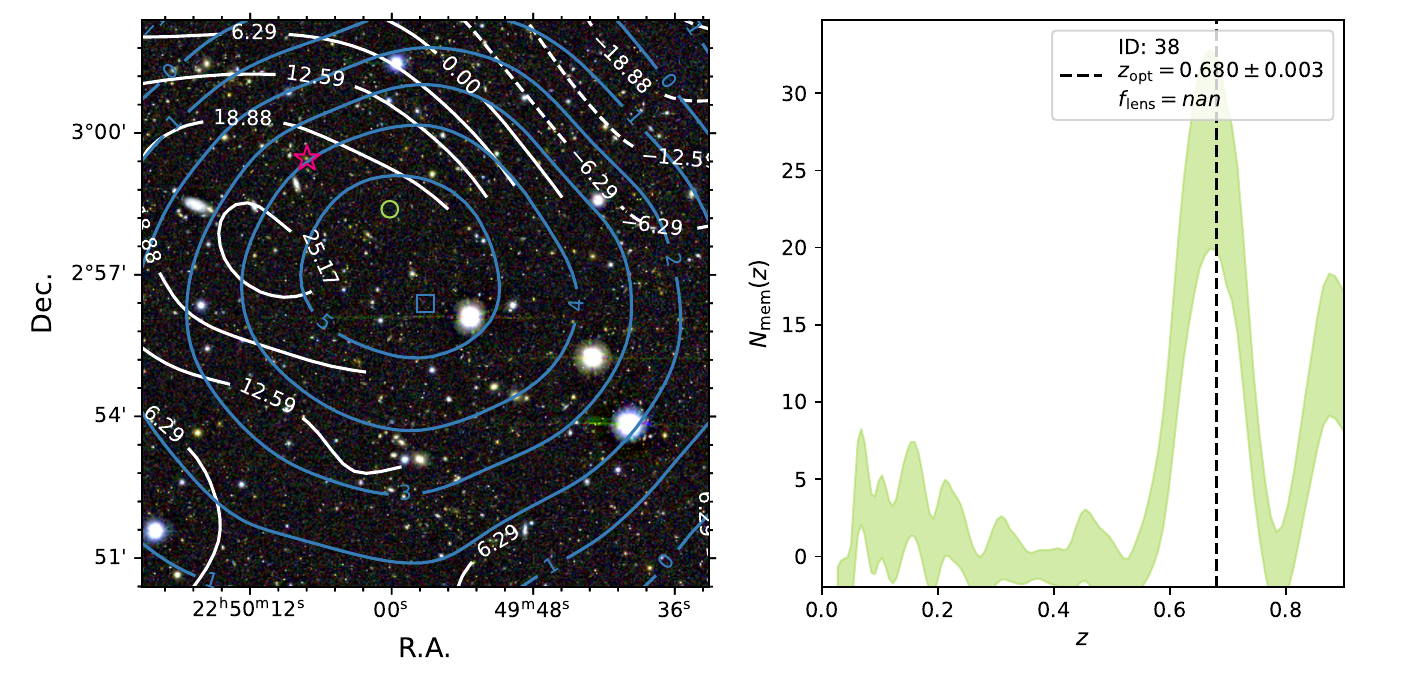}}
\resizebox{0.245\textwidth}{!}{\includegraphics[scale=1]{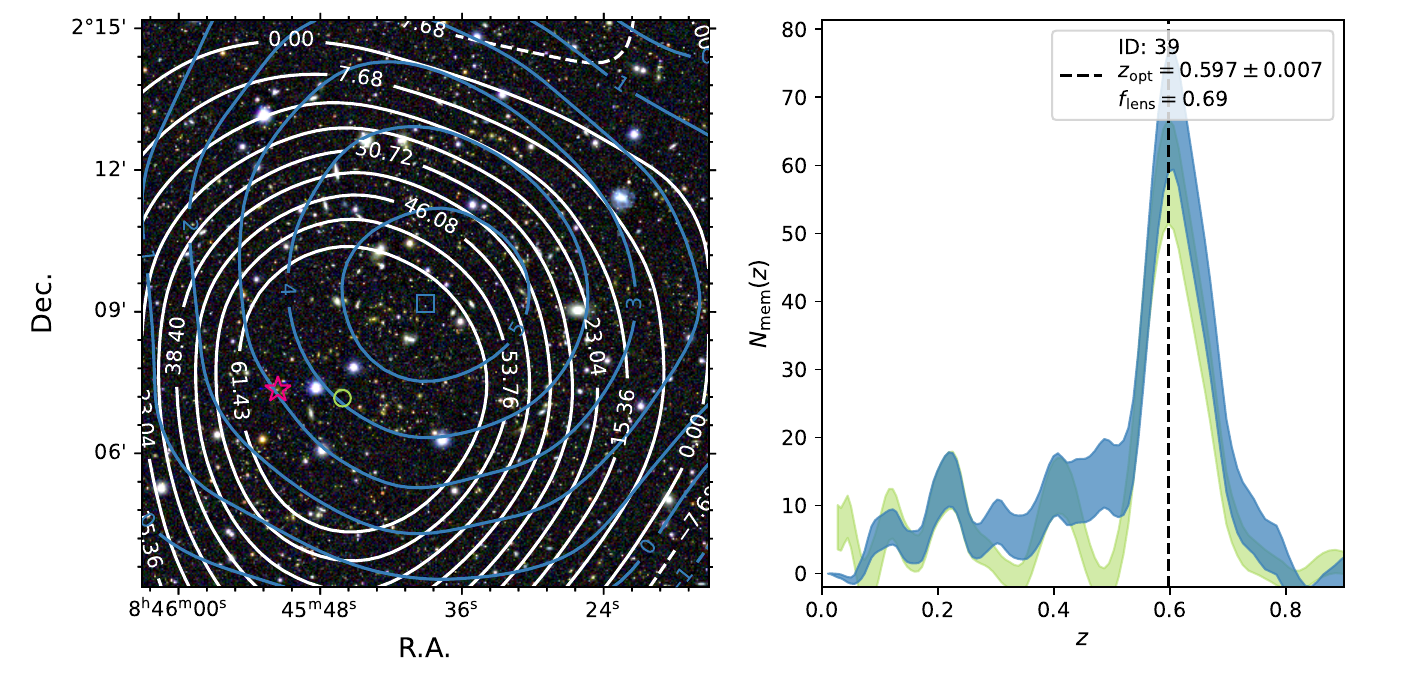}}
\caption{
The cutout RGB images of the individual WL shear-selected clusters.
Each subplot is made following the same manner as in Figure~\ref{fig:rgb_example}.
}
\label{fig:optical_images_0}
\end{figure*}
\begin{figure*}
\centering
\resizebox{0.245\textwidth}{!}{\includegraphics[scale=1]{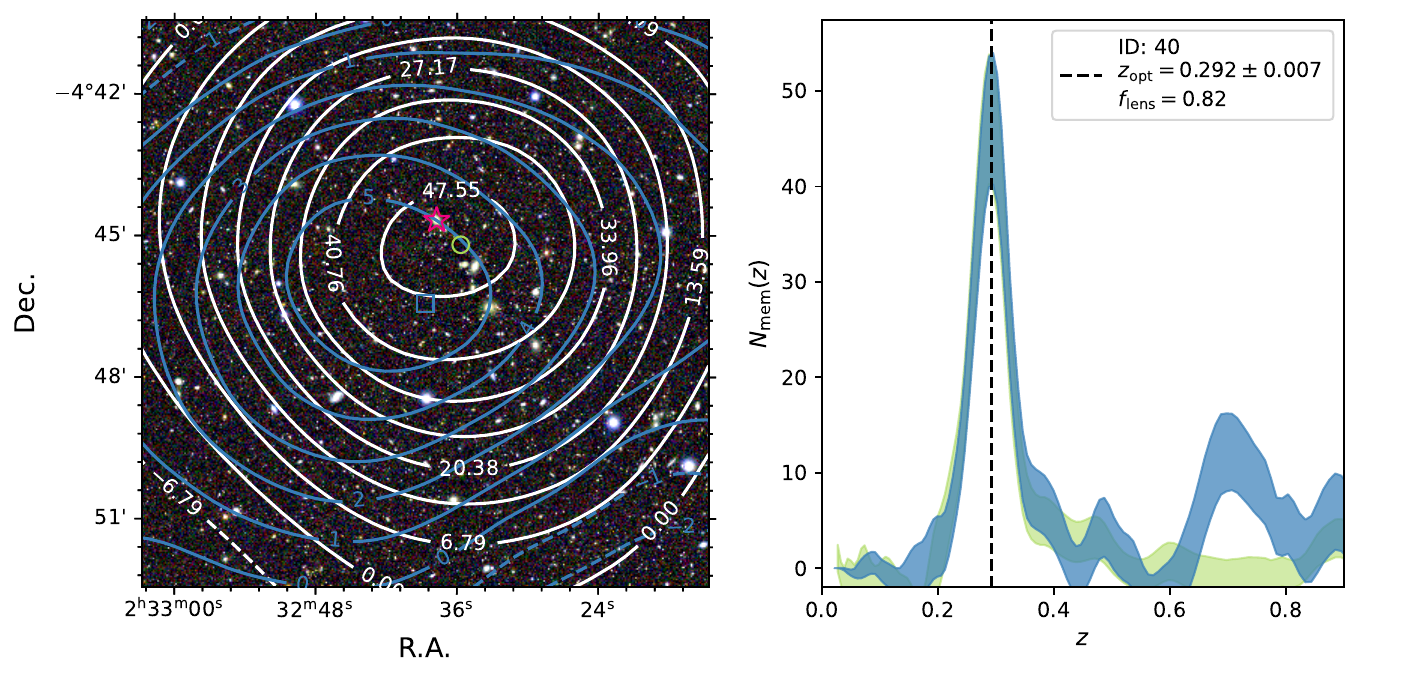}}
\resizebox{0.245\textwidth}{!}{\includegraphics[scale=1]{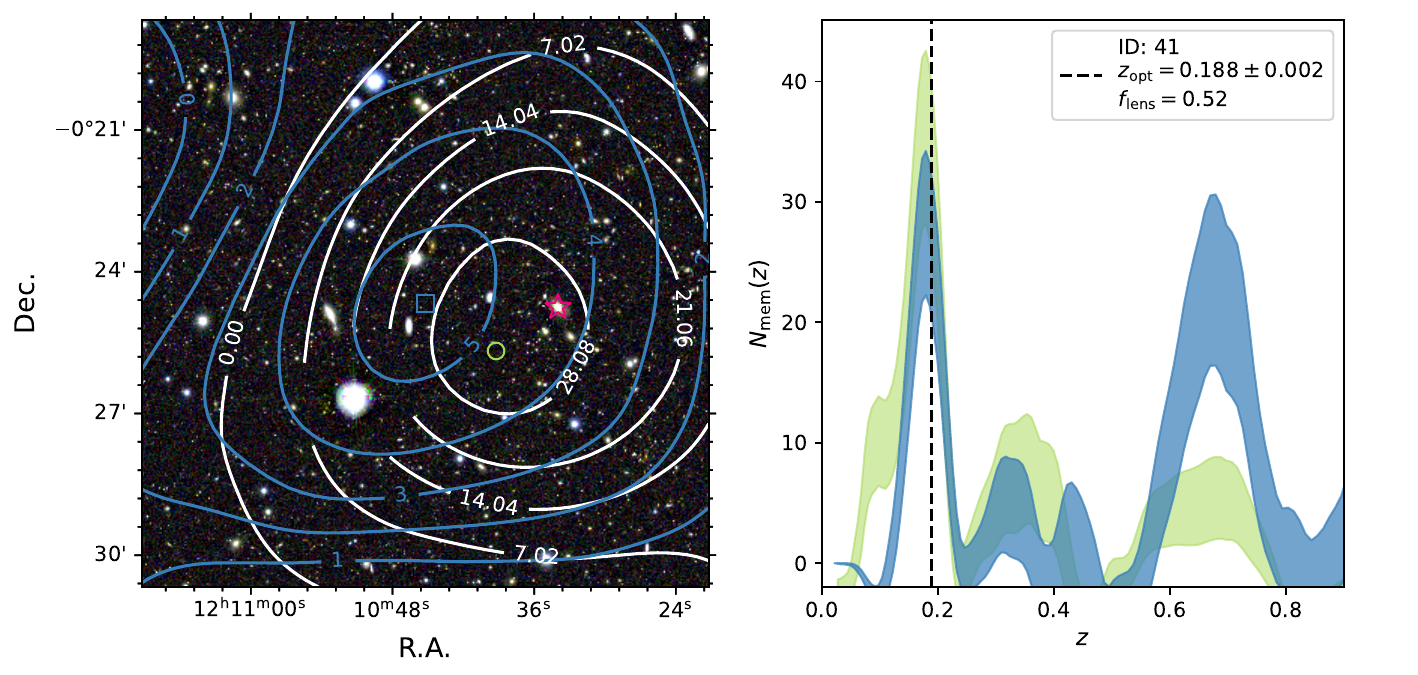}}
\resizebox{0.245\textwidth}{!}{\includegraphics[scale=1]{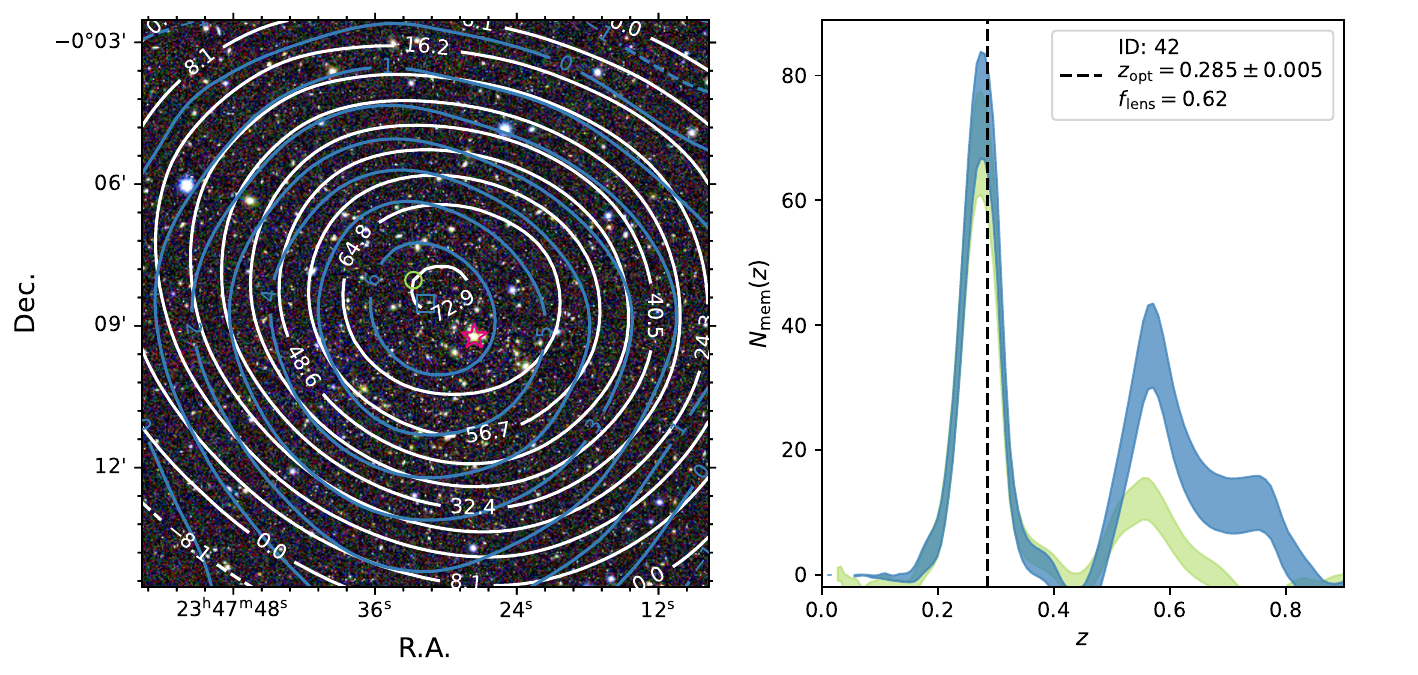}}
\resizebox{0.245\textwidth}{!}{\includegraphics[scale=1]{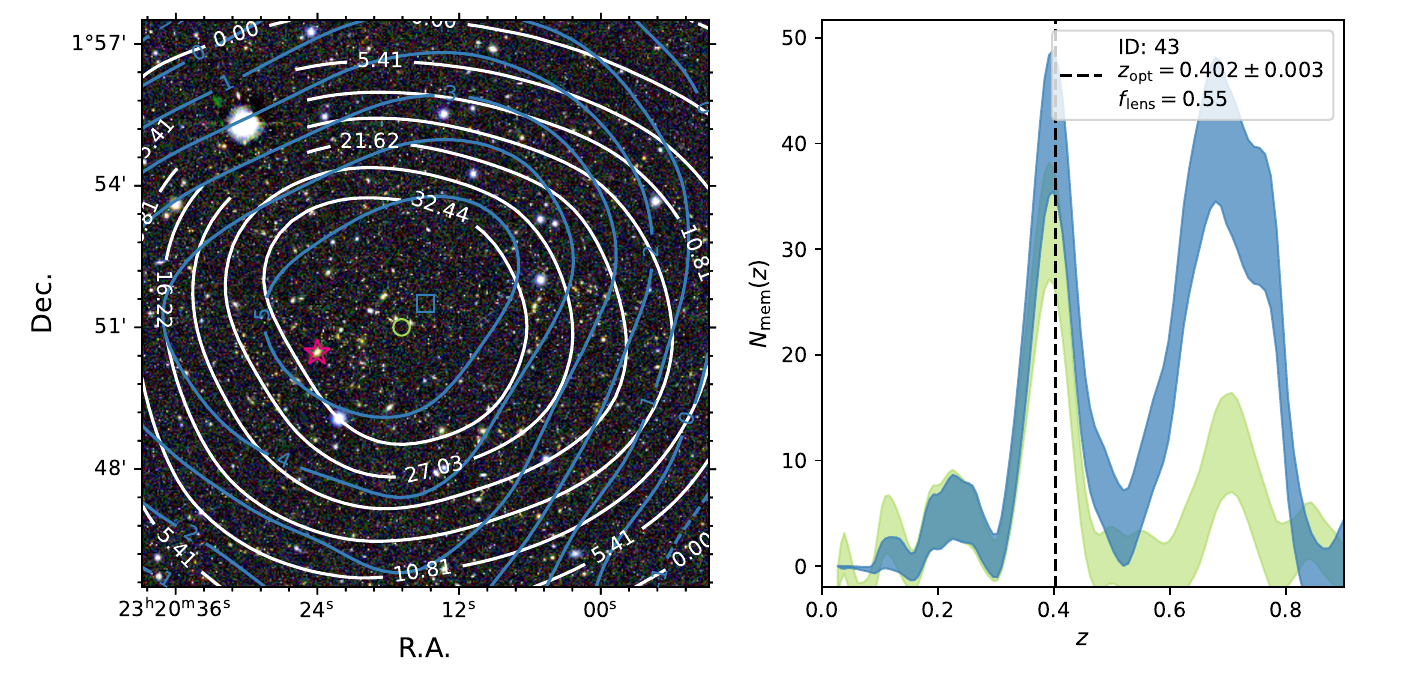}}
\resizebox{0.245\textwidth}{!}{\includegraphics[scale=1]{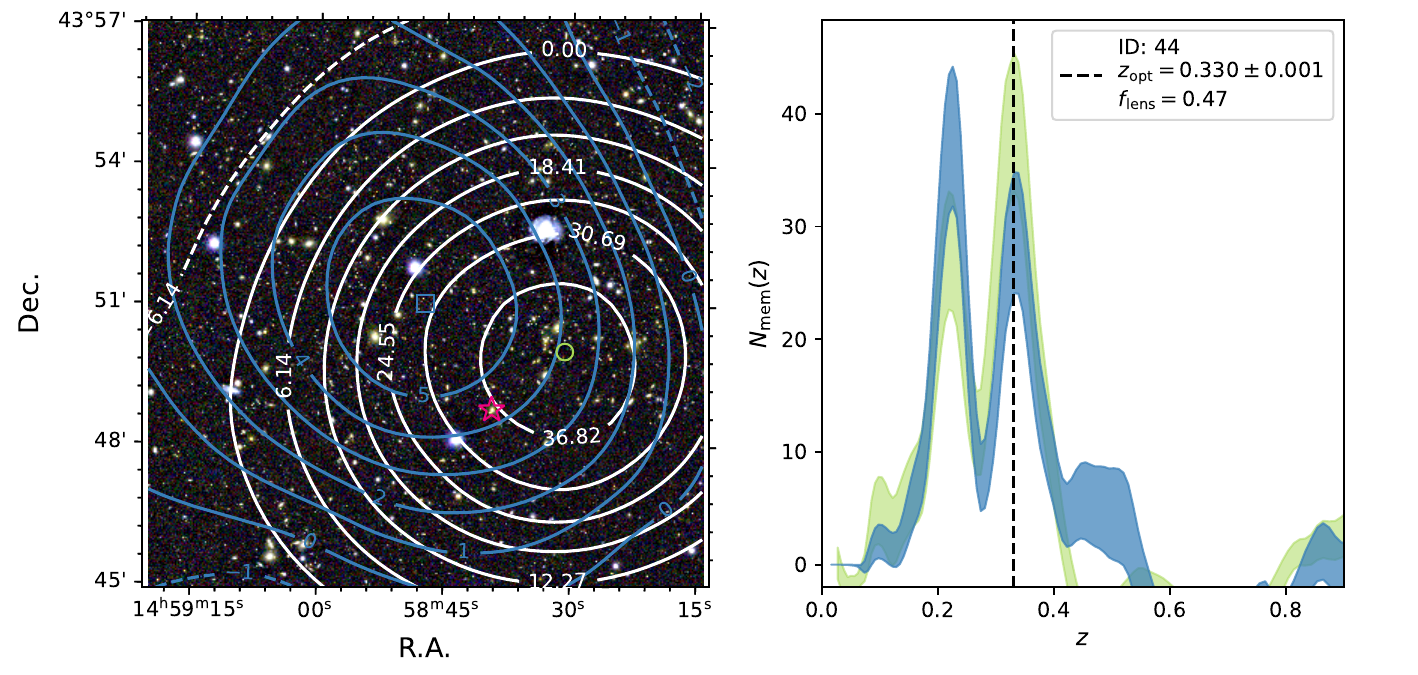}}
\resizebox{0.245\textwidth}{!}{\includegraphics[scale=1]{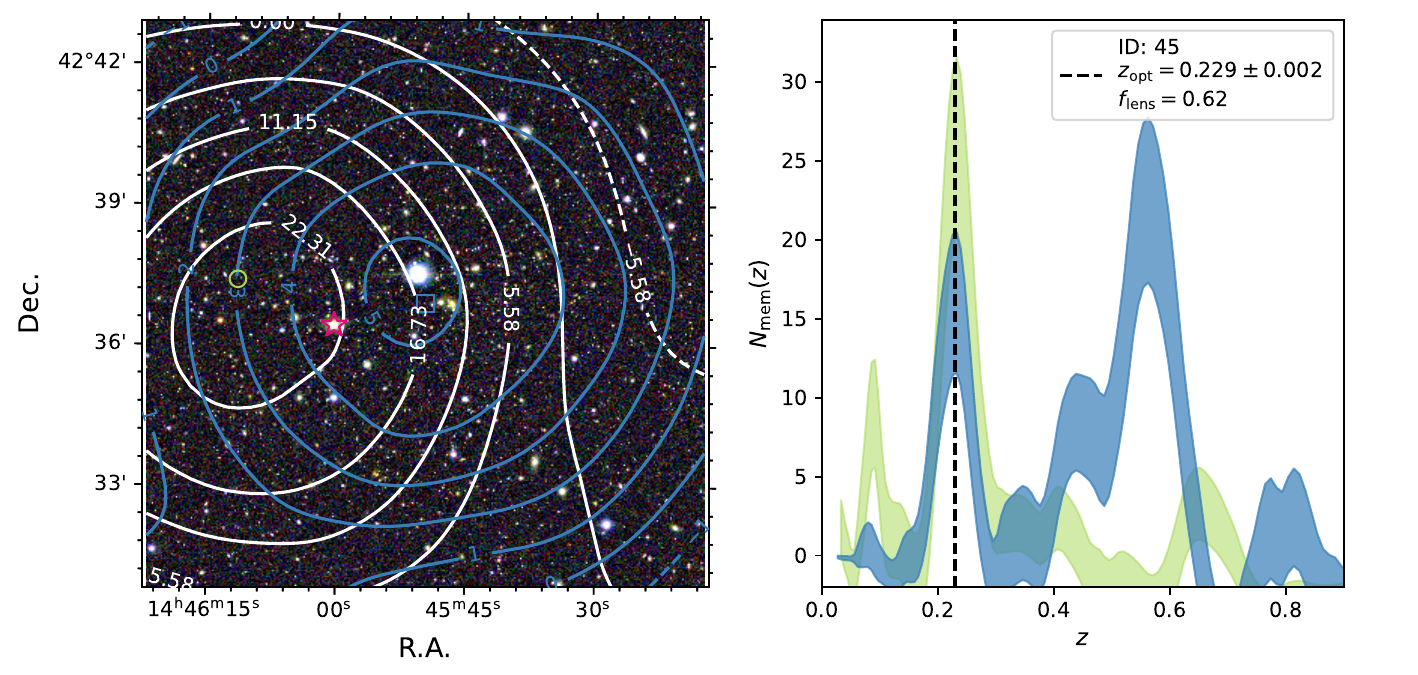}}
\resizebox{0.245\textwidth}{!}{\includegraphics[scale=1]{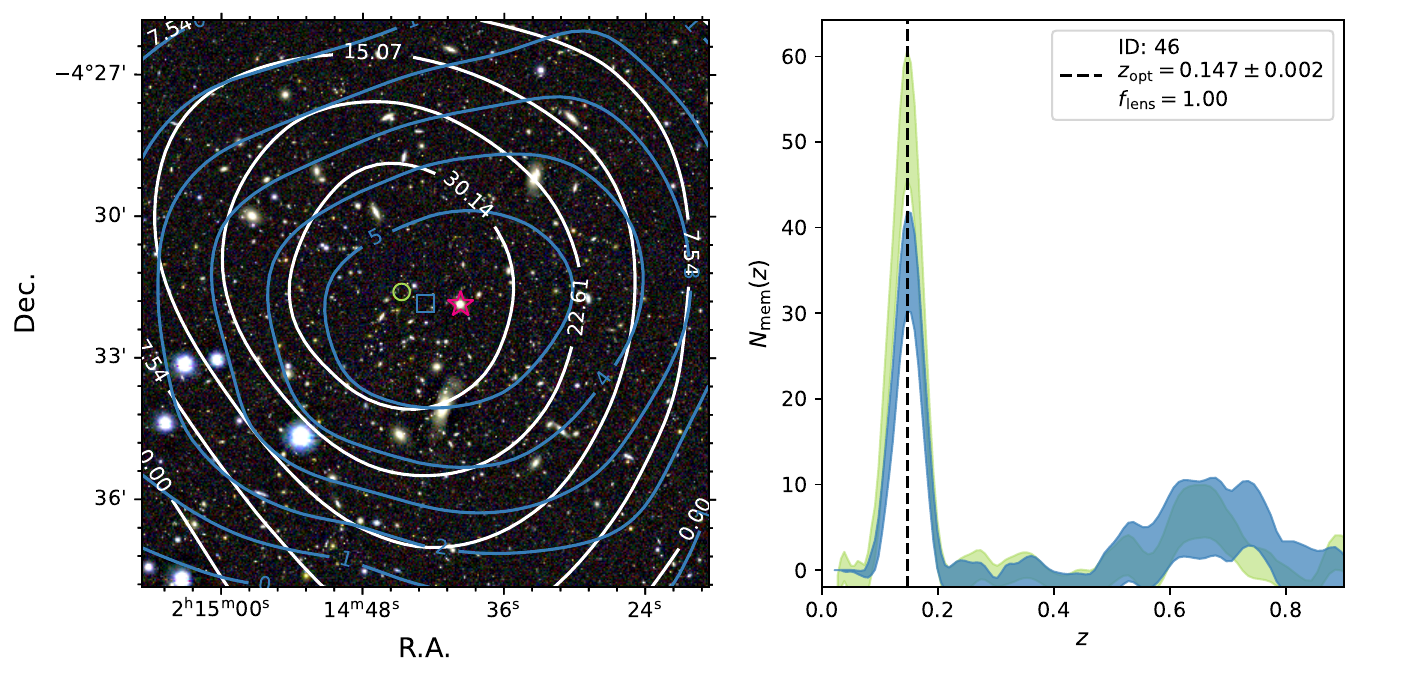}}
\resizebox{0.245\textwidth}{!}{\includegraphics[scale=1]{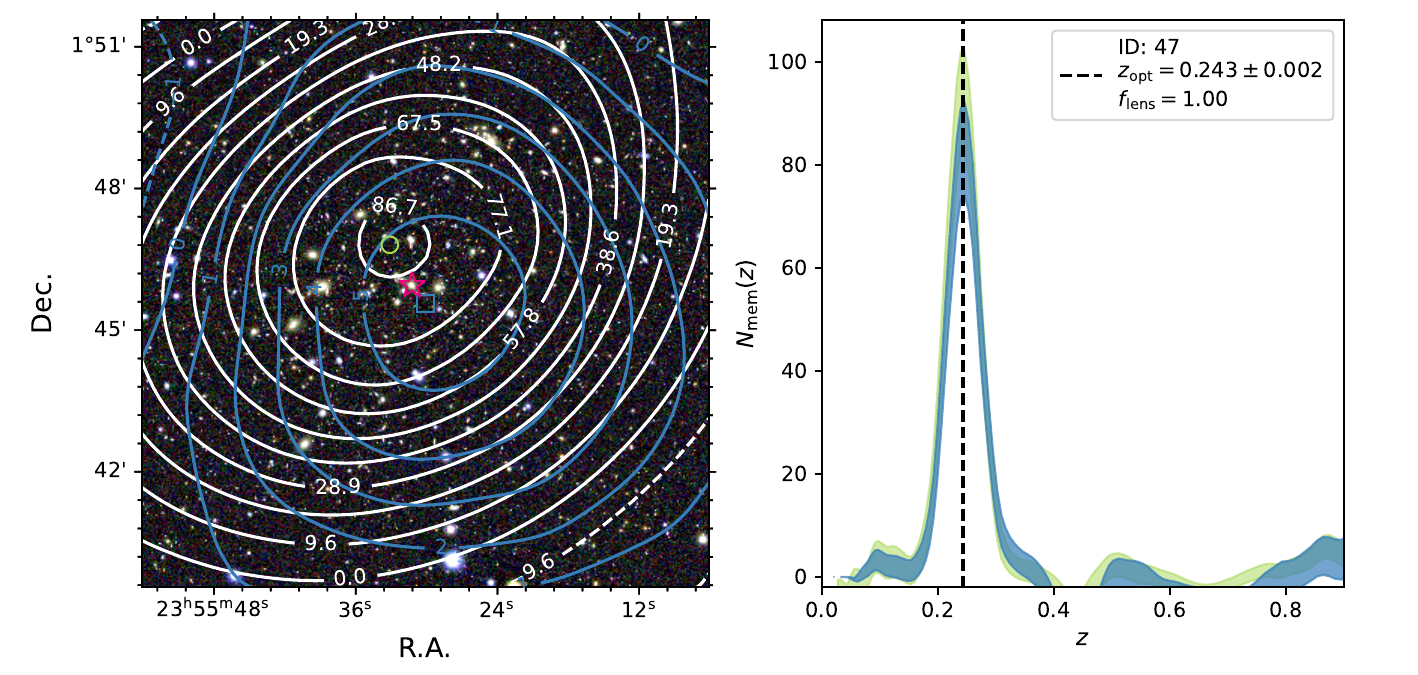}}
\resizebox{0.245\textwidth}{!}{\includegraphics[scale=1]{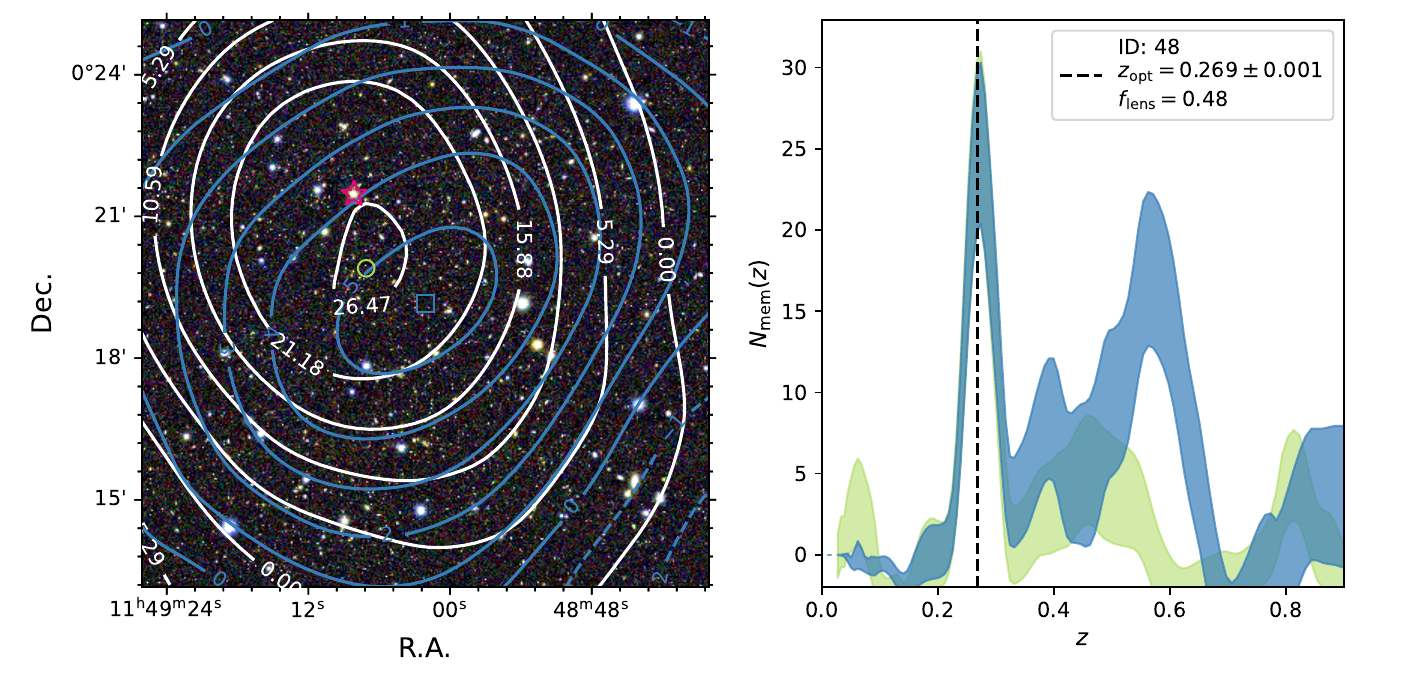}}
\resizebox{0.245\textwidth}{!}{\includegraphics[scale=1]{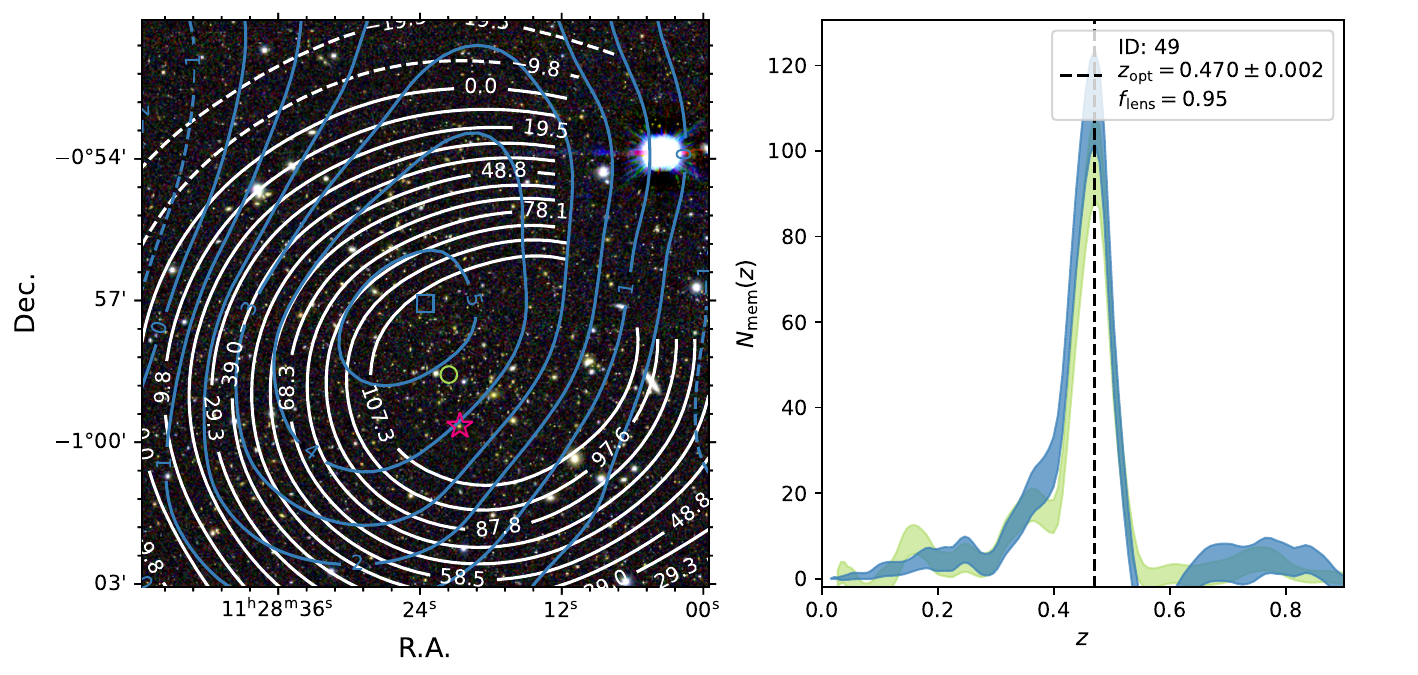}}
\resizebox{0.245\textwidth}{!}{\includegraphics[scale=1]{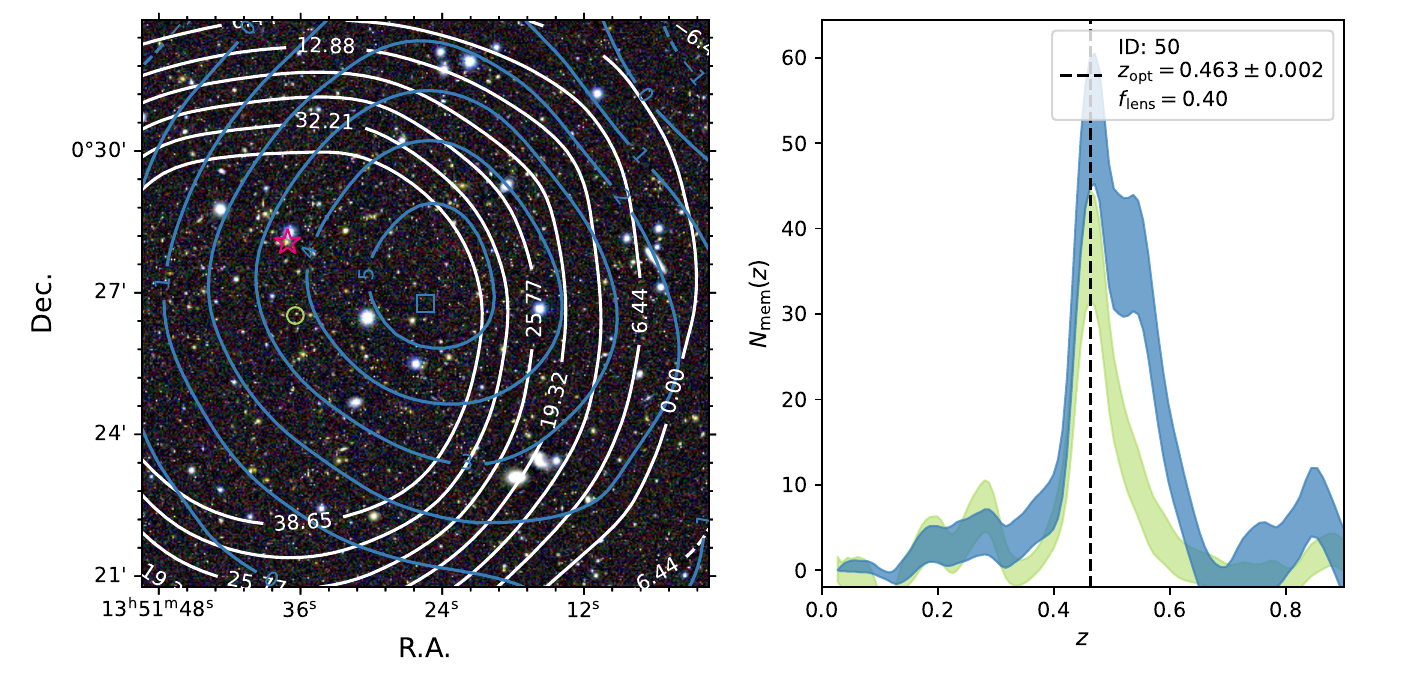}}
\resizebox{0.245\textwidth}{!}{\includegraphics[scale=1]{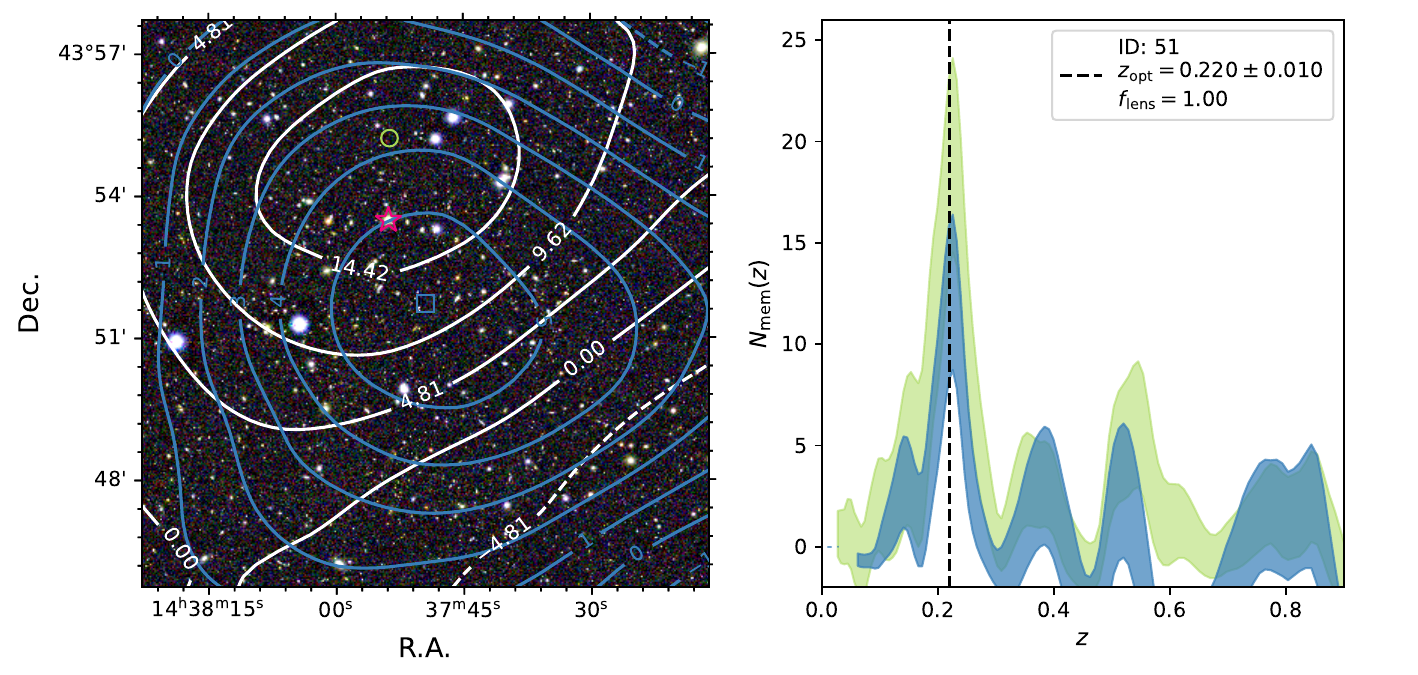}}
\resizebox{0.245\textwidth}{!}{\includegraphics[scale=1]{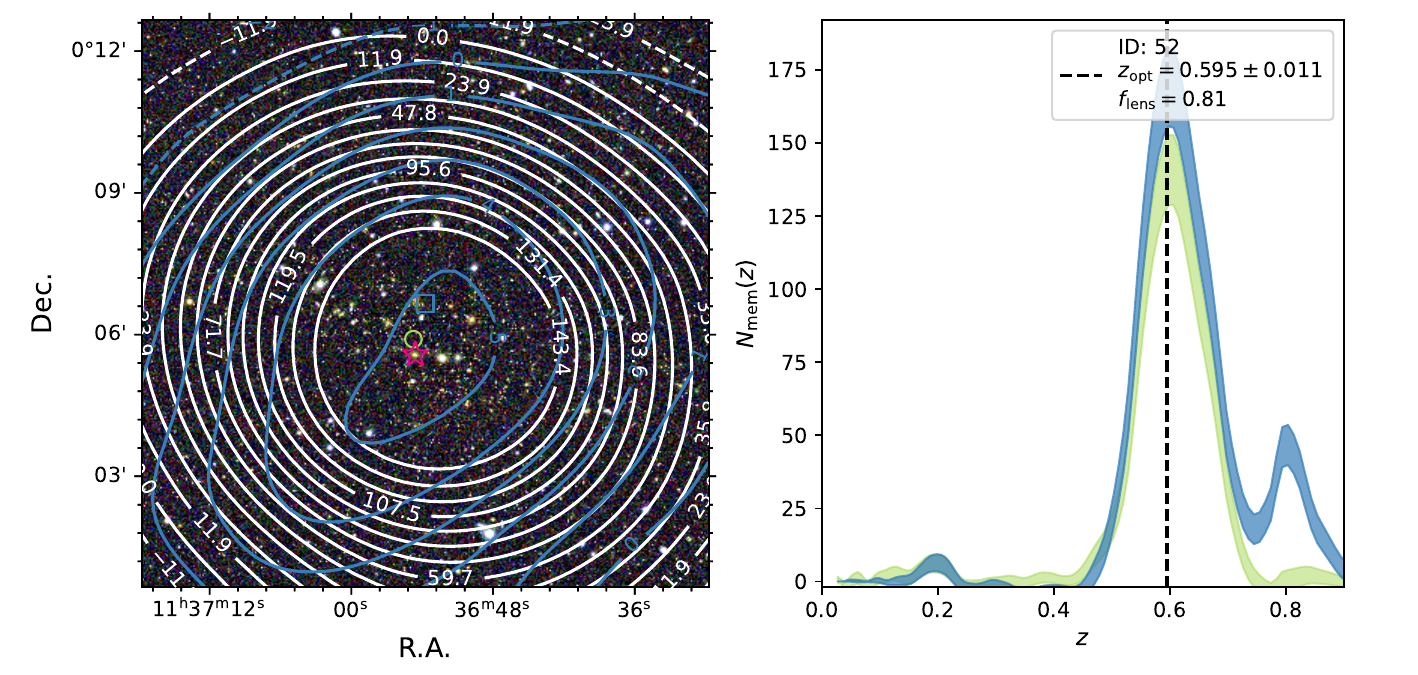}}
\resizebox{0.245\textwidth}{!}{\includegraphics[scale=1]{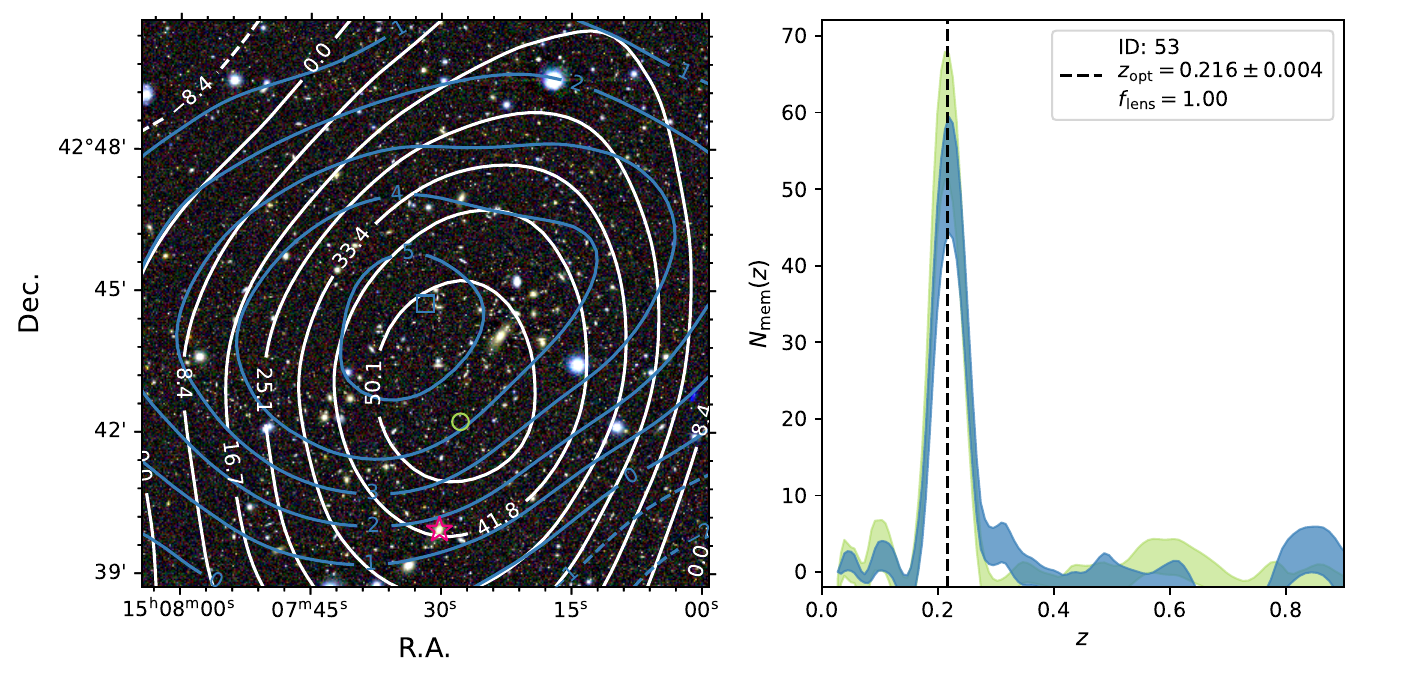}}
\resizebox{0.245\textwidth}{!}{\includegraphics[scale=1]{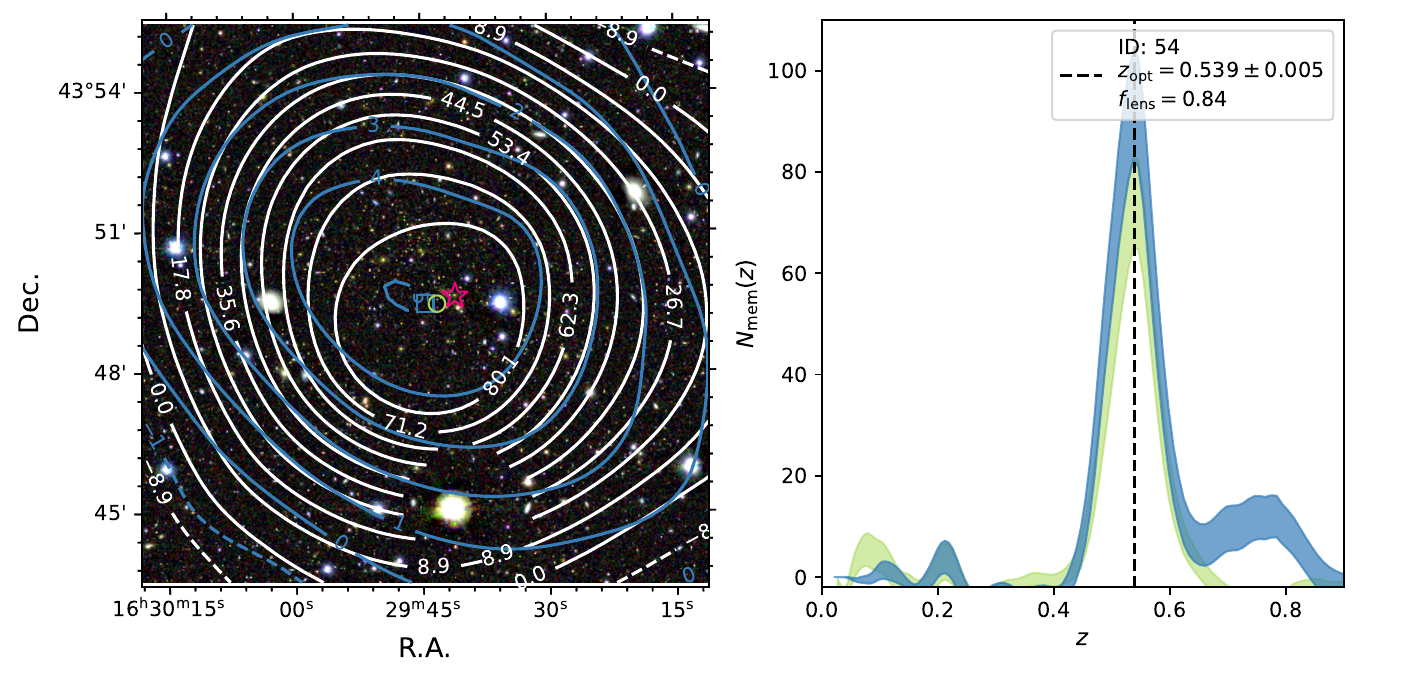}}
\resizebox{0.245\textwidth}{!}{\includegraphics[scale=1]{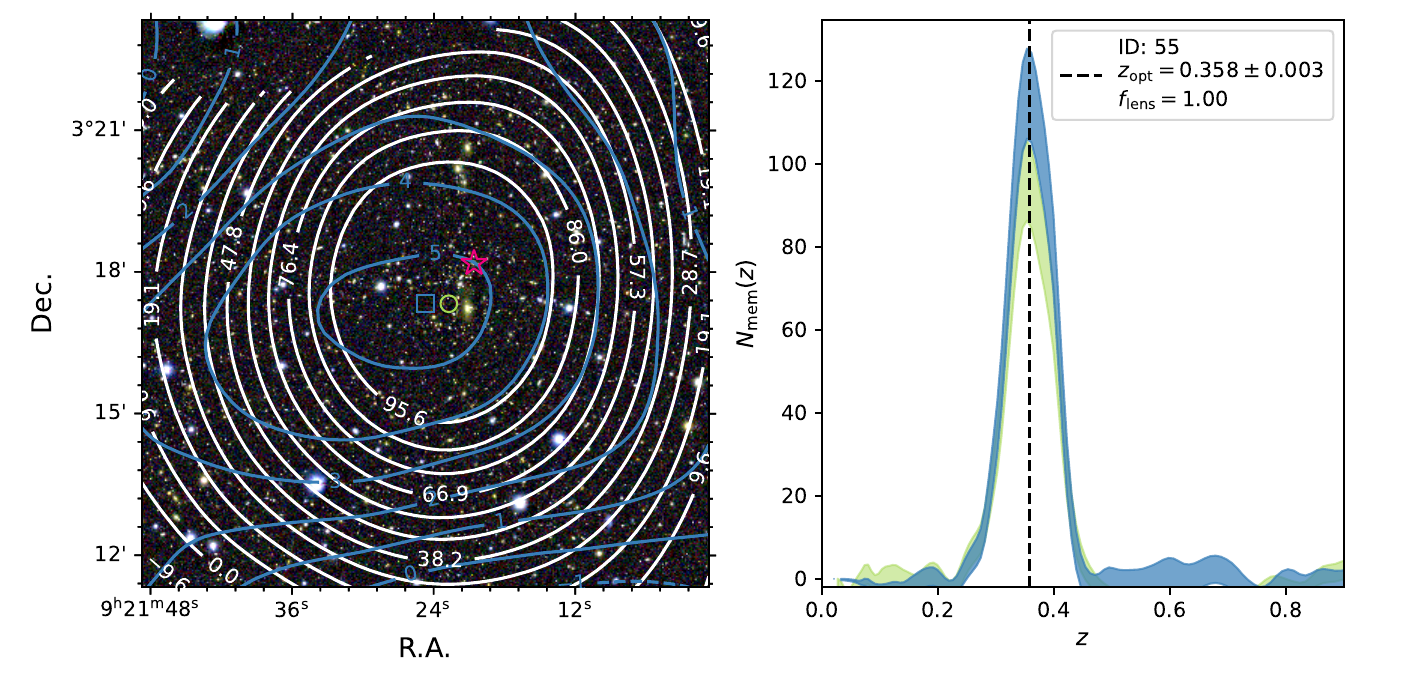}}
\resizebox{0.245\textwidth}{!}{\includegraphics[scale=1]{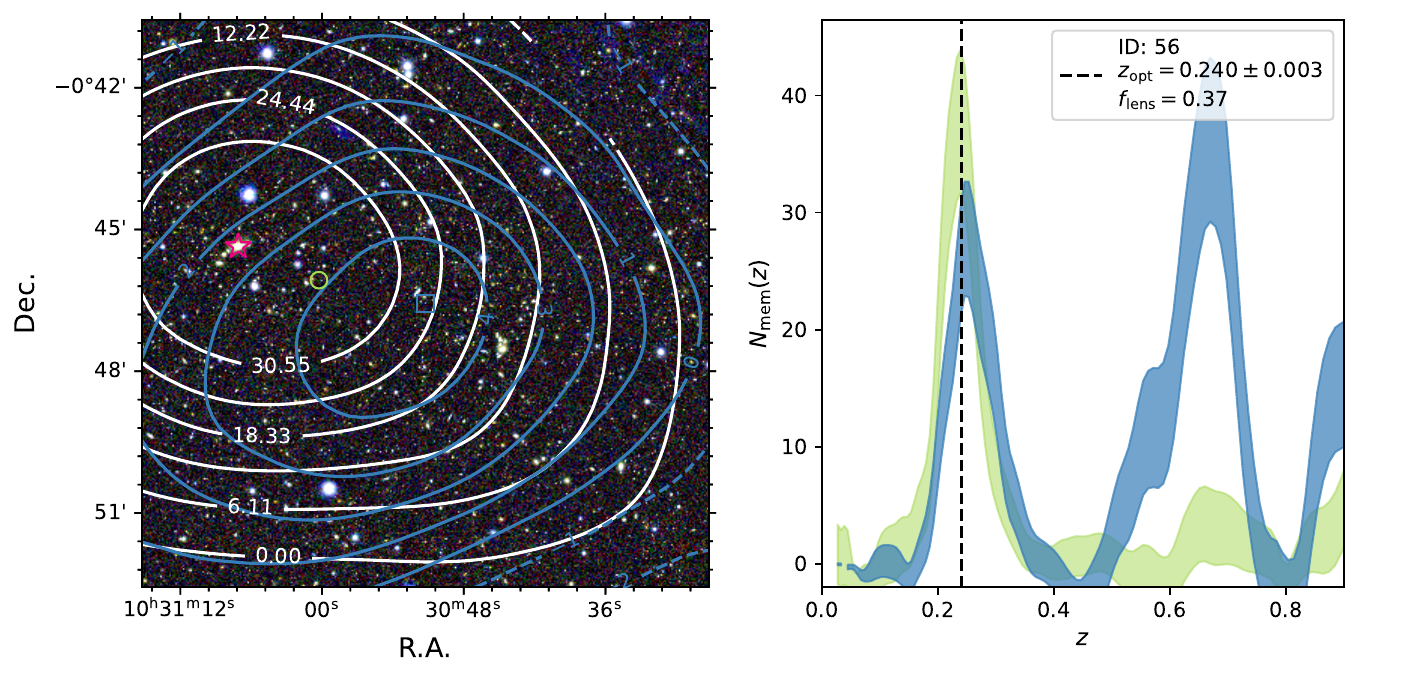}}
\resizebox{0.245\textwidth}{!}{\includegraphics[scale=1]{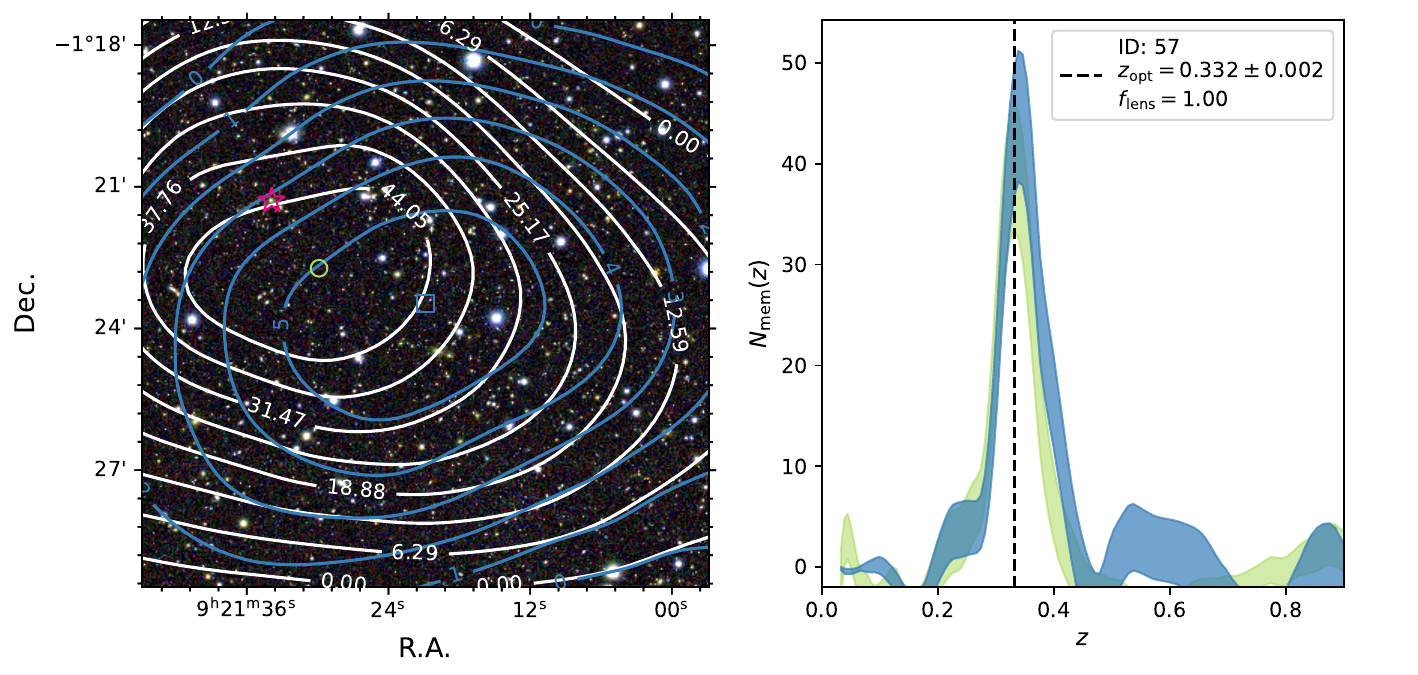}}
\resizebox{0.245\textwidth}{!}{\includegraphics[scale=1]{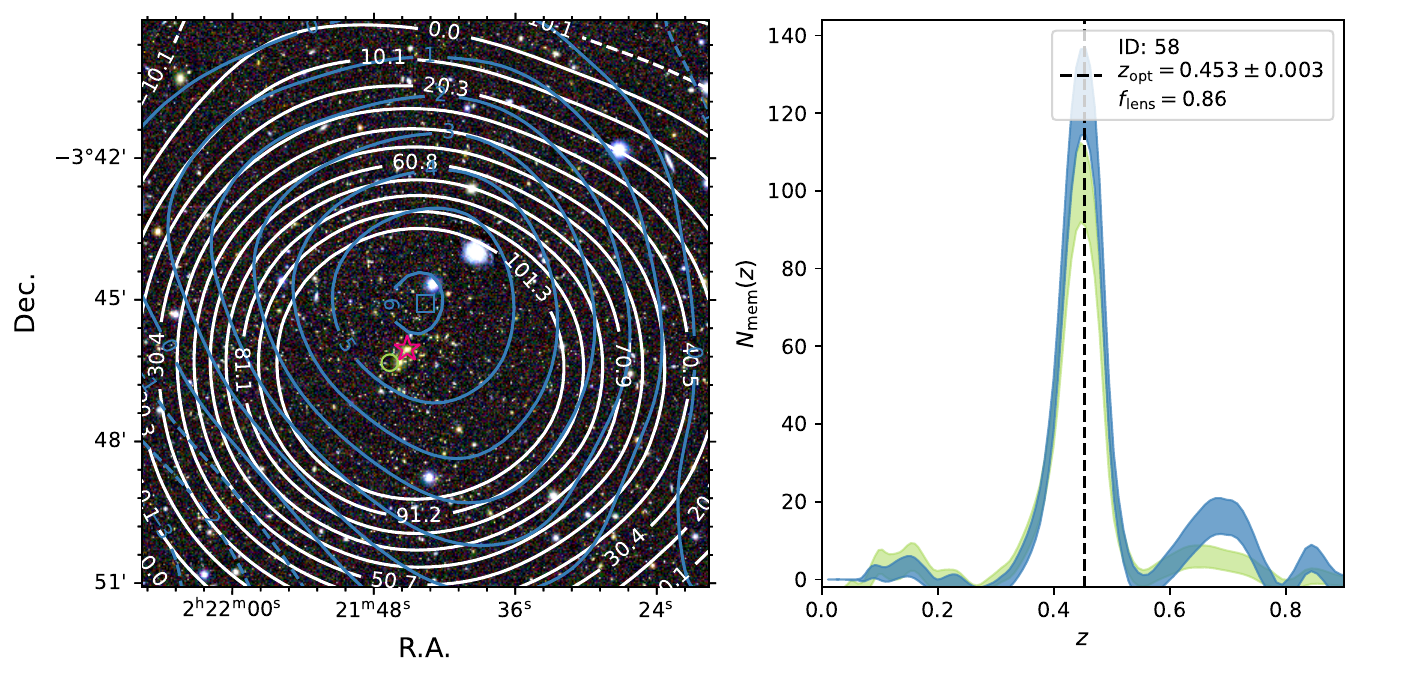}}
\resizebox{0.245\textwidth}{!}{\includegraphics[scale=1]{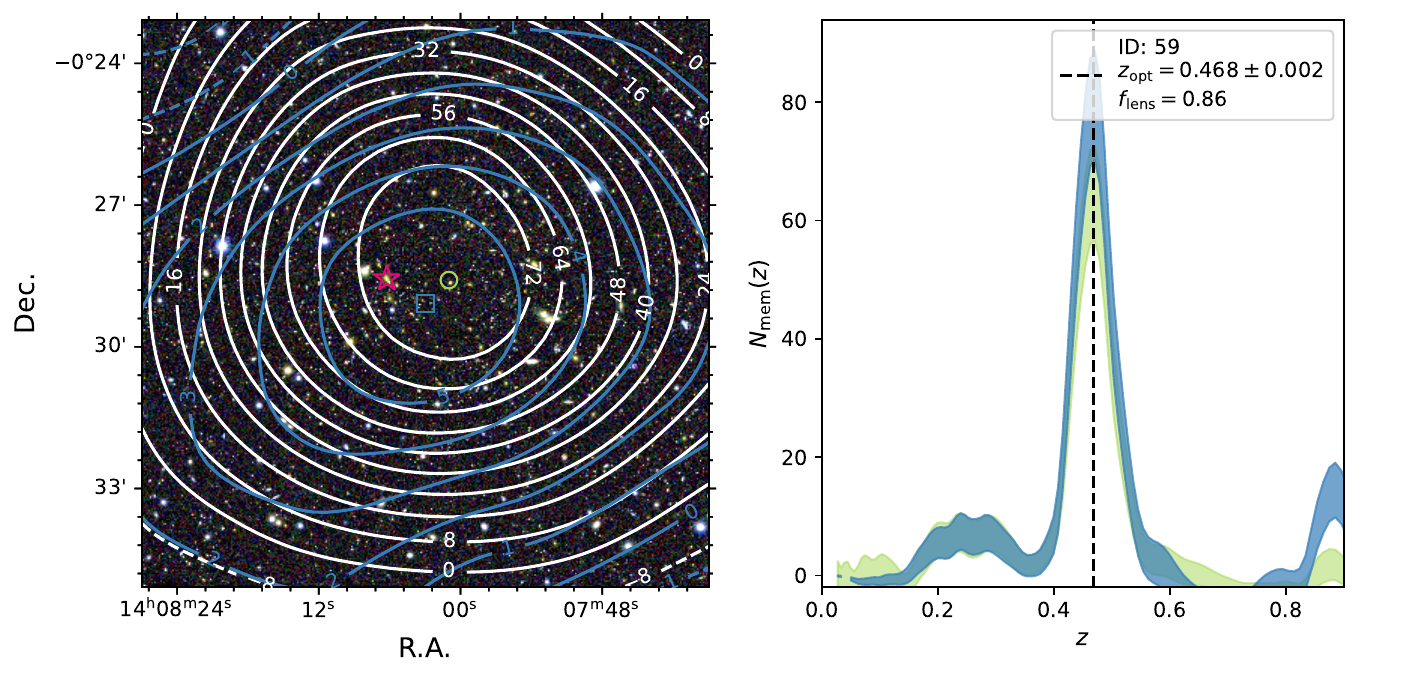}}
\resizebox{0.245\textwidth}{!}{\includegraphics[scale=1]{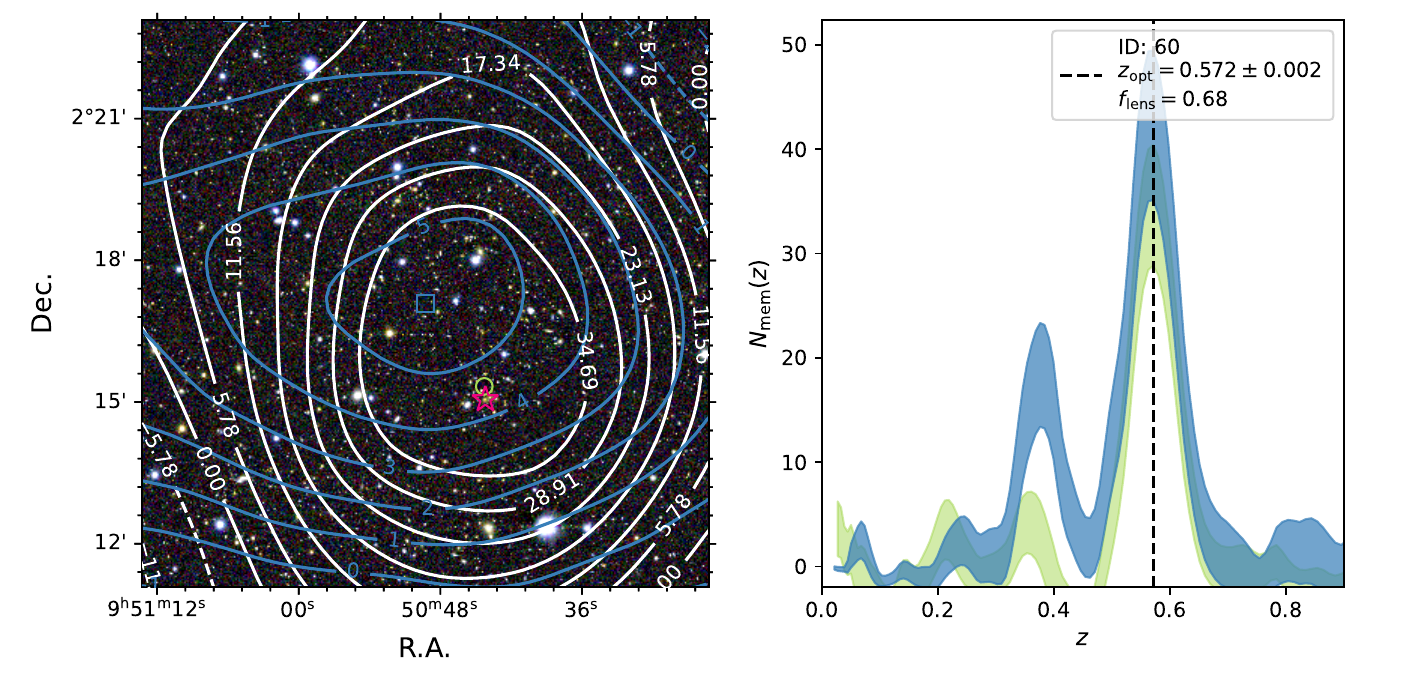}}
\resizebox{0.245\textwidth}{!}{\includegraphics[scale=1]{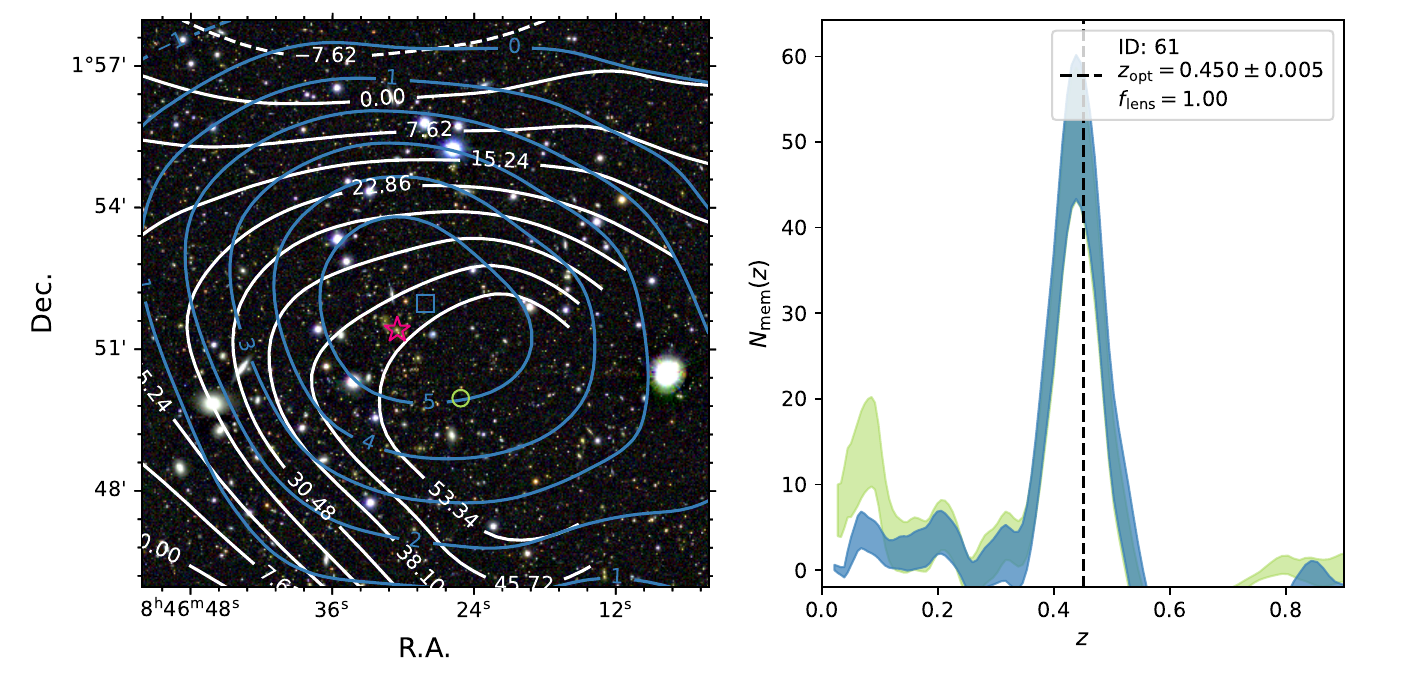}}
\resizebox{0.245\textwidth}{!}{\includegraphics[scale=1]{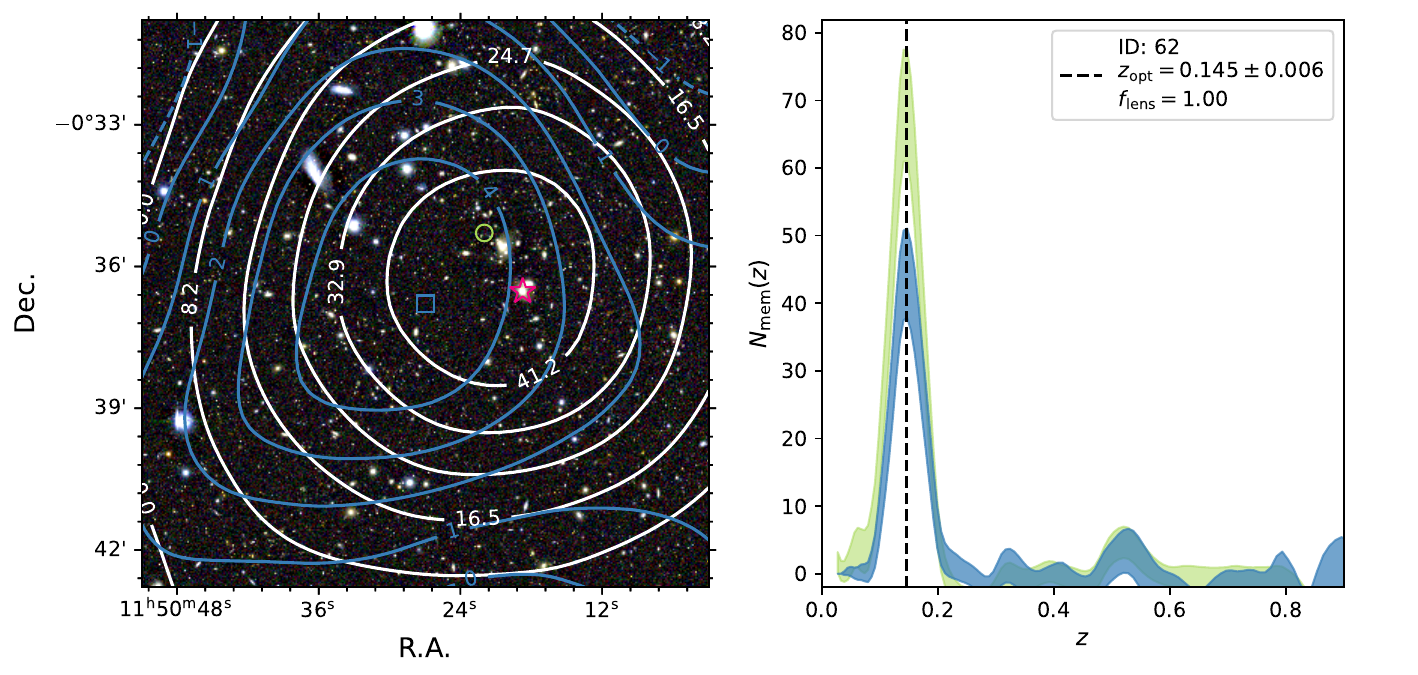}}
\resizebox{0.245\textwidth}{!}{\includegraphics[scale=1]{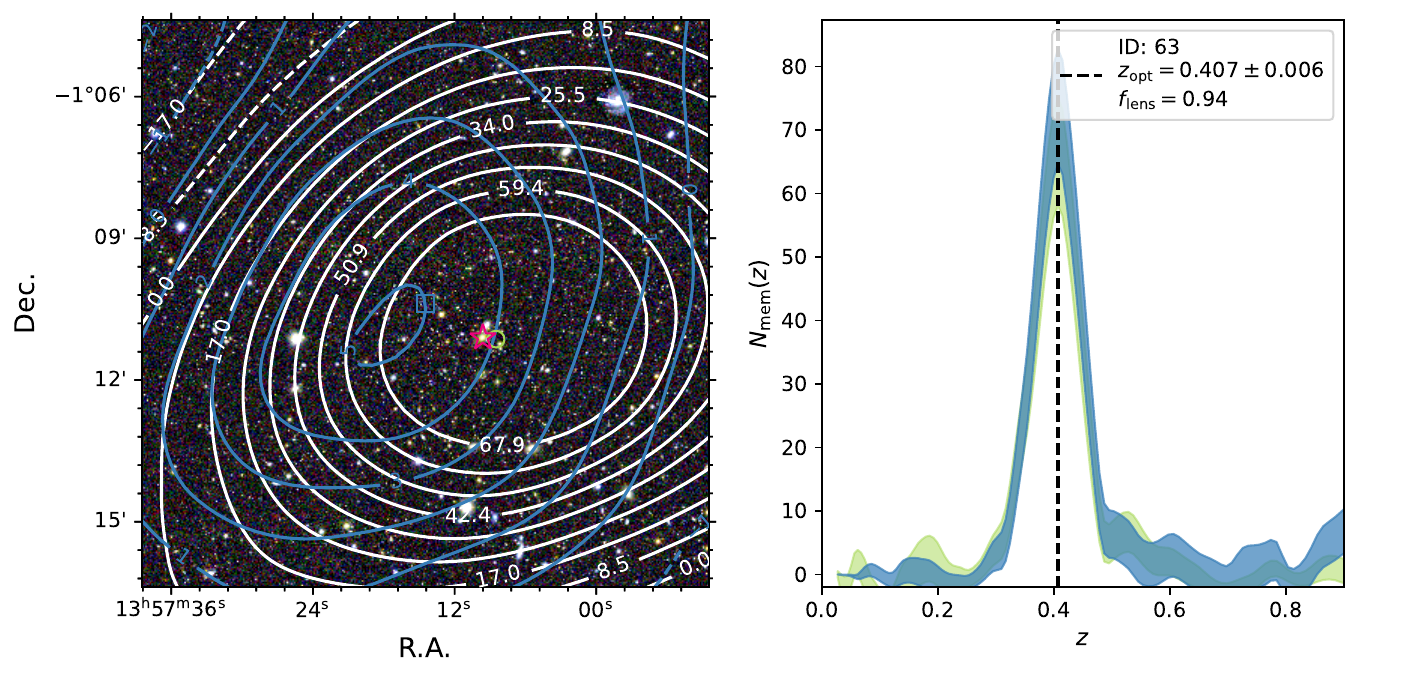}}
\resizebox{0.245\textwidth}{!}{\includegraphics[scale=1]{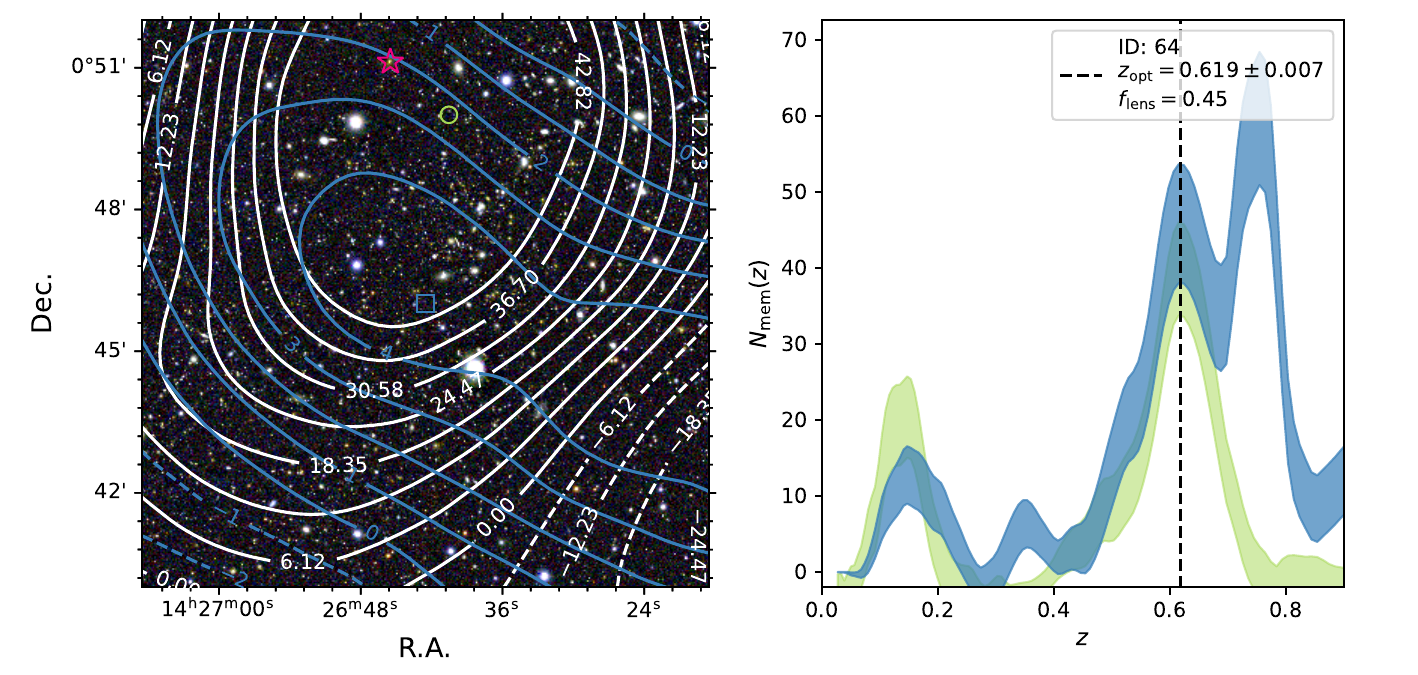}}
\resizebox{0.245\textwidth}{!}{\includegraphics[scale=1]{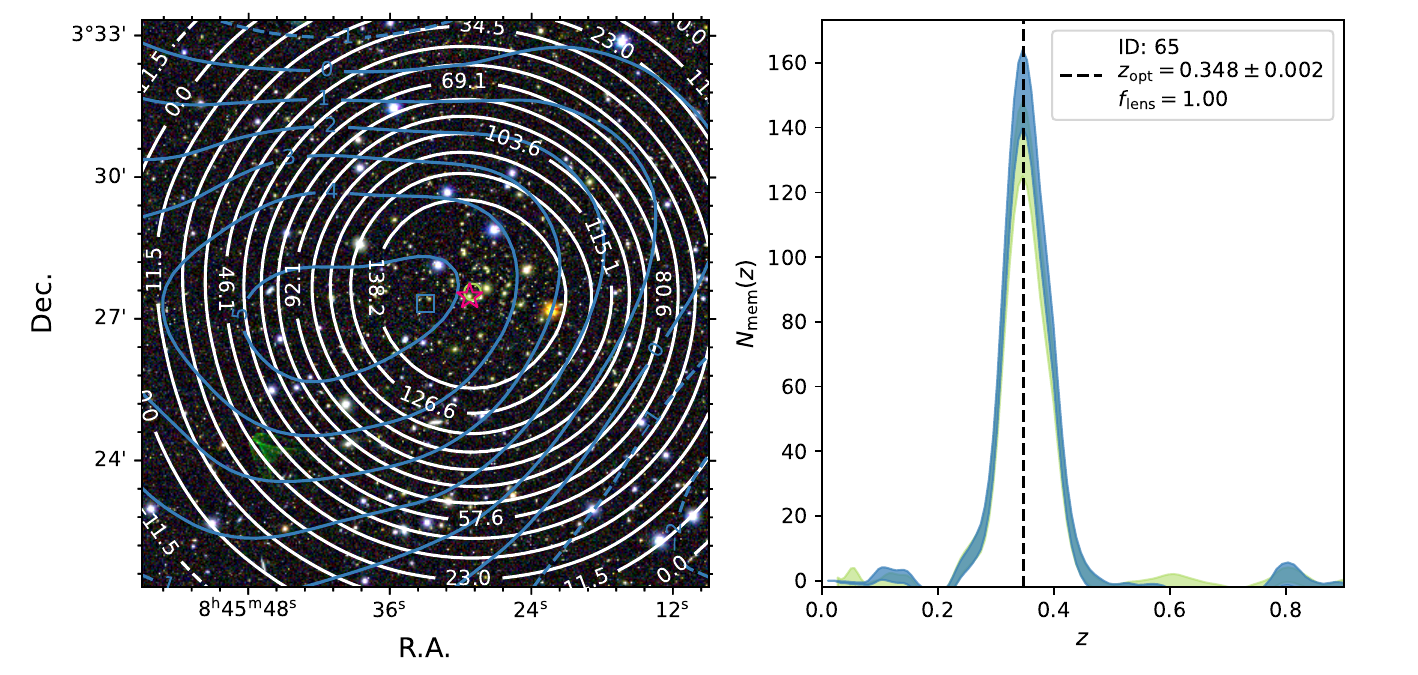}}
\resizebox{0.245\textwidth}{!}{\includegraphics[scale=1]{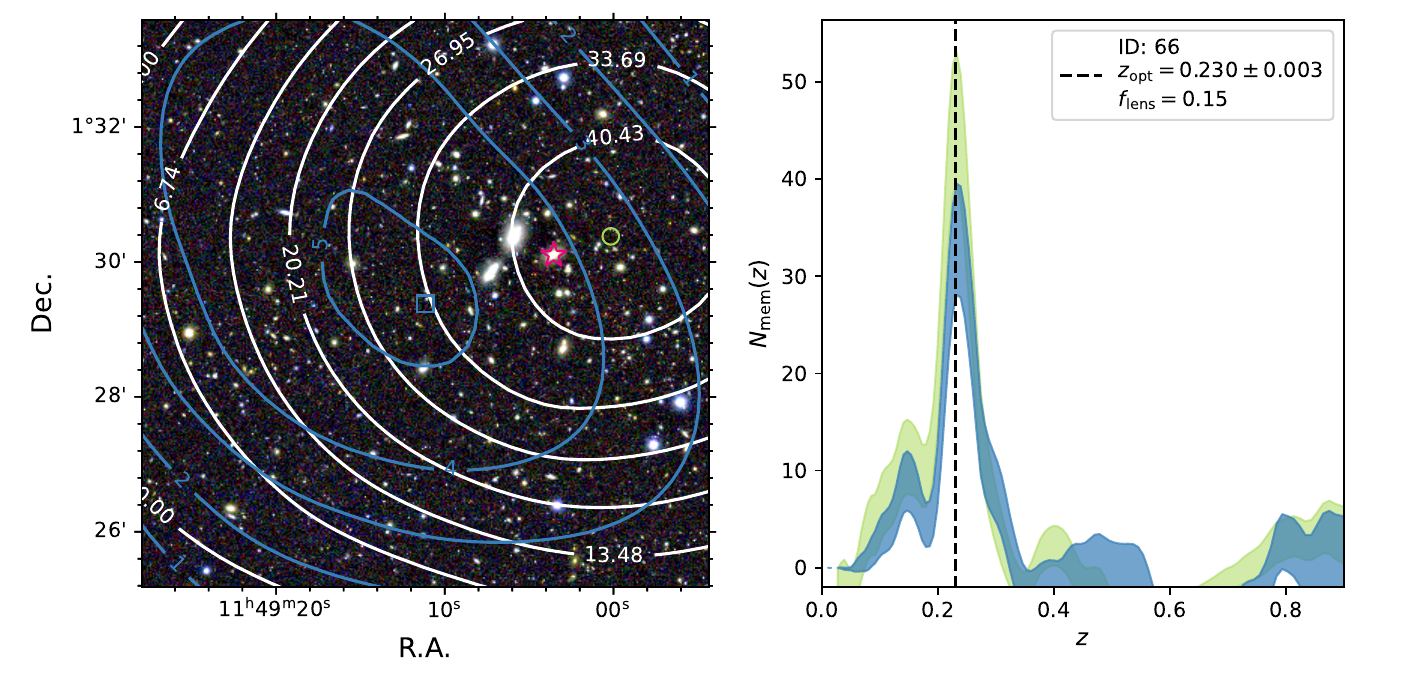}}
\resizebox{0.245\textwidth}{!}{\includegraphics[scale=1]{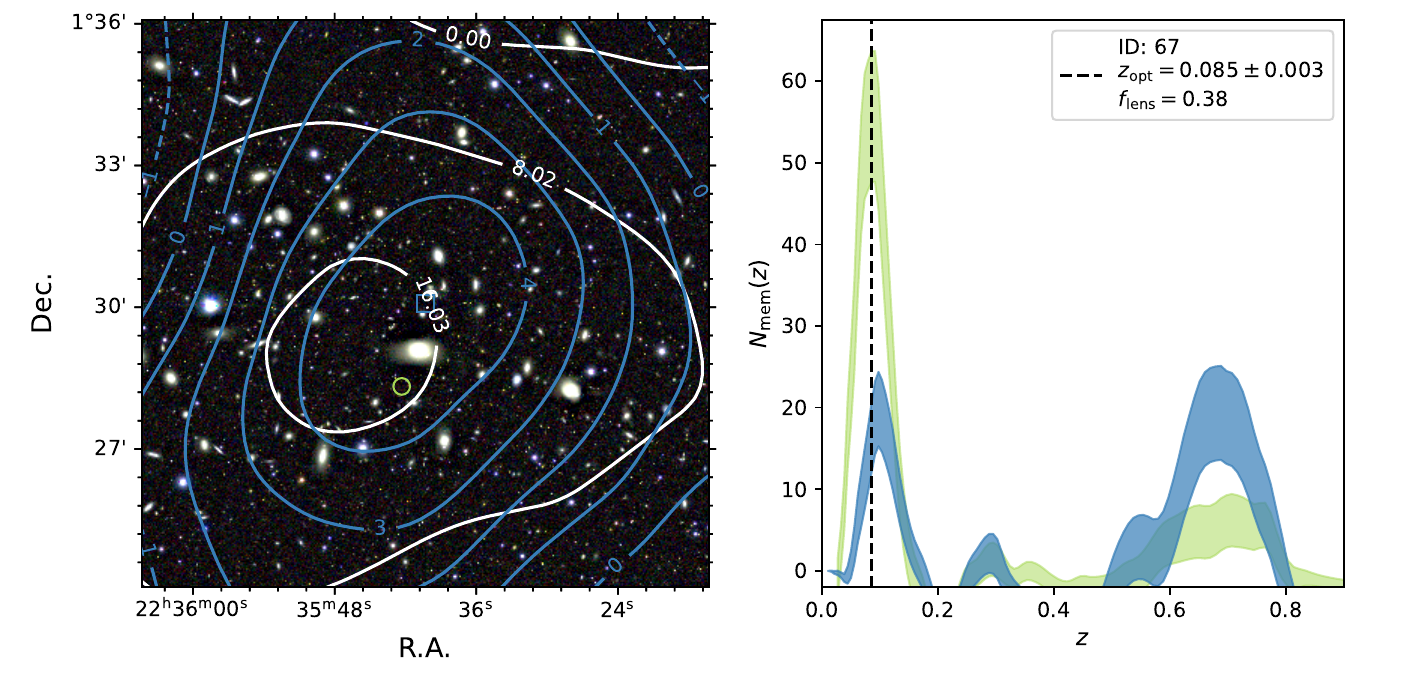}}
\resizebox{0.245\textwidth}{!}{\includegraphics[scale=1]{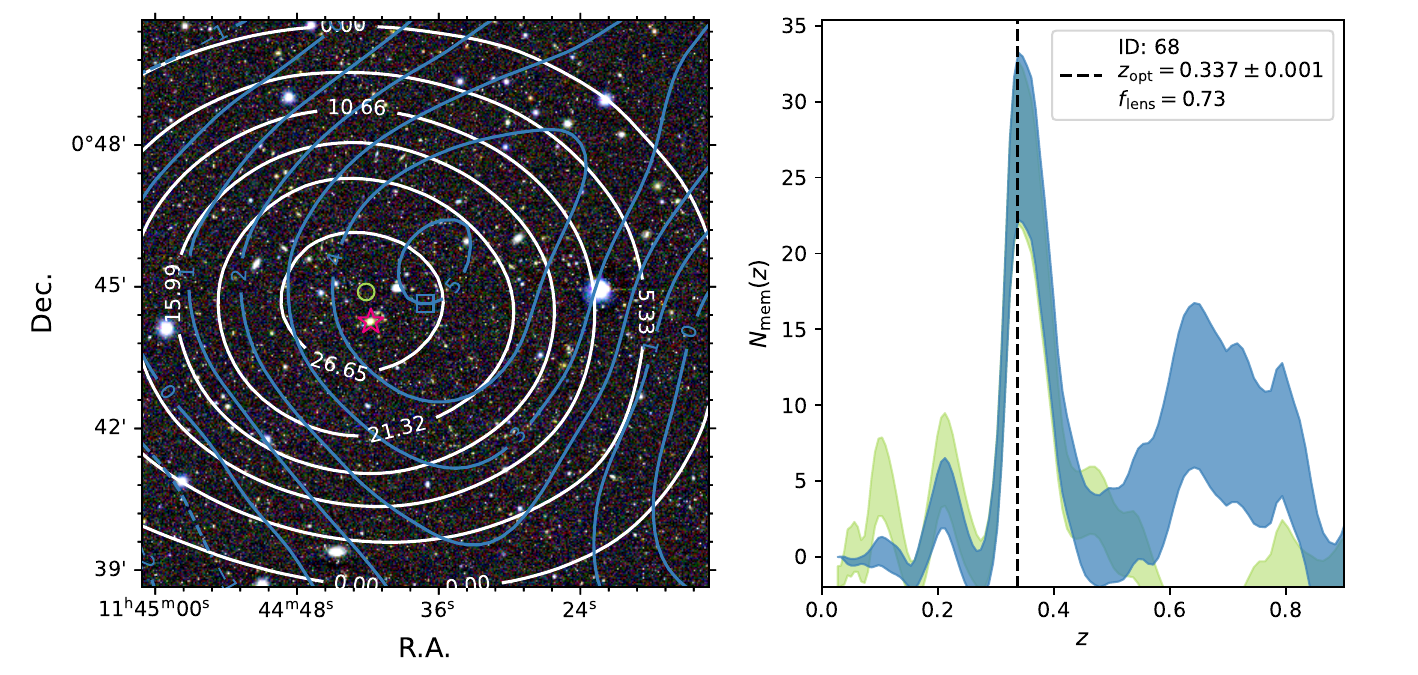}}
\resizebox{0.245\textwidth}{!}{\includegraphics[scale=1]{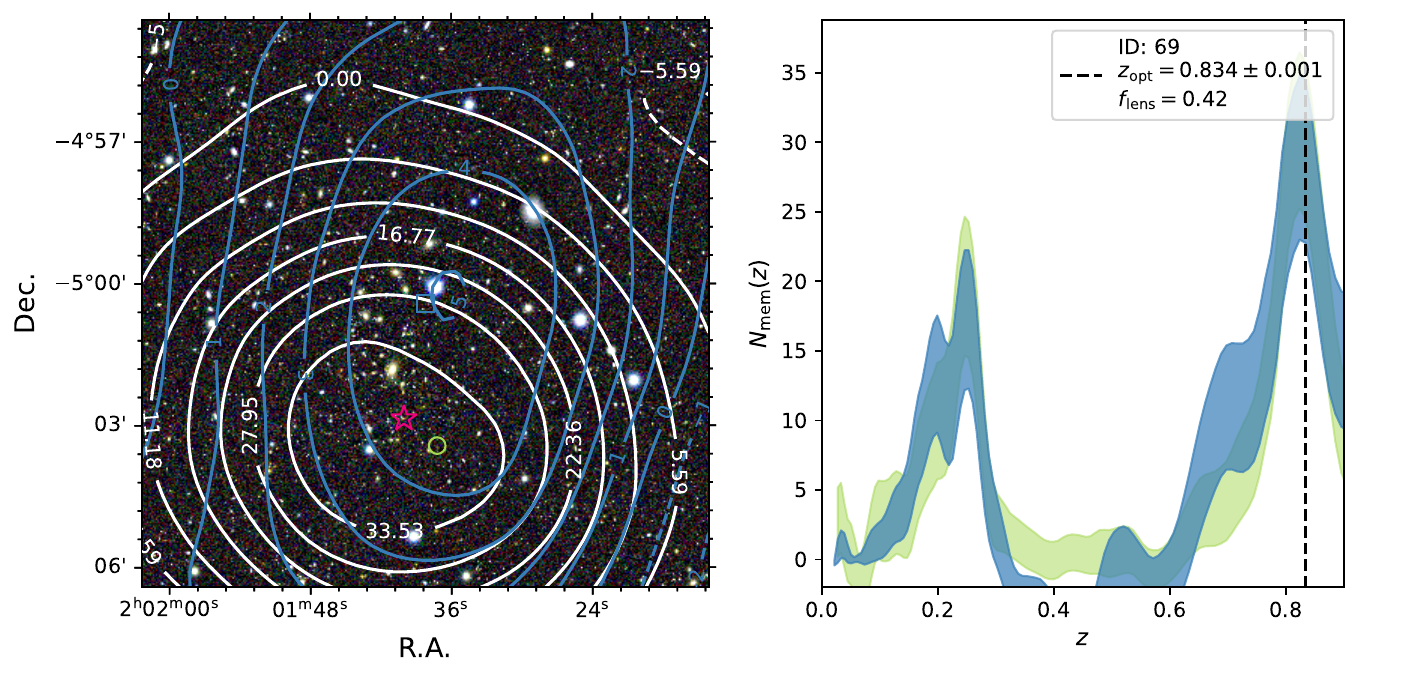}}
\resizebox{0.245\textwidth}{!}{\includegraphics[scale=1]{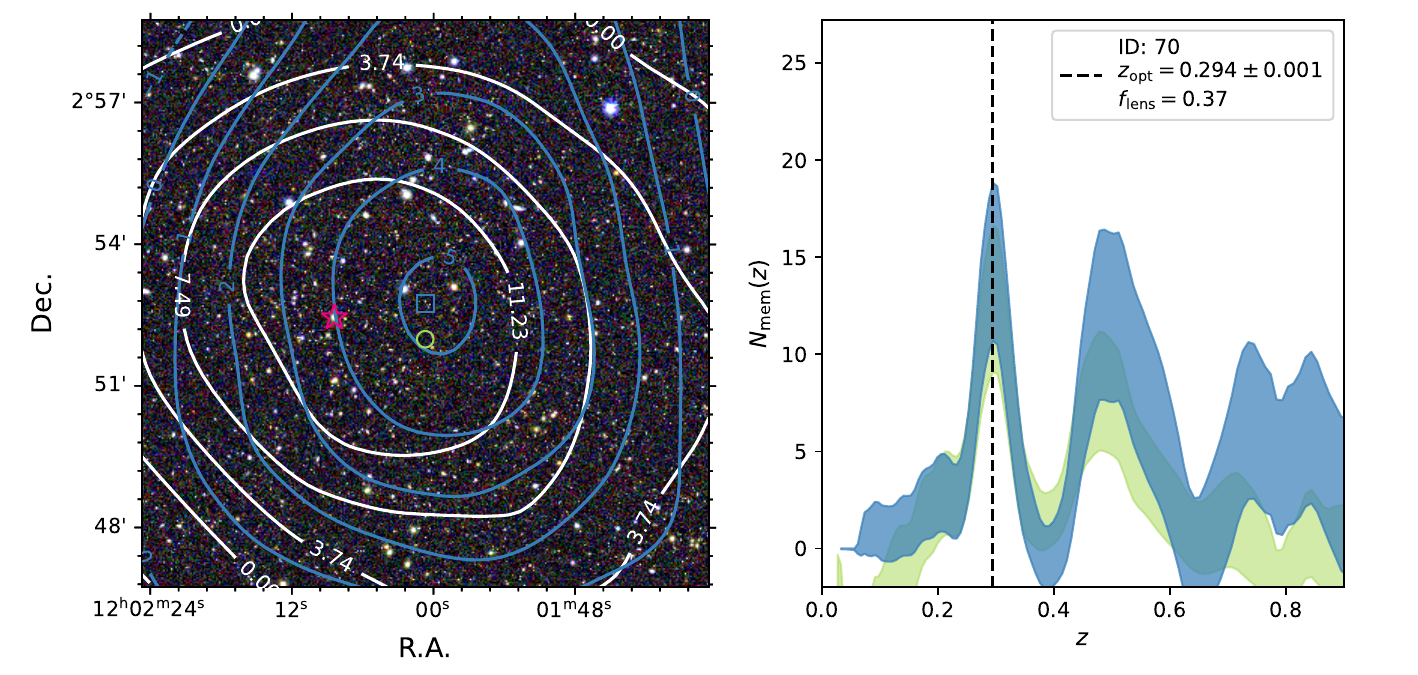}}
\resizebox{0.245\textwidth}{!}{\includegraphics[scale=1]{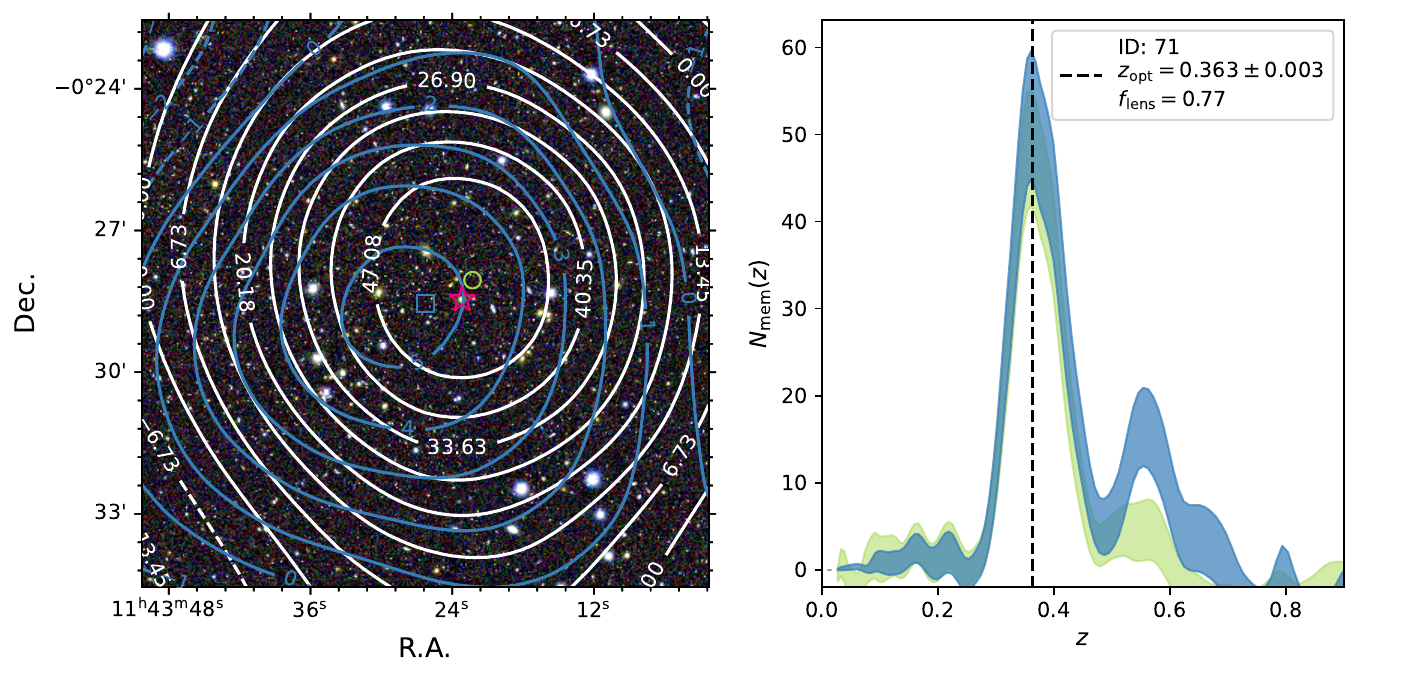}}
\resizebox{0.245\textwidth}{!}{\includegraphics[scale=1]{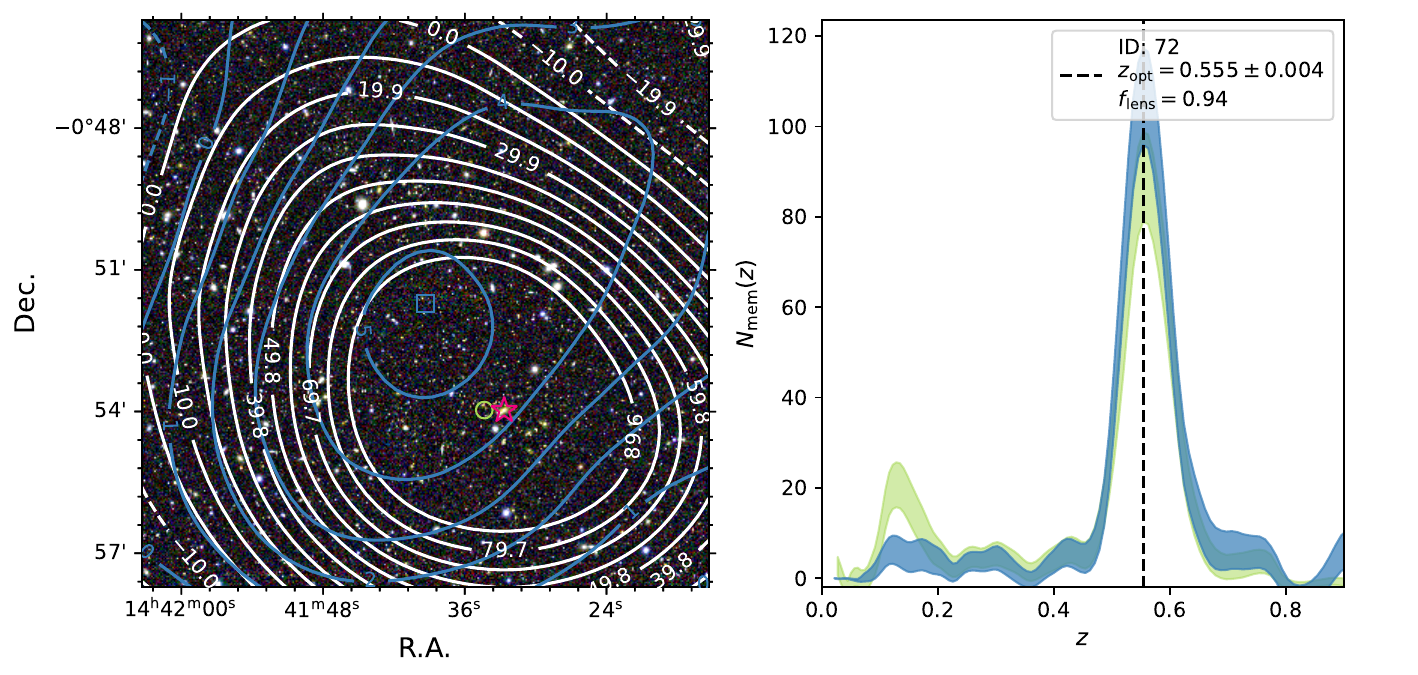}}
\resizebox{0.245\textwidth}{!}{\includegraphics[scale=1]{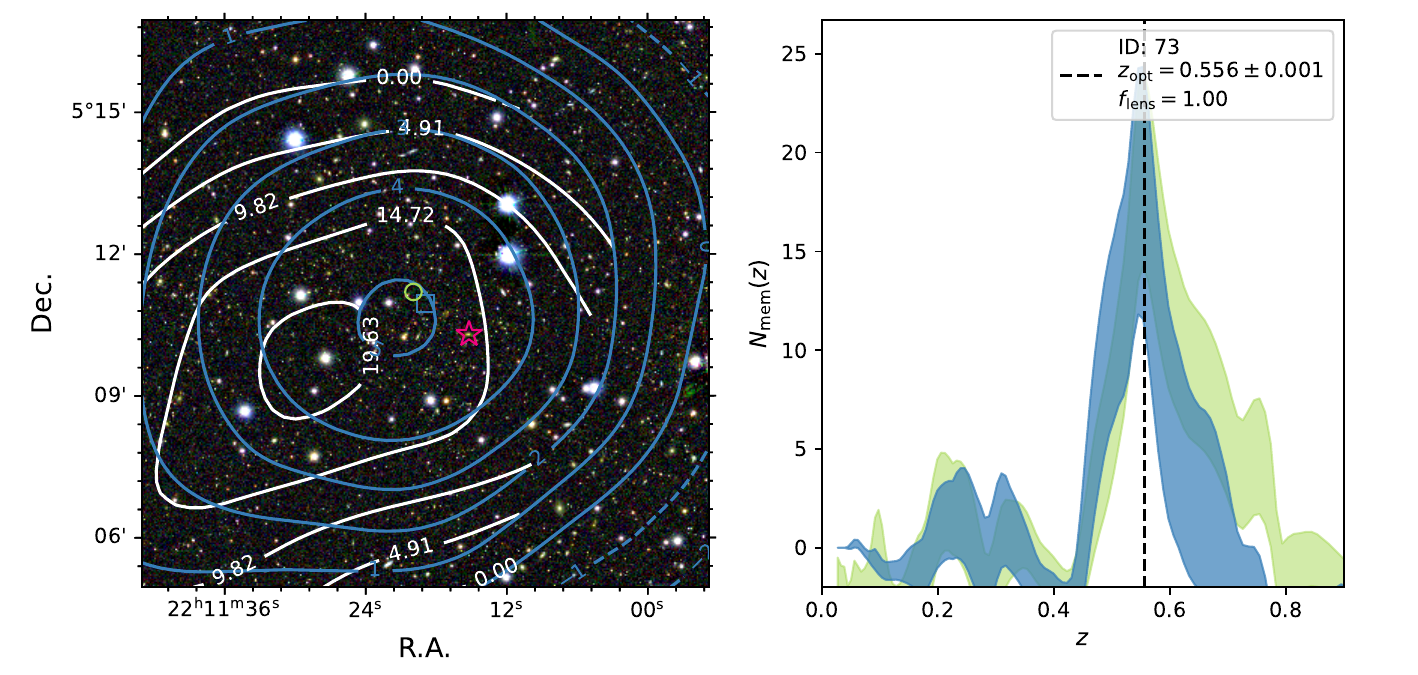}}
\resizebox{0.245\textwidth}{!}{\includegraphics[scale=1]{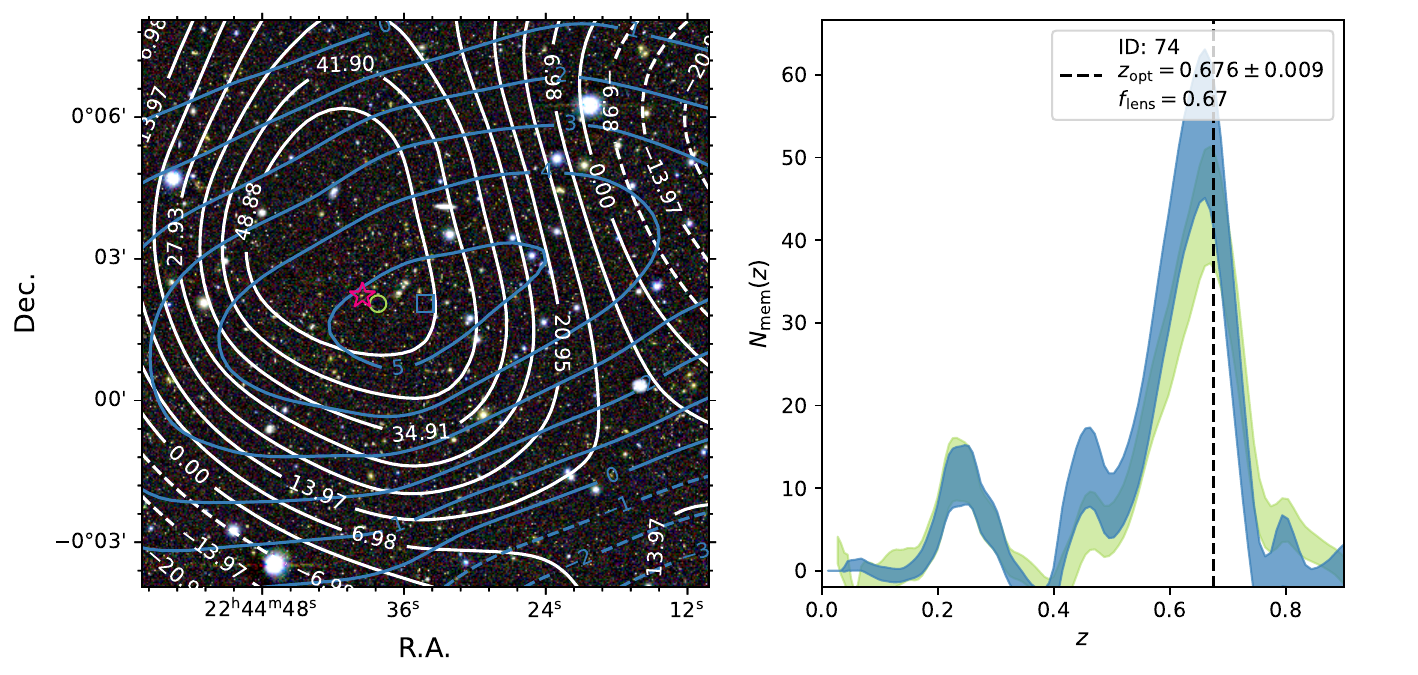}}
\resizebox{0.245\textwidth}{!}{\includegraphics[scale=1]{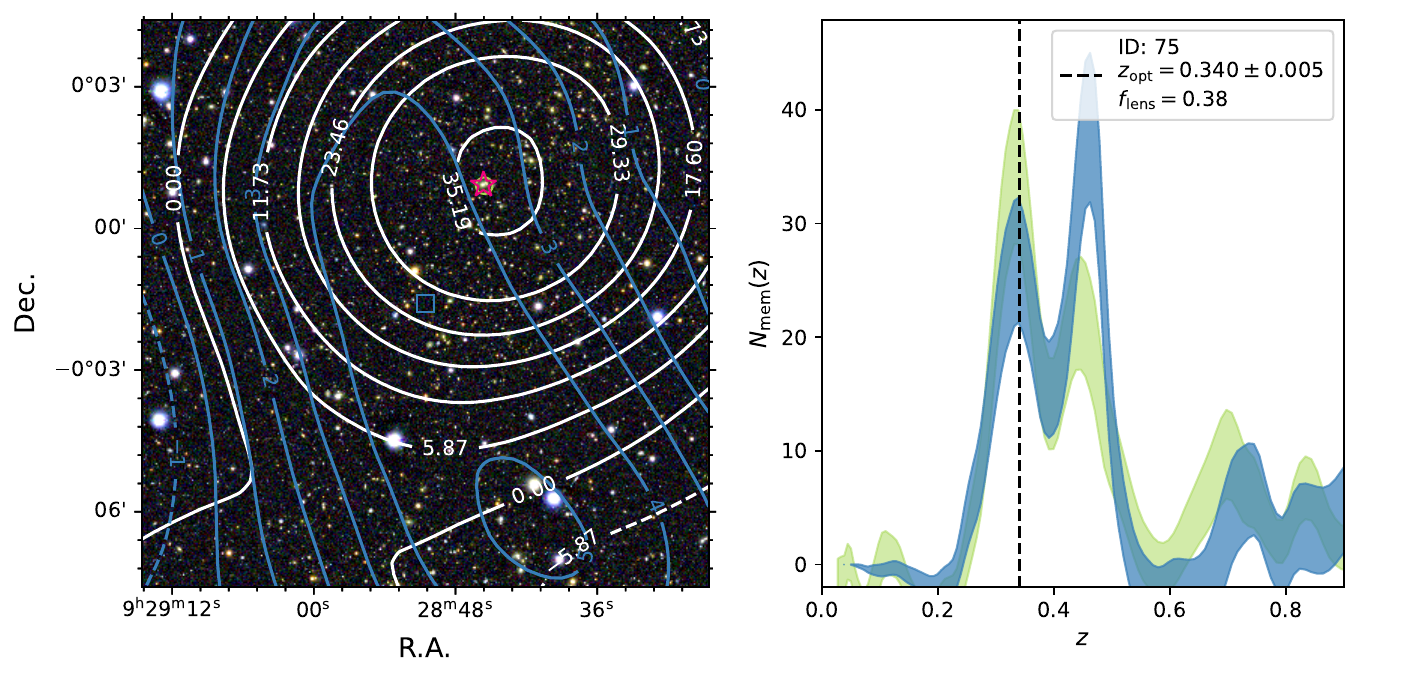}}
\resizebox{0.245\textwidth}{!}{\includegraphics[scale=1]{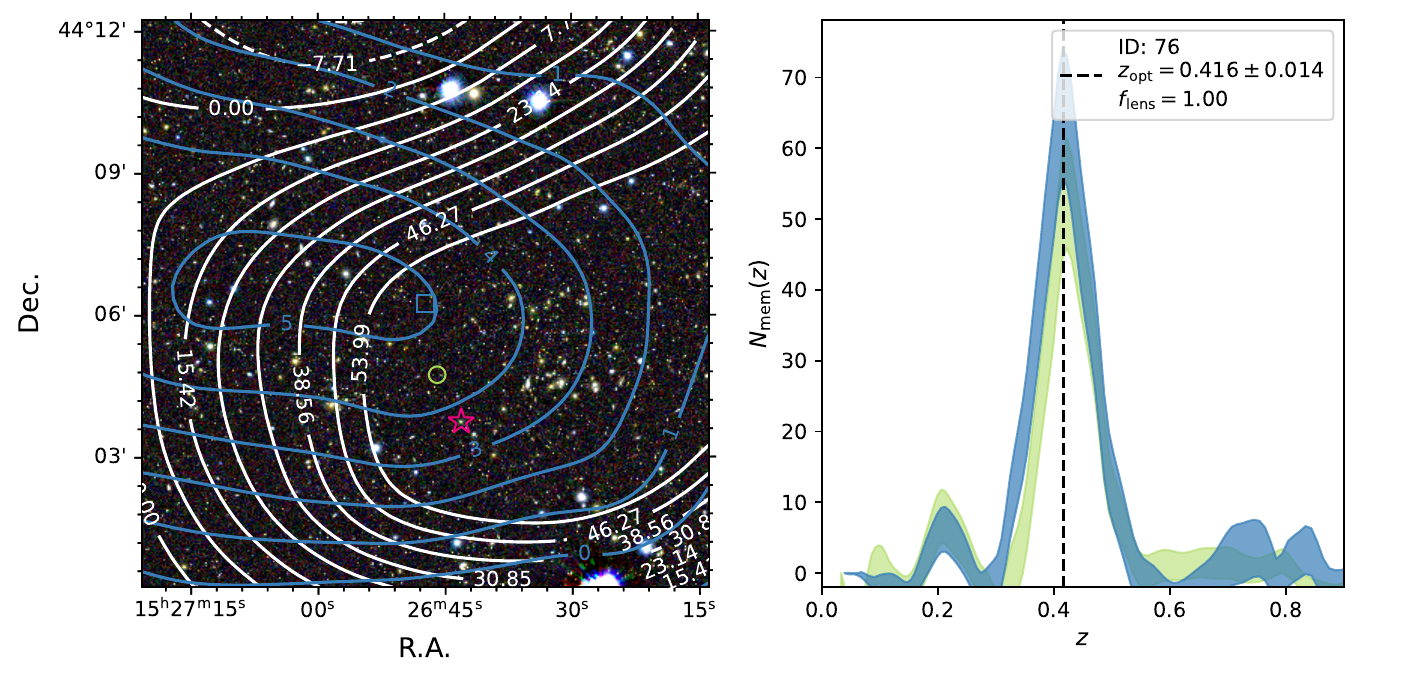}}
\resizebox{0.245\textwidth}{!}{\includegraphics[scale=1]{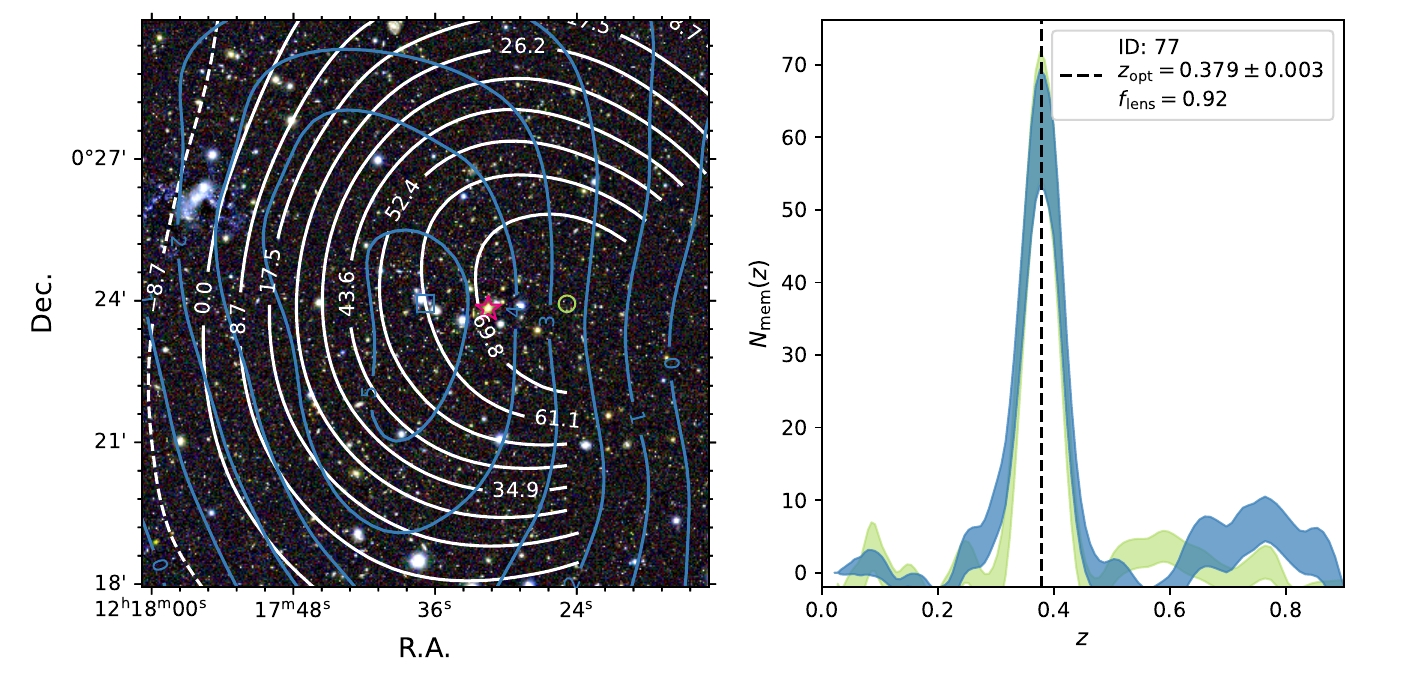}}
\resizebox{0.245\textwidth}{!}{\includegraphics[scale=1]{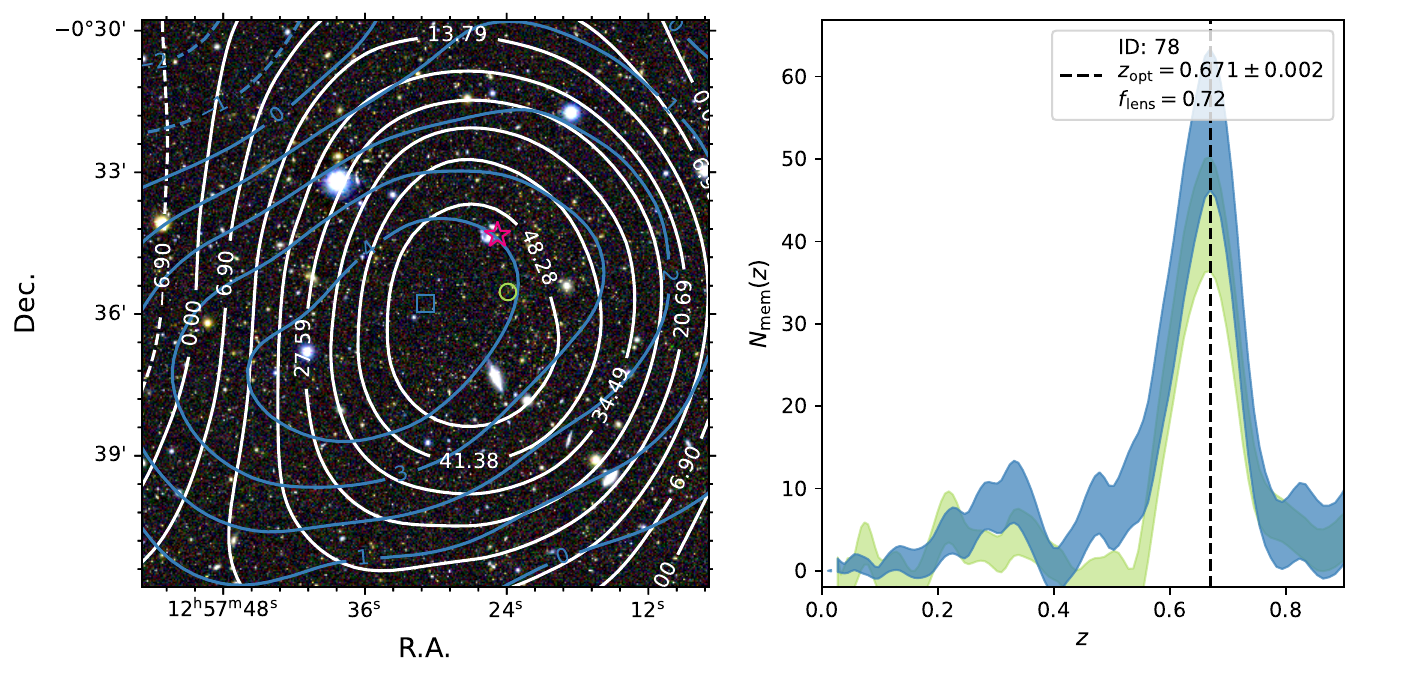}}
\resizebox{0.245\textwidth}{!}{\includegraphics[scale=1]{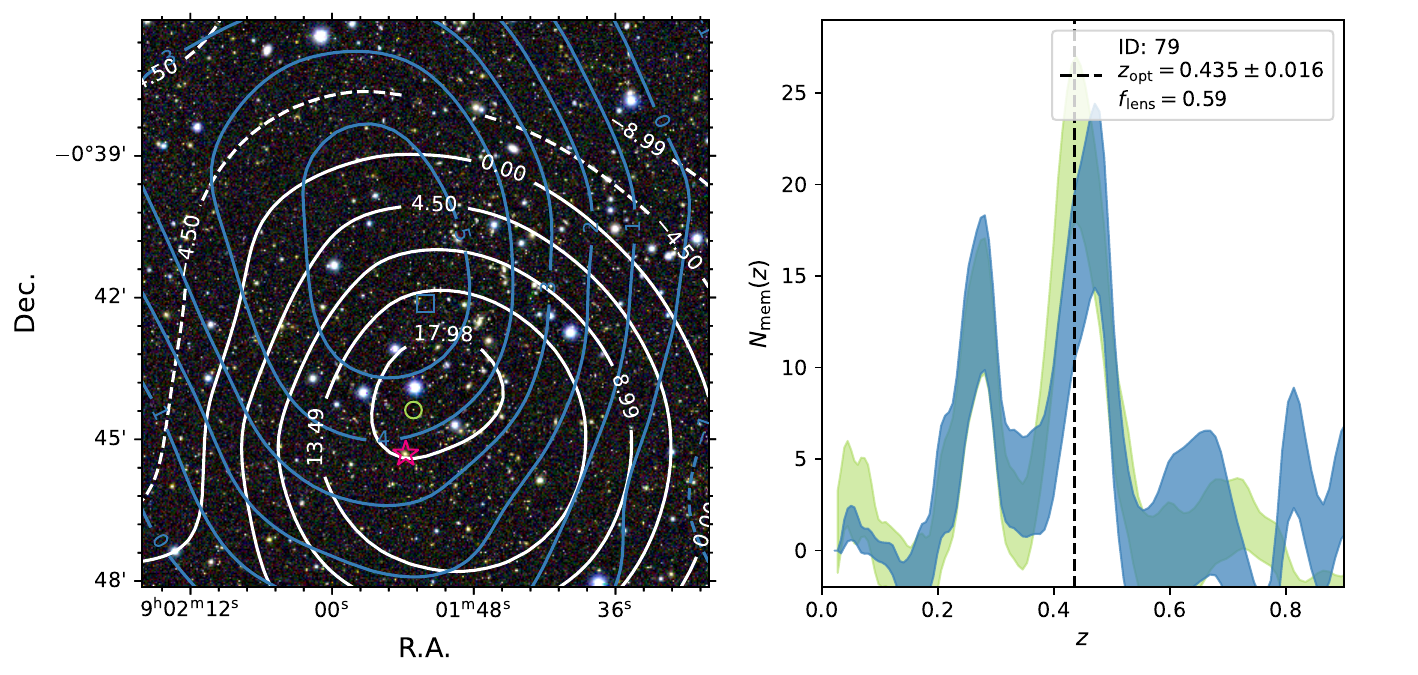}}
\caption{Continued from Figure~\ref{fig:optical_images_0}.}
\label{fig:optical_images_1}
\end{figure*}
\begin{figure*}
\centering
\resizebox{0.245\textwidth}{!}{\includegraphics[scale=1]{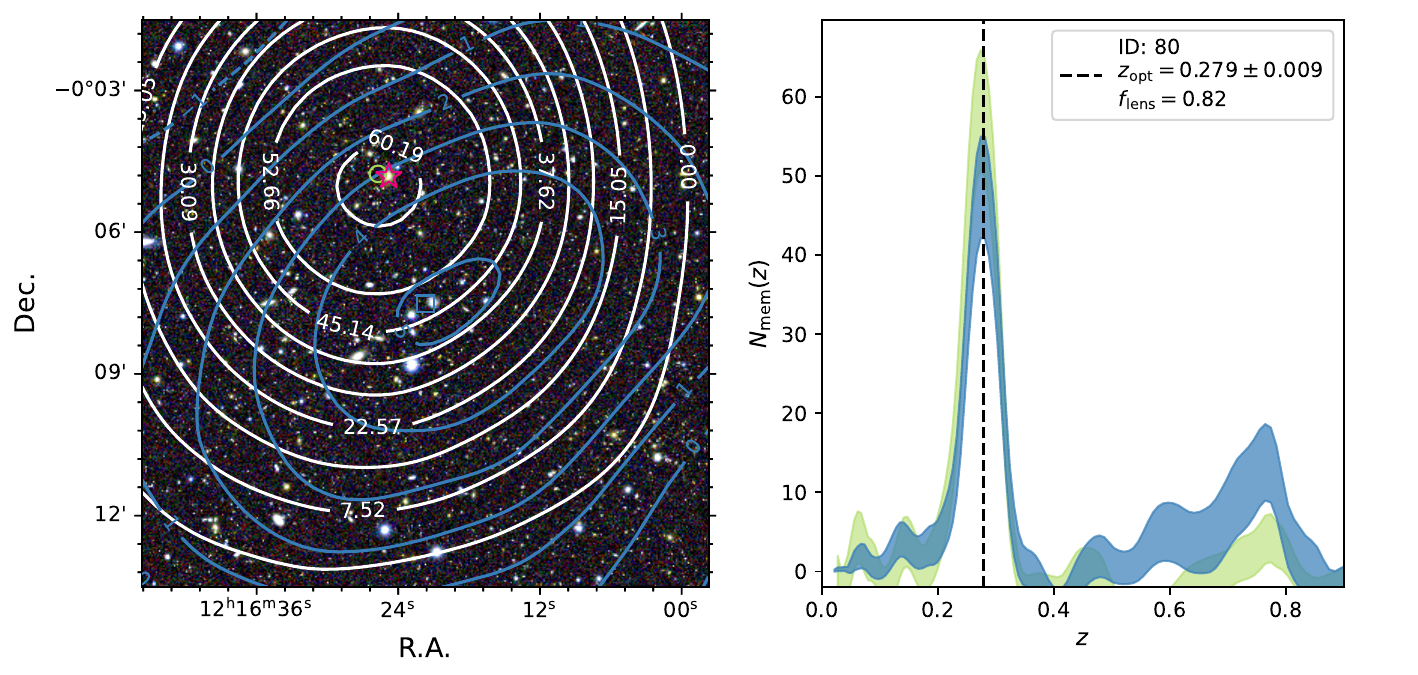}}
\resizebox{0.245\textwidth}{!}{\includegraphics[scale=1]{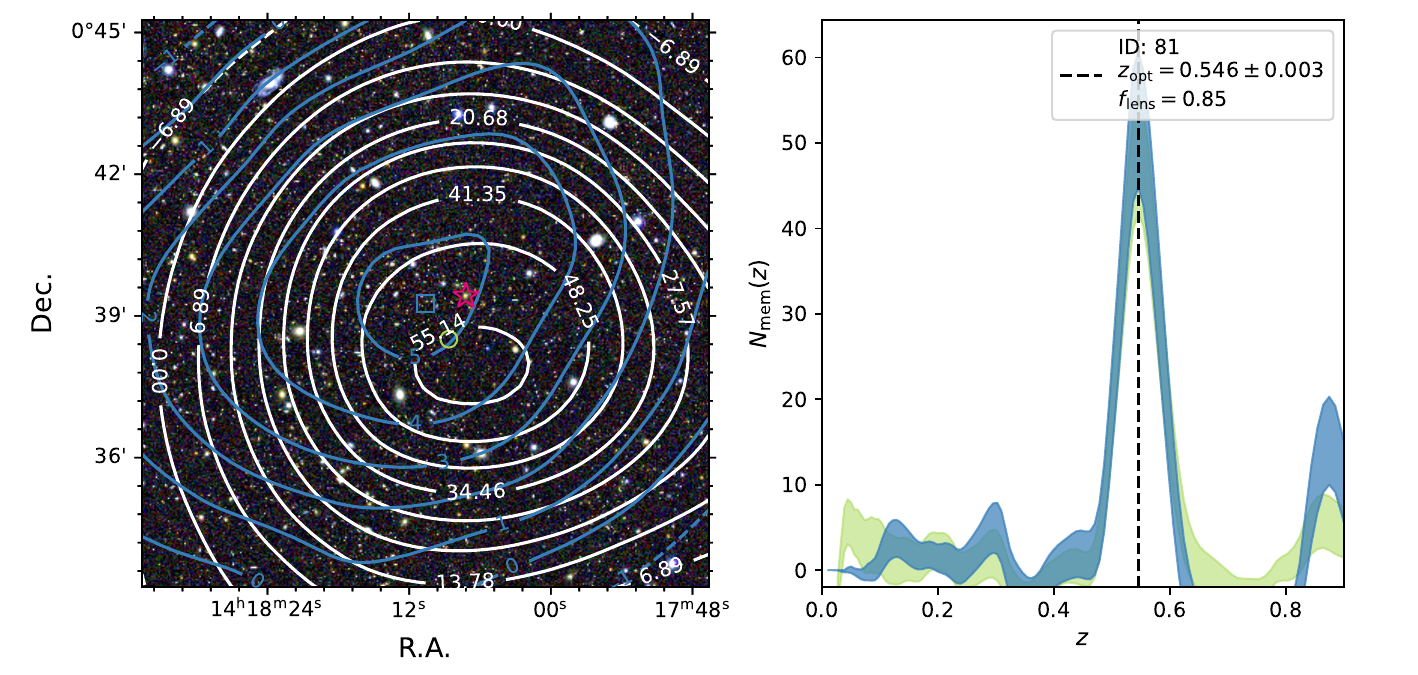}}
\resizebox{0.245\textwidth}{!}{\includegraphics[scale=1]{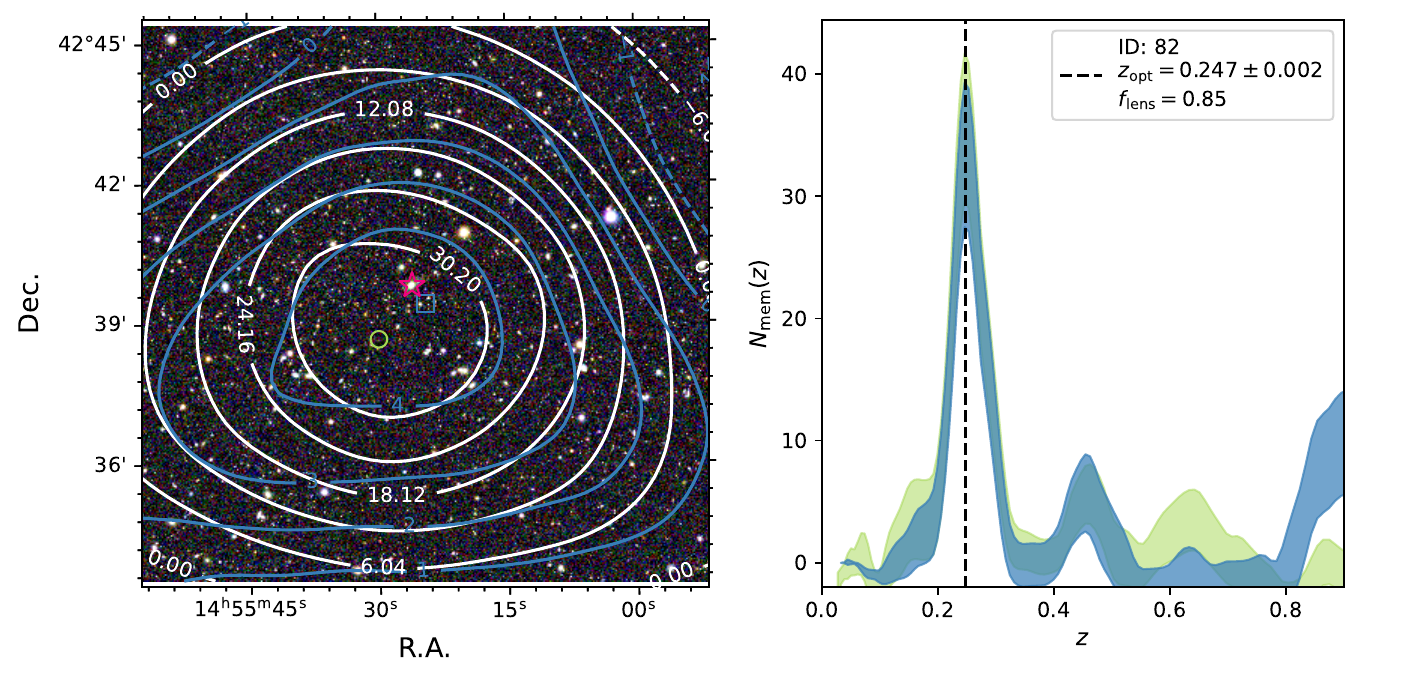}}
\resizebox{0.245\textwidth}{!}{\includegraphics[scale=1]{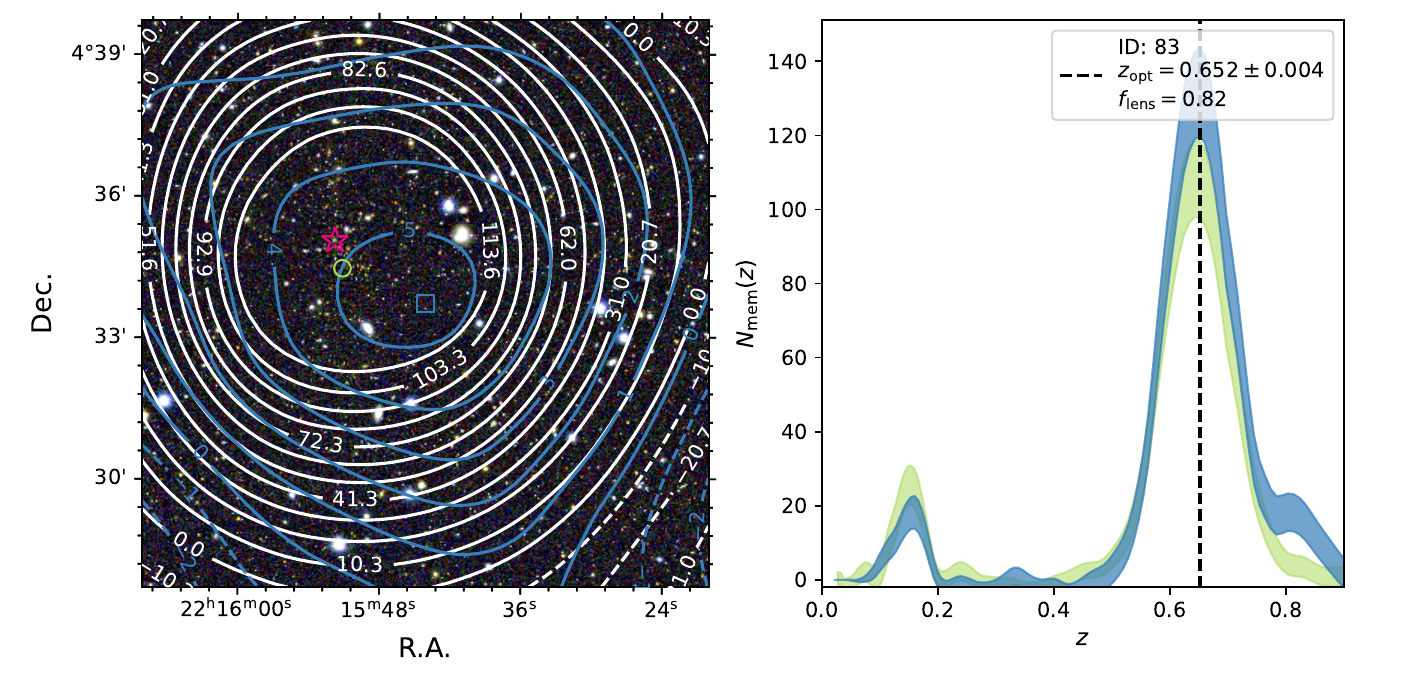}}
\resizebox{0.245\textwidth}{!}{\includegraphics[scale=1]{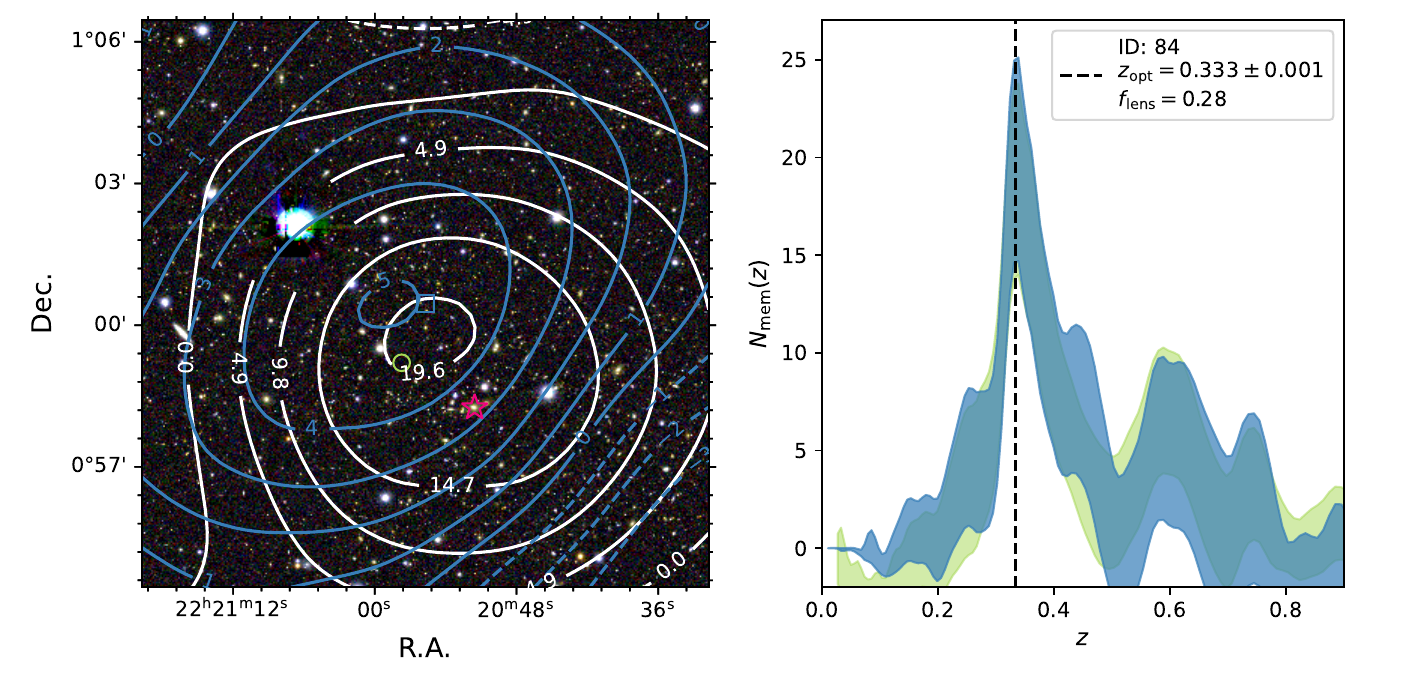}}
\resizebox{0.245\textwidth}{!}{\includegraphics[scale=1]{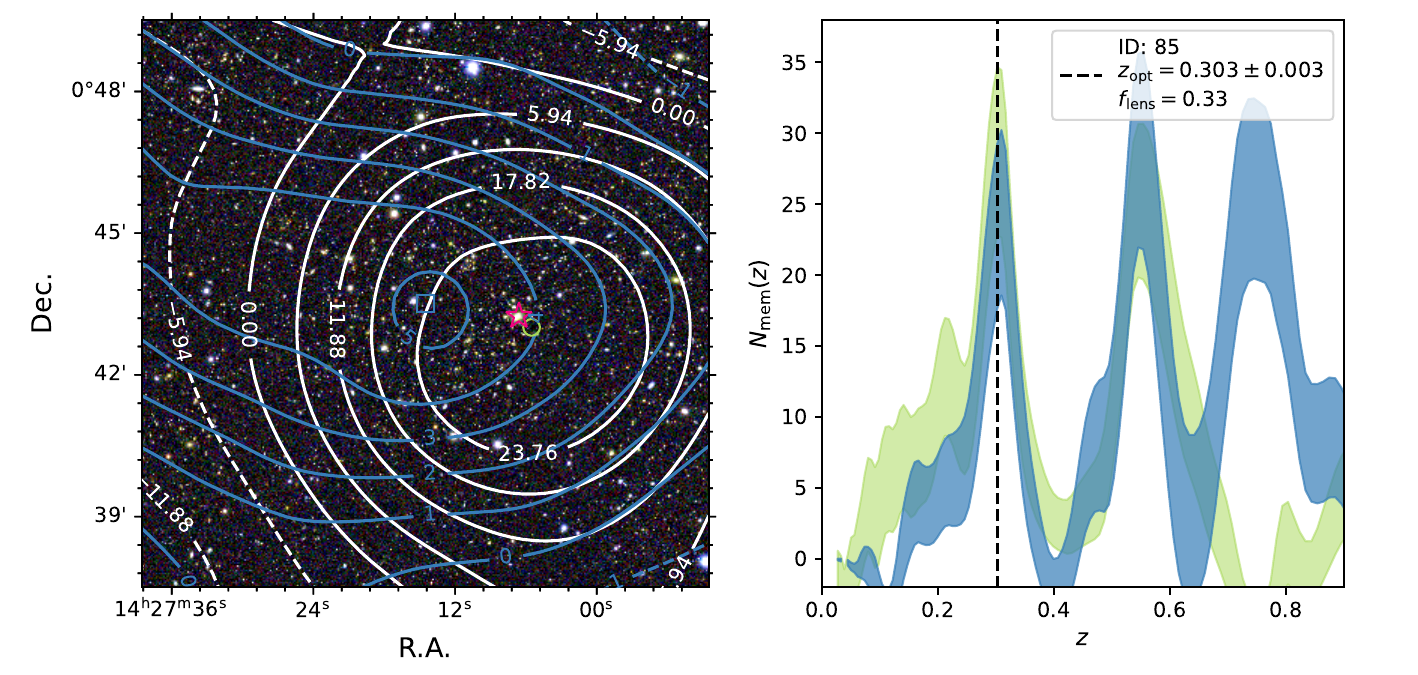}}
\resizebox{0.245\textwidth}{!}{\includegraphics[scale=1]{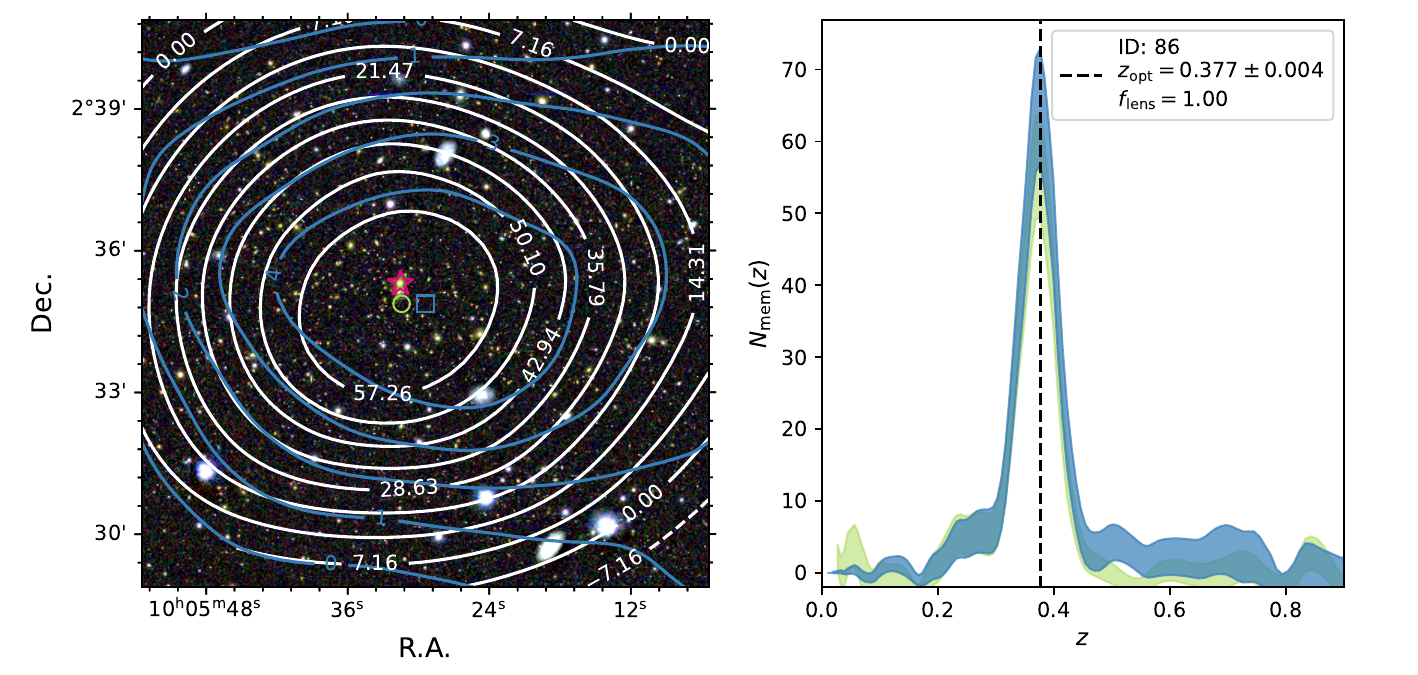}}
\resizebox{0.245\textwidth}{!}{\includegraphics[scale=1]{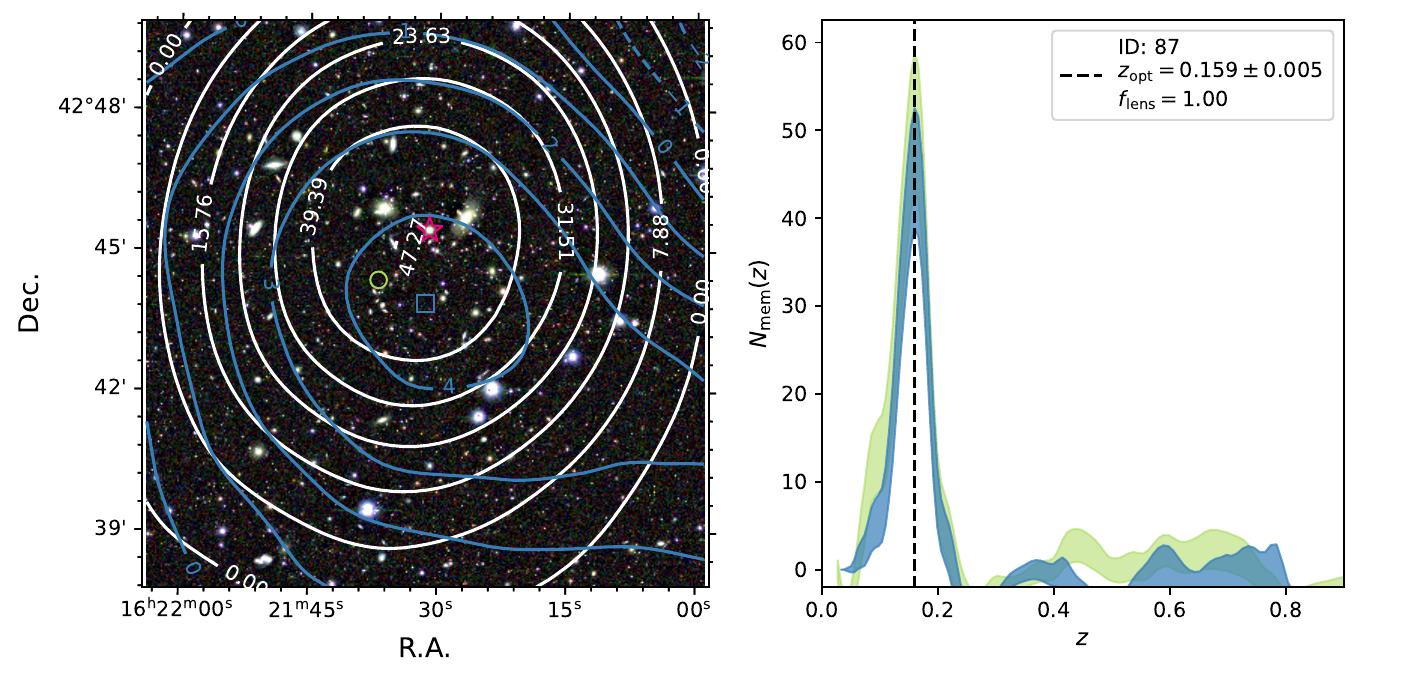}}
\resizebox{0.245\textwidth}{!}{\includegraphics[scale=1]{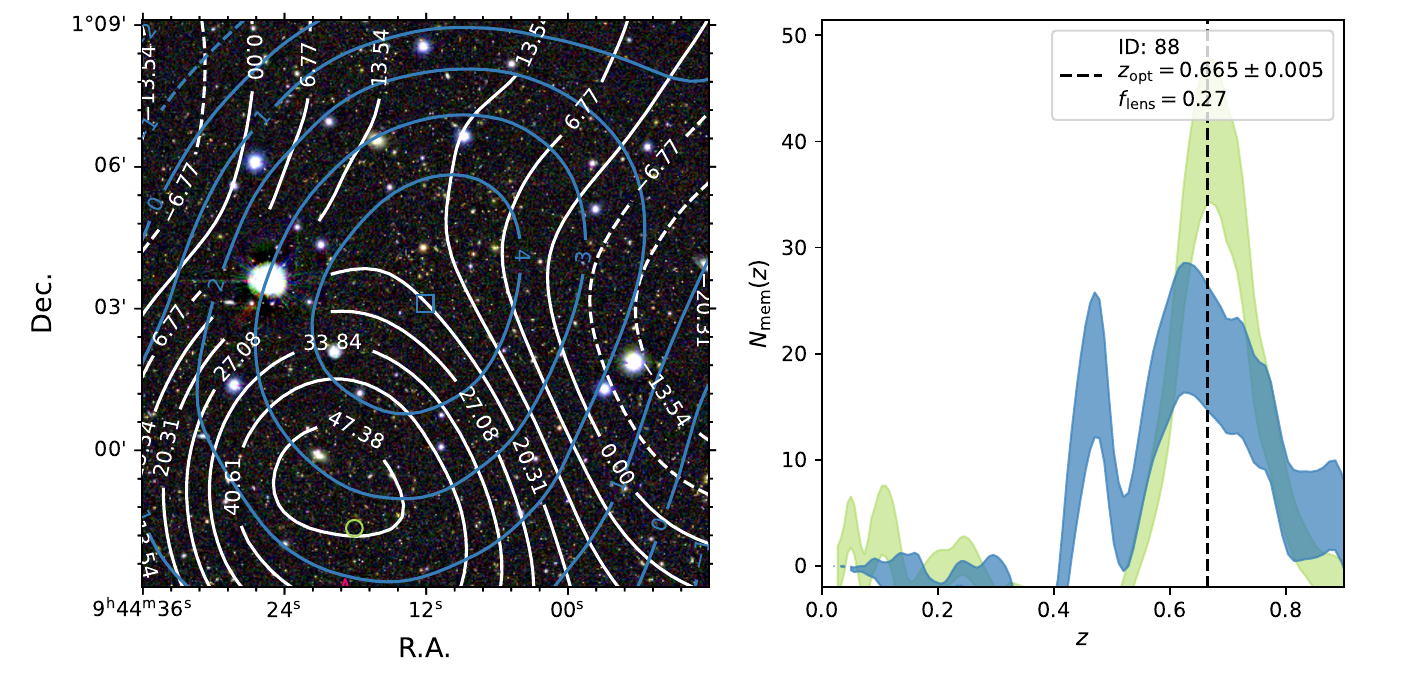}}
\resizebox{0.245\textwidth}{!}{\includegraphics[scale=1]{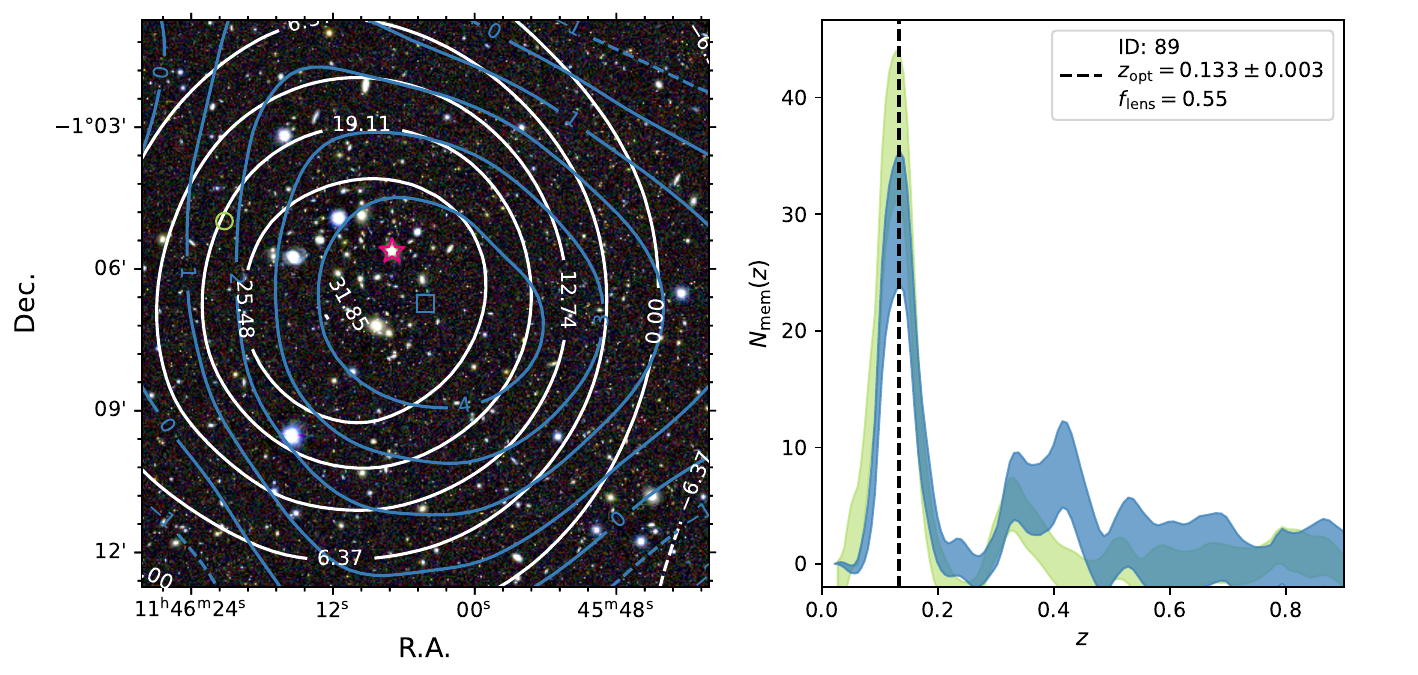}}
\resizebox{0.245\textwidth}{!}{\includegraphics[scale=1]{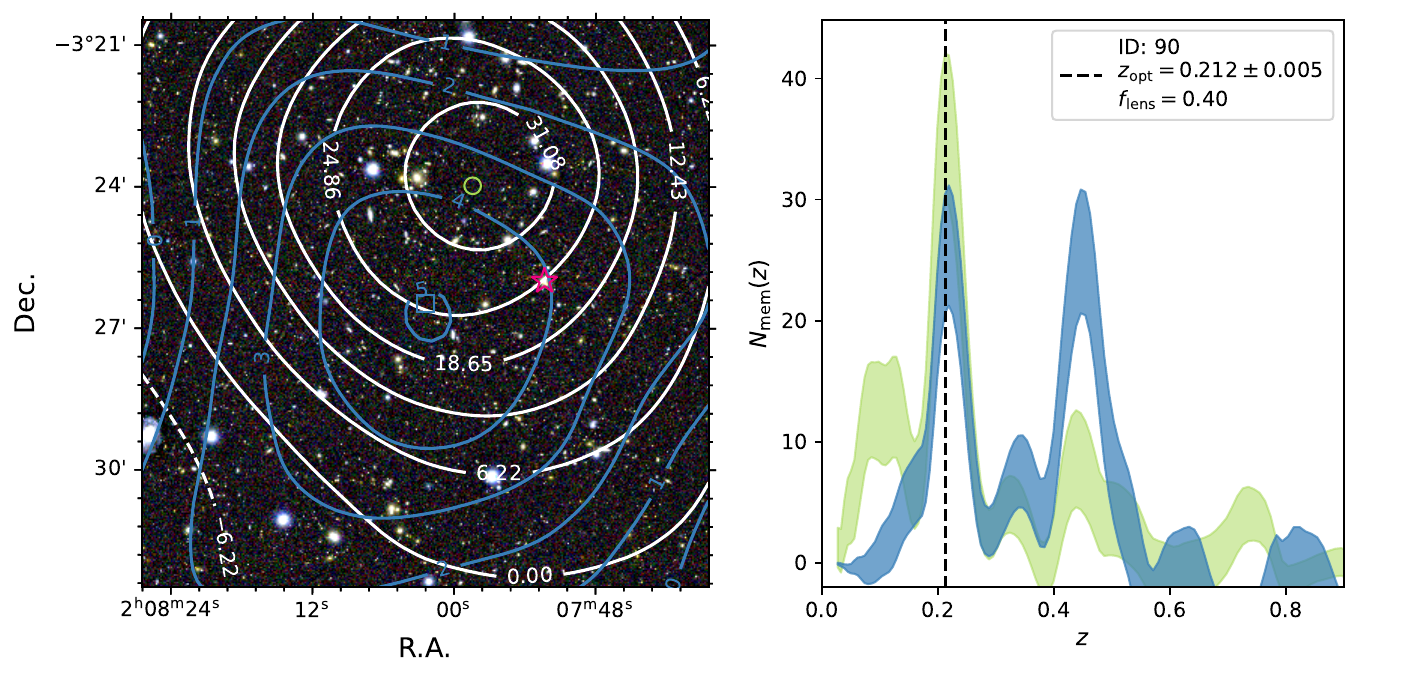}}
\resizebox{0.245\textwidth}{!}{\includegraphics[scale=1]{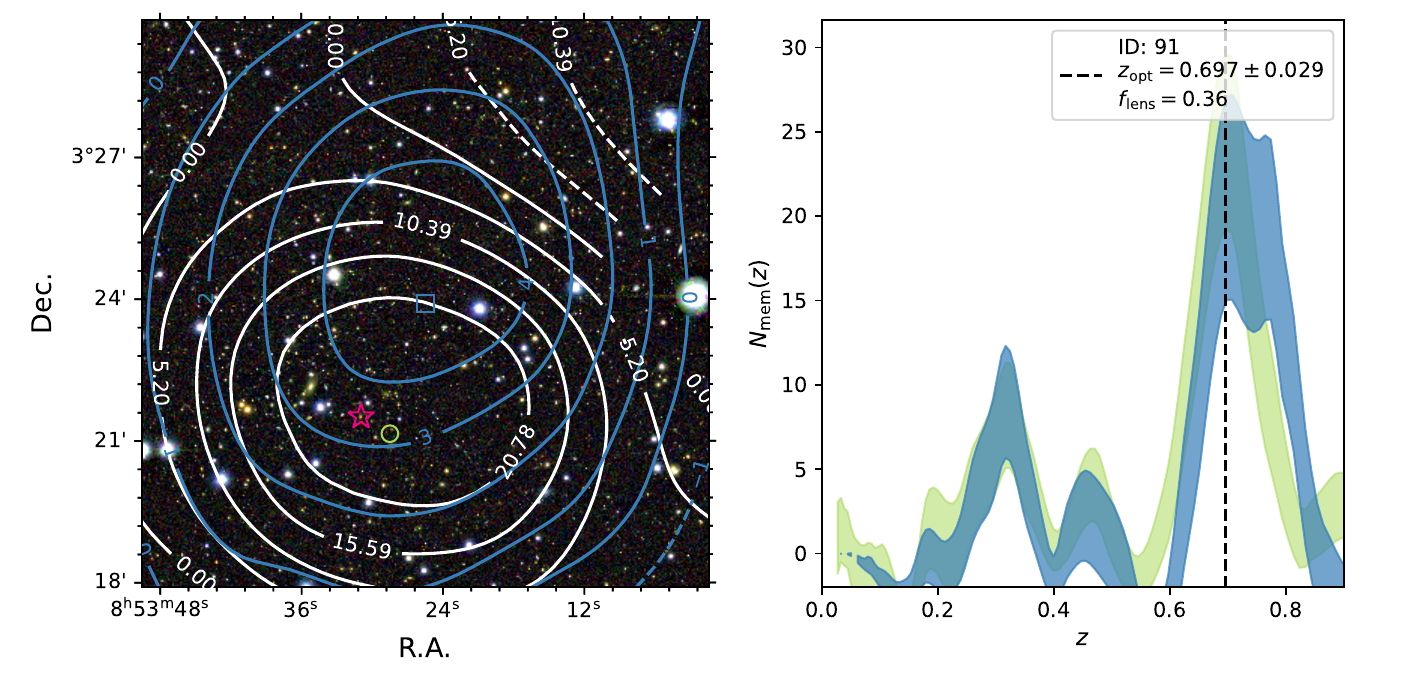}}
\resizebox{0.245\textwidth}{!}{\includegraphics[scale=1]{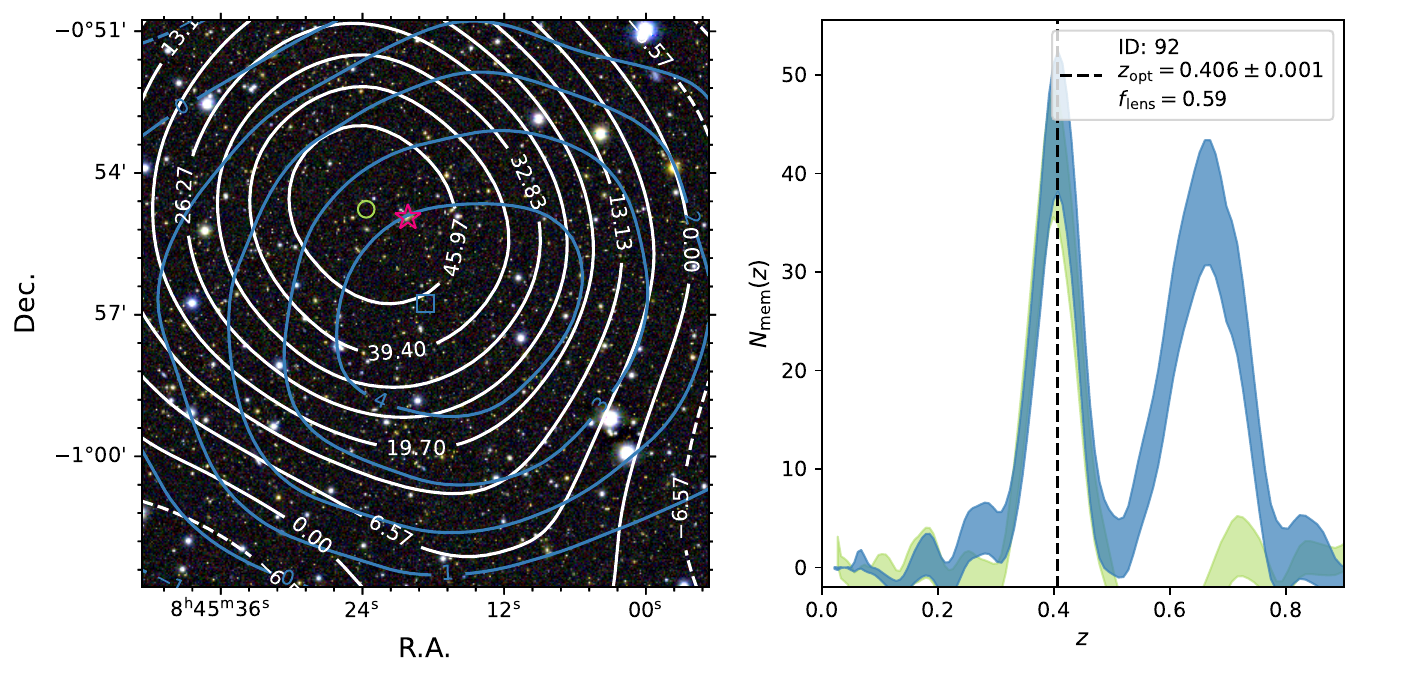}}
\resizebox{0.245\textwidth}{!}{\includegraphics[scale=1]{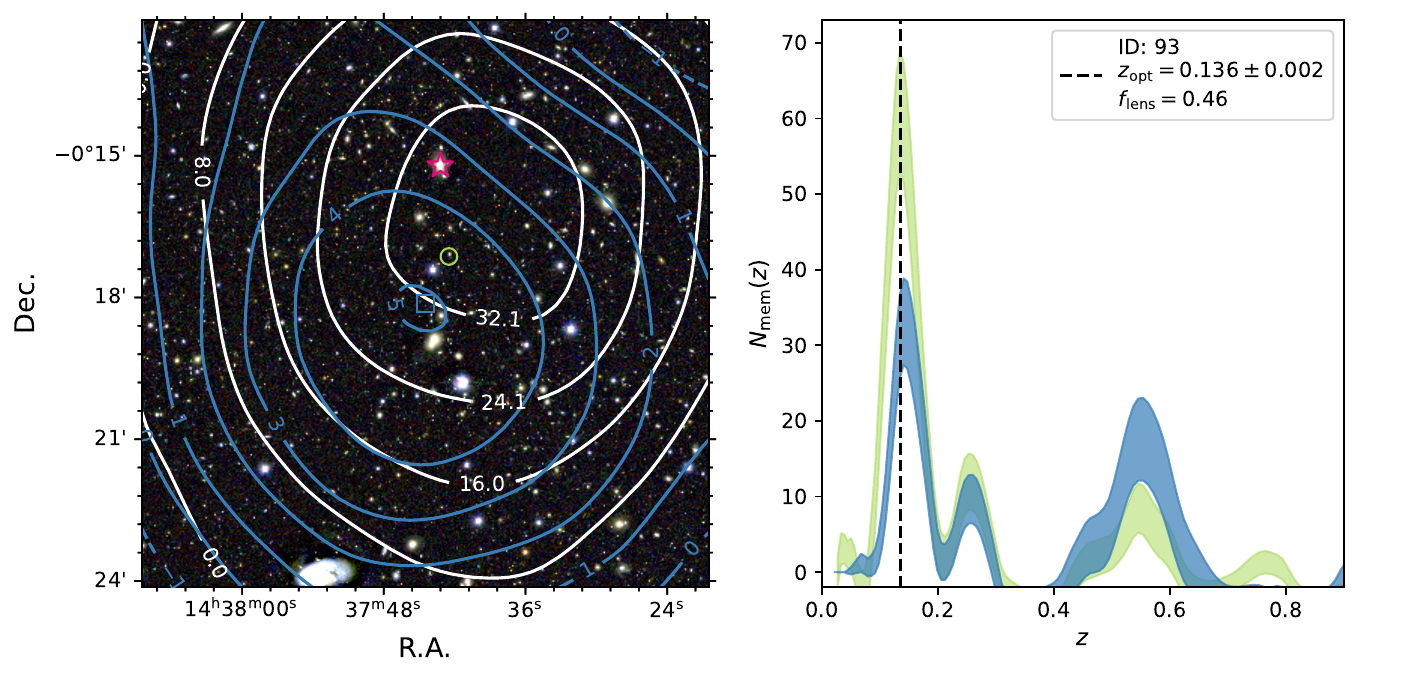}}
\resizebox{0.245\textwidth}{!}{\includegraphics[scale=1]{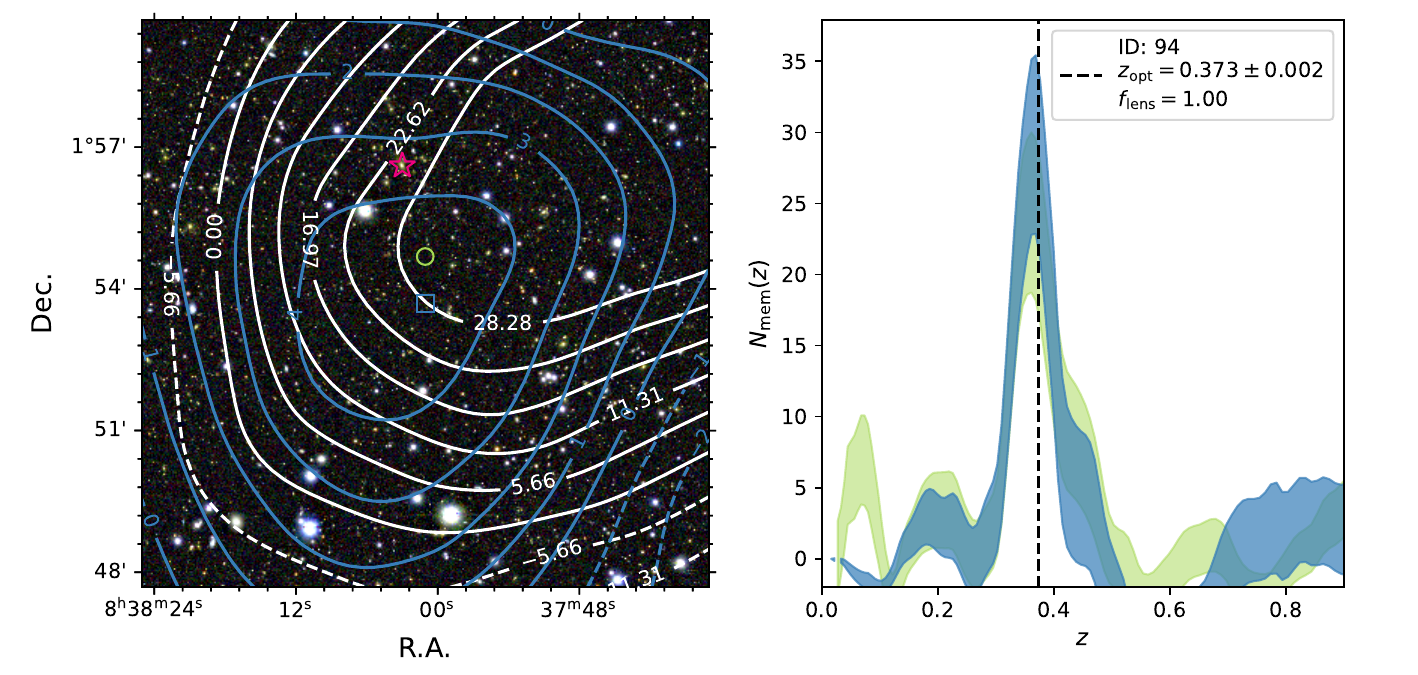}}
\resizebox{0.245\textwidth}{!}{\includegraphics[scale=1]{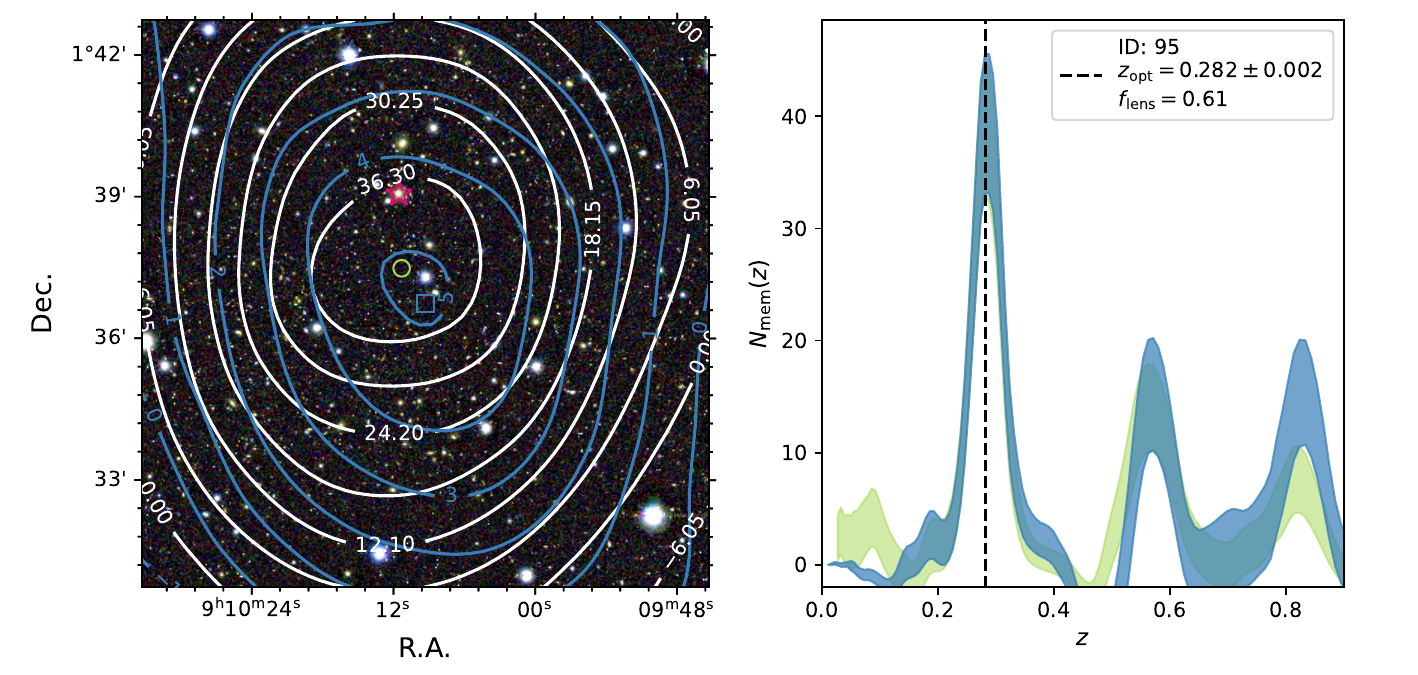}}
\resizebox{0.245\textwidth}{!}{\includegraphics[scale=1]{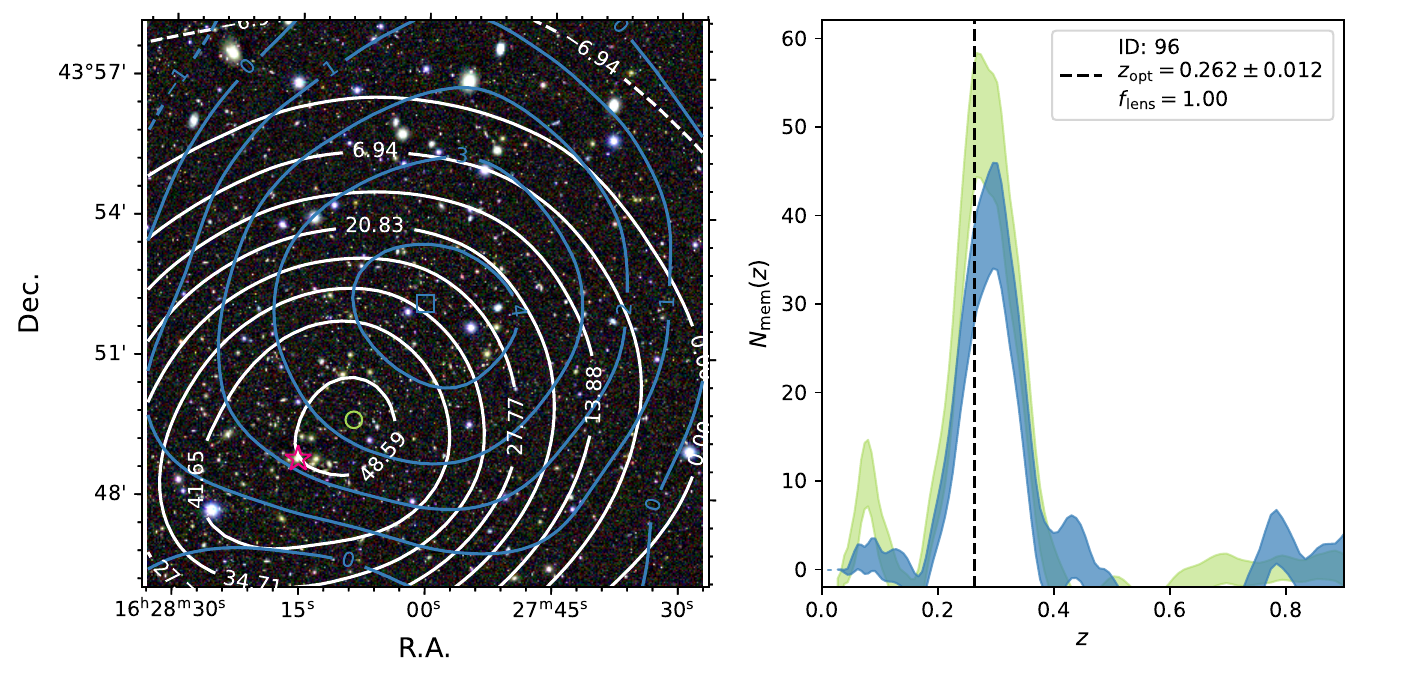}}
\resizebox{0.245\textwidth}{!}{\includegraphics[scale=1]{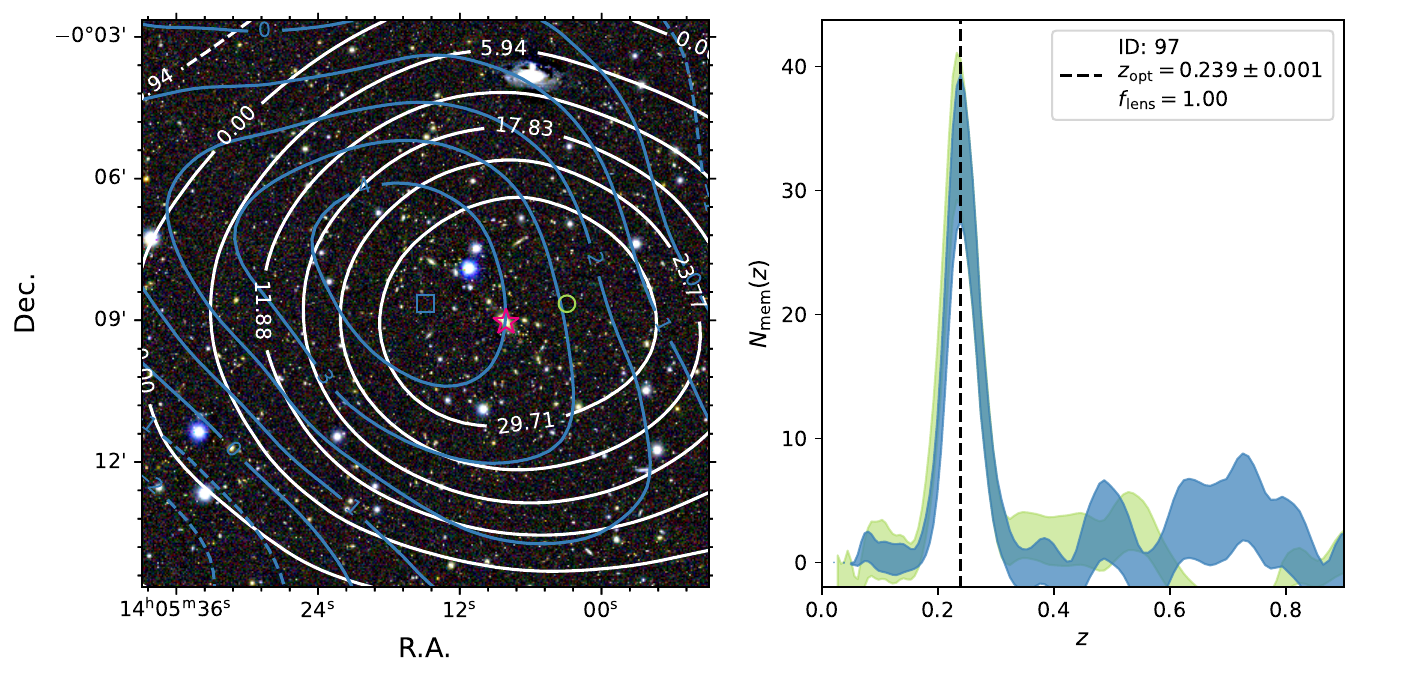}}
\resizebox{0.245\textwidth}{!}{\includegraphics[scale=1]{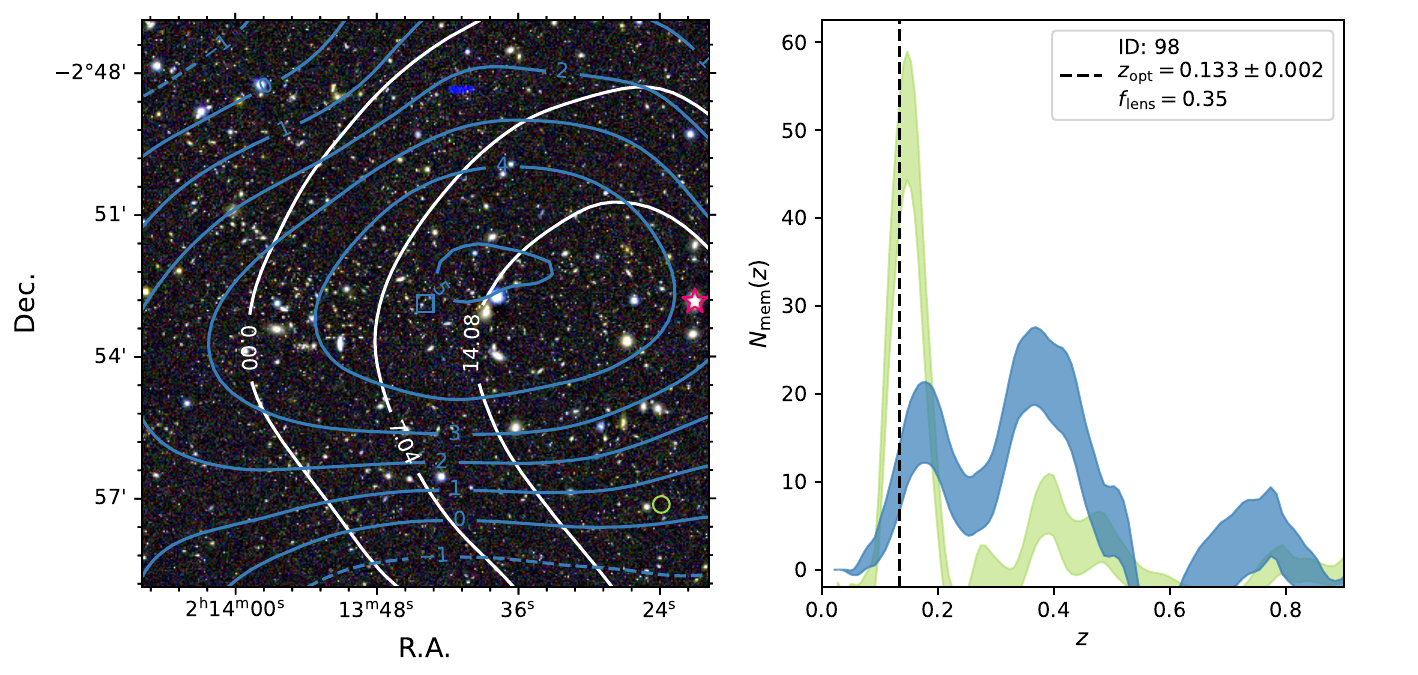}}
\resizebox{0.245\textwidth}{!}{\includegraphics[scale=1]{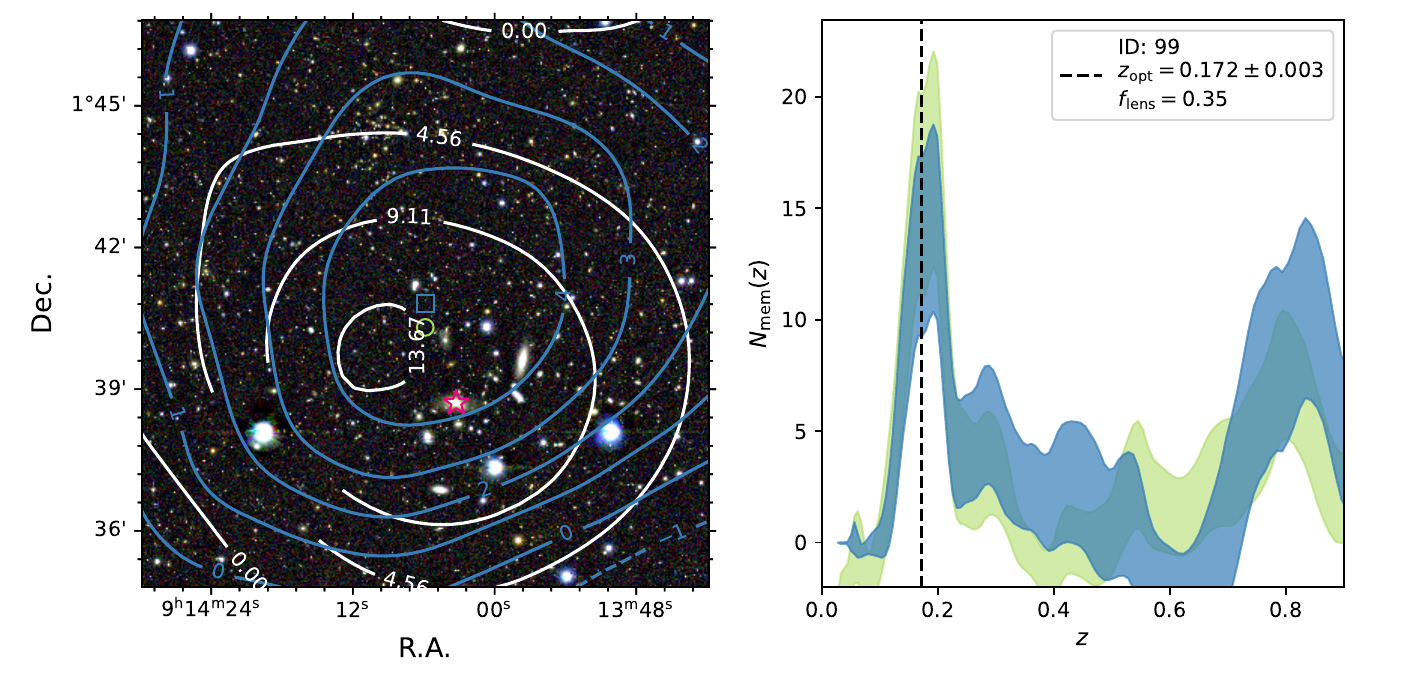}}
\resizebox{0.245\textwidth}{!}{\includegraphics[scale=1]{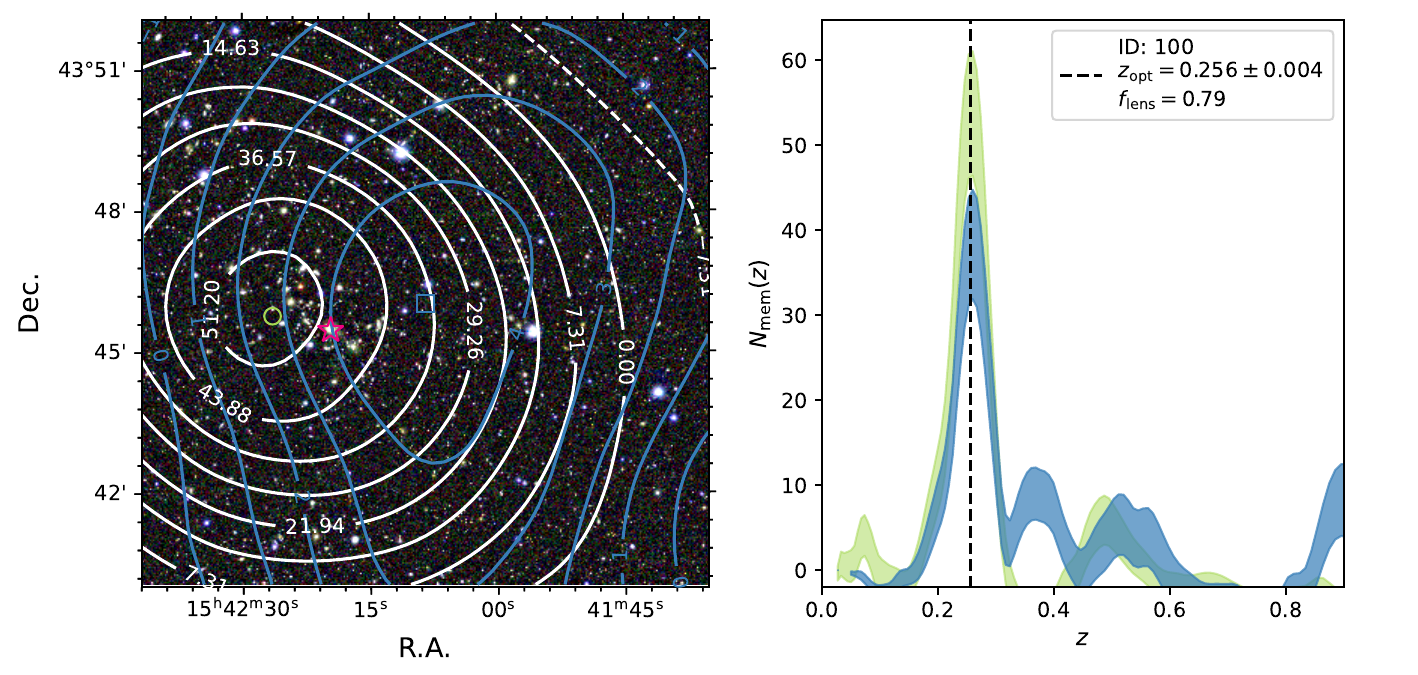}}
\resizebox{0.245\textwidth}{!}{\includegraphics[scale=1]{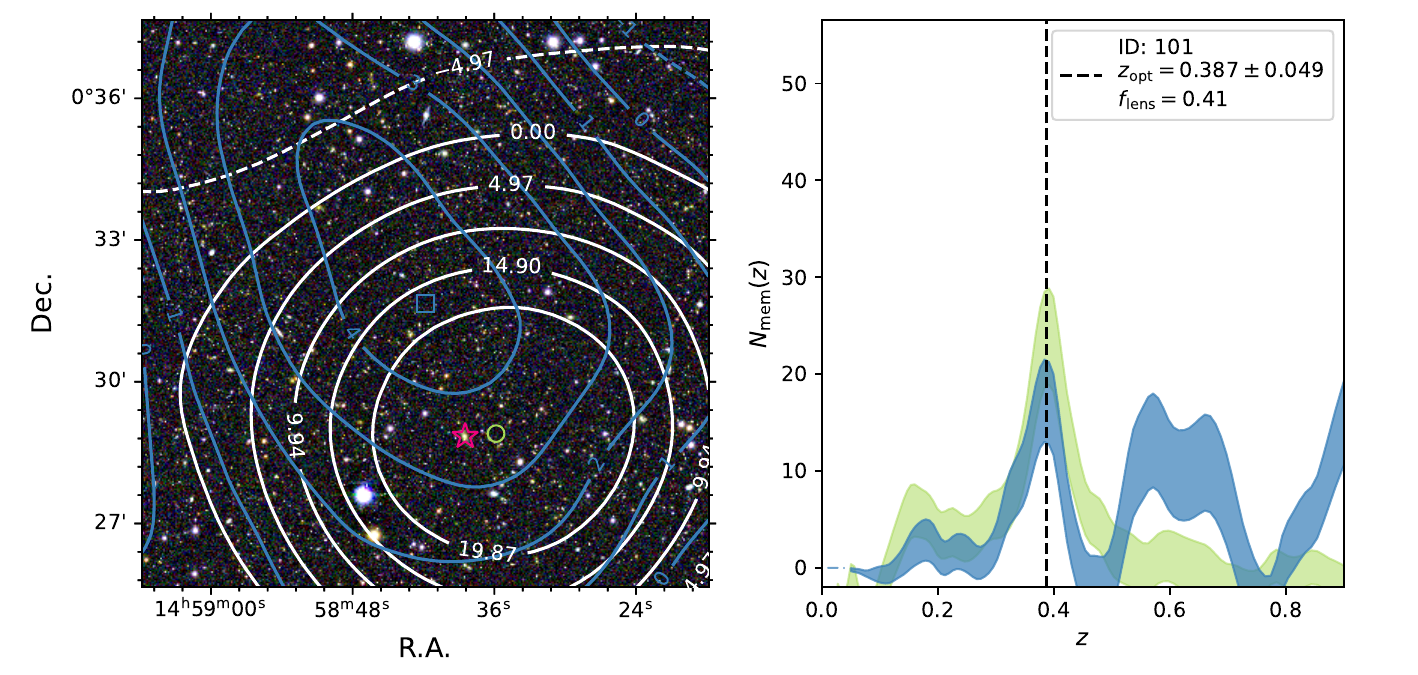}}
\resizebox{0.245\textwidth}{!}{\includegraphics[scale=1]{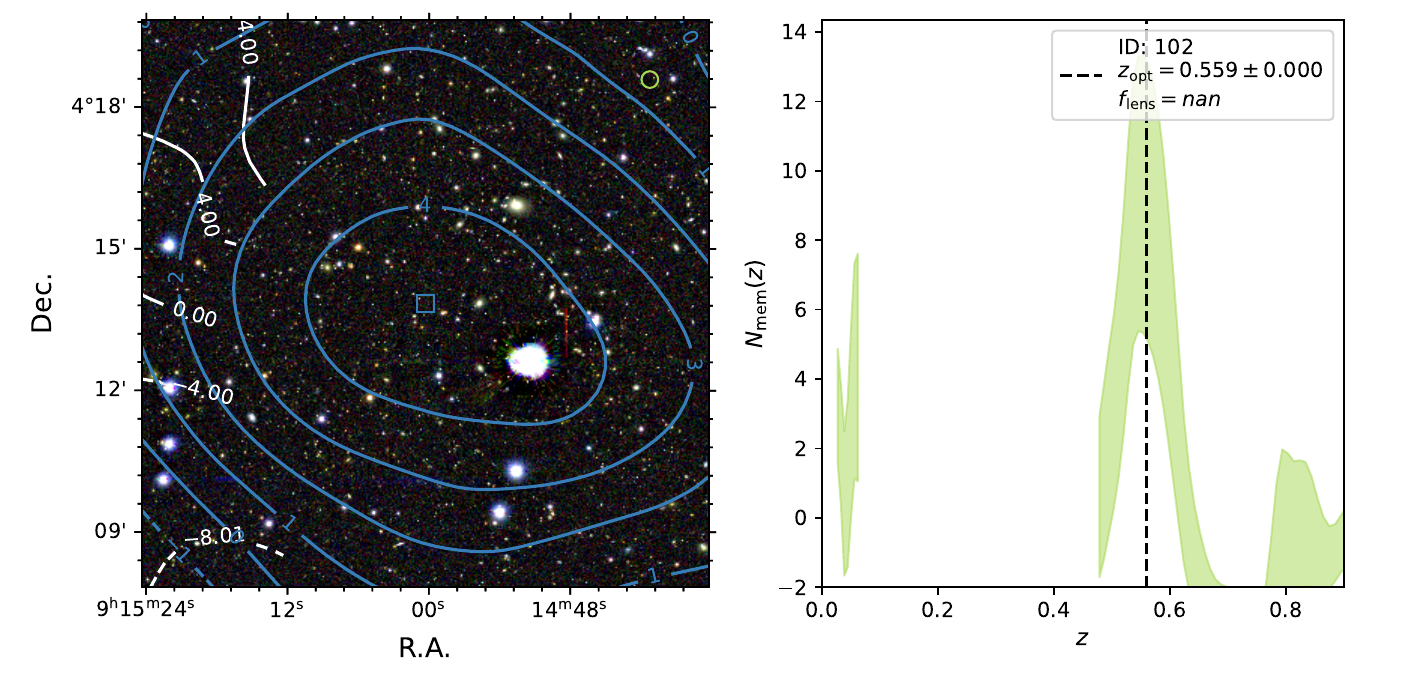}}
\resizebox{0.245\textwidth}{!}{\includegraphics[scale=1]{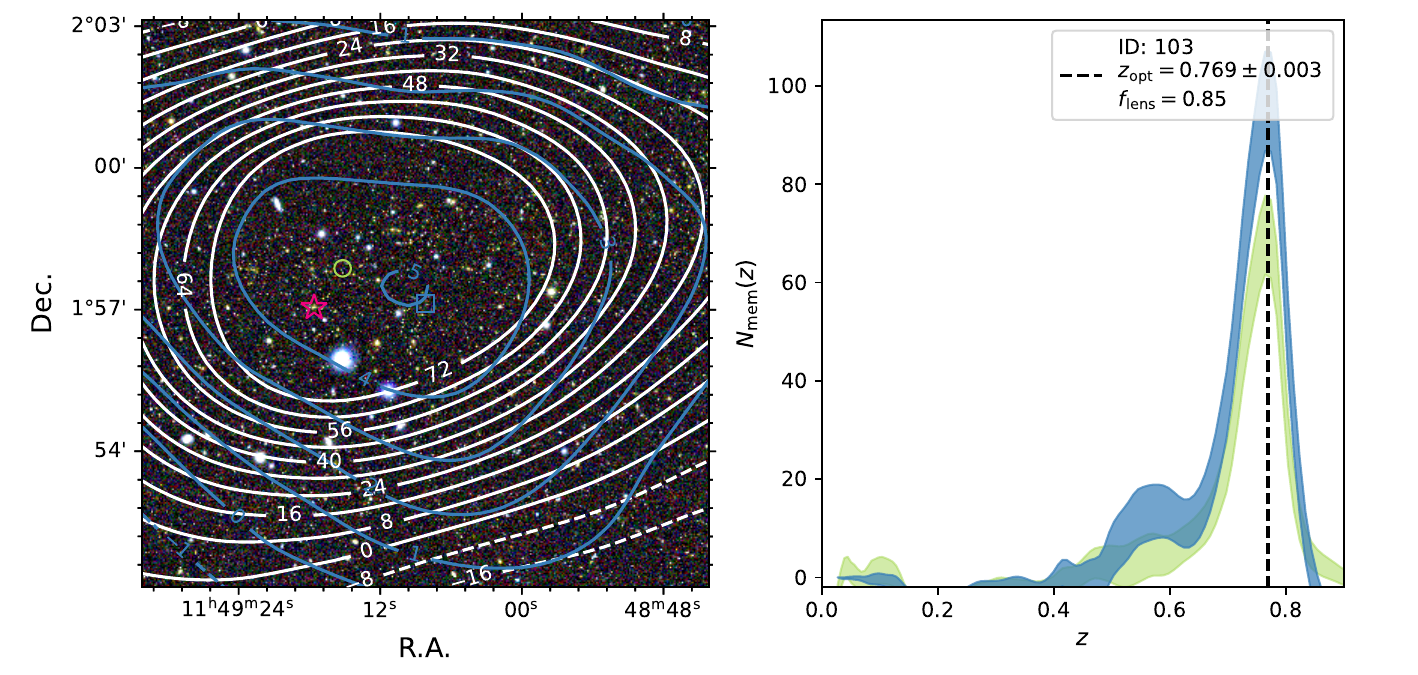}}
\resizebox{0.245\textwidth}{!}{\includegraphics[scale=1]{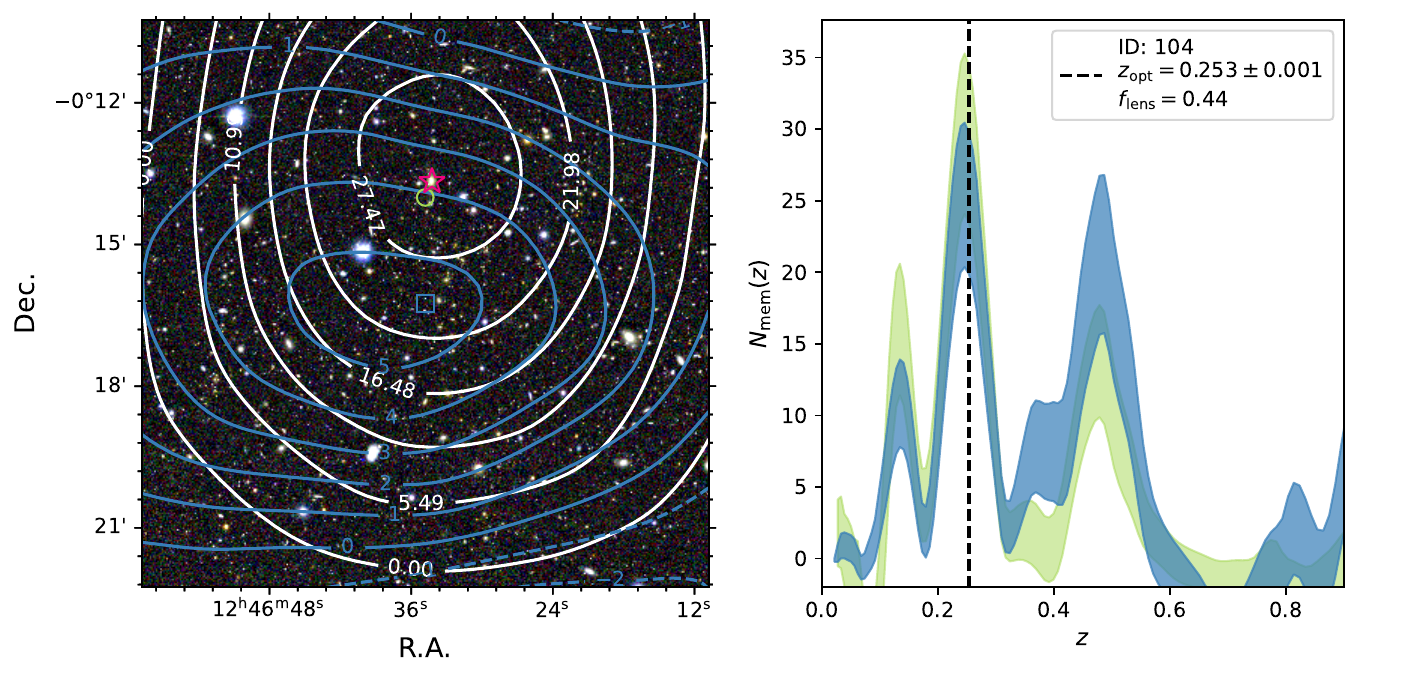}}
\resizebox{0.245\textwidth}{!}{\includegraphics[scale=1]{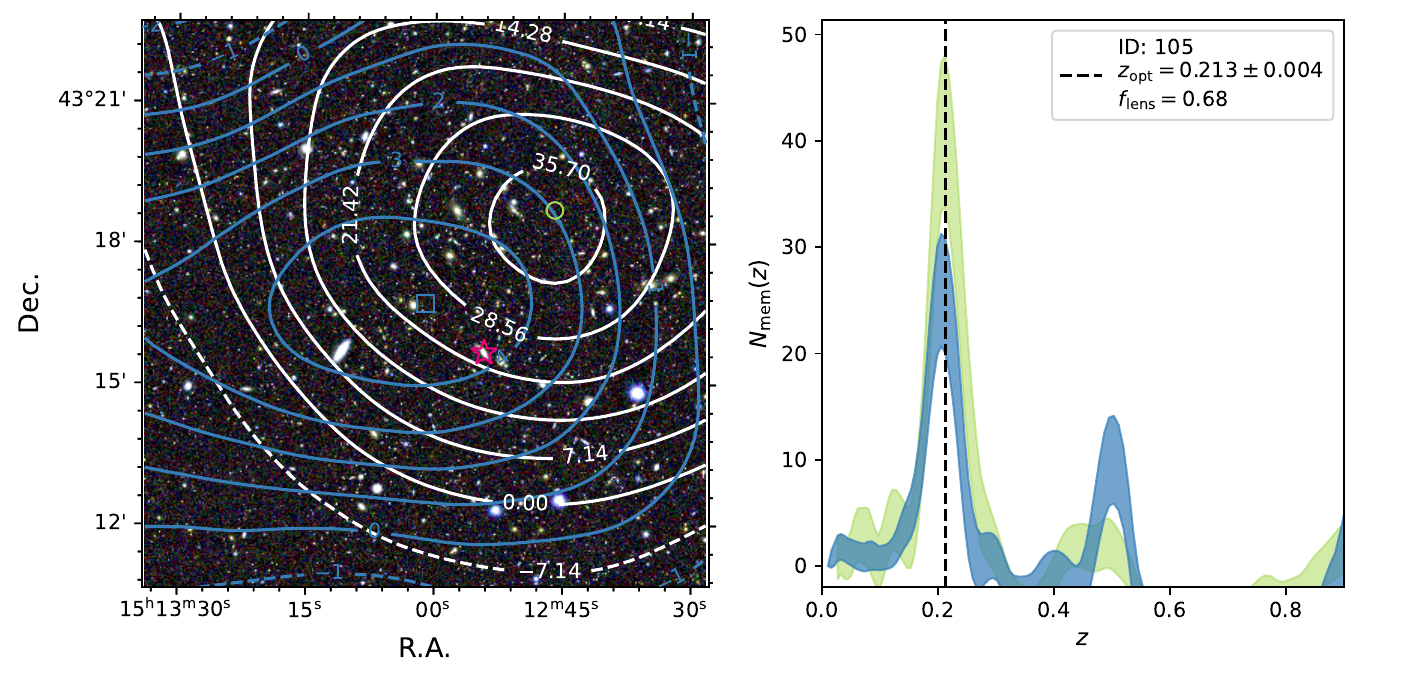}}
\resizebox{0.245\textwidth}{!}{\includegraphics[scale=1]{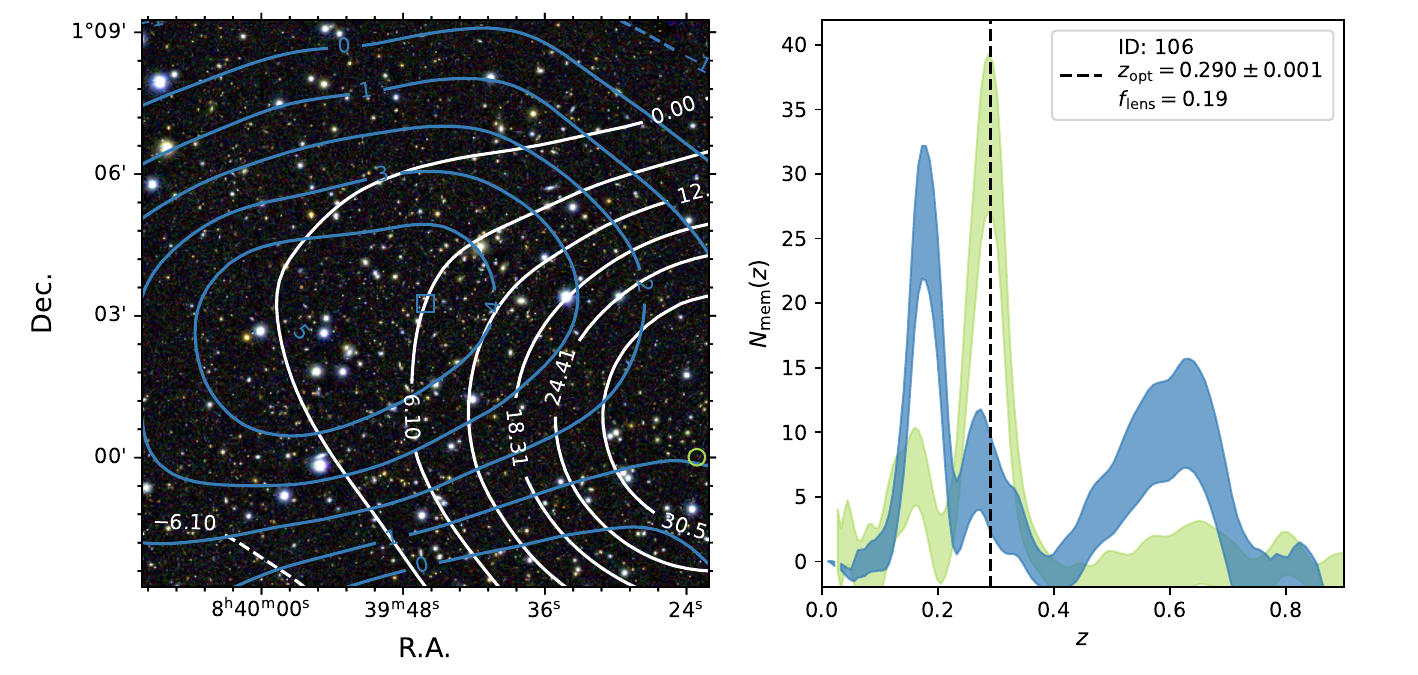}}
\resizebox{0.245\textwidth}{!}{\includegraphics[scale=1]{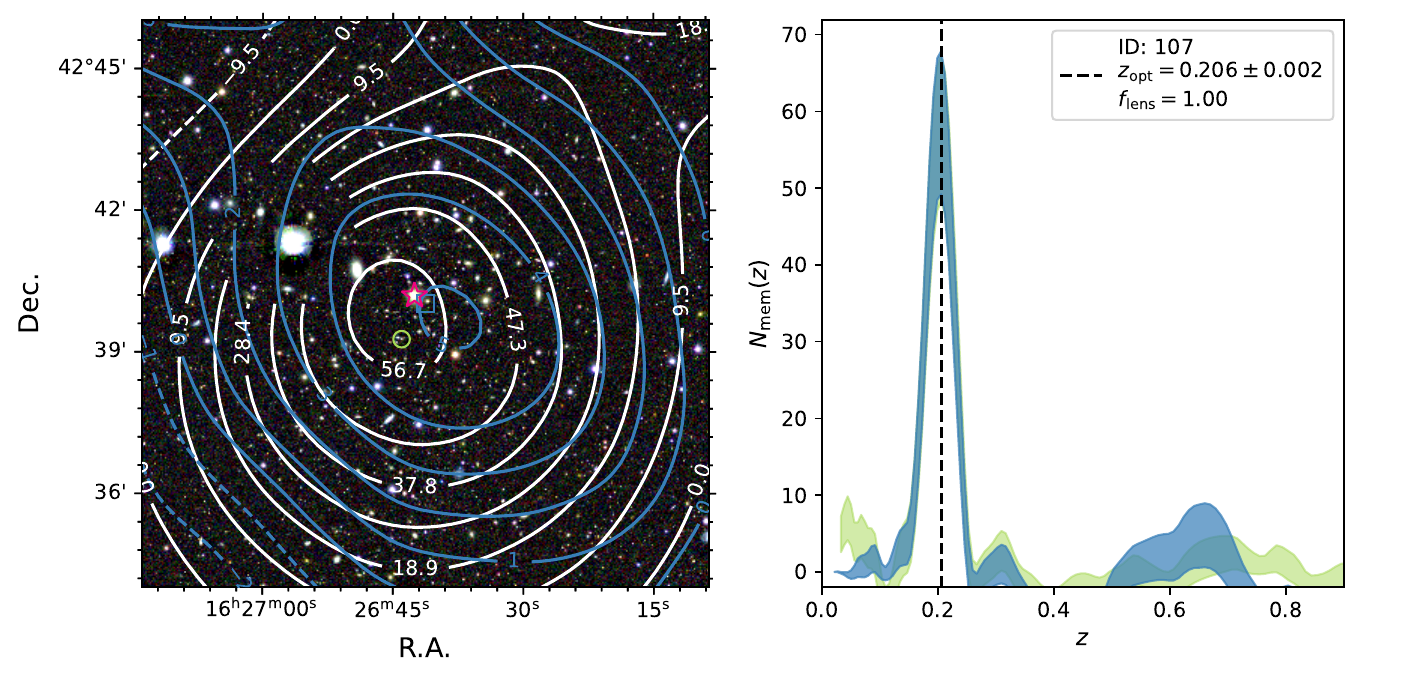}}
\resizebox{0.245\textwidth}{!}{\includegraphics[scale=1]{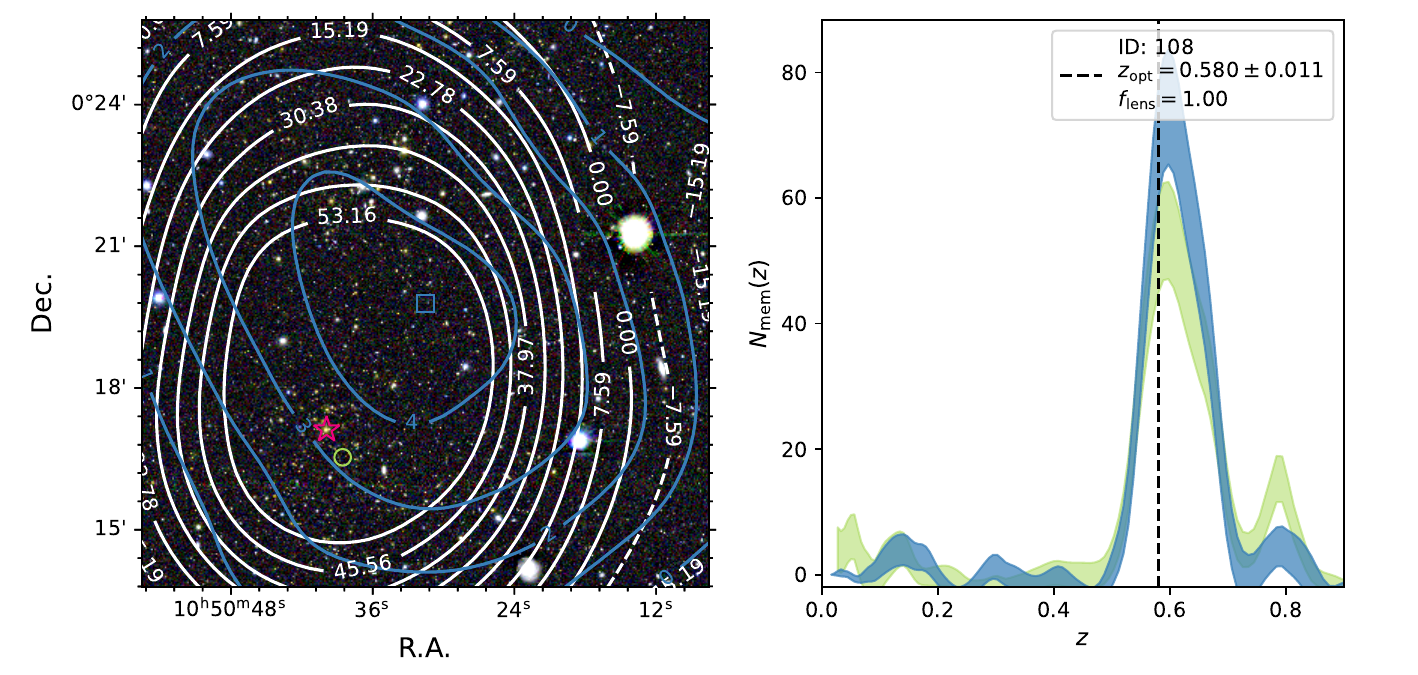}}
\resizebox{0.245\textwidth}{!}{\includegraphics[scale=1]{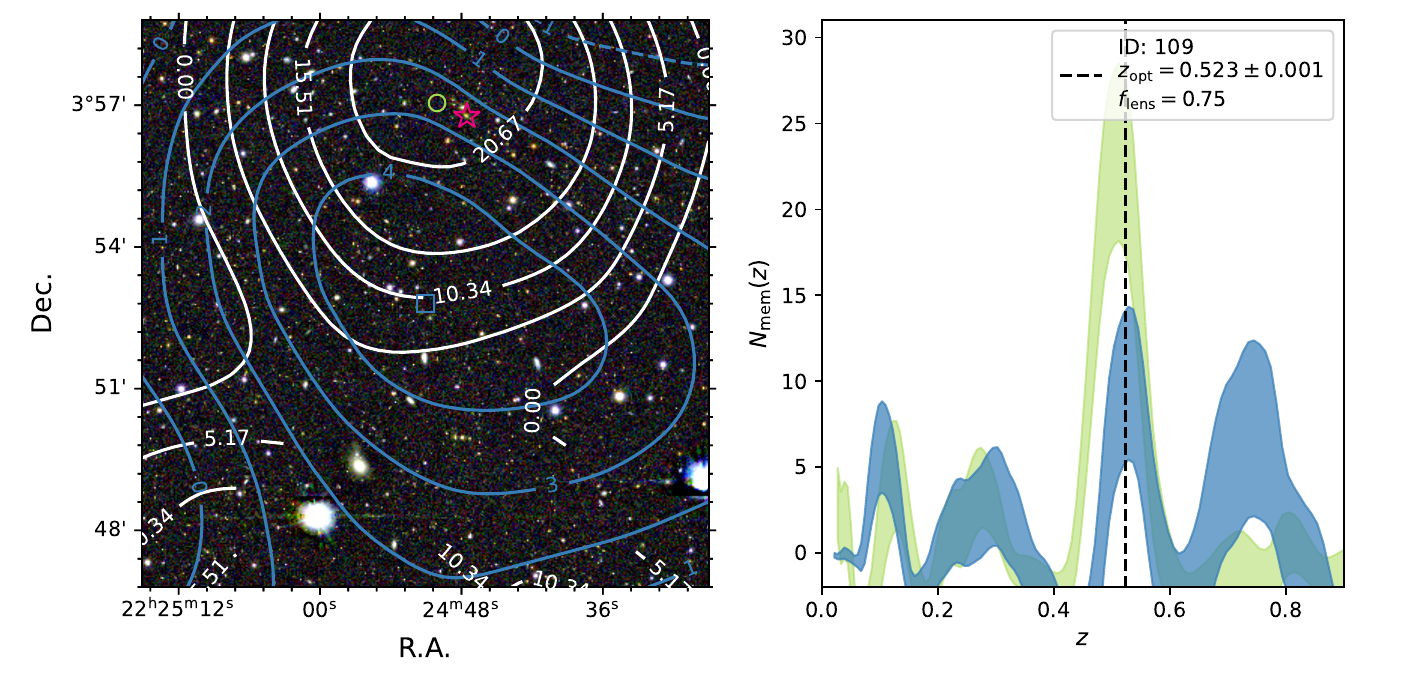}}
\resizebox{0.245\textwidth}{!}{\includegraphics[scale=1]{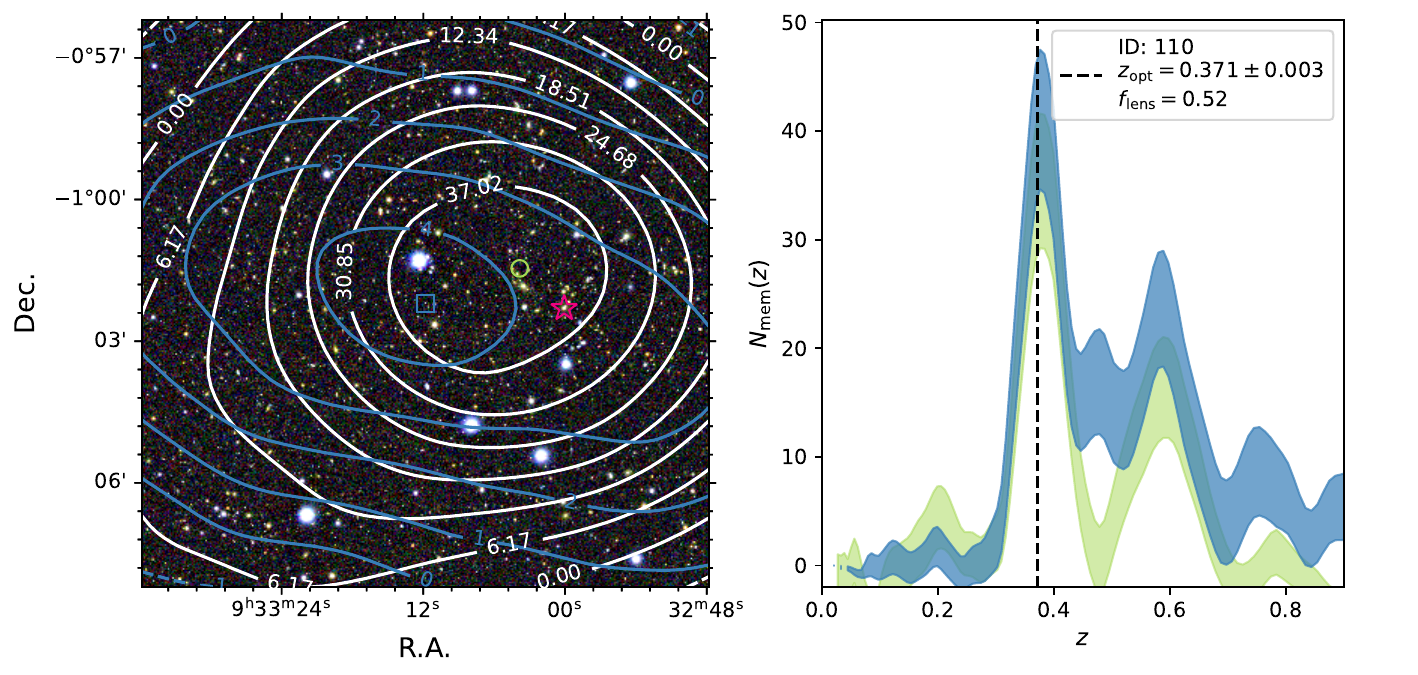}}
\resizebox{0.245\textwidth}{!}{\includegraphics[scale=1]{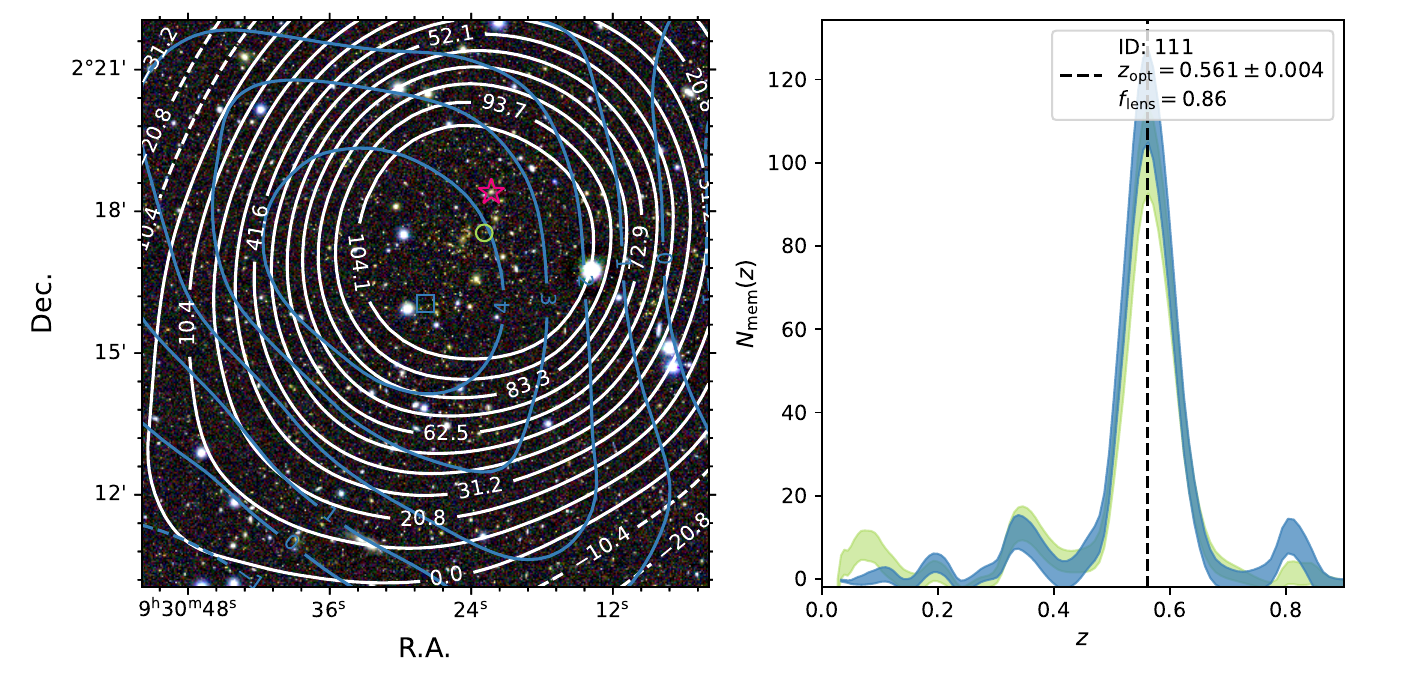}}
\resizebox{0.245\textwidth}{!}{\includegraphics[scale=1]{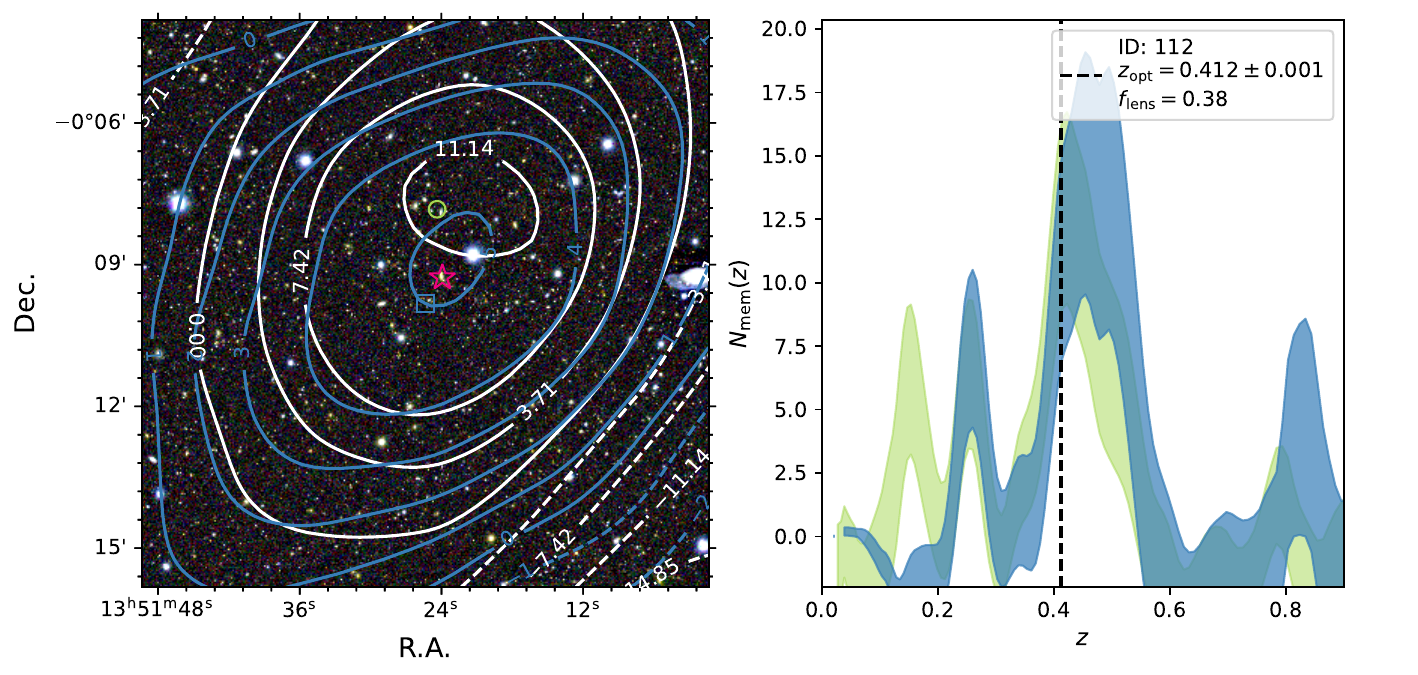}}
\resizebox{0.245\textwidth}{!}{\includegraphics[scale=1]{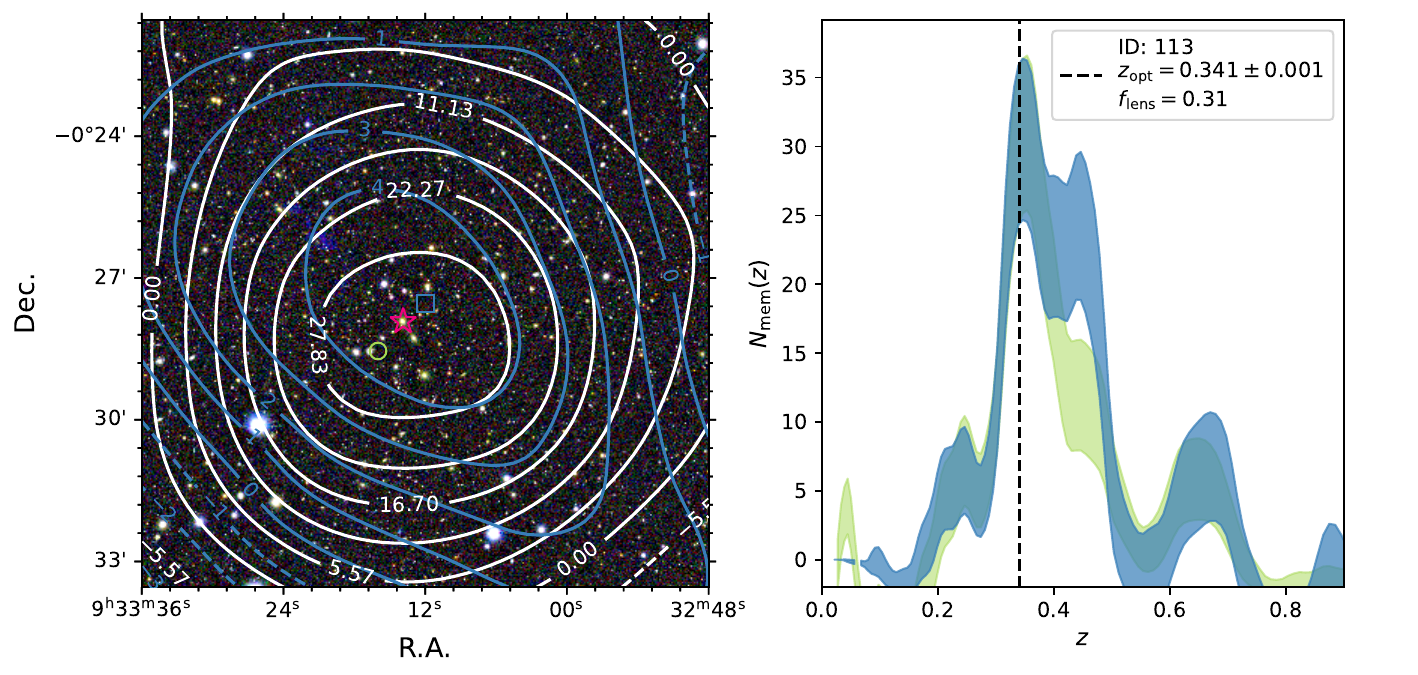}}
\resizebox{0.245\textwidth}{!}{\includegraphics[scale=1]{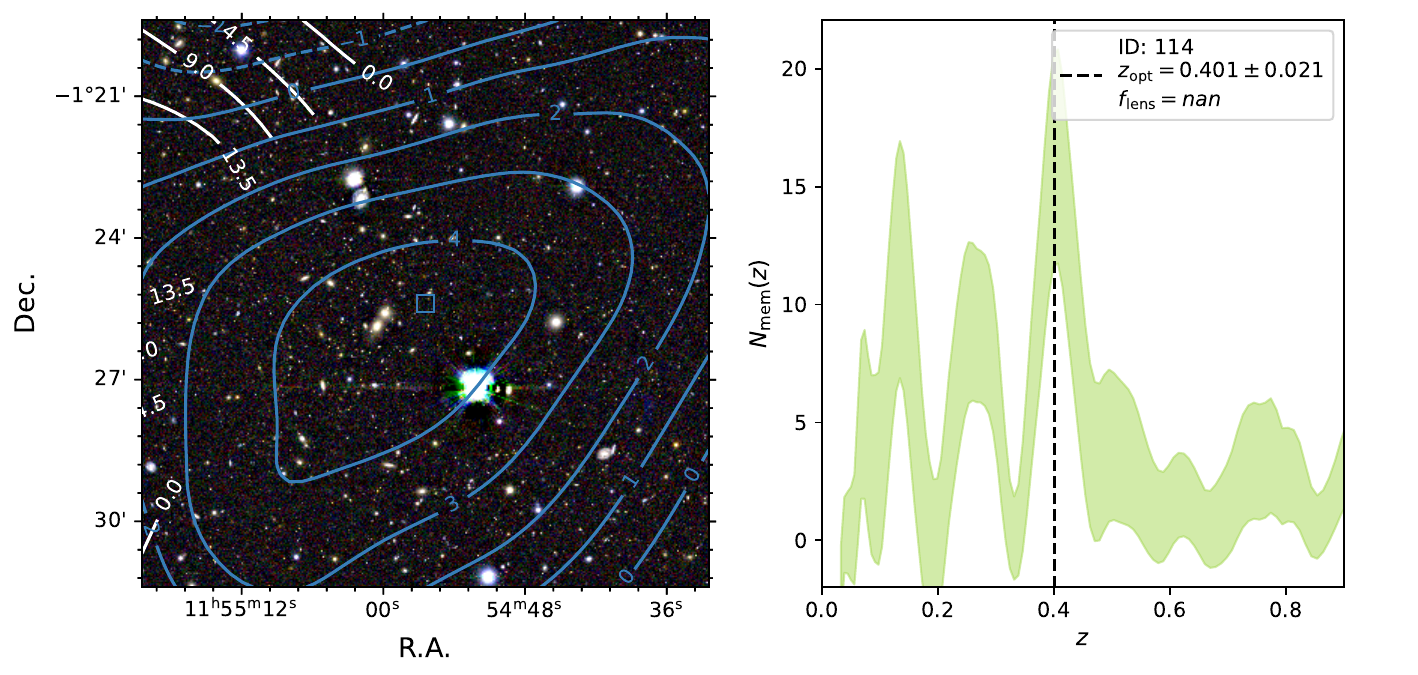}}
\resizebox{0.245\textwidth}{!}{\includegraphics[scale=1]{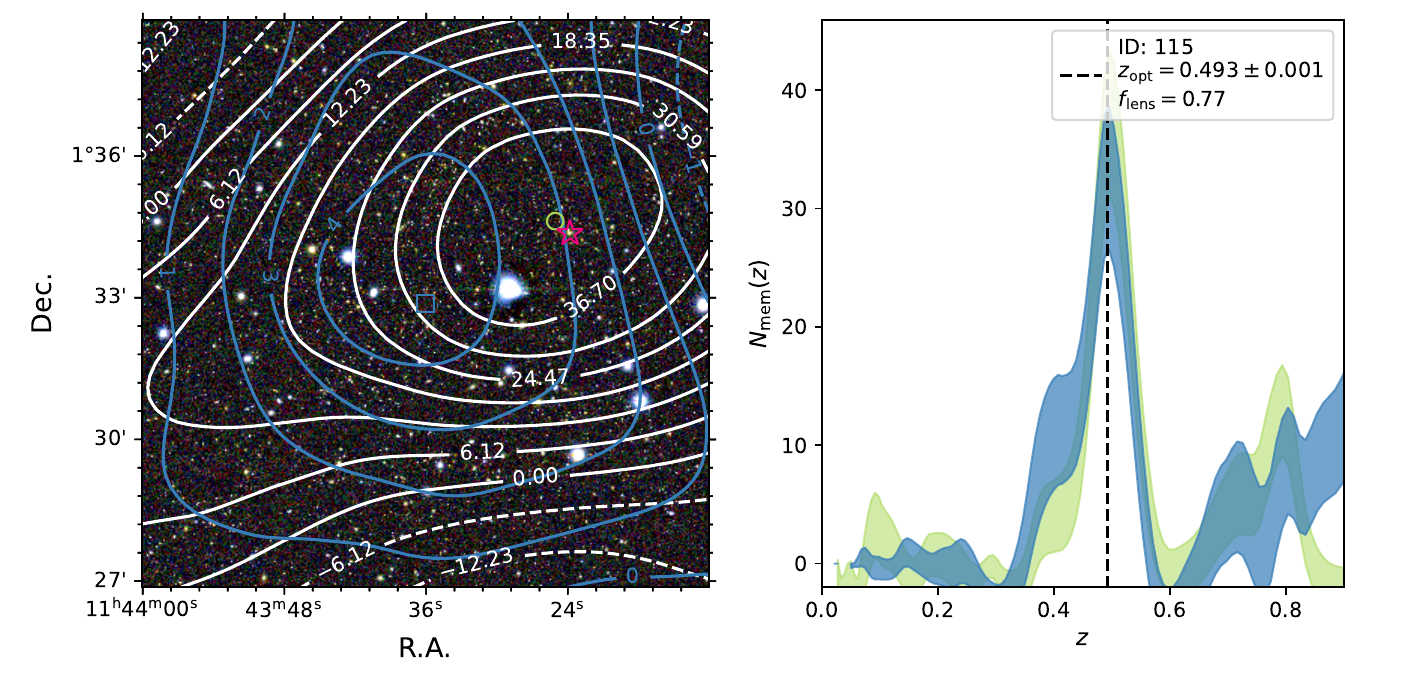}}
\resizebox{0.245\textwidth}{!}{\includegraphics[scale=1]{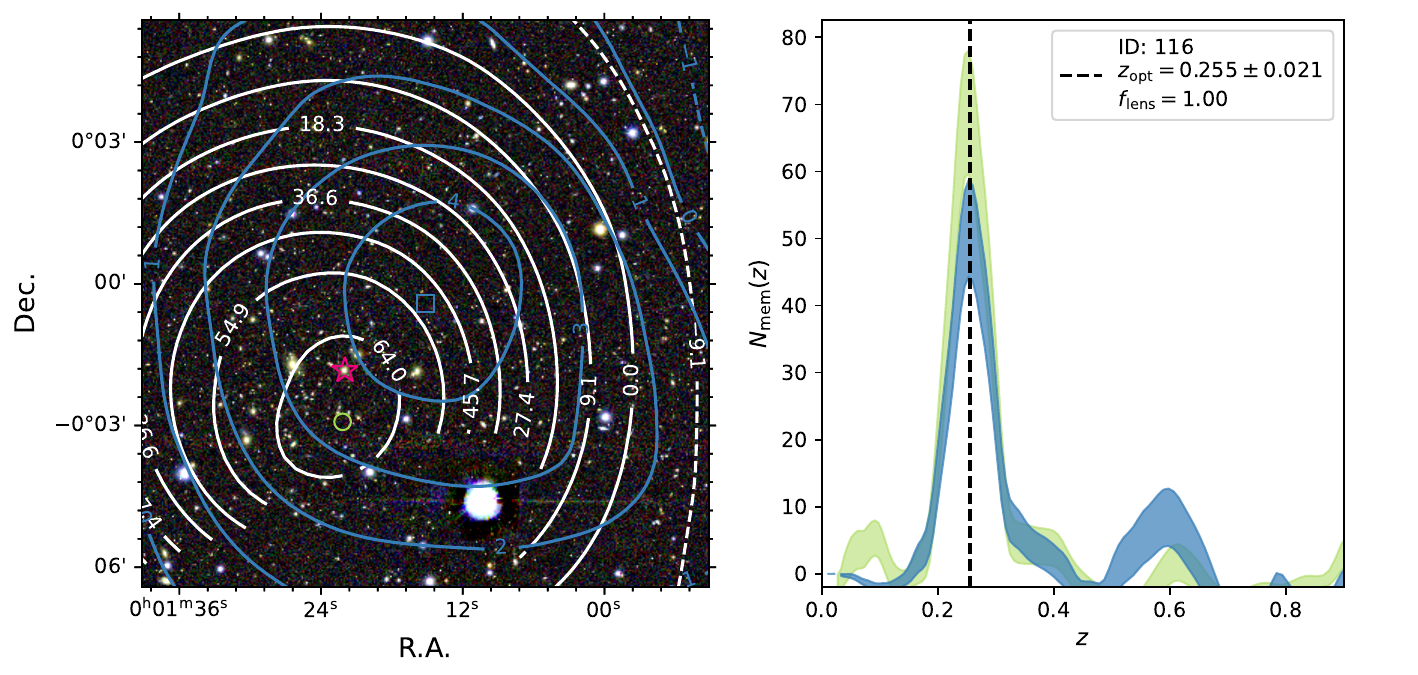}}
\resizebox{0.245\textwidth}{!}{\includegraphics[scale=1]{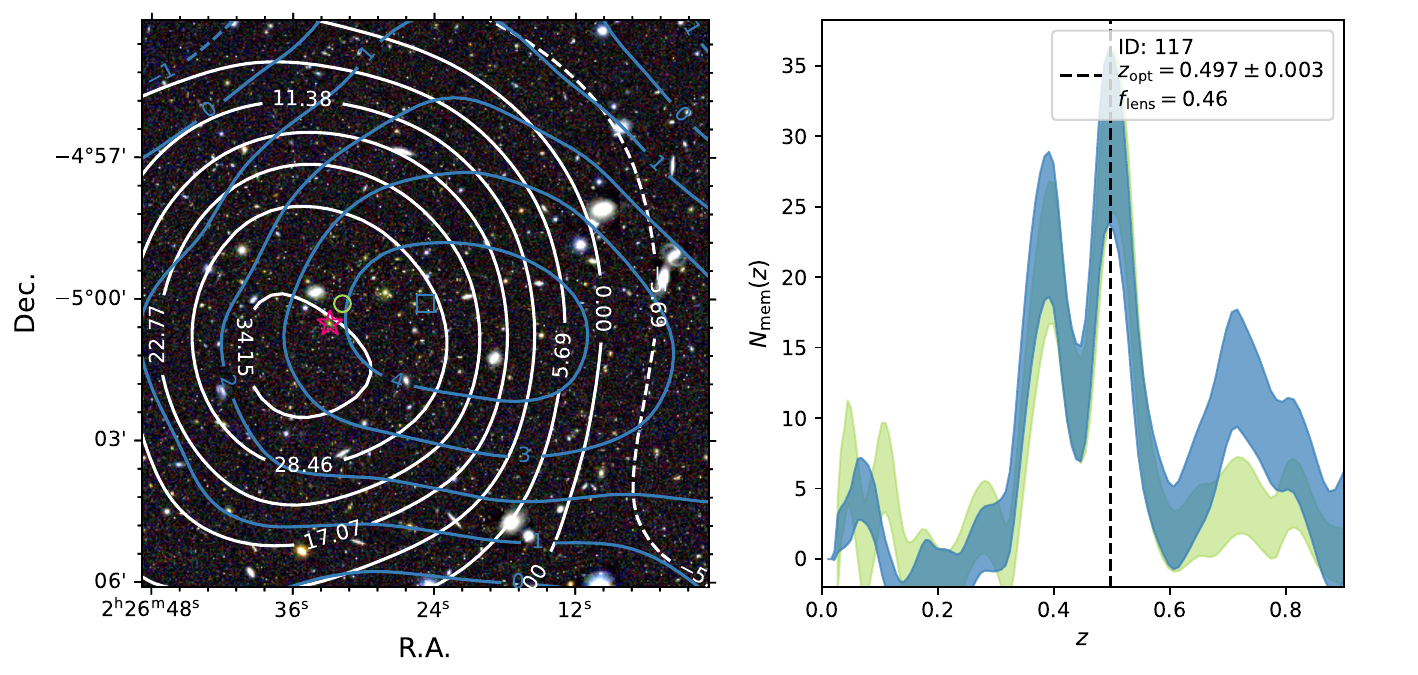}}
\resizebox{0.245\textwidth}{!}{\includegraphics[scale=1]{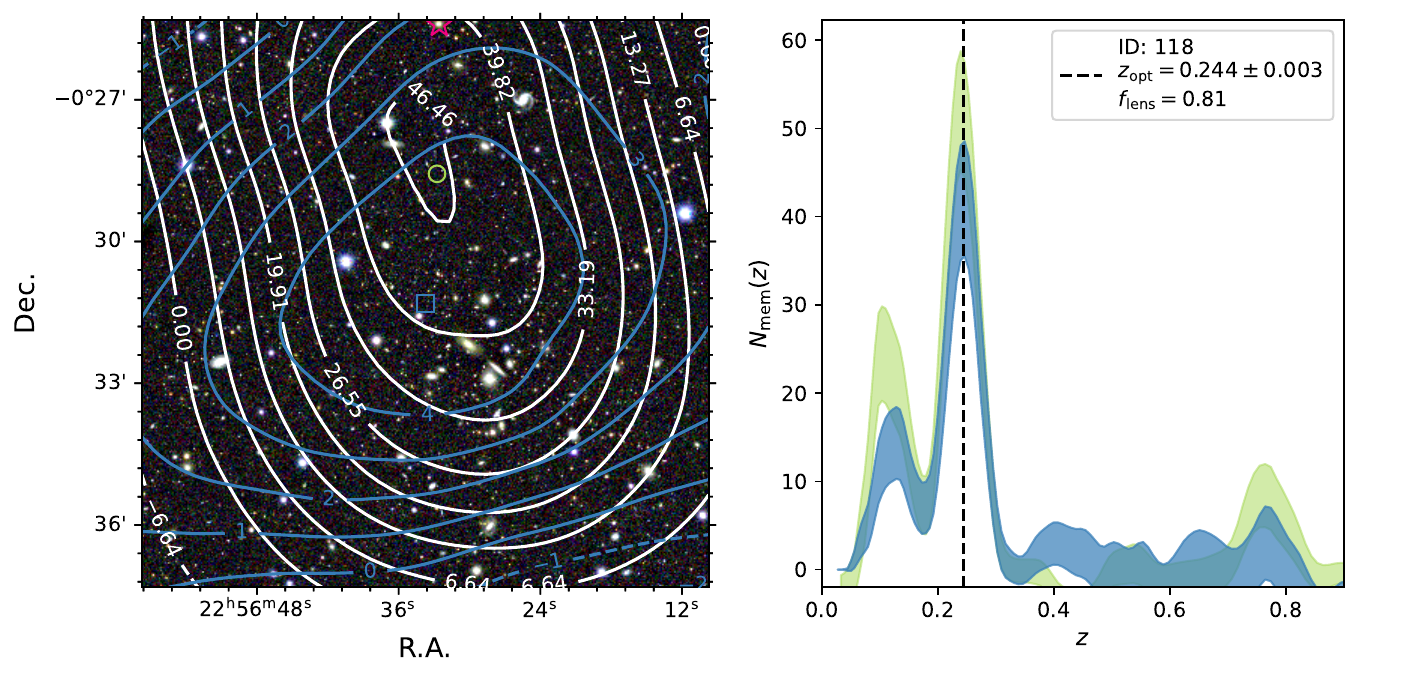}}
\resizebox{0.245\textwidth}{!}{\includegraphics[scale=1]{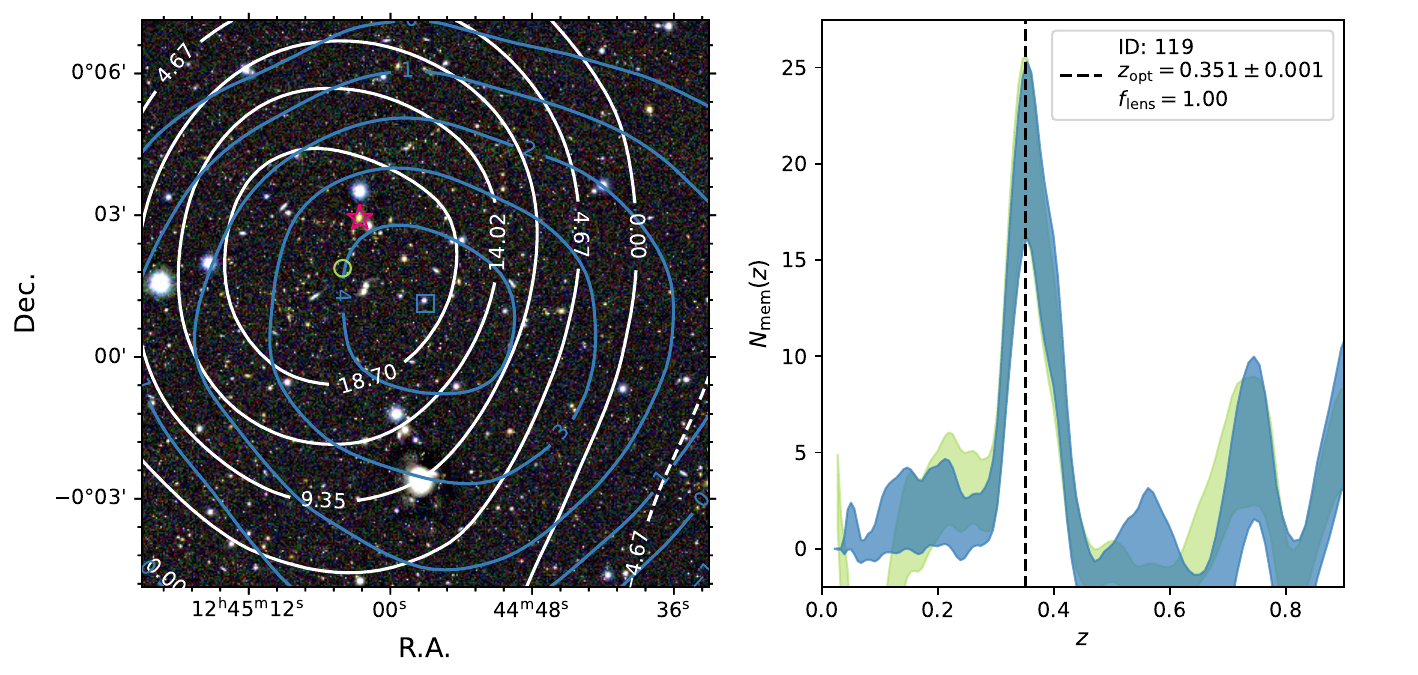}}
\caption{Continued from Figure~\ref{fig:optical_images_1}.}
\label{fig:optical_images_2}
\end{figure*}
\begin{figure*}
\centering
\resizebox{0.245\textwidth}{!}{\includegraphics[scale=1]{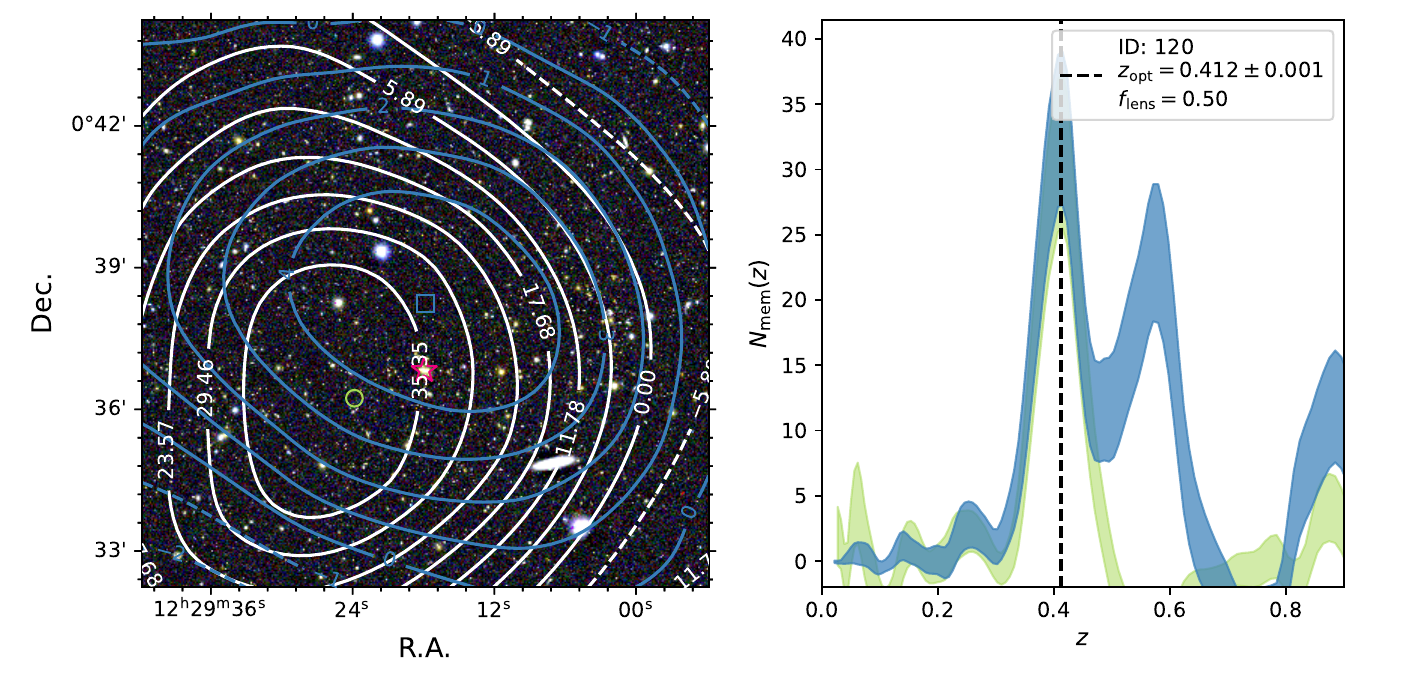}}
\resizebox{0.245\textwidth}{!}{\includegraphics[scale=1]{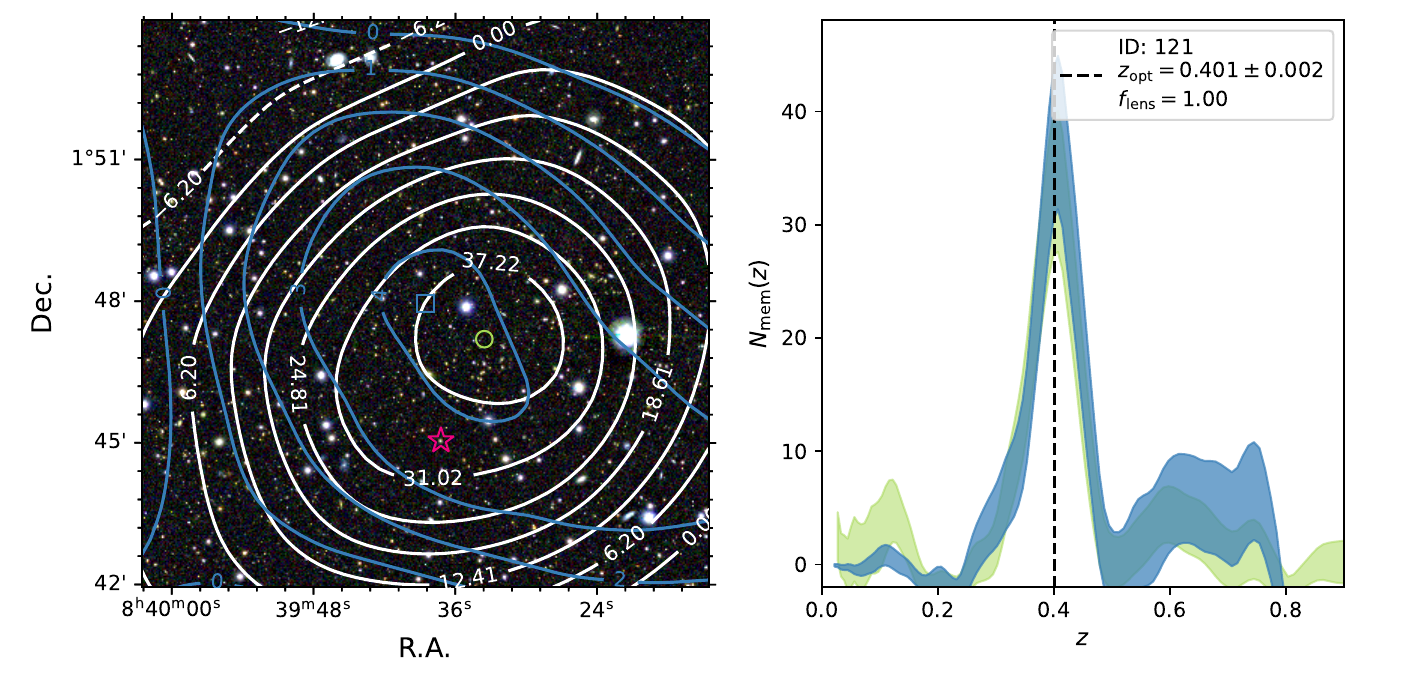}}
\resizebox{0.245\textwidth}{!}{\includegraphics[scale=1]{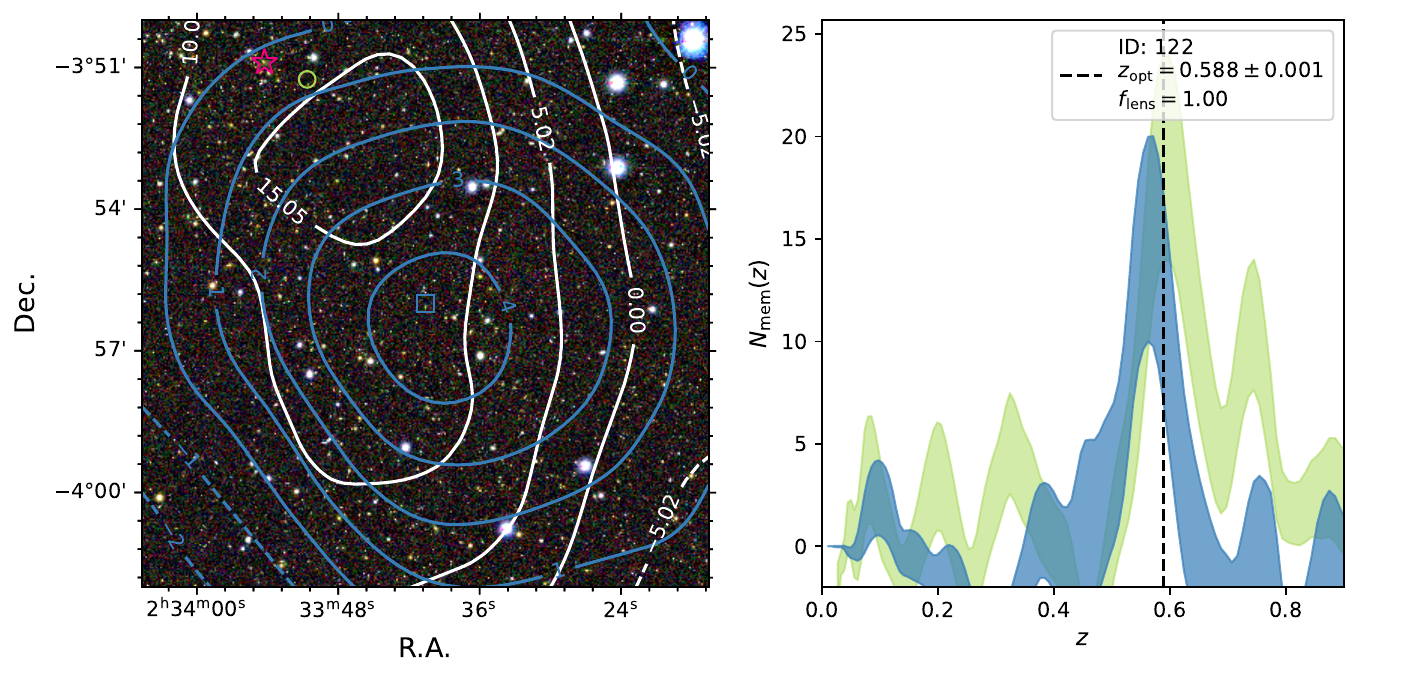}}
\resizebox{0.245\textwidth}{!}{\includegraphics[scale=1]{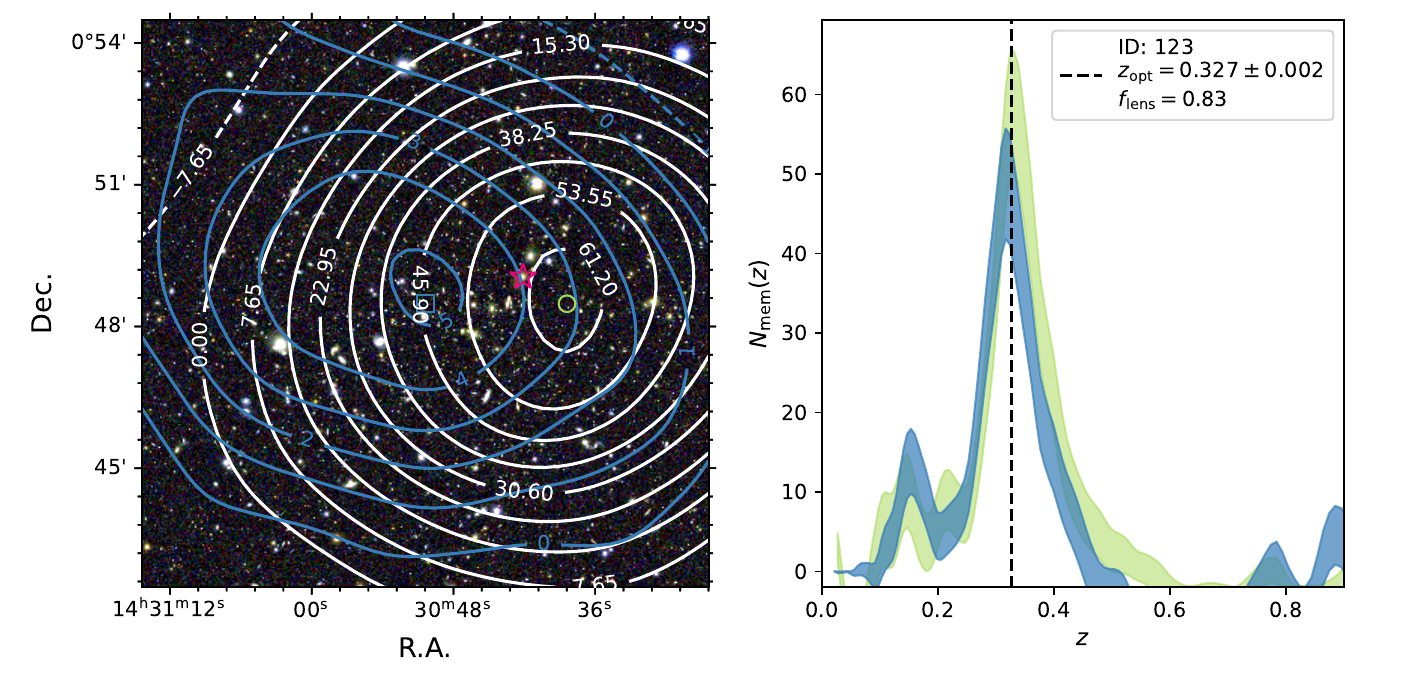}}
\resizebox{0.245\textwidth}{!}{\includegraphics[scale=1]{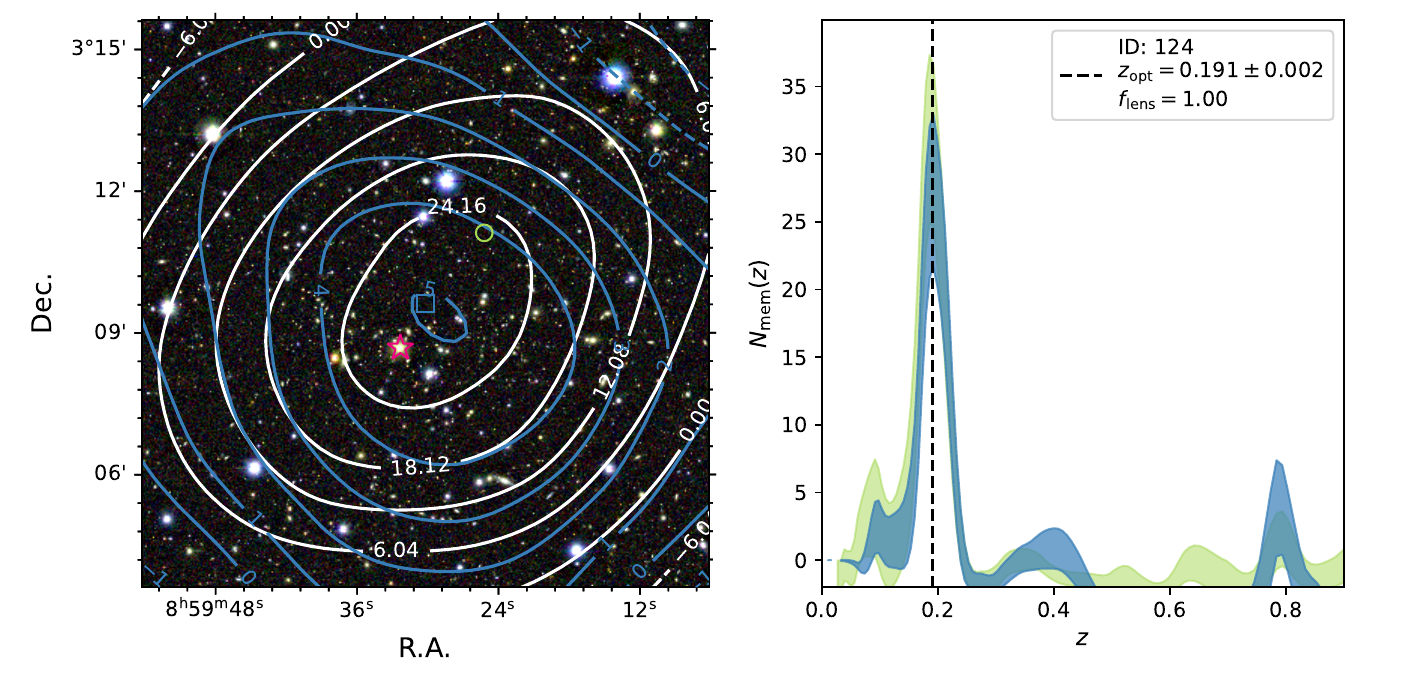}}
\resizebox{0.245\textwidth}{!}{\includegraphics[scale=1]{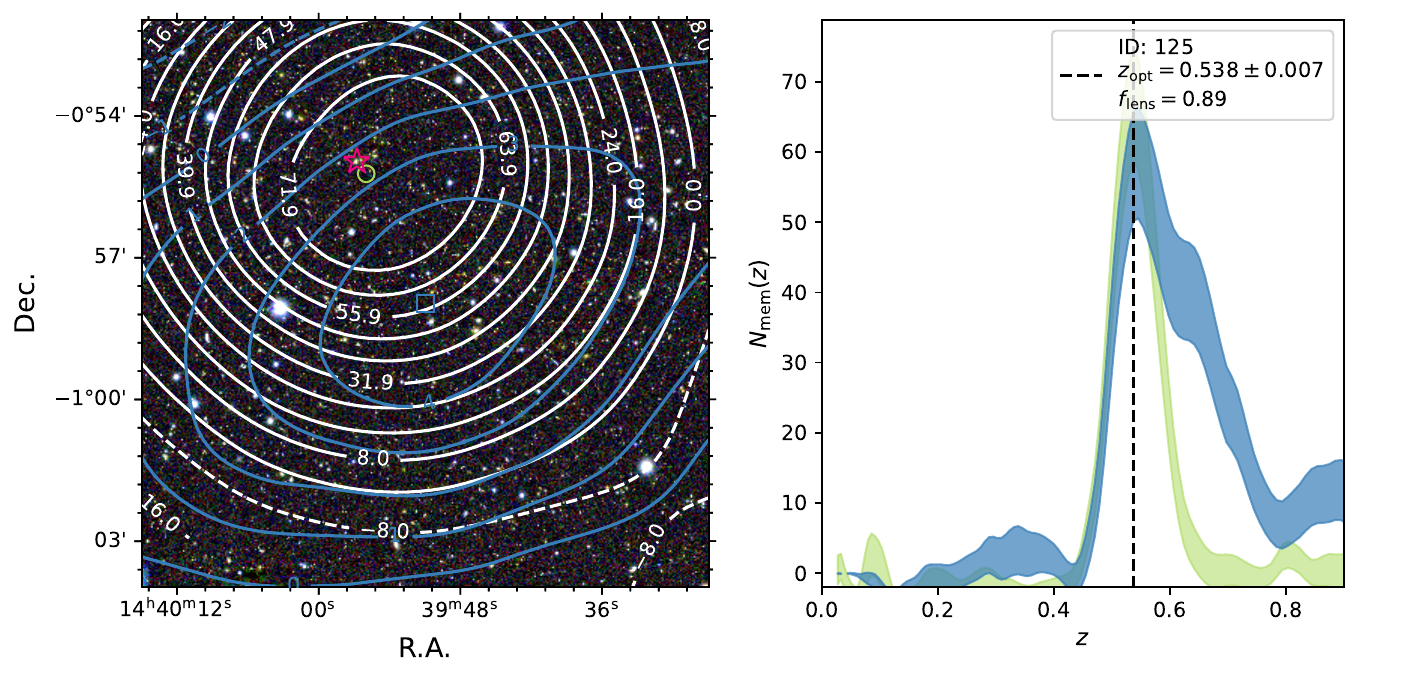}}
\resizebox{0.245\textwidth}{!}{\includegraphics[scale=1]{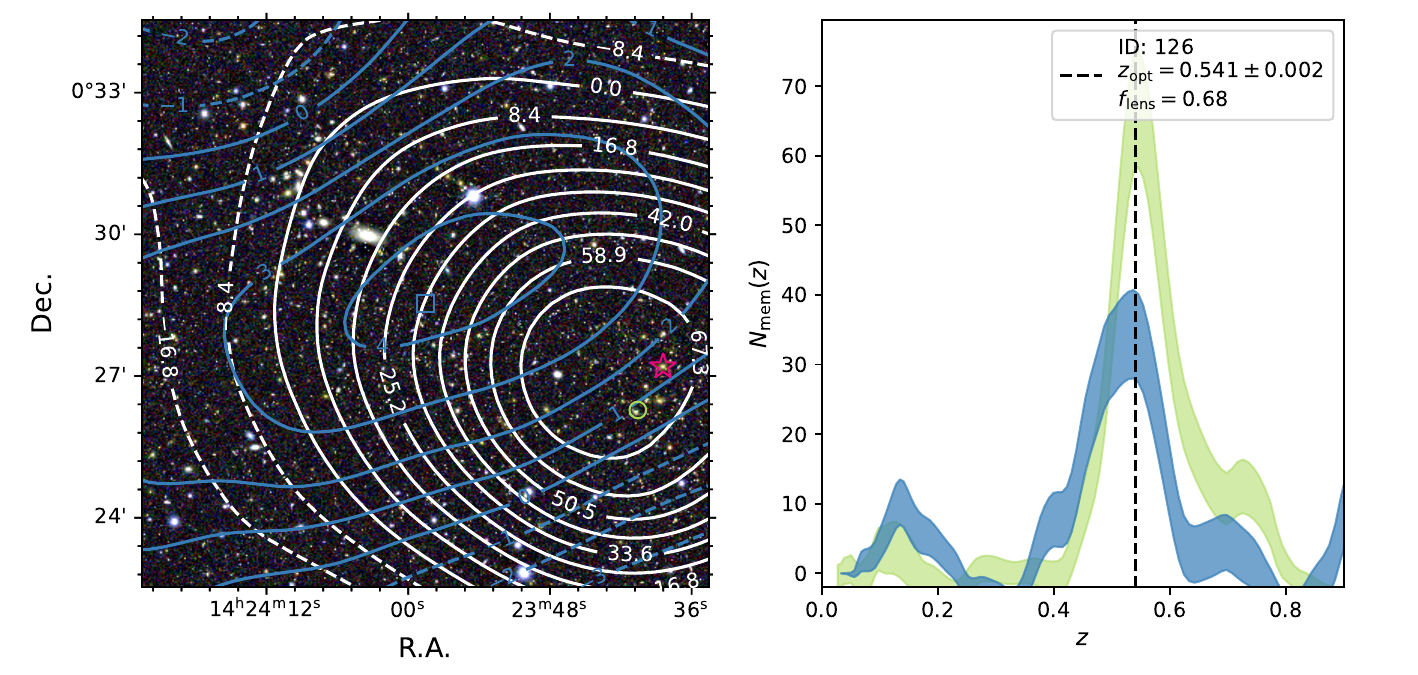}}
\resizebox{0.245\textwidth}{!}{\includegraphics[scale=1]{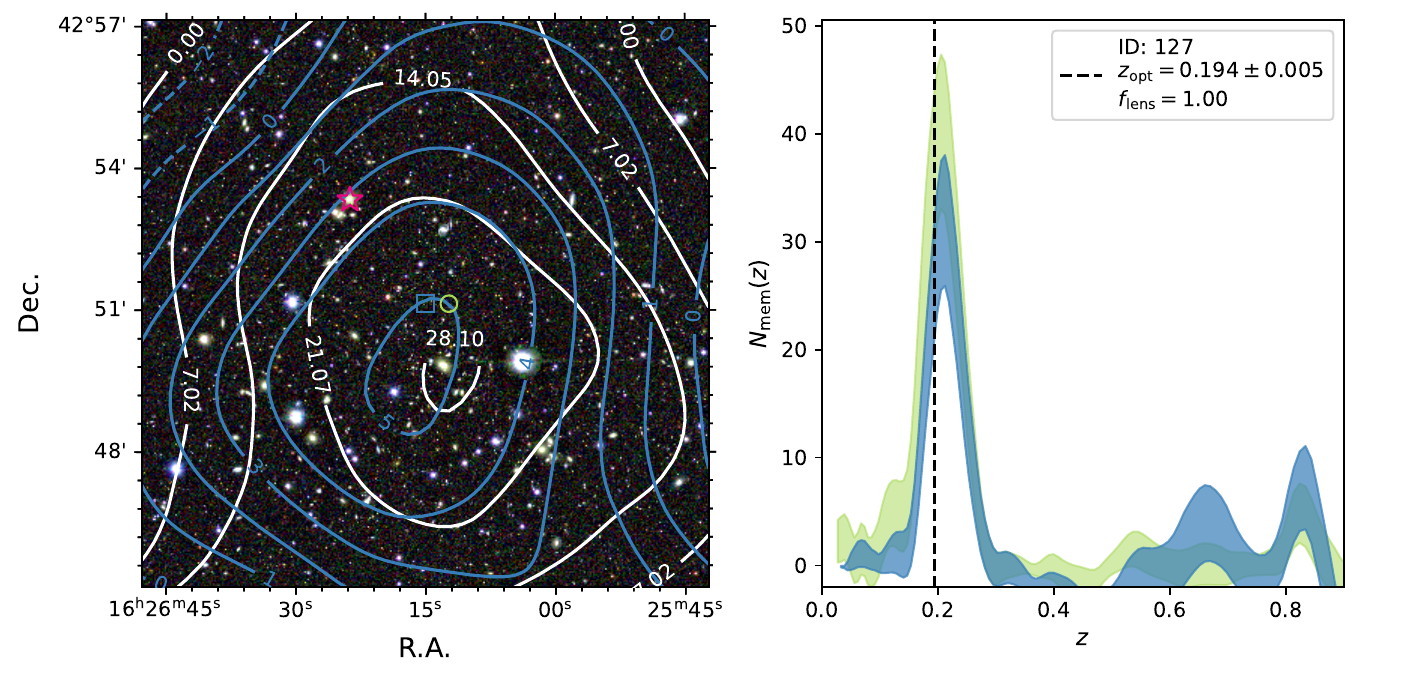}}
\resizebox{0.245\textwidth}{!}{\includegraphics[scale=1]{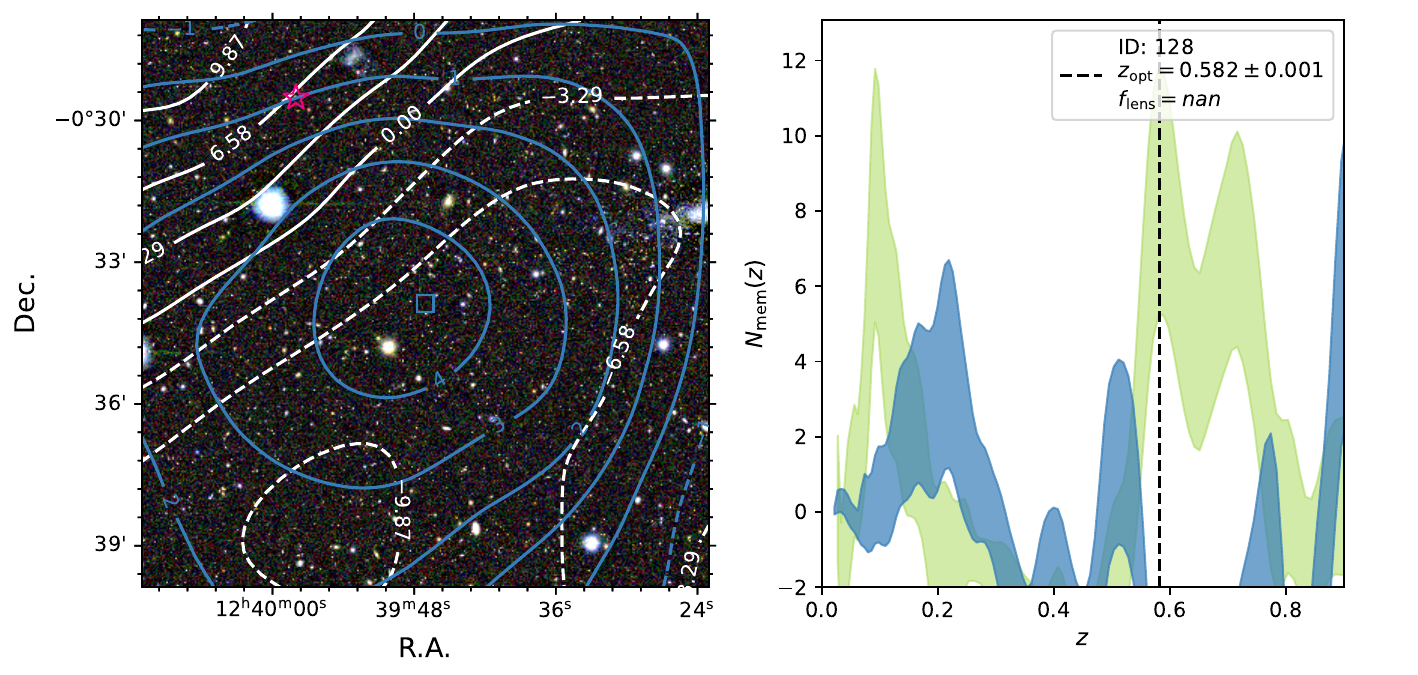}}
\caption{Continued from Figure~\ref{fig:optical_images_2}.}
\label{fig:optical_images_3}
\end{figure*}
%


\label{lastpage}
\end{document}

%% file: newcommands.tex
\newcommand{\lcdm}{\ensuremath{\Lambda\textrm{CDM}}}

\newcommand{\omegam}{\ensuremath{\Omega_{\mathrm{m}}}}
\newcommand{\omegal}{\ensuremath{\Omega_{\Lambda}}}

\newcommand{\Hnow}{\ensuremath{H_{0}}}

\newcommand{\sigmaeight}{\ensuremath{\sigma_{8}}}
\newcommand{\seight}{\ensuremath{S_{8}}}

\newcommand{\w}{\ensuremath{w}}

\newcommand{\Msunh}{\ensuremath{h^{-1}\mathrm{M}_{\odot}}}
\newcommand{\Mpc}{\ensuremath{\mathrm{Mpc}}}
\newcommand{\Mpch}{\ensuremath{h^{-1}\mathrm{Mpc}}}

\newcommand{\Rfiveoo}{\ensuremath{R_{500\mathrm{c}}}}
\newcommand{\Mfiveoo}{\ensuremath{M_{500\mathrm{c}}}}

\newcommand{\Rtwooo}{\ensuremath{R_{200\mathrm{c}}}}
\newcommand{\Mtwooo}{\ensuremath{M_{200\mathrm{c}}}}
\newcommand{\ctwooo}{\ensuremath{c_{200\mathrm{c}}}}
\newcommand{\redshift}{\ensuremath{z}}
\newcommand{\mass}{\ensuremath{M}}
\newcommand{\dif}{\ensuremath{\mathrm{d}}}
\newcommand{\angstrom}{\textup{\AA}}

\newcommand{\mstar}{\ensuremath{m_{\ast}}}

\newcommand{\erosita}{\emph{eROSITA}}

\newcommand{\rich}{\ensuremath{N_{\mathrm{mem}}}}
\newcommand{\richti}{\ensuremath{N_{\mathrm{TI20}}}}
\newcommand{\richcamira}{\ensuremath{N_{\mathrm{CAMIRA}}}}
\newcommand{\richopt}{\ensuremath{N_{\mathrm{opt}}}}

\newcommand{\zcl}{\ensuremath{z_{\mathrm{cl}}}}
\newcommand{\zs}{\ensuremath{z_{\mathrm{src}}}}

\newcommand{\bwl}{\ensuremath{b_{\mathrm{WL}}}}
\newcommand{\mwl}{\ensuremath{M_{\mathrm{WL}}}}

\newcommand{\snr}{\ensuremath{\nu}}
\newcommand{\numin}{\ensuremath{\nu_{\mathrm{min}}}}
\newcommand{\mkappahat}{\ensuremath{\hat{M}_{\kappa}}}
\newcommand{\mkappa}{\ensuremath{M_{\kappa}}}
\newcommand{\thetas}{\ensuremath{\theta_{\mathrm{s}}}}

\newcommand{\rdd}{\ensuremath{R}}

\newcommand{\comp}{\ensuremath{\mathcal{C}}}

\newcommand{\Awl}{\ensuremath{A_{\mathrm{WL}}}}
\newcommand{\Bwl}{\ensuremath{B_{\mathrm{WL}}}}

\newcommand{\gammawl}{\ensuremath{\gamma_{\mathrm{WL}}}}
\newcommand{\sigmawl}{\ensuremath{\sigma_{\mathrm{WL}}}}

\newcommand{\magnitude}{\ensuremath{m}}
\newcommand{\gmag}{\ensuremath{g}}
\newcommand{\rmag}{\ensuremath{r}}
\newcommand{\imag}{\ensuremath{i}}
\newcommand{\zmag}{\ensuremath{z}}
\newcommand{\ymag}{\ensuremath{y}}
\newcommand{\wmem}{\ensuremath{w_{\mathrm{mem}}}}
\newcommand{\skyloc}{\ensuremath{\vect{\theta}}}
\newcommand{\wlcenter}{\ensuremath{\vect{\theta}_{\mathrm{WL}}}}
\newcommand{\optcenter}{\ensuremath{\vect{\theta}_{\mathrm{opt}}}}

\newcommand{\flens}{\ensuremath{f_{\mathrm{lens}}}}
\newcommand{\photz}{\ensuremath{z_{\mathrm{phot}}}}
\newcommand{\specz}{\ensuremath{z_{\mathrm{spec}}}}